\documentclass[
reprint,
amsmath,amssymb,
 aps,
prx,
]{revtex4-2}
\usepackage{tikz}
\usepackage{comment}
\usepackage{braket}
\usepackage{graphicx}
\usepackage{dcolumn}
\usepackage{bm}
\usepackage{hyperref}
\usepackage{CJKutf8}
\newcommand{\frmint}[0]{\mathcal{D}\bar\psi\mathcal{D}\psi}
\usepackage{xcolor}

\newcommand*\circled[1]{\tikz[baseline=(char.base)]{
            \node[shape=circle,draw,inner sep=1pt] (char) {#1};}}
\makeatletter
\def\l@subsection#1#2{}
\def\l@subsubsection#1#2{}
\makeatother

\begin{document}
\title{Geometric landscape annealing as an optimization principle \\ underlying the coherent Ising machine}
\author{Atsushi Yamamura (\begin{CJK}{UTF8}{ipxm}山村 篤志\end{CJK})}
\author{Hideo Mabuchi}
\author{Surya Ganguli}
\affiliation{Dept. of Applied Physics, Stanford University, Stanford, CA}
\date{\today}

\begin{abstract}
Given the fundamental importance of combinatorial optimization across many diverse application domains, there has been widespread interest in the development of unconventional physical computing architectures that can deliver better solutions with lower resource costs. These architectures embed discrete optimization problems into the annealed, analog evolution of nonlinear dynamical systems. However, a theoretical understanding of their performance remains elusive, unlike the cases of simulated or quantum annealing. We develop such understanding for the coherent Ising machine (CIM), a network of optical parametric oscillators that can be applied to any quadratic unconstrained binary optimization problem. Here we focus on how the CIM finds low-energy solutions of the Sherrington-Kirkpatrick spin glass. As the laser gain is annealed, the CIM interpolates between gradient descent on the soft-spin energy landscape, to optimization on coupled binary spins. By exploiting spin-glass theory, we develop a detailed understanding of the evolving geometry of the high-dimensional CIM energy landscape as the laser gain increases, finding several phase transitions, from flat, to rough, to rigid. Additionally, we develop a cavity method that provides a precise geometric interpretation of supersymmetry breaking in terms of the response of a rough landscape to specific perturbations. We confirm our theory with numerical experiments, and find detailed information about critical points of the landscape. Our extensive analysis of phase transitions provides theoretically motivated optimal annealing schedules that can reliably find near-ground states. This analysis reveals geometric landscape annealing as a powerful optimization principle and suggests many further avenues for exploring other optimization problems, as well as other types of annealed dynamics, including chaotic, oscillatory or quantum dynamics.  
\end{abstract}
\maketitle
\tableofcontents

\section{Introduction}
\subsection{Theory for Coherent Ising Machines}
Combinatorial optimization~\cite{Papadimitriou1998-hj} is a key enabler of performance in diverse application domains, including for example, machine learning, robotics, chip design, operations research, and manufacturing.  Thus the co-development of algorithms and hardware that can provide better solutions with lower consumption of resources such as time and energy could substantially impact many fields.  Promising recent demonstrations of unconventional hardware architectures have ignited broad interest in physics-based approaches to solving NP-hard problems, in which combinatorial optimization over discrete variables is embedded in the analog evolution of nonlinear dynamical systems~\cite{MohseniMcMahonByrnes2022, kalinin2022computational,syed2023physics, molnar2020accelerating, dutta2020ising}. This interplay between discrete optimization and analog evolution spawns a rich new field of research based on fresh foundations to complement more traditional approaches. While benchmarking experiments have established high-performance scaling of physics-based approaches to the regime of $10^5$ optimization variables~\cite{Honjo2021}, scant theory exists for extrapolating future prospects for unconventional architectures or analyzing their strengths and weaknesses relative to mainstream heuristics.

In this article, we develop substantial components of a theoretical framework for the Coherent Ising Machine (CIM)~\cite{inagakiCoherentIsingMachine2016, mcmahonFullyProgrammable100spin2016, yamamotoCoherentIsingMachines2020, yamamotoCoherentIsingMachines2017, wangCoherentIsingMachine2013, yamamuraQuantumModelCoherent2017, strinati2021all}, an unconventional physical optimization architecture based on coupled optical parametric oscillators (OPOs). The CIM may be understood as a heuristic solver for the Ising ground-state problem, which is to identify the spin configuration $\{s_i\}_{i=1,2,\cdots,N}$ with spin variables $s_i = \pm 1$ that minimizes the Ising Hamiltonian $\mathcal{H}=-\frac{1}{2} \sum_{i,j} J_{ij} s_{i} s_{j}$, where $J_{ij}$ is an $N\times N$ symmetric matrix. This problem, also known as quadratic unconstrained binary optimization (QUBO), is known to be NP-hard \cite{barahonaComputationalComplexityIsing1982}. Indeed, many optimization problems, including partitioning, covering, packing, matching, clique finding, graph coloring, minimum spanning trees, and the traveling salesman problem, can be mapped to a corresponding QUBO problem with only polynomial overhead~\cite{lucasIsingFormulationsMany2014}. The central role of OPOs as building blocks makes the CIM architecture especially interesting within the broader field of physics-based optimization, as comprehensive quantitative models for OPO networks can be constructed in ways that interpolate between classical and quantum operating regimes (as a function, {\it e.g.}, of linear decoherence rates relative to coherent nonlinear dynamical rates~\cite{Jankowski2023}). This makes CIM theory a fertile setting for exploring how novel information dynamics that emerge in the classical-quantum crossover~\cite{Yanagimoto2023} may impact optimization performance. But the first step in this program must be to establish a baseline understanding of classical CIM mechanics, against which quantum differences can be highlighted. In this article we begin to draw this classical baseline.

The CIM approaches QUBO by relaxing the binary Ising spins to continuous soft spins. Each OPO functions as a relaxed analog state (soft-spin) with a continuous state variable $x$, exposed to a double-well energy potential $E_I(x, a):= \frac{1}{4}x^4 - \frac{a}{2}x^2$. Here, a particular laser gain parameter $a$ controls the depth of the two wells. As the gain parameter increases, each OPO becomes strongly confined in one of the wells, effectively functioning as a binary spin $s_i$. At very large gain, there are $2^N$ minima in the energy landscape, and the global minimum corresponds to the ground state of the Ising Hamiltonian \cite{wangCoherentIsingMachine2013}. To locate a global minimum, the CIM anneals the gain, by first optimizing the energy of the soft-spin network at a low gain, where the energy landscape is convex, and then adiabatically increasing the gain parameter until each soft-spin starts to exhibit behavior akin to a binary spin. Such optimization mechanisms, which continually reshape the energy landscape starting from a trivial form, have also been suggested in various other contexts such as mean-field annealing \cite{bilbroOptimizationMeanField1989} (a deterministic approximation of simulated annealing), annealed stochastic gradient descent \cite{chaudhariEnergyLandscapeDeep2017} in the context of deep neural networks, and topology trivialization \cite{fyodorovTopologyTrivializationLarge2014} in the random landscape literature.

Numerous numerical and experimental benchmarks have shown that this landscape annealing approach can achieve high performance \cite{Honjo2021,mcmahonFullyProgrammable100spin2016, inagakiCoherentIsingMachine2016,haribaraPerformanceEvaluationCoherent2017,
hamerlyExperimentalInvestigationPerformance2019}. However, to the best of our knowledge, no theoretical analysis for this performance has been established. This is in stark contrast to other well-recognized annealing algorithms, such as simulated annealing and quantum annealing, which are known to successfully find the optimum given a sufficiently slow annealing schedule \cite{vincentSimulatedAnnealingProof1994, kadowakiQuantumAnnealingTransverse1998}. Interestingly, the CIM may fail to find the ground state for certain frustrated instances, even if the annealing speed is appropriately slow  \cite{wangCoherentIsingMachine2013}.  This is believed to stem from the amplitude heterogeneity of the soft spins, which makes the mapping from Ising energy to the soft-spin network's energy less precise when the gain is not substantial enough. Indeed, right after the landscape becomes non-convex, the global minimum of the energy landscape lies along the eigenvector of the $J$-matrix with the minimum eigenvalue, generally different from the true Ising ground state configuration \cite{leleuCombinatorialOptimizationUsing2017}.
This amplitude heterogeneity issue has been discussed, and a few methods have been proposed to mitigate its effect \cite{leleuCombinatorialOptimizationUsing2017,leleuDestabilizationLocalMinima2019}.
While the amplitude heterogeneity initially compels the Ising machine to find the eigenvector rather than a global minimum, as we further ramp up the gain, the signs of soft-spin variables $x_i$ successively flip, leading to a continuous decrease in Ising energy. These configuration adjustments enhance the Ising machine, making it a robust Ising optimizer rather than just a simple linear solver. To understand how the CIM state evolves with the landscape annealing process, we need to understand the changes in the energy landscape as the gain increases.

This type of question has been extensively investigated for simulated annealing and quantum annealing, especially with purely random instances. In the former case, we generally observe a phase transition from the paramagnetic phase to the spin-glass phase as we cool down the system \cite{mezardSpinGlassTheory1986}. In the spin-glass phase, free energies of different thermodynamic states are generally crossing successively, and the equilibrium states at two slightly different temperatures can be dramatically different \cite{brayChaoticNatureSpinGlass1987,krzakalaChaoticTemperatureDependence2002,rizzoChaosTemperatureSherringtonKirkpatrick2003}. This phenomenon, known as temperature chaos, obstructs simulated annealing from finding the global minimum within a time span linear in the system size \cite{krzakalaFollowingGibbsStates2010}. Quantum annealing presents similar properties; it undergoes a  phase transition from a quantum paramagnetic phase to a spin-glass phase or many-body localized phase as we cease the quantum fluctuation \cite{laumannManybodyMobilityEdge2014}. In systems with local interactions, energy level crossings of low energy states occur in the localized phase, and it takes exponential time to follow the ground state due to the small overlap of those localized states \cite{foiniSolvableModelQuantum2010,altshulerAndersonLocalizationMakes2010} (see \cite{bapstQuantumAdiabaticAlgorithm2013} for reviews on this topic.)

To our knowledge, such analysis has not been applied to soft-spin networks and hardware like the coherent Ising machine. In this paper, we focus on purely random instances, corresponding in the Ising setting to the Sherrington-Kirkpatrick (SK) spin glass \cite{mezardSpinGlassTheory1986}, and we examine phase transitions the geometry of the CIM energy landscape.  We discover significant phase transitions in the energy landscape as well as evidence for potential level crossings within a particular phase. Furthermore, we demonstrate that these phase transitions are intimately tied to the annealing schedule and optimization performance. In addition to contributing to a type of baseline theory that can eventually be used to study the impact of increasingly quantum OPO behavior in CIM-type architectures, our analysis may also be useful for exploring the potential utility of non-degenerate oscillatory OPO dynamics~\cite{Roy2021} for evading landscape obstacles within the QUBO setting (see discussion). Such studies will be the subject of future work, but our results here provide essential foundations.

The structure of this paper is as follows: After discussing how our work on CIM theory connects with statistical physics results based on related technical approaches, we review in Section~\ref{sec:overview} the classical formulation of the CIM as a soft-spin network as well as the structure of the energy landscape in both the small and large gain regimes, in the case of random connectivity matrices corresponding to the SK spin glass.  We furthermore derive a theory delineating the dependence of the curvature of the landscape, quantified through the Hessian eigenspectrum, on where one is located in the landscape.  This dependence is critical in all following sections, given the CIM energy landscape possesses no special symmetries. In Section~\ref{sec:performance} we demonstrate numerically that the CIM performs well in finding a near ground state solution of the SK spin glass, using an optimal annealing schedule for the laser gain that we derive using our subsequent landscape analysis; it significantly outperforms a spectral algorithm, and finds a solution that is within about $1\%$ of the true intensive ground state energy.  

In Section~\ref{sec:evolving} we begin our geometric landscape annealing analysis by performing a supersymmetry breaking replica calculation to derive detailed predictions about the structure and organization of critical points of the CIM energy landscape and how it evolves as the laser gain is increased. In Section~\ref{sec:cavity-method} we re-derive these results by developing a novel supersymmetry (SUSY) breaking cavity method, thereby providing considerable geometric insight into the meaning of SUSY breaking in terms of extreme landscape reactivity to external perturbations.  In Section~\ref{sec:numericaltests} we further analyze our replica and cavity theory predictions and compare them to numerical explorations of the CIM energy landscape, finding an excellent match between theory and numerical experiments.  In Section~\ref{sec:fullrsb} we derive a supersymmetric but full replica symmetry breaking theory of global minima of the CIM energy landscape and further confirm the predictions of this theory in numerical experiments. Together, Section~\ref{sec:numericaltests} and \ref{sec:fullrsb}  provide matching theory and experiments for the typical energy, distance from the origin, and Hessian eigenspectra of saddle points, local minima, and global minima as a function of laser gain, and reveal a sequence of important phase transitions in the landscape geometry which we summarize in a phase diagram in Section~\ref{sec:phase-diagram}. 

In Section~\ref{sec:relationship}, we relate the phase transitions in the landscape geometry to the performance of the CIM as a function of the annealing schedule, and explain how these phase transitions suggest the optimal annealing schedule employed earlier in Section~\ref{sec:performance} to obtain good CIM performance for the SK spin glass.  We end with a discussion and future directions in Section \ref{sec:conclusion}. Finally, we provide self-contained and detailed derivations in our Supplementary Material.

\subsection{Statistical physics context}

The fundamental problem of understanding how the high dimensional geometry of even the classical CIM energy landscape evolves with increasing laser gain poses several interesting challenges from the perspective of random landscape theory, which has a rich history involving the analysis of several models, including for example, TAP free energy landscapes~\cite{
brayMetastableStatesSpin1980, 
brayMetastableStatesSolvable1981,
aspelmeierComplexityIsingSpin2004,   
crisantiSpinGlassComplexity2004, 
parisiSupersymmetryBreakingComputation2004, 
crisantiComplexityMeanfieldSpinglass2005, 
rizzoTAPComplexityCavity2005, 
aspelmeierFreeenergyLandscapesDynamics2006,  
fanTAPFreeEnergy2020, 
mullerMarginalStatesMeanfield2006}
, random Gaussian fields~\cite{
fyodorovComplexityRandomEnergy2004, 
fyodorovCountingStationaryPoints2005, 
brayStatisticsCriticalPoints2007, 
fyodorovReplicaSymmetryBreaking2007, 
fyodorovStatisticalMechanicsSingle2008} 
and spherical spin glasses~\cite{
cavagnaQuenchedComplexityMeanfieldpspin1999, 
cavagnaStationaryPointsThoulessAndersonPalmer1998, 
crisantiComplexitySherringtonKirkpatrickModel2003, 
crisantiComplexitySphericalMathsfp2003, 
fyodorovHighDimensionalRandomFields2013, 
auffingerComplexityRandomSmooth2013,  
auffingerRandomMatricesComplexity2013, 
rosComplexEnergyLandscapes2019, 
rosComplexityEnergyBarriers2019, 
beckerGeometryEnergyLandscapes2020}.  
Here we situate our work within this prior context. 

To describe the evolving CIM landscape geometry, we seek to describe changes in the number, location, energy, Hessian eigenspectrum, and local susceptiblity of various critical points, including typical saddle points, local minima, and global minima.  We apply a combination of the Kac-Rice method~\cite{
fyodorovComplexityRandomEnergy2004,
adler2007random}, 
replica theory~\cite{
mezardSpinGlassTheory1986}, 
random matrix theory~\cite{
Potters2020-ta}, and supersymmetry~\cite{
cavagnaFormalEquivalenceTAP2003, 
crisantiComplexitySherringtonKirkpatrickModel2003,
annibaleRoleBecchiRouet2003,
annibaleSupersymmetricComplexitySherringtonKirkpatrick2003,
annibaleCoexistenceSupersymmetricSupersymmetrybreaking2004,
parisiSupersymmetryBreakingComputation2004,
cavagnaCavityMethodSupersymmetrybreaking2005,
rizzoTAPComplexityCavity2005} to derive an analytic theory of the organization of critical points in the CIM energy landscape as a function of laser gain.  Prior theoretical studies of the geometry of critical points in continuous high-dimensional random landscapes focused on simplified settings in which symmetry played a crucial role in carrying forward calculations.  For example, in the case of random Gaussian fields~\cite{
fyodorovComplexityRandomEnergy2004, 
fyodorovCountingStationaryPoints2005, 
brayStatisticsCriticalPoints2007, 
fyodorovReplicaSymmetryBreaking2007, 
fyodorovStatisticalMechanicsSingle2008} 
and spherical spin glasses~\cite{
cavagnaQuenchedComplexityMeanfieldpspin1999, 
cavagnaStationaryPointsThoulessAndersonPalmer1998, 
crisantiComplexitySherringtonKirkpatrickModel2003, 
crisantiComplexitySphericalMathsfp2003, 
fyodorovHighDimensionalRandomFields2013, 
auffingerComplexityRandomSmooth2013,  
auffingerRandomMatricesComplexity2013, 
rosComplexEnergyLandscapes2019, 
rosComplexityEnergyBarriers2019, 
beckerGeometryEnergyLandscapes2020}  
, translational and spherical symmetry respectively were crucial. The reason symmetry has greatly simplified past calculations is that, as we will see below, the combination of the Kac-Rice and replica methods require an analysis of how the Hessian eigenspectrum of the energy landscape depends on the location $\mathbf{x}$ within the landscape. When strong translational or spherical symmetries are present, the Hessian eigenspectrum becomes independent of location and the problem of computing properties of critical points can be reduced to computing properties of the spectrum of a single random Hessian matrix.  The TAP free energy landscape on other hand does not possess such simple symmetries, but does have a nongeneric property, namely that the Hessian eigenspectrum of typical critical points has a bulk that is gapped away from the origin, apart from a single zero eigenvalue \cite{aspelmeierComplexityIsingSpin2004, rizzoTAPComplexityCavity2005}, which again simplifies certain analyses as described below. 

In contrast, as we shall see below, the CIM energy landscape possesses neither translational nor spherical symmetry, and its Hessian eigenspectra extend continuously to zero, even for local minima.  All of this necessitates a more involved analysis of the relationship between the Hessian eigenspectra of critical points and their location in the CIM energy landscape. One of the contributions of this article from the perspective of random landscape theory is to provide an analysis of how the Hessian eigenspectrum depends on location in a scenario in which no strong symmetries are present. Intriguingly, in the case of the CIM we find a simple connection from location to Hessian eigenspectrum through Dyson's Brownian motion \cite{Dyson1962-vn}.  We furthermore provide a framework for incorporating this dependence into the combined Kac-Rice and replica methods to analytically derive the organization of critical points in the CIM energy lansdscape for arbitrary laser gains. Such a framework could be broadly useful for other random landscape problems.

Our work also sheds new light on the geometric meaning of supersymmetry (SUSY) breaking, which is one approach to analyzing random landscape geometries \cite{
annibaleCoexistenceSupersymmetricSupersymmetrybreaking2004,
annibaleRoleBecchiRouet2003,
annibaleSupersymmetricComplexitySherringtonKirkpatrick2003,
cavagnaCavityMethodSupersymmetrybreaking2005,
cavagnaFormalEquivalenceTAP2003, 
crisantiComplexitySherringtonKirkpatrickModel2003,
parisiSupersymmetryBreakingComputation2004,
rizzoTAPComplexityCavity2005}.
The reason SUSY can emerge in random landscape analysis is that the Kac-Rice formula can be expressed in terms of a partition function integral over bosonic degrees of freedom related to the location $\mathbf{x}$ as well as fermionic degrees of freedom, which, when integrated alone, yield the determinant of the Hessian of the energy landscape. This integral possesses a supersymmetry (SUSY) that exchanges bosonic and fermionic degrees of freedom. When the integral is computed via the saddle point method, the correct saddle point can sometimes break SUSY, and therefore yield nonzero SUSY breaking order parameters. Given the abstract nature of this calculation, the fundamental geometric meaning of SUSY breaking and the resultant nonzero order parameters has often remained mysterious in general settings.  

Prior work has derived geometric interpretations of SUSY breaking in limited settings \cite{mullerMarginalStatesMeanfield2006} using modifications of the cavity method \cite{cavagnaCavityMethodSupersymmetrybreaking2005,rizzoTAPComplexityCavity2005} that take into account the possibility that critical points may have Hessian eigenspectra with a {\it single} zero mode corresponding to a single flat direction in the energy landscape, with the rest of the bulk spectrum gapped away from the origin.  Indeed \cite{mullerMarginalStatesMeanfield2006} showed that the presence of this single flat direction indicates SUSY breaking, and the SUSY breaking order parameters for local minima are related to the inner product between the location of the minimum and the flat direction.  This analysis suffices for the TAP free energy landscape of the SK model which is known to have such an isolated single flat direction, or soft mode, around local minima \cite{aspelmeierComplexityIsingSpin2004, rizzoTAPComplexityCavity2005}. However, as we will see below, this is not the case for the CIM energy landscape, in which typical critical points can have a continuous Hessian spectral density extending to $0$ indicating an extensive number of near-flat directions about such critical points. 

Another main contribution of our work is to not only derive the properties of critical points using the Kac-Rice formula combined with the SUSY-breaking replica method, but also derive a generalized cavity method for the SUSY-breaking phase.  We demonstrate the generalized cavity and replica methods yield identical results, but our novel SUSY-breaking cavity method yields important geometric insights into the meaning of SUSY breaking in more general scenarios than previously derived. Importantly, unlike prior work, our cavity method can handle Hessian eigenspectra whose spectral density extends continuously to zero, indicating a critical point that is marginally stable, with extensively many soft modes, corresponding to the small eigenvalues. These soft modes are highly susceptible to perturbations of the landscape.  Our cavity method shows that SUSY breaking coincides with the presence of exponentially many such marginally stable, soft critical points with high susceptibility to perturbations. In such a scenario, a small change in the landscape can induce bifurcations in these exponentially many critical points, resulting in exponentially more or fewer critical points. Moreover, we show that the nonzero SUSY-breaking order parameters quantitatively reflect the exponential reactivity of the number of critical points of the energy landscape to specific perturbations. Thus our work provides a new, general, and quantitative geometric interpretation of SUSY breaking in terms of the extreme reactivity of the landscape stemming from exponentially many marginally stable critical points.       

Thus overall we see that the general analysis of a physical analog computing device for solving random discrete combinatorial optimization problems, even in the classical limit, yields an incredibly rich theoretical picture that interfaces with numerous branches of physics and mathematics, including the replica method, the cavity method, supersymmetry breaking, random matrix theory, Dyson's Brownian motion, and the geometry of random landscapes.  This rich picture serves as an interesting foundational baseline for analyzing how the classical to quantum transition may aid in optimization, in a physically implementable device. 

\section{The overview of the CIM and its adiabatic evolution}
\label{sec:overview}

Our fundamental problem of interest is to find ground states of the Ising energy function, given by 
\begin{equation}
    E_{\operatorname{Ising}}(\mathbf{s}) = \frac{1}{2} \sum_{i,j=1}^N J_{ij} s_i s_j,
    \label{eq:Ising}
\end{equation}
where each $s_i = \pm 1$ is a binary spin. The reason for this is many optimization problems can be cast as Ising optimization problems for a given choice of spin-connectivity $J_{ij}$ \cite{lucasIsingFormulationsMany2014}. However, we will focus in particular on one generic ensemble of optimization problems in which $J_{ij}$ are chosen to be iid zero mean random Gaussian variables with variance $\frac{1}{N}$.  This is known as the Sherrington-Kirpatrick spin glass \cite{mezardSpinGlassTheory1986}.

\subsection{A model of the coherent Ising machine}

We consider a model of the CIM as a network of $N$ soft spins, each of which is described by a scalar $x_i\in \mathbb{R}$ ($i=1,2,\cdots, N$), corresponding to the $x$-quadrature of a degenerate OPO. The total energy of the network is given by 
\begin{equation}
    E_{\operatorname{tot}}(\mathbf{x}) = \sum_{i=1}^N E_I(x_i) + \frac{1}{2} \sum_{i,j=1}^N J_{ij} x_i x_j,
    \label{eq:Etot}
\end{equation}
where $E_I(x)$ is a single site energy function governing the dynamics of a single OPO and $J_{ij}$ reflects the symmetric network connectivity between the OPOs.  

While many of our derivations apply to arbitrary internal energy functions $E_I(x)$ that are bounded from below, we will focus our comparisons to numerics using the particular internal energy function   
\begin{equation}
    E_I(x) = \frac{1}{4}x^4 - \frac{a}{2}x^2,
    \label{eq:EsingleOPO}
\end{equation}
which governs the dynamics of each individual OPO in the CIM. Here  $a$ is an important effective laser gain parameter that controls the overall shape of the internal energy of individual OPOs. 
Note that $a$ reflects a balancing between the linear dissipation and the gain of the CIM system.
Therefore, it can be negative when the dissipation is stronger.
For $a<0$, $E_I(x)$ is convex with a single minimum at $x=0$.  But as $a$ increases beyond $0$ to become positive, the single OPO energy landscape undergoes a pitchfork bifurcation wherein the minimum at $x=0$ becomes a local maximum and two new minima appear at $x=\pm \sqrt{a}$, both with energy $E_I = -\frac{1}{4}a^2$. This corresponds to a symmetric double well potential in which the wells move further out and become deeper and sharper as $a$ increases, leading to stronger confinement of the soft-spins around $x=\pm \sqrt{a}$.   

The simplified dynamics of the CIM at zero temperature and fixed gain $a$ can be described as gradient descent dynamics \cite{wangCoherentIsingMachine2013,leleuCombinatorialOptimizationUsing2017}
\begin{equation}
    \tau \frac{dx_i}{dt} = - \frac{dE_{\operatorname{tot}}(\mathbf{x})}{dx_i}.
    \label{eq:graddescent}
\end{equation}
We will work in units of time in which the intrinsic CIM time-scale $\tau=1$.
The CIM is typically operated by annealing the gain $a$ as follows \cite{yamamotoCoherentIsingMachines2017}.
First, the gain parameter $a$ is large and negative, so that the initial CIM state is prepared near the origin $\mathbf{x}=0$, corresponding to all OPOs approximately in their vacuum state.  
Then the gain $a$ is slowly increased over time, while the OPOs simultaneously undergo their natural gradient descent dynamics in \eqref{eq:graddescent}.  Finally, at a large enough gain $a$, the OPO states $x_i$ are measured and their signs $s_i = \operatorname{sign}{x_i}$ are interpreted as a binary spin configuration, which ideally would achieve a very low Ising energy in the original Ising energy minimization problem of interest in \eqref{eq:Ising}.

This typical annealing of the gain $a$ leads to several questions.  First, how and why does annealing $a$ lead to a final answer with low Ising energy? Second, what determines a good annealing schedule and at what value of $a$ should we stop annealing?  In this work, we take a high dimensional geometric perspective to these questions, by seeking to understand the changing structure of $E_{\operatorname{tot}}(\mathbf{x})$ in \eqref{eq:Etot} as $a$ increases.  

In particular, as $a$ increases, a sequence of bifurcations in the geometry of the high dimensional energy landscape $E_{\operatorname{tot}}(\mathbf{x})$ takes place.  
In each such bifurcation, new critical points (i.e. points where the gradient $\nabla E_{\operatorname{tot}}(\mathbf{x})$ vanishes) are either created or destroyed.  Additionally, at bifurcations, the index of a critical point can change, where the index is defined to be the number negative eigenvalues of the Hessian matrix of second derivatives of $E_{\operatorname{tot}}(\mathbf{x})$, evaluated at the critical point.  We seek to understand, at each value of $a$, the high dimensional geometry of $E_{\operatorname{tot}}(\mathbf{x})$ by analyzing where critical points of a given index lie in terms of their typical energies and their typical locations in $\mathbf{x}$ space.  An elucidation of this changing high dimensional geometry provides insights into the functional optimization advantage gained by annealing the laser gain in the CIM.  Furthermore it suggests properties of good annealing schedules for $a$.

\subsection{Energy landscape geometry at extremal gains}

As a warmup to understanding the high dimensional geometry of $E_{\operatorname{tot}}(\mathbf{x})$ for arbitrary $a$, we first focus on two extremal regimes: small $a \ll 0$ and large $a \gg 0$. 

\subsubsection{The small laser gain regime: the CIM computes a spectral approximation to the Ising problem}
\label{subsubsection:smallgain}
For $a \ll 0$, we expect the energy landscape to be convex, with the only minimum occurring at $\mathbf{x}=0$.  As $a$ increases, the landscape will first become nonconvex, by definition, when the Hessian matrix $H(\mathbf{x})$ at any location $\mathbf{x}$ first acquires a negative eigenvalue.  The elements of this $N$ by $N$ Hessian matrix are given by 
\begin{equation}
    H(\mathbf{x})_{ij} =
        \frac{\partial^2 E_{\operatorname{tot}}}
             {\partial x_i \partial x_j} = 
             H^I(\mathbf{x})_{ij} + J_{ij},
    \label{eq:Hessian}
\end{equation}
where
\begin{equation}
    H^I(\mathbf{x})_{ij} = \partial^2 E_I(x_i) \delta_{ij},
    \label{eq:diagonalHessian}
\end{equation}
is the diagonal contribution to the Hessian coming from the internal single site OPO energy function $E_I(x)$ alone.
To determine both the smallest $a$ and the location $\mathbf{x}$ at which the first negative eigenvalue of $H(\mathbf{x})$ can occur, we lower bound the eigenvalues of $H(\mathbf{x})$ for all $\mathbf{x}$ as follows. 

First, note that since  $H(\mathbf{x}) = H^I(\mathbf x) + J$, and that the minimum eigenvalue $\lambda_{\operatorname{min}}$ of a symmetric matrix is a concave function of its matrix elements, we have, by Jensen's inequality
\begin{align}
\lambda_{\operatorname{min}}(H(\mathbf{x})) & \geq & 
\lambda_{\min}(H^I)
& + \lambda_{\min}(J) \nonumber \\
& = & \min_i \partial^2 E_I(x_i) & + \lambda_{\min}(J) \nonumber \\
& = & \min_i 3x_i^2 - a & + \lambda_{\min}(J).
\label{eq:lowerboundlammin}
\end{align}
In the last line we have used the specific form of the single OPO energy function in \eqref{eq:EsingleOPO}. Then a sufficient condition for $\lambda_{\operatorname{min}}(H(\mathbf{x}))$ to be nonnegative is that its lower bound   \eqref{eq:lowerboundlammin} is also nonnegative. This yields the sufficient (but not necessary) condition that if 
$a \leq \min_i 3x_i^2 + \lambda_{\min}(J)$
at any spin configuration $\mathbf{x}$ then $E_{\operatorname{tot}}(\mathbf{x})$ is convex at $\mathbf{x}$. The contrapositive then implies that if $E_{\operatorname{tot}}(\mathbf{x})$ violates convexity at any fixed location $\mathbf{x}$, because the Hessian obeys $\lambda_{\operatorname{min}}(H(\mathbf{x})) < 0$, then we must have $a > \min_i 3x_i^2 + \lambda_{\min}(J)$.  This is a necessary (but not sufficient) condition for $E_{\operatorname{tot}}(\mathbf{x})$ to be nonconvex at $\mathbf{x}$.    

As $a$ increases, this inequality is first satisfied at the origin $\mathbf{x}=0$,
yielding the result that the origin is the first place where the Hessian $H(\mathbf{x})$ acquires a negative eigenvalue. Moreover this occurs when $a$ crosses $\lambda_{\min}(J)$.  
Since the Hessian at the origin is simply $H(\mathbf{0}) = -a I + J$, the associated eigenvector of this Hessian is simply the minimal eigenvector $\mathbf{v}_{\min}$ of $J$ 
which solves the variational problem 
\begin{equation} 
\mathbf{v}_{\min} = \operatorname{argmin}_{\{\mathbf{v}|\mathbf{v}^T\mathbf{v}=1\}} \mathbf{v}^T J \mathbf{v}. 
\label{eq:spectralopt}
\end{equation}
As $a$ increases beyond $\lambda_{\min}(J)$, the first nonconvex behavior of $E_{\operatorname{tot}}(\mathbf{x})$ is a pitchfork bifurcation where the minimum at $\mathbf{x}=0$ becomes an index $1$ saddle with a single negative curvature direction along $\mathbf{v}_{\min}$, and two new minima appearing that are closely aligned to $\pm \mathbf{v}_{\min}$.  If one simply computes the signs of the spin configuration $\mathbf{x}$ in these minima, then one obtains an Ising configuration given by $s_i = \operatorname{sign}(\mathbf{v}_{\min})$ where $\mathbf{v}_{\min}$ is the solution to \eqref{eq:spectralopt}. This is known as the spectral approximation to the Ising energy minimization problem in \eqref{eq:Ising}.  Thus for small $a$ just above $\lambda_{\min}(J)$, the CIM computes the spectral approximation.  We will see below that increasing $a$ can improve upon this spectral solution by finding Ising spin configurations with energy lower than that of the spectral solution. 

In summary, our analysis above yields the following picture.  For any fixed value of $a$, $E_{\operatorname{tot}}(\mathbf x)$ can only be nonconvex in the region obeying  $\min_i 3x_i^2 < a-\lambda_{\min}(J)$ (a necessary condition for nonconvexity).  Contrapositvely, if $\min_i 3x_i^2 \geq a-\lambda_{\min}(J)$ then $E_{\operatorname{tot}}(\mathbf{x})$ must be convex at $\mathbf{x}$ (a sufficient condition for convexity).

\subsubsection{The large laser gain regime: the CIM global minimum coincides with the Ising global minimum}
\label{subsubsection:largegain}
In the absence of the connectivity $J$, the $N$ spins decouple and the energy landscape of \eqref{eq:Etot} and \eqref{eq:EsingleOPO} has $3^N$ critical points given by 
\begin{equation}
     x_i = \sqrt{a} s_i \quad \text{where} \quad s_i \in \{-1,0,+1\}.
     \label{eq:decoupledcp}
 \end{equation}  
 Thus in the absence of connectivity $J$, the scale of the soft spins $x_i$ grows as the square root of gain $a$. If we work with rescaled variables $x'_i := a^{-\frac{1}{2}}x_i$ which remain $O(1)$ as $a$ becomes large, the total energy in \eqref{eq:Etot} and \eqref{eq:EsingleOPO} can be written as 
 \begin{equation}
     a^{-2}E_{\operatorname{tot}} = \sum_i \frac{1}{4}x'^4_i - \frac{1}{2} x'^2_i + \frac{1}{2a} \sum_{i,j} J_{ij} x'_i x'_j.
 \label{eq:Etotrescaled}    
 \end{equation}
 This shows that for large $a \gg \lambda_{\operatorname{max}}(J)$, the effect of the connectivity $J$ on the geometry of the energy landscape can be treated as a weak perturbation of the decoupled landscape in which $J=0$.  Therefore it is useful to first understand this simple decoupled energy landscape. 

 In this landscape with $3^N$ critical points given by \eqref{eq:decoupledcp}, 
 the Hessian matrix $H$ of each critical point is diagonal, 
 with each diagonal element either:  \circled{1} taking the value $-a$ for every ``uncommitted" spin sitting at the saddle point $x_i=0$ of the double well potential in \eqref{eq:EsingleOPO},
 or  \circled{2} taking the value $2a$ for every ``committed" spin sitting at a minimum $x_i=\pm \sqrt{a}$ of the double well potential.  
 Thus the intensive index $r$ of each critical point, defined as the fraction of negative eigenvalues of $H$, simply corresponds to the fraction of uncommitted spins in the critical point.  
 Since each uncommitted (committed) spin contributes internal energy $E_I=0$ ($E_I = -\frac{1}{4}a^2$) in \eqref{eq:EsingleOPO}, the energy of every critical point is determined by its index $r$ via 
 \begin{equation}
     E_{\operatorname{tot}} = -\frac{N}{4}(1-r)a^2. 
 \end{equation}
 Thus a saddle point's energy decreases linearly with its index. 
 
 However, the introduction of the connectivity $J$ breaks the energy degeneracy between all critical points of the same index.  Applying perturbation theory in the small parameter $1/a$ to \eqref{eq:Etotrescaled}, shows that each critical point of the decoupled landscape in \eqref{eq:decoupledcp} moves to
 \begin{equation}
    x_i =  \sqrt{a}s_i - (3s_i^2-1)^{-1}a^{-\frac{1}{2}} h^0_i +  O(a^{-3/2}),
    \label{eq:perturbedcp}
\end{equation}  
where $h^0_i = \sum_j J_{ij} s_j$ is the field on spin $i$ before the perturbation. Inserting \eqref{eq:perturbedcp} into \eqref{eq:Etot} and \eqref{eq:EsingleOPO} shows that the energy of each critical point at large $a$ is given by 
\begin{equation}
    E_{\operatorname{tot}} =  -\frac{N}{4}(1-r)a^2 + \frac{a}{2} \sum_{i,j}J_{ij} s_i s_j + O(a^{0}).
    \label{eq:CIM_Ising}
 \end{equation}
 Thus, to leading order in $a$, the term breaking the degeneracy of critical points in the decoupled landscape is proportional to the Ising energy in \eqref{eq:Ising}. This implies that at large $a$, the sign configuration of the global minimum of the CIM energy function in \eqref{eq:Etot} and \eqref{eq:EsingleOPO} is equal to that of the global minimum of the Ising energy function in \eqref{eq:Ising}. 

Additionally, the Hessian $H(\mathbf{x})$ in \eqref{eq:Hessian} at a critical point $\bf{x}$ in \eqref{eq:perturbedcp} takes the form $H = H^I(\mathbf{x}) + J$ where $H^I(\mathbf{x})$ is diagonal with elements 
\begin{equation}
    H^I_{ii} = 
    \begin{cases}
       2a - 3h^0_i + O(1/a) & \text{for } s_i = \pm 1, \\           
       -a + O(1/a) & \text{for } s_i=0.
    \end{cases}       
    \label{eq:perturbedHI}
\end{equation}
The eigenvalue spectrum of this Hessian, in the case where $J$ is the random Gaussian connectivity of the SK model, can be understood using the random matrix theory of the next subsection, which will also form a basis for many subsequent analyses.  

\subsection{A theory of Hessian eigenspectra in the CIM with an SK spin glass connectivity}

The eigenvalue distribution of the Hessian $H(\mathbf x)$ in \eqref{eq:Hessian} and \eqref{eq:diagonalHessian} will play a key role in this work.  
Here we provide a theory for the spectrum of $H(\mathbf{x})$, at any spin configuration $\mathbf{x}$, when $J_{ij}$ is a rotationally invariant symmetric Wigner random matrix with i.i.d elements distributed as 
\begin{equation}
    J_{ij} = J_{ji} \sim
    \begin{cases}
        \mathcal{N}(0, {g^2}/{N})  & \text{for }i\neq j  \\  \mathcal{N}(0, 2{g^2}/{N}) & \text{for }i=j,
    \end{cases}  
    \label{eq:Jdistrib}
\end{equation}
where $\mathcal{N}(\mu, \sigma^2)$ denotes a Gaussian distribution with mean $\mu$ and variance $\sigma^2$. This connectivity corresponds to the SK spin glass in \eqref{eq:Ising}.  Because of the fundamental importance of the eigenvalue distribution of $H(\mathbf{x})$ in understanding the high dimensional geometry of the CIM energy landscape, we discuss this spectral distribution in the next two subsections in two different ways: first, in a conceptual way, as the outcome of a Dyson's Brownian motion with initial condition determined by $\mathbf{x}$, and second, in a computationally tractable manner  in terms of a self-consistent formula involving the resolvent of $H(\mathbf{x})$. Finally, in the third subsection, we apply this random matrix theory to analytically calculate the Hessian eigenspectra of CIM critical points at large $a$ and verify our formula by comparing to numerics.       
In the following, we set the connectivity variance parameter $g$ in \eqref{eq:Jdistrib} to $1$ without loss of generality, because the case of $g\neq1$ can be reduced to $g=1$ through the rescaling $\mathbf{x}\to \sqrt{g}\mathbf{x}$ and  $a\to ga$.  Note that for $g=1$, the eigenvalue spectrum of $J$ follows the well-known Wigner semicircular law with minimum/maximum eigenvalues given by $\lambda_{\min}(J) \approx -2$ and $\lambda_{\max}(J) \approx +2$ \cite{madanlalRandomMatrices2004}. 

\subsubsection{From the distribution of spins to Hessian eigenspectra through Dyson's Brownian motion}

Now at any spin configuration $\mathbf{x}$ for which the diagonal elements $H^I_{ii}(x_i) = \partial^2 E_I(x_i)$ are large relative to the elements of $J_{ij}$, one can compute the eigenvalues of $H(\mathbf{x})$ through first order perturbation theory, treating $J$ as perturbation to $H^I$ in \eqref{eq:Hessian}. This yields an approximate expression for the eigenvalues $\lambda_i$ of $H(\mathbf{x})$ given by 
\begin{equation}
    \lambda_i = H^I_{ii} + J_{ii} 
    + \sum_{j\neq i} \frac{J_{ij}^2}{H^I_{ii} - H^I_{ij}}.
    \label{eq:HessianEigPert}
\end{equation}
This expression is applicable for example, when $\mathbf{x}$ corresponds to a critical point of the CIM energy landscape at large $a$, where each $x_i$ in \eqref{eq:perturbedcp} is $O(\sqrt{a})$, and therefore each $H^I_{ii}$ in \eqref{eq:perturbedHI} is $O(a)$.  

However, at smaller $a$, when critical points are closer to the origin, the perturbative expression in \eqref{eq:HessianEigPert} may not be accurate. One can go beyond this perturbation theory by exploiting the fact that $H(\mathbf{x})$ is the sum of a fixed matrix $H^I$ and a Wigner matrix. This sum can be thought of as the outcome of a white noise driven diffusion process in the space symmetric matrices running from time $t=0$ to $t=g$ starting from the initial condition $H^I(\mathbf{x})$ and ending at $H(\mathbf{x})$.  This diffusion process on symmetric matrices in turn induces the well known Dyson's Brownian motion on the corresponding eigenvalues \cite{Dyson1962-vn,Potters2020-ta}, described by the stochastic differential equation
\begin{equation}
d\lambda_i = \sqrt{\frac{2}{N}} dW_{ii} + \frac{1}{N} \sum_{j\neq i} \frac{dt}{\lambda_i - \lambda_j},
\label{eq:dysonmotion}
\end{equation}
where $dW_{ii}$ is a standard white noise process. This stochastic evolution has a physical interpretation in which each $\lambda_i$ can be thought of as a Coulomb charge in the complex plane, confined to the real axis, feeling a deterministic, repulsive $2D$ Coulomb force from all the other charges $\lambda_j$, in addition to an independent stochastic drive.  If this Brownian motion is initialized at $t=0$ so that $\lambda_i(0) = H^I_{ii}(x_i)$, and is run up to time $t=g$, then the resulting eigenvalue distribution,
\begin{equation}
\rho_H(\lambda) \equiv  \frac{1}{N} \sum_{i=1}^N \delta(\lambda-\lambda_i(g))
\label{eq:hessdist}
\end{equation}
will, at large $N$, converge to the eigenvalue distribution of $H(\mathbf{x})$ in \eqref{eq:Hessian} with $J_{ij}$ distributed as in \eqref{eq:Jdistrib}.

Thus Dyson's Brownian motion provides an elegant and intuitive understanding of the relationship between a spin configuration $\mathbf{x}$ and the eigenvalue distribution of the Hessian $H(\mathbf{x})$: simply initialize a set of $N$ charges at the positions $H^I_{ii} = \partial^2E_I(x_i)$ and allow them to diffuse under \eqref{eq:dysonmotion} for a time $g$. However, this does not by itself provide an analytic method for computing the final outcome of the diffusion in \eqref{eq:hessdist}. 

\subsubsection{From the distribution of spins to the Hessian eigenspectra through the resolvent}
\label{subsubsec:resolvent}

In Section S-App.II, we provide a direct replica calculation of the Hessian eigenspectrum $\rho_H(\lambda)$ of $H(\mathbf{x})$ as a function of the distribution of spins at $\mathbf{x}$, defined as
\begin{equation}
    P_x(x) \equiv  \frac{1}{N} \sum_{i=1}^N \delta(x-x_i).
    \label{eq:xspindist}
\end{equation}
Our replica calculation yields a self-consistent equation for the resolvent of $H(\mathbf{x})$. In general, the resolvent of any $N$ by $N$ symmetric matrix $H$ is defined as
\begin{equation}
    R(z) = \frac{1}{N} \operatorname{Tr}\frac{1}{H-z},
\end{equation}
where $z \in \mathbb{C}$ is a complex scalar. One can recover the eigenvalue density $\rho_H(\lambda)$ from the resolvent $R(z)$ via the inversion formula 
\begin{equation}
    \rho_H(\lambda)=\lim _{\varepsilon \rightarrow 0^{+}} \frac{R(\lambda-i \varepsilon)-R(\lambda+i \varepsilon)}{2\pi i}.
    \label{eq:inversionformula}
\end{equation}
For $H(\mathbf{x})$, our replica-based self-consistent equation for its resolvent, when $g=1$, is given by (see Section S-App.I for a derivation), 
\begin{equation}
    R(z) = \int \frac{P_x(x)}{\partial^2 E_I(x)-z-R(z)} dx,
    \label{eq:resolventeqn}
\end{equation}
where $P_x(x)$ is the distribution of spins in \eqref{eq:xspindist}. This result agrees with Pastur's self-consistent equation for the resolvent of the sum of a fixed matrix and a Wigner matrix  \cite{pasturSpectrumRandomMatrices1972}.

Thus we obtain a simple calculational framework to obtain the Hessian eigenspectrum at any spin configuration $\mathbf{x}$: \circled{1} insert the distribution of spins $P_x(x)$ in \eqref{eq:xspindist} into the self-consistent equation for the resolvent $R(z)$ in \eqref{eq:resolventeqn}, \circled{2} solve this equation to find $R(z)$, and \circled{3} insert this solution into the inversion formula in  \eqref{eq:inversionformula} to obtain the Hessian eigenvalue distribution $\rho_H(\lambda)$.  The result will be equivalent to the distribution in \eqref{eq:hessdist} at time $t=1$ obtained by running Dyson's Brownian motion in \eqref{eq:dysonmotion} starting from the initial distribution of $H^I_{ii}$ induced by the distribution of $P_x(x)$ under the map $x \rightarrow \partial^2 E_I(x)$.    

\subsubsection{Hessian eigenspectra of critical points at laser gain}

Given this random matrix theory, we now return to the large gain regime in Section~\ref{subsubsection:largegain} to compute the Hessian eigenspectra of critical points of the form in \eqref{eq:perturbedcp}. 
In a typical index $r$ critical point, a fraction $r$ of the spins (before the perturbation by $J$) take the uncommitted value $s_i=0$, 
while the remaining fraction $1-r$ takes the committed values $s_i = \pm 1$ with equal probability. 
Moreover the field $h^0_i = \sum_j J_{ij}s_j$ in \eqref{eq:perturbedcp},
which perturbs the critical point after introducing the SK connectivity in \eqref{eq:Jdistrib} is, at large $N$, a zero mean Gaussian random variable with variance $(1-r)$, originating from the fraction $1-r$ of nonzero committed spins.  Thus the distribution of the diagonal elements in $H^I(\mathbf{x})$ in \eqref{eq:perturbedHI} is given by
\begin{equation}
    p_{H^I}(h) = r \delta(h+a) + (1-r)\mathcal{N}(2a, 9(1-r)). 
    \label{eq:pHI}
\end{equation}
This corresponds to a mixture of a $\delta$-function at $-a$ with weight $r$ coming from the  uncommitted spins, and a Gaussian centered at $2a$ with weight $1-r$ coming from the committed spins. The variance of $9(1-r)$ arises from the amplification of $h^0_i$ by a factor of $3$ in \eqref{eq:perturbedHI}. 

This initial distribution then undergoes Dyson's Brownian motion in \eqref{eq:dysonmotion} to yield the full distribution $\rho_H(\lambda)$ of $H(\mathbf{x})$. Alternatively, we can make the change of variables from $x$ to $h = \partial^2 E_I(x)$ in \eqref{eq:resolventeqn} to obtain a self-consistent equation $R_H(z)$ in terms of $p_{H^I}(h)$: 
\begin{equation}
R(z) = \int \frac{p_{H^I}(h)}{h-z-R(z)} dh.
\label{eq:resovlenth}
\end{equation}
We can then solve this equation (numerically) and insert the solution into \eqref{eq:inversionformula} to obtain $\rho_H(\lambda)$.  

\begin{figure}[!htbp]
    \includegraphics[width=\linewidth]{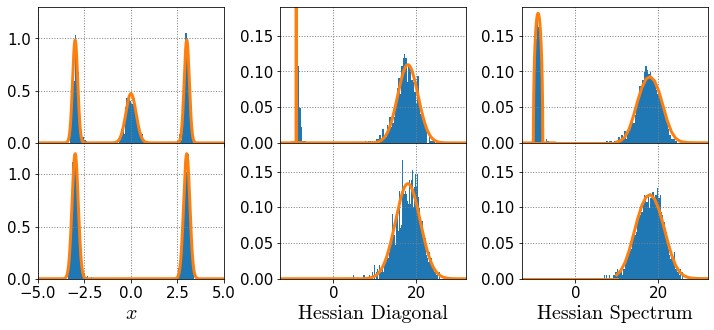} 
    \caption{{\bf Distribution of spins, Hessian diagonal elements and Hessian Spectrum.} The upper (or lower) panels showcase the distribution of OPO amplitudes $x$ (left) Hessian diagonal elements (middle), and Hessian eigenvalues (right) corresponding to a typical critical point (or a typical local minimum) with a large gain $a=9$. The empirical distributions portrayed as blue histograms are obtained with a system size of $N=10^3$. The orange curves in the left and middle figures are obtained with the perturbation theory in  \eqref{eq:perturbedcp} and \eqref{eq:perturbedHI}. The distributions of Hessian diagonal elements in the middle panels diffuse via Dyson's Brownian motion in \eqref{eq:dysonmotion} to generate the Hessian eigenspectrum in the right panels. The orange curves in the right panel are obtained from solutions of  \eqref{eq:resovlenth}.} 
    \label{fig:dyson-brownian}
\end{figure}

We calculated the Hessian eigenspectrum in this fashion both for typical critical points with index $r=1/3$, and for typical minima with index $r=0$, finding an excellent match with direct numerical searches for such critical points at a finite system size of $N=10^3$ and at large laser gain $a=9$ (Fig. \ref{fig:dyson-brownian}). Some features of the outcome of Dyson's Brownian motion in going from $p_{H^I}(h)$ in \eqref{eq:pHI} to $\rho_H(\lambda)$ are readily apparent in Fig. \ref{fig:dyson-brownian}. For example, at large $a$ for a typical critical point with index $r=1/3$, the charges start in two far apart clumps in \eqref{eq:pHI}, with a delta-function at $-a$ and a Gaussian at $2a$. Thus these two distant charge clumps do not interact strongly with each other in the diffusion. However each clump itself expands under the repulsive diffusion. The delta function expands into a Wigner semicircle, still centered at $-a$, while the Gaussian expands a bit more, largely retaining its shape and remaining centered at $2a$ (Fig. \ref{fig:dyson-brownian} top). 

\section{The performance of geometric landscape annealing for the SK spin glass}
\label{sec:performance}

We have seen in Section~\ref{subsubsection:smallgain} that at small gains $a$ just above $\lambda_{\operatorname{min}}(J)$, the CIM global minimum computes the spectral solution in \eqref{eq:spectralopt}, which is not of direct interest. 
On the other hand in Section~\ref{subsubsection:largegain}, at large gain $a \gg \lambda_{\operatorname{max}}(J)$, we have seen that the CIM global minimum computes the Ising energy minimization, which {\it is} of direct interest. 
However, our analysis of the energy landscape at large laser gain in Section~\ref{subsubsection:largegain} reveals a complex landscape with exponentially many local minima and saddle points of all indices.  
Thus direct gradient descent in the large laser gain energy landscape of the CIM is unlikely to find the CIM global minimum (as we verify below in Section~\ref{sec:relationship}).
Therefore to understand how the CIM solves optimization problems by annealing the laser gain, a key first step is to understand how the geometry of the landscape changes from small to large gain.

In particular, we would like to understand in general how the first local minimum to occur, which is aligned along the lowest eigenvector $\mathbf{v}_{\min}$ in \eqref{eq:spectralopt}, changes as the laser gain is increased. There are several possibilities. 
The first is that this local minimum is continuously connected to one of the CIM global minima as we increase the gain to large values. In this case, annealing will find the global minimum. The second possibility is that the first local minimum to appear as the gain increases is continuously connected to a higher energy CIM local minimum at large gain. In this case annealing will not find the CIM global minimum.  A third possibility is that this first minimum may disappear through a saddle-node bifurcation, and then slowly annealed gradient descent will flow to another nearby minimum, which in-turn can exhibit these same $3$ possibilities. 

It is an exceedingly difficult problem to analytically predict, in advance of geometric landscape annealing, which of the these possibilities will occur for any large, fixed connectivity matrix $J$. One would have to map out the entire bifurcation structure of critical points as $a$ increases.  Moreover, one would have to analytically derive the CIM ground state energy for that connectivity $J$ at large $a$ and compare it to the energy of all critical points that are continuously connected through bifurcations to the first minimum to appear along $\mathbf{v}_{\min}$ near the origin.  All of this is more complex than simply performing geometric landscape annealing itself.  

We circumvent these difficulties by not analyzing any fixed connectivity $J$, but rather analyzing typical CIM behavior in random Gaussian connectivities $J$ in \eqref{eq:Jdistrib} corresponding to an SK spin glass. For this problem we can use techniques from the statistical mechanics of quenched disorder to analytically calculate the CIM ground state energy at arbitrary laser gain $a$, as well as the location, Hessian eigenstructure, and energy levels of critical points of any index. We can compare these quantities to numerical simulations of geometric landscape annealing to provide insights into its operation.     

\begin{figure}
    \includegraphics[width=\linewidth]{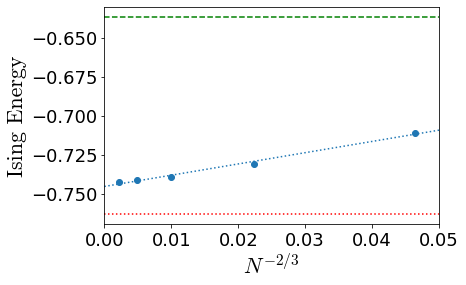}
    \caption{{\bf Ising energy of the final state obtained by the annealing process for random instances.} We simulated the annealing process of the soft-spin network with many instances for each system size $N$, and plot the medians as blue dots. The blue dotted line is the linear regression against $N^{-2/3}$. This scaling comes from the SK model's finite scaling \cite{aspelmeierFinitesizeCorrectionsSherrington2008}. The y-intercept of this line represents the reachable lowest Ising energy by the dynamics under the large-$N$ limit. The horizontal red dotted line is the SK model's ground state energy in the large-$N$ limit $E_{g} \sim -0.763$. The horizontal green dashed line is $E_{\operatorname{sp}}=-2/\pi$, the energy of spin configuration obtained by rounding the principal eigenvector of $J$. The annealing schedule used here is given by \eqref{eq:schedule} with $\tau=10^2$, $a_{\max} = 0.0$, and $a(0) = \lambda_{\min}$. The number of sampled instances are $300, 300, 100, 100, 20, 5$ for $N=10^2, 3\times 10^2, 10^3,$ $ 3\times 10^3, 10^4$, respectively.} 
    \label{fig:ising-energy-by-cim}
\end{figure}

First, we assess the performance of CIM geometric landscape annealing as we increase the laser gain $a$.  
We performed numerical simulations of geometric landscape annealing with several system sizes $N$ by the integration of \eqref{eq:graddescent}.
During the integration, we slowly increased the gain parameter $a$ starting from $\lambda_{\min}(J)$ to achieve the best performance for each system size $N$.
In Fig.(\ref{fig:ising-energy-by-cim}), the blue dots represent the medians of the final Ising energy obtained by the simulations with several instances, and the blue dotted line is the linear regression of those points against $N^{-2/3}$. This scaling comes from the finite-size scaling of the SK model's ground state energy \cite{aspelmeierFinitesizeCorrectionsSherrington2008}. The y-intercept of this blue line represents the reachable lowest energy by the annealing dynamics in the large-$N$ limit. We call this energy $E_{\text{anneal}}$. The horizontal red dotted line is the theoretically obtained ground state energy $E_{g}\sim-0.763$ in the large-$N$ limit \cite{crisantiAnalysisEnsuremathInfty2002, schmidtReplicaSymmetryBreaking2008}, and the horizontal green dashed line is the energy obtained from the Ising spin configuration by rounding the principal eigenvector, which yields the known value  $E_{\operatorname{sp}}=-2/\pi$ \cite{montanariOptimizationSherringtonKirkpatrickHamiltonian2019}.  We can see that the energy $E_{\operatorname{anneal}}\sim -0.75$ obtained by the annealing process in the large-$N$ limit is much lower than $E_{\operatorname{sp}}$. This means that the first minimum to appear along the eigenvector $\mathbf{v}_{\min}$, must undergo multiple sign flips induced by further bifurcations as $a$ increases. 

Remarkably, these bifurcations substantially lower the Ising energy found by the CIM, making it very close to the actual ground state Ising energy of the SK model. 
In the remainder of this paper we will study the changing geometry of the CIM energy landscape to understand how geometric annealing of this landscape empowers the performance of the CIM in finding low Ising energy solutions.

\section{The evolving energy landscape geometry}
\label{sec:evolving}
To address the questions raised above, we first analytically derive a formula for the typical number $\mathcal{N}(r,e |J)$ of critical points of a given intensive index $r$ and energy $e$. In order to average over the connectivity $J$ we will work with the complexity $\Sigma(r,e |J)$ of critical points, which is defined via the relation
\begin{equation}
\mathcal{N}(r,e | J) = e^{N\Sigma(r,e | J)}.
\end{equation}
This complexity can be formally written as a sum over all critical points
\begin{equation}
    \Sigma(r,e| J) = \frac{1}{N}\log\sum_{\alpha\in\operatorname{Crt}(E)} \delta(I(\mathbf{x}^\alpha) - r) \delta(E(\mathbf{x}^\alpha) - e),
\end{equation}
where $E(\mathbf{x}) = \frac{1}{N} E_{\operatorname{tot}}(\mathbf{x})$ is the intensive energy, and $\operatorname{Crt(E)}$ denotes the set of all critical points of $E(\mathbf{x})$.  

Unlike the potentially exponentially large number $\mathcal{N}(r,e | J)$ itself, which could fluctuate across random samples of $J$, we expect the complexity function $\Sigma(r,e| J)$ to be self-averaging with respect to $J$.  
This means that typical values of $\Sigma(r,e| J)$ for random samples of $J$ concentrate closely around the sample average $\Sigma(r,e) \equiv \Braket{\Sigma(r,e| J)}_J$, where $\Braket{\cdot}_J$ denotes an average over $J$.
Furthermore, in order to compute this sample averaged complexity $\Sigma(r,e)$, we will first compute the sample average of the grand potential $\Omega(\beta,\mu| J)$, defined as
\begin{equation}
    -\beta\Omega(\beta, \mu | J) = \frac{1}{N}\log\sum_{\alpha\in \operatorname{Crt}(E)}
        e^{-\beta E(\mathbf{x}^\alpha) + \mu \mathcal{I}(\mathbf{x}^\alpha)},
    \label{eq:grandpotdef}
\end{equation}
where $\mathcal{I}(x^\alpha)\in[0,1]$ denotes the intensive index of the critical point $\mathbf{x}^\alpha$, i.e., $\mathcal{I}(\mathbf{x^\alpha}) = I(\mathbf{x^\alpha})/N$.
If one can compute the sample averaged grand potential $\Omega(\beta, \mu) \equiv \Braket{\Omega(\beta, \mu | J)}_J$, then one can recover the average complexity $\Sigma(e,r)$ via Legendre transform 
\begin{equation}
    \Sigma(e,r) = \mathrm{inf}_{\beta,\mu} \left[ \beta e -\mu r  - \beta\Omega(\beta, \mu)\right].
    \label{eq:legendre}
\end{equation}
Here the effective inverse temperature $\beta$ and energy density $e$ form a Legendre dual pair, as does the chemical potential $\mu$ and the intensive index $r$. Indeed the typical values of $e$ and $r$ that dominate in the sum in \eqref{eq:grandpotdef} are those that achieve the infimum in \eqref{eq:legendre}. If the infimum does not occur at a boundary, the typical $e$ and $r$ are related to $\beta$ and $\mu$ through
\begin{equation}
    e = \frac{\partial}{\partial \beta} \left[\beta\Omega(\beta, \mu)\right]
    \quad
    r = - \frac{\partial}{\partial \mu} \left[\beta\Omega(\beta, \mu)\right].
    \label{eq:legendreinf}
\end{equation}

Finally, we compute the sample averaged grand potential $\Omega(\beta,\mu)$ via the replica trick \cite{moroneReplicaTheorySpin2014}, i.e.,
\begin{eqnarray}
    -\beta\Omega(\beta,\mu) 
    &=& \frac{1}{N} \Braket{\log(Z)}_J \nonumber \\
    &=& \frac{1}{N} \lim_{n\to 0} \frac{1}{n} \log \Braket{Z^n}_J,
\label{eq:grandpotrep}
\end{eqnarray}
where $Z$ is the partition function 
\begin{equation}
    Z = \sum_{\alpha\in \operatorname{Crt}(E)}e^{- \beta E(\mathbf{x}^{\alpha}) 
        + \mu \mathcal{I}(\mathbf{x^\alpha})}.
    \label{eq:partitionfunc}
\end{equation}

Below we will apply the Kac-Rice formula to \eqref{eq:grandpotrep} to compute $\Omega(\beta,\mu)$. This replica based calculation, which is given in full detail in Section S-II,  involves introducing both bosonic degrees of freedom (replicated soft spins $\mathbf{x}^a$ for $a=1,\dots,n$) as well as fermionic degrees of freedom whose integral computes the determinant of the Hessian which arises in the Kac-Rice formula below.  The resulting integrals possess both replica symmetry, involving permutations of the replicas, as well as supersymmetry, involving exchanges of bosonic and fermionic degrees of freedom. Such a SUSY based framework has also been used in a variety of works~\cite{
annibaleCoexistenceSupersymmetricSupersymmetrybreaking2004,
annibaleRoleBecchiRouet2003,
annibaleSupersymmetricComplexitySherringtonKirkpatrick2003,
cavagnaCavityMethodSupersymmetrybreaking2005,
cavagnaFormalEquivalenceTAP2003, 
crisantiComplexitySherringtonKirkpatrickModel2003,
parisiSupersymmetryBreakingComputation2004,
rizzoTAPComplexityCavity2005}.
These integrals can be solved via a saddle point approximation, and the order parameters, whose extremal values determine the saddle point, can either exhibit or break replica symmetry or supersymmetry. Which pattern of symmetry breaking occurs or not depends on the particular values of the inverse temperature $\beta$, chemical potential $\mu$, and gain $a$ considered. 

In the following, we will consider three regimes in detail. First, we will consider $\beta=\mu=0$, corresponding to a white average in \eqref{eq:partitionfunc} in which all critical points are equally weighted.  This white average yields information about the typical behavior of a randomly chosen critical point, regardless of its energy or index.  We will find that in this regime, at the saddle point order parameters, replica symmetry always holds, but SUSY is preserved at low laser gain $a$, while it is broken at large laser gain. The order parameters at the saddle point yield information about the distribution of spins and Hessian eigenvalues at typical critical points.  
    
The second regime we will consider is $\beta=0$ and $\mu\rightarrow -\infty$. This concentrates the sum over critical points onto those with vanishing intensive index, independent of their energy. This corresponds to a sum over all minima (we refer to critical points with zero {\it intensive} index as minima). We will find a similar pattern for typical minima as we did for typical critical points: order parameters at the saddle point exhibit replica symmetry, but can break SUSY depending on the laser gain.  The order parameters, distribution of spins, and Hessian eigenvalues for $\beta=0$ and arbitrary $\mu$ are given in Section~\ref{sec:replica-cal} in the case of replica symmetry and broken SUSY. 

Unfortunately, the geometric interpretation of these replica calculations, and in particular the geometric meaning of broken SUSY in terms of the original energy landscape, is unclear. Because the geometric interpretation of SUSY breaking is a subject of considerable interest, 
we provide in Section~\ref{sec:cavity-method} a completely different derivation of the results of Section~\ref{sec:replica-cal} using the cavity method instead of the supersymmetric method (see Section S-IV for a detailed cavity derivation). 
This derivation yields a new interpretation of nonzero supersymmetry breaking order parameters as signaling a high sensitivity of the complex energy landscape to small changes in external fields.  
We discuss in particular the case of typical critical points in  Section~\ref{sec:typical-crt} and the typical minima in  Section~\ref{sec:typical-min} and successfully match our theoretical predictions with numerical experiments. 

In Section~\ref{sec:global-min} we move on to the case of the global minima, corresponding to the regime $\beta \rightarrow \infty$ in \eqref{eq:partitionfunc}.  We will find that the global minima of the energy landscape occur at significantly lower energies than that of typical local minima for large values of the laser gain. Therefore, as in other spin glass problems, replica symmetry is broken.  We provide an analysis of the global minima through two methods: \circled{1} a replica-based calculation of the grand potential (see Section~S-II.G) or \circled{2} a calculation of the free energy (see Section~S-III), and we demonstrate their equivalence.  

Finally in Section \ref{sec:phase-diagram}, we summarize and describe the significant phase transitions we can observe in the energy landscape due to successive SUSY and replica symmetry breaking, and their geometric consequences. 

\subsection{The replica-based calculation}
\label{sec:replica-cal}
In our setting, the Kac-Rice formula (see \cite{adler2007random} for an introduction) enables us to convert the sum of any function $F(\mathbf x)$ over all critical points $\mathbf{x}^\alpha$ for $\alpha \in \operatorname{Crt}(E)$ of a landscape $E(\mathbf{x})$, into an integral over the entire domain $\mathbf{x} \in \mathbb{R}^N$ of the landscape. It is given by
\begin{equation}
    \sum_{{\alpha} \in \operatorname{Crt}(E)} F(\mathbf{x}^\alpha) = \int \prod_{i=1}^N\left[dx_i\ \delta(\partial_i E(\mathbf{x}))\right] |\det H(\mathbf{x})| F(\mathbf{x}),
    \label{eq:kacrice}
\end{equation}
where $H(\mathbf{x})$ is the Hessian of $E$ at $\mathbf{x}$. Here the $\delta$-functions in \eqref{eq:kacrice} localize the integral to critical points of $E(\mathbf{x})$ as desired, while the absolute value of the Hessian determinant $|\det H(\mathbf{x})|$ corresponds to the Jacobian of the change of variables from $x_i$ to $y_i = \partial_i E(\mathbf{x})$. Indeed performing this change of variables on the right hand side of \eqref{eq:kacrice} and then integrating recovers the left hand side. 

Now applying the Kac-Rice formula in \eqref{eq:kacrice} to the partition function in \eqref{eq:partitionfunc} yields
\begin{equation}
    Z = \int \prod_{i=1}^N\left[dx_i\ \delta(\partial_i E(\mathbf{x}))\right] |\det H(\mathbf{x})|
        e^{-\beta E(\mathbf{x}) + \mu \mathcal{I}(\mathbf{x})}.
\label{eq:partitionfunckac}
\end{equation}
Then inserting \eqref{eq:partitionfunckac} into \eqref{eq:grandpotrep} provides the starting point for the replica and supersymmetry based calculation of the sample-averaged grand potential. A detailed derivation is given in Section ~S-II. The final answer, at a replica symmetric, annealed level with $\beta=0$, but with broken supersymmetry is given by
\begin{widetext}
\begin{equation}
    -\beta\Omega(0,\mu) = -\frac{1}{2}(Cq + A^2) - A t+\log\frac{1}{\sqrt{2\pi q}}\int dx |\partial^2 E_I(x)-t|\exp\left( -\frac{1}{2q}h(x)^2 + \frac{A}{q} xh(x) + \frac{1}{2} \frac{qC - A^2}{q}x^2+\mu\bar {\mathcal{I}}(x)\right)
    \label{eq:complexity-replica}
\end{equation}
Additionally, beyond the sample averaged grand potential, we consider the sample averaged distribution of spins in an ensemble of critical points, defined as
\begin{equation}
    P(x)= \Braket{Z^{-1}\sum_{\alpha\in \operatorname{Crt}(E)} e^{-\beta E_I(\mathbf{x}^\alpha) + \mu\mathcal{I}(\mathbf{x}^\alpha)}  \left(\frac{1}{N}\sum_{i=1}^{N} \delta(x-x^\alpha_i)\right)}_J, 
    \label{eq:defpx}
\end{equation}
where $Z$ is given in \eqref{eq:partitionfunc}. We derive a formula for this distribution in Section~S-II.B. In the case of $\beta=0$ and arbitrary $\mu$, which is relevant for typical critical points ($\mu=0$) and typical minima ($\mu \rightarrow -\infty)$, the answer is 
\begin{equation}
P(x)\propto |\partial^2 E_I(x)-t| \exp\left( -\frac{1}{2q}h(x)^2 + \frac{A}{q} xh(x) + \frac{1}{2} \frac{qC - A^2}{q}x^2+\mu\bar {\mathcal{I}}(x)\right).
    \label{eq:Px-replica}
\end{equation}
\end{widetext}
Here $\overline{\mathcal{I}}(x)$ is given by
\begin{equation}
    \overline{\mathcal{I}}(x) = \Theta(-(\partial^2 E_I(x)-t)),
    \label{eq:mean-field-index}
\end{equation}
where $\Theta$ is the Heaviside step function, and
\begin{equation}
    h(x) = \partial E_I(x) - tx.
\end{equation}

The formulas for the sample averaged grand potential in \eqref{eq:complexity-replica} and distribution of spins in \eqref{eq:Px-replica} depend on four order parameters, $q,t,A$ and $C$ which satisfy the following self-consistent equations arising from extremizing the grand potential in \eqref{eq:complexity-replica}:
\begin{eqnarray}
\left\{
  \begin{aligned}
  & q = \braket{x^2}\\
  & t= \Braket{\frac{1}{\partial^2 E_I(x) - t}}\\
  & A = \frac{\braket{xh(x)}}{2q} - \frac{t}{2}\\
  & C = -q^{-1} + q^{-2}\Braket{h^2(x)} - 2q^{-2}A\Braket{xh(x)} +q^{-1}A^2.
  \end{aligned}\right.
  \nonumber\\
  \label{eq:self-consistent-replica}
\end{eqnarray}
Here $\Braket{\cdot}$ denotes an average with respect to the distribution $P(x)$ in \eqref{eq:Px-replica}. We note that self-consistent solutions with nonzero values for the order parameters $A$ and $C$ correspond to broken supersymmetry \cite{crisantiComplexitySherringtonKirkpatrickModel2003, crisantiSpinGlassComplexity2004}.

Finally, with knowledge of the typical distribution of spins $P(x)$ in an ensemble of critical points, we can obtain the typical distribution of Hessian eigenvalues by inserting $P(x)$ in \eqref{eq:Px-replica} into \eqref{eq:resolventeqn}, solving for the resolvent $R(z)$, and inserting this solution into \eqref{eq:inversionformula} to obtain $\rho_H(\lambda)$ for any $\beta=0$, $\mu$, and $a$. Note that we here assume that the correlation between $H^I(\mathbf{x})$ and $J$ is negligible for any critical point $\mathbf{x}$. 

Importantly, we note that the last self-consistent equation for $t$ in \eqref{eq:self-consistent-replica} is equivalent to the self-consistent equation for $R(z)$ at $z=0$ in \eqref{eq:resolventeqn}. 
But more precisely,
while the resolvent $R(z)$ of a large random symmetric matrix $H$ 
is not well defined at any point $\lambda$ 
on the real axis where the eigenvalue distribution $\rho_H(\lambda)$ is nonzero, 
$R(z)$ {\it is} defined on the complex plane near the real axis for $z=\lambda+i\varepsilon$ 
with arbitrarily small $\varepsilon$. 
Thus we can define the complex number $t_R + i t_I = R(0+i\varepsilon)$ for a small $\varepsilon$.  
By the inversion formula in \eqref{eq:inversionformula}, $t_I$ is nonzero if and only if the Hessian eigenvalue density $\rho_H(0)$ is nonzero.  
We will see empirically that $\rho_H(0)$ is very close to $0$. 
Therefore assuming $t_I=0$, $t$ in \eqref{eq:self-consistent-replica} should be properly be thought of as $t_R = R(0+i\varepsilon)$.  
On the otherhand, if the Hessian eigenvalue density $\rho_H(\lambda)$ at the origin $\lambda=0$ were nonzero, one would have to self-consistently solve for another order parameter $t_I$.  
The full self-consistent equations for all $5$ order parameters $q, t_R, t_I, A$ and $C$ are given in Section~S-II.F. 
However, to match numerics below, we will only need to find approximate self-consistent solutions to \eqref{eq:self-consistent-replica} assuming that $t_I=0$, or equivalently $\rho_H(0)=0$.

In summary, the replica analysis provides an efficient calculational framework to obtain key information about the number and properties of typical critical points ($\mu=0$) and typical minima ($\mu \rightarrow -\infty$), as well as critical points of any index $r$ related to $\mu$ through Legendre duality in \eqref{eq:legendreinf}.  The procedure is as follows: \circled{1} solve the self-consistent equations for the order parameters in \eqref{eq:self-consistent-replica}; \circled{2} insert them into \eqref{eq:Px-replica} to obtain the typical distribution of spins $P(x)$ at a critical point; \circled{3} insert $P(x)$ into \eqref{eq:resolventeqn} and \eqref{eq:inversionformula} to obtain the typical distribution of Hessian eigenvalues $\rho_H(\lambda)$; \circled{4} insert the formula for the grand potential $\Omega(\beta,\mu)$ in \eqref{eq:grandpotrep} into the Legendre transform in \eqref{eq:legendre} to obtain the complexity $\Sigma(e,r)$ at typical energy $e$ and index $r$ given by  
\eqref{eq:legendreinf}. 

Finally, we note that for typical critical points and typical minima, if we only wish to compute the grand potential at $\beta=0$, we can still compute the typical energy $e$ of critical points without using the first Legendre dual relation in \eqref{eq:legendreinf}. We do this by noting that any critical point $\mathbf{x}$ of \eqref{eq:Etot} obeys $\partial E_I(x_i) + h_i = 0$ where $h_i \equiv \sum_j J_{ij} x_j$.  This implies that at any critical point $\mathbf{x}$, the normalized intensive energy obeys the special relation 
\begin{eqnarray}
    E(\mathbf{x}) &=& \frac{1}{N} \sum_{i=1}^N \left[ E_I(x_i) + \frac{1}{2} x_i h_i \right] \nonumber \\
                  &=& \frac{1}{N} \sum_{i=1}^N \left[ E_I(x_i) - \frac{1}{2} x_i \partial E_I(x_i) \right].
\end{eqnarray}  
This site decoupled expression for the energy allows us to calculate the typical energy $e$ at critical points  directly from the typical distribution of spins $P(x)$ in \eqref{eq:Px-replica} via
\begin{equation}
    e = \int dx \, P(x) \left[ E_I(x) - \frac{1}{2} x \partial E_I(x)  \right].
    \label{eq:eformula}
\end{equation}

Similarly, the typical intensive index $r$ can be calculated, without resorting to the second Legendre dual relation in \eqref{eq:legendreinf}, by directly using the typical Hessian eigenvalue distribution $\rho_H(\lambda)$ obtained from \eqref{eq:inversionformula} using the distribution of $P(x)$ in \eqref{eq:Px-replica} inserted into the formula for $R(z)$ in \eqref{eq:resolventeqn}.  In terms of this $\rho_H(\lambda)$, $r$ is simply 
\begin{equation}
    r = \int_{-\infty}^0 d\lambda \, \rho_H(\lambda).
    \label{eq:rformula}
\end{equation}

Overall, these results yield a complete characterization of the typical energy $e$, index $r$, grand potential $\Omega(0,\mu)$, complexity $\Sigma(e,r)$, distribution of spins $P(x)$ and distribution of Hessian eigenvalues $\rho_H(\lambda)$ of both typical critical points and typical minima.  We will successfully confirm these theoretical predictions with numerical simulations below in Section~\ref{sec:numericaltests}. But first, we provide an alternate derivation of these results by developing a novel cavity method.

\section{A geometric interpretation of supersymmetry breaking via a generalized cavity method}
\label{sec:cavity-method}

While the replica based calculation above provides detailed information about critical points, the form of the answers are difficult to understand. For example, why does the grand potential $\beta \Omega$
in \eqref{eq:grandpotrep}, the distribution of spins $P(x)$ in \eqref{eq:Px-replica}, and the self-consistent equations for the order parameters in \eqref{eq:self-consistent-replica} take the forms that they do?  Moreover, what is the geometric meaning of the order parameters, especially the SUSY breaking order paramters $A$ and $C$? In essence, what is the qualitative difference between high dimensional energy landscapes described by broken SUSY versus preserved SUSY? To obtain answers to these questions, we developed a new generalized version of the cavity method and demonstrate the equivalence between our generalized cavity method and replica derivations (see Section~S-IV for a detailed derivations).  However, our generalized cavity method yields considerable conceptual insights into the replica results as well as a geometric interpretation of SUSY breaking.

\subsection{The naive cavity method}

We first take a naive approach to the cavity method, which we will see is appropriate when SUSY is preserved. The cavity method in general for many mean-field systems involves: \circled{1} analyzing the effect of adding a single new degree of freedom to a system (called a cavity system because it excludes the new degree of freedom), \circled{2} describing how the cavity system responds to the new degree of freedom, often using simple perturbation theory under the assumption that the single new degree of freedom exerts a small effect on the large cavity, and \circled{3} quantifying how the response of the cavity exerts a backreaction onto the new degree of freedom as it comes to equilibrium with the cavity system.  The backreaction of the cavity onto the new degree of freedom depends on certain order parameters associated with the cavity.  The cavity method then yields self-consistent equations for these order parameters assuming the cavity system without the new degree of freedom, and the full system with the new degree of freedom, have the same order parameters, due to the existence of a thermodynamic limit.

For example, in the context of the CIM, critical points of any index (not necessarily energy minima), obey the gradient equations

\begin{equation}
    \partial E_I(x_i) + \sum_{j=1}^{N-1}J_{ij} x_j  = 0 \quad \text{for } i=1,\dots,N-1.
    \label{eq:cavitysystem}
\end{equation}
Here this corresponds to a cavity system with only $N-1$ spins.  Next we introduce a new spin $x_0$ coupled to the cavity system via new random coupling constants $\{J_{0i}\}_{i=1,\dots,N-1}$. The gradient equations for $x_1,\dots x_N$ in the presence of the new spin in the full system become 
\begin{equation}
    \partial E_I(x_i) + \sum_{j=1}^{N-1}J_{ij} x_j + J_{i0} x_0 = 0,
    \label{eq:cavity+force}
\end{equation}
while the new spin, after it equilibriates with the cavity must obey
\begin{equation}
    \partial E_I(x_0) + \sum_{i=1}^{N-1}J_{0i} x_i = 0.
    \label{eq:cavitynewspin}
\end{equation}
The cavity method relates the critical point solutions of the full system in \eqref{eq:cavity+force} and \eqref{eq:cavitynewspin} to the critical point solutions of the cavity system in \eqref{eq:cavitysystem}. 
In particular, let $x_i^{/0}$ 
for $i=1,\dots,N-1$ be a critical point of the cavity system in the {\it absence} 
of spin $0$. 
Thus $x_i^{/0}$ is a solution to \eqref{eq:cavitysystem} for all $i=1,\dots,N-1$. 
Now when the new spin $0$ is brought into contact with the cavity and held at a {\it fixed} value $x_0$, the cavity will react to the new spin so as to solve the modified equations \eqref{eq:cavity+force}, which are simply equivalent to the original cavity equations \eqref{eq:cavitysystem} plus a small perturbative term $J_{i0}x_0$ that is $O(\frac{1}{\sqrt{N}})$. 

Assuming the effect of the new spin $x_0$ on the cavity is small, one can solve \eqref{eq:cavity+force} using perturbative linear response theory, by Taylor expanding the first two terms about $x_i = x_i^{/0}$ and using the fact that $x_i^{/0}$ satisfies \eqref{eq:cavitysystem}. The resulting approximate linear response of the cavity to the new spin $x_0$ (i.e. approximate solution to \eqref{eq:cavity+force}) is 
\begin{equation}
x_i = x_i^{/0} - \sum_{j=1}^N H_{ij}^{-1}(\mathbf{x}^{/0}) J_{j0} x_0.
\label{eq:cavityresponse}
\end{equation}
Here $H_{ij}^{-1}(\mathbf{x}^{/0})$ is the inverse Hessian of the cavity system evaluated at its critical point $\mathbf{x}^{/0}$ {\it before} the new spin $x_0$ is introduced.  
As usual, this inverse Hessian acts as a linear susceptibility matrix $\chi = H^{-1}(\mathbf{x}^{/0})$ that translates the force $J_{j0} x_0$ exerted by the new spin into the response of the cavity from $x_i^{/0}$ to $x_i$ in \eqref{eq:cavityresponse}.

Now with \eqref{eq:cavity+force} solved perturbatively via the cavity response in \eqref{eq:cavityresponse} for arbitrary $x_0$, we must next find the equilibrium value of $x_0$ that generates an approximate critical point of the full system by inserting \eqref{eq:cavityresponse} into \eqref{eq:cavitynewspin}, obtaining 
\begin{equation}
    \partial E_I(x_0) - \sum_{i,j=1}^N J_{0i} H_{ij}^{-1}(\mathbf{x}^{/0}) J_{j0} x_0 + \sum_{i=1}^N J_{0i} x_i^{/0} = 0. 
    \label{eq:mfgradient}
\end{equation}
Here the final term
\begin{equation}
    h^{/0} \equiv - \sum_{i=1}^N J_{0i} x_i^{/0}
    \label{eq:cavityfield}
\end{equation}
is the cavity field that the cavity would have exerted on the new spin had it not reacted to the new spin at all, and remained at configuration $\mathbf{x}^{/0}$.  The second term takes into account the reaction of the cavity to $x^0$ through the force $J_{j0}x^0$, and its resultant backreaction on the new spin through the connections $J_{0i}$.  This is an example of an Onsager backreaction type term \cite{mezardSpinGlassTheory1986}. 

Now both the cavity field and the backreaction term depend on the cavity system through two simple order parameters.  First, note that $x_i^{/0}$, a critical point of the cavity system in the absence of the new spin $x_0$, is necessarily independent of the new connectivity $J_{0i}$, which is not a part of the cavity system. Thus we can apply the central limit theorem to conclude that $h^{/0}$ in \eqref{eq:cavityfield} is a random Gaussian variable distributed as $\mathcal{N}(0,q)$ where the variance $q$ is an order parameter given by
\begin{equation}
    q = \frac{1}{N-1} \sum_{i=1}^{N-1} (x^{/0}_i)^2
    = \frac{1}{N} \sum_{i=0}^{N-1} \left(x_i\right)^2.
      \label{eq:orderparamq}
\end{equation}
Here we have assumed that the order parameter $q$ is self-averaging and is the same in both the cavity system and the full system, at large $N$. Similarly, we assume that the Onsager-back reaction term is self-averaging and we replace it with its average over the connectivity in \eqref{eq:Jdistrib}, yielding a second order parameter $t$ which we assume is the same both in the cavity and the full system: 
\begin{equation}
    t = \frac{1}{N-1} \operatorname{Tr} H^{-1}(\mathbf{x}^{/0})
      = \frac{1}{N} \operatorname{Tr} H^{-1}(\mathbf{x}).
    \label{eq:orderparamt}
\end{equation}
While $q$ is the squared length of a critical point, $t$ is the trace of a critical point's linear susceptibility matrix to small external forces. 

With the definition of the cavity field $h^{/0}$ in \eqref{eq:cavityfield} and the order parameters $q$ in \eqref{eq:orderparamq} and $t$ in \eqref{eq:orderparamt}, the solution(s) of $x_0$ in \eqref{eq:mfgradient} are in one to one correspondence with critical points of a mean-field energy function 
\begin{equation}
    E_{\operatorname{MF}}[h](x) \equiv 
    E_I(x) -\frac{1}{2} t x^2 - hx, 
    \label{eq:E_MF}
\end{equation}
where the random external cavity field $h \sim \mathcal{N}(0,q)$.  Here we have dropped the index $0$ from both the new spin $x_0$ and its cavity field $h^{/0}$, because under the random mean field connectivity in \eqref{eq:Jdistrib}, there is nothing special about removing and adding back spin $0$.  We could have done this for any spin $x_i$, yielding its own cavity field $h^{/i}$ which is also distributed as $\mathcal{N}(0,q)$. Moreover, each cavity field $h^{/i}$ in the absence of $x_i$ is independent of any other cavity field $h^{/j}$ in the absence of $x_j$.  Therefore the empirical distribution of spins $x_i$ across the index $i$, defined as $P(x) = \frac{1}{N}\sum_{i=1}^N \delta(x-x_i)$, can be obtained, in the large $N$ limit as
\begin{equation}
    P(x) \propto \Braket{\sum_{x_*\in\mathrm{Crt}(E_{\operatorname{MF}}[h])}\delta(x-x_*)}_{h},
    \label{eq:Px_cavitySUSY}
\end{equation}
where $\mathrm{Crt}(E_{\operatorname{MF}}[h])$ denotes the set of critical points of the function $E_{\operatorname{MF}}[h]$ in \eqref{eq:E_MF}, and $\Braket{\cdot}_h$ denotes an average with respect to the Gaussian cavity field $h \sim \mathcal{N}(0,q)$. The normalization factor in \eqref{eq:Px_cavitySUSY} is simply the mean number of critical points in the random ensemble of mean field energy functions $E_{\operatorname{MF}}[h]$. 

Now, with the distribution of spins $P(x)$ in a typical critical point in hand, we can derive self-consistent equations for the order parameters. In particular, it is clear that $q$ in \eqref{eq:orderparamq} is simply the second moment of $P(x)$, yielding the self-consistent equation
\begin{equation}
    q = \int dx \, x^2 P(x).
    \label{eq:qselfcons}
\end{equation}
Furthermore, $t$ in \eqref{eq:orderparamt} is simply the mean of the diagonal elements of the inverse Hessian. The Hessian of the mean field energy function is given by $H(x) = \partial^2 E_I(x) - t$ and is independent of the cavity field $h$.  Taking the average of its inverse yields the self-consistent equation
\begin{equation}
    t = \int dx \, \frac{P(x)}{\partial^2 E_I(x) - t}.
    \label{eq:tselfcons}
\end{equation}
Together, \eqref{eq:E_MF}, \eqref{eq:Px_cavitySUSY}, \eqref{eq:qselfcons}, and \eqref{eq:tselfcons} constitute a theoretical prediction for the distribution of spins in a typical critical point (i.e. the special case of $\beta=\mu=0$ in \eqref{eq:defpx}).  Interestingly, the cavity result appealingly and intuitively replaces the problem of summing over critical points in a large $N$ dimensional system (i.e. \eqref{eq:Etot} and \eqref{eq:defpx} with $\mu=\beta=0$) with the problem of summing over critical points in a random ensemble of $1$ dimensional systems (i.e. \eqref{eq:E_MF} and \eqref{eq:Px_cavitySUSY}).  

\subsection{Equivalence of the naive cavity method with the supersymmetric replica method}

We next show that these cavity results are exactly equivalent to those of the replica method in the further special case where SUSY is preserved (i.e. $A=C=0$).  We can demonstrate the equivalence of the cavity result for $P(x)$ in \eqref{eq:Px_cavitySUSY} 
 with the replica result for $P(x)$ in \eqref{eq:Px-replica} with $\mu=A=C=0$ as follows.  First we can apply the Kac-Rice formula in \eqref{eq:kacrice} to \eqref{eq:Px_cavitySUSY}, and perform the resulting integral over $x_*$ which simply fixes it to $x$, yielding
 \begin{equation}
 P(x) \propto \Braket{\delta(\partial E_I(x) - tx - h) |\partial^2 E_I(x) - t|}_{h}.
 \end{equation}
Then performing the integral over $h$ fixes it to be $h(x)=\partial E_I(x) - tx$, and recalling that $\Braket{\cdot}_h$ denotes an average w.r.t. the Gaussian distribution $\mathcal{N}(0,q)$, we obtain 
\begin{equation}
P(x)\propto |\partial^2 E_I(x)-t| \exp\left(-\frac{h(x)^2}{2q}\right), 
\label{eq:Pxcav_rep_susy}
\end{equation}
where $h(x) = \partial E_I(x) - tx$ is the external field $h$ required to make $x$ a critical point of the mean field energy function in \eqref{eq:E_MF}. Thus the distribution of spins $P(x)$ in \eqref{eq:Pxcav_rep_susy}, and therefore in \eqref{eq:Px_cavitySUSY}, is entirely equivalent to the replica expression for $P(x)$ when $\mu=A=C=0$.  

Moreover, given this equivalence of $P(x)$, the cavity-derived self-consistent equations for the order parameters $q$ in \eqref{eq:qselfcons} and $t$ in \eqref{eq:tselfcons} are entirely equivalent to the first two self-consistent equations derived via the replica method in \eqref{eq:self-consistent-replica}.  Thus overall, the naive cavity method recovers the results of the {\it supersymmetric} solution, but cannot account for supersymmetry breaking. 

\subsection{Beyond the naive cavity method: accounting for supersymmetry breaking}

Why does the naive cavity method only recover the replica results in the case of preserved SUSY -- i.e. \eqref{eq:Px-replica} when $A=C=0$ and the first two equations in \eqref{eq:self-consistent-replica}? Here we resolve this issue as well as generalize to nonzero $\mu$. The key idea is that the naive cavity method makes an implicit assumption about the nature of the perturbative reaction of the cavity system to the addition of a single new spin in \eqref{eq:cavityresponse}.  In particular, this account of the reaction assumes that the only effect of adding a new spin $x_0$ is to move every critical point of the cavity system a small amount to generate a critical point of the full system.  Thus it is assumed that critical points of the cavity system and the full system are in one-to-one correspondence with each other. 

This assumption is likely to be valid if the Hessian matrix $H(\mathbf{x})$ has an eigenvalue distribution $\rho_H(\lambda)$ which vanishes in a finite region about $\lambda=0$.  Because the susceptibility matrix $\chi$ is the inverse Hessian, such a gap in the Hessian spectrum yields a non-degenerate, structurally stable critical point that is unlikely to undergo a bifurcation or change its index upon the addition of a new spin. However, if the Hessian spectral density $\rho_H(\lambda)$ extends continuously to $\lambda=0$, such a critical point is degenerate with extremely soft modes, and the addition of a single spin could cause it to either disappear or bifurcate to create additional critical points. If the landscape has exponentially many critical points whose typical Hessian eigenspectra are gapless, then the addition of a single new spin $x_0$ could lead to exponentially more or fewer critical points of any given index, depending on the realization of the couplings $J_{i0}$ and the value of $x_0$ at its own equilibrium.  This extreme reactivity of the landscape to the addition of a single spin, marked by an exponential change in the number of critical points, is a fundamental possibility that is not accounted for by the naive cavity method. 

We provide a generalized cavity method that can account for this extreme reactivity. We provide a detailed derivation in  Section~S-IV. Here we simply outline the key ideas and intermediate results. Our generalized cavity method starts from the expression for the grand potential  
\begin{equation}
    - \beta \Omega = \frac{1}{N} \operatorname{ln} \Braket{Z}_J,
    \label{eq:grandpotgencav}
\end{equation}
where $Z$ is the partition given in \eqref{eq:partitionfunckac}.  Thus we start from an annealed approximation.  Next, because of the critical importance of the presence of soft modes in the energy landscape in the vicinity of critical points, corresponding to eigenvectors of the Hessian with small eigenvalues, we soften the $\delta$ functions of the gradient in \eqref{eq:partitionfunckac} and replace them with Gaussians via  
\begin{equation}
    \delta(\partial_iE(\mathbf{x})) \rightarrow 
    \sqrt{\frac{\gamma}{\pi}} e^{-\gamma (\partial_i E(\mathbf{x}))^2}.
    \label{eq:softgauss}
\end{equation}
We work at finite $\gamma$ throughout the calculation, taking $\gamma \rightarrow \infty$ at the end.  A finite $\gamma$ crucially allows the partition function $Z$ in \eqref{eq:partitionfunckac} to receive contributions not only from critical points, but also from the geometry of the landscape in the vicinity of critical points, including the nature of the nonzero gradient in the neighborhood of each critical point. 

Next we split the degrees of freedom $\mathbf{x}$ into that of a cavity system $\mathbf{x}^{/0}$ with components $x^{/0}_i$ for $i=1,\dots,N-1$ and a single spin $x_0$. Mirroring this split, we would like to express the grand potential of the full system in \eqref{eq:grandpotgencav} in terms of the grand potential of the cavity system $\mathbf{x}^{/0}$ (taking into account the effect of the new spin on it) and an effective mean field grand potential of the new spin $x_0$ (taking into account the effect of the cavity on it in terms of certain cavity fields and order parameters). Achieving this decomposition {\it prima facie} poses several challenges because $x_0$ and $\mathbf{x}^{/0}$ appear intricately coupled in the expressions for the Hessian determinant $|\det H(\mathbf{x})|$
and the Hessian index $\mathcal{I}(\mathbf{x})$ in $Z$ in \eqref{eq:partitionfunckac}. 
Despite this seemingly intricate coupling, we can show that, upon averaging over the random choice of coupling $J_{i0}$ between the cavity $\mathbf{x}^{/0}$ and the new spin $x_0$, 
the interaction between them depends on the cavity system $\mathbf{x}^{/0}$ {\it only} 
through the mean cavity susceptibility order parameter $t$, defined in \eqref{eq:orderparamt}. 

In particular, for the Hessian determinant, we show in Section~S-IV.B.1 that after averaging over $J_{i0}$,
\begin{equation}
    |\det H(\mathbf{x})| = |\partial^2 E_I(x_0) - t| |\det H(\mathbf{x}^{/0})|.
    \label{eq:decomphess}
\end{equation}
The first term is nothing other than the absolute value of the Hessian of the mean field energy function $E_{MF}$ in \eqref{eq:E_MF} evaluated at $x=x_0$, while the second term is the same Hessian determinant for the cavity system. 

Similarly, for the index of the Hessian, we show in Section~S-IV.B.2 that after averaging over $J_{i0}$, 
\begin{equation}
    \mathcal{I}(\mathbf{x}) = \overline{\mathcal{I}}(x_0) + \mathcal{I}(\mathbf{x}^{/0}).
    \label{eq:decompind}
\end{equation}
Here $\overline{\mathcal{I}}(x_0)$ is defined in \eqref{eq:mean-field-index} and can be interpreted simply as the index of the mean-field energy function $E_{MF}$ in \eqref{eq:E_MF} evaluated at $x=x_0$. Thus remarkably, the index of the full system is simply the sum of the index of the mean field system and the cavity system, on average.  

Now assuming formulas \eqref{eq:decomphess} and \eqref{eq:decompind} are self-averaging (i.e. they also hold to high accuracy for typical random choices of $J_{i0}$), we can substitute these formulas into \eqref{eq:partitionfunckac}, thereby achieving a partial decomposition of the full partition function $Z$ into that of a cavity system of size $N-1$ and a mean field system of size $1$, coupled so far only through the cavity susceptibility order parameter $t$ in \eqref{eq:orderparamt}. 

However, to fully complete this decomposition, we must also account for interactions between the cavity system $\mathbf{x}^{/0}$ and the single spin $x_0$ through the Gaussian softening in \eqref{eq:softgauss} of the $\delta$ functions in \eqref{eq:partitionfunckac}.  We show in Section~S-IV.B.3 and S-IV.B.4 that these interactions are mediated precisely by two fields:
\begin{eqnarray}
    \bar h &\equiv& -\mathbf{J}_{0} \cdot \mathbf{x}^{/0} \nonumber \\
    \bar z &\equiv& -\gamma \mathbf{J}_{0} \cdot \nabla E(\mathbf{x}^{/0}).
    \label{eq:barhz}
\end{eqnarray} 
Here $\mathbf{J}_{0}$ is the $N-1$ dimensional coupling vector between $x_0$ and the cavity system $\mathbf{x}^{/0}$. Note that $\bar h$ and $\bar z$ are jointly Gaussian distributed with a $2$ by $2$ covariance matrix that depends on cavity order parameters specified by inner products of $\mathbf{x}^{/0}$ and $\nabla E(\mathbf{x}^{/0})$. In particular, $\bar h$ is simply the Gaussian cavity field that already appears in the naive cavity method in \eqref{eq:cavityfield} with variance $q$ in \eqref{eq:orderparamq}.  

But most importantly, $\bar z$ is a new cavity field that only appears in our generalized cavity method, and plays a fundamental role in accounting for the extreme sensitivity of the landscape to the addition of a new spin $x_0$.  In particular, as detailed in Section~S-IV.B.4, the field $\bar z$ couples the new spin $x_0$ to the cavity system through an exponential modification of the partition function $Z$ in \eqref{eq:partitionfunckac} via a multiplicative factor $\exp(x_0 \bar z)$.  Given the form of $\bar z$ in \eqref{eq:barhz}, this means that if the coupling vector $\mathbf{J}_{0}$ were aligned to the cavity gradient $\nabla E(\mathbf{x}^{/0})$ in the vicinity of a typical critical point, so that $\bar z$ is negative, then the partition function $Z$ would be exponentially enhanced (diminished) if $x_0$ were to assume larger negative (positive) values. Conversely, if $\mathbf{J}_{0}$ were anti-aligned to the cavity gradient $\nabla E(\mathbf{x}^{/0})$ so that $\bar z$ were positive, then $Z$ would be exponentially enhanced (diminished) if $x_0$ were to assume larger positive (negative) values. The end result of the field $\bar z$ is then to exponentially reweight the distribution of spins in the mean-field theory of a single spin $x_0$ according to the exponential weight that different values of $x_0$ exert on the cavity partition function, and therefore on the grand potential and the complexity. Thus while the usual cavity field $\bar h$ exerts a force on the new spin $x_0$ through an energy term $-\bar h x_0$ in the mean-field energy function $E_{MF}$ in \eqref{eq:E_MF}, we will see that the new field $\bar z$ yields an entropic force on the new spin $x_0$ through the proliferation or destruction of exponentially many critical points in the cavity system for different values of $x_0$.  

Now in order to take the $\gamma \rightarrow \infty$ limit, it is useful not to work directly with the fields $\bar h$ and $\bar z$, but to perform a change of variables (detailed in Section~S-IV.B.4) to $h$ and $z$ which remain jointly Gaussian distributed with density $P(h,z)$ given by 
\begin{equation}
    P(h,z) \sim 
    \mathcal N \left(
    \left[\begin{matrix}
        0 \\
        0 
    \end{matrix}\right],
    \left[\begin{matrix}
        q & A \\
        A & C
    \end{matrix}\right]
    \right).
    \label{eq:phz}
\end{equation}
Here the covariance parameters at finite $\gamma$ are given by 
\begin{eqnarray}
    &&q = \frac{1}{N-1}|\mathbf{x}^{/0}|^2\nonumber\\
    &&A = \frac{2}{N-1}\gamma \nabla E(\mathbf{x}^{/0})\cdot \mathbf{x}^{/0} - t\nonumber\\
    &&C = \frac{4}{N-1}\gamma^2 |\nabla E(\mathbf{x}^{/0})|^2 - 2\gamma,
    \label{eq:orderparamqAC}
\end{eqnarray}
and correspond to cavity order parameters involving inner products of $\mathbf{x}^{/0}$ and $\nabla E(\mathbf{x}^{/0})$ in the vicinity of critical points.  

Now with the definition of the cavity order parameters  $q$, $A$, $C$ in \eqref{eq:orderparamqAC} and $t$ in \eqref{eq:orderparamt}, as well as the Gaussian fields $h$ and $z$ with distribution $P(h,z)$ in \eqref{eq:phz}, we can achieve a decomposition of the partition function $Z$ in \eqref{eq:partitionfunckac}, and therefore of the grand potential $\Omega$ in \eqref{eq:grandpotgencav}, into a cavity system $\mathbf{x}^{/0}$ and a single spin $x_0$. However, there is one remaining issue: the resultant grand potential of the cavity system has a mismatched variance; the size of the cavity system is $N-1$ while the variance of its connectivity in \eqref{eq:Jdistrib} for $g=1$ is $\frac{1}{N}$.  Given the potentially extreme reactivity of the energy landscape, we cannot ignore this mismatch. Indeed, to obtain self-consistent equations for the order parameters $q,A,C$ and $t$ of the full system, and we must analyze the susceptibility of the grand potential in response to small changes in the variance of its connectivity. We perform this analysis in Section~S-IV.B.6, obtaining a simple formula for this susceptibility in terms of the cavity order parameters:
\begin{eqnarray}
    \left.\frac{d\Omega(g)}{dg}\right|_{g=1} &=& \frac{1}{2}(qC + A^2) +A t.
\label{eq:cavitysuscvar}
\end{eqnarray}
Here, $\Omega(g)$ denotes the grand potential of the full system with a general variance parameter $g$ in \eqref{eq:Jdistrib}.  

Finally, putting everything together and taking the $\gamma \rightarrow \infty$ limit, we find (see Section~S-IV.B.6 for details) that the grand potential in \eqref{eq:grandpotgencav}, or equivalently the annealed connectivity average of the grand potential in \eqref{eq:grandpotdef} (in the special case of $\beta=0$ relevant to typical critical points and minima) is given by 
\begin{equation}
    -\beta \Omega(0, \mu) = 
     \underset{(q,A,C,t)}{\mathrm{ext}} 
     \left\{
     - \, \left.\frac{d\Omega(g)}{dg}\right|_{g=1}
     - \, \Omega_{\operatorname{MF}}
     \right\}.
\label{eq:grandpotgencavity}
\end{equation}
Here the first term is a simple function of the order parameters given in \eqref{eq:cavitysuscvar} while the second term is the mean-field grand potential $\Omega_{\operatorname{MF}}$ of a single spin given by    
\begin{equation}
     - \Omega_{\operatorname{MF}} =  
      \log \Braket{Z_{\operatorname{MF}}[h,z]}_{h,z},
\label{eq:grandpotmf}
\end{equation}
where $\Braket{\cdot}_{h,z}$ denotes an average over the Gaussian distribution $P(h,z)$ of cavity fields in \eqref{eq:phz}, and $Z_{\operatorname{MF}}[h,z]$ denotes the mean field partition function of a single spin in the presence of cavity fields $h$ and $z$, given by 
\begin{equation}
     Z_{\operatorname{MF}}[h,z] =  
     \sum_{x\in\mathrm{Crt}(E_{\operatorname{MF}}[h])}e^{xz+\mu \bar{\mathcal{I}}(x)}.
\label{eq:partfuncmf}
\end{equation}
Here, as above, $\mathrm{Crt}(E_{\operatorname{MF}}[h])$ denotes the set of critical points of the mean-field energy function $E_{\operatorname{MF}}[h](x)$ in \eqref{eq:E_MF}, and the mean-field index function  $\bar{\mathcal{I}}(x)$ defined in \ref{eq:mean-field-index} is simply the index of $E_{\operatorname{MF}}$ evaluated at $x$. Finally, our generalized cavity method computes the distribution of spins in a typical critical point, defined in \eqref{eq:defpx},  to be (see Section~S-IV.C for details) 
\begin{equation}
    P(x) \propto \Braket{\sum_{x_*\in\mathrm{Crt}(E_{\operatorname{MF}}[h])}e^{xz+\mu\bar{\mathcal{I}}(x)}\delta(x-x_*)}_{h,z}.
    \label{eq:Px_cavity}
\end{equation}

Appealingly, \eqref{eq:grandpotmf}, \eqref{eq:partfuncmf} and \eqref{eq:Px_cavity} all correspond to the problem of counting critical points in a random ensemble of $1$ dimensional systems with mean-field energy functions  $E_{\mathrm{MF}}[h]$ subject to a random external field cavity field $h$, in addition to a random exponential factor $e^{xz}$ that reweights both the partition function $Z_{\operatorname{MF}}[h,z]$ in \eqref{eq:partfuncmf} and the spin distribution $P(x)$ in \eqref{eq:Px_cavity}.  Notably, when the variance $C$ of  $z$, and therefore the covariance $A$ between $h$ and $z$ is $0$, the reweighting factor $e^{xz}$ plays no role, and $P(x)$ in \eqref{eq:Px_cavity} reduces to the prediction of the naive replica method in \eqref{eq:Px_cavitySUSY} (when $\mu=0$).  The origin of this reweighting factor for nonzero $A$ and $C$, as summarized above and described in detail in Section~S-IV.B, arises from entropic effects in the cavity system due to exponential changes in the number of critical points, depending on the value $x$ of an added spin and the random alignment $z$ of its coupling vector to the cavity system gradient near critical points. This entropic effect of the cavity on the new spin is encapsulated in the mean-field theory of the new spin simply through the reweighting factor $e^{xz}$.  

Finally, we can {\it directly} obtain self-consistent equations for the order parameters $q,A,C$ and $t$ through our generalized cavity method, without resorting to replicas (see Section~S-IV.C for details).  The self-consistent equations for $q$ and $t$ are identical in form to those obtained in the naive cavity method in \eqref{eq:qselfcons} and \eqref{eq:tselfcons} respectively, with the sole difference being that the distribution of spins $P(x)$ obtained in the naive cavity method in \eqref{eq:Px_cavitySUSY} is replaced with the reweighted distribution of spins $P(x)$ obtained in the generalized cavity method in \eqref{eq:Px_cavity}.  The generalized cavity method also enables us to find self-consistent equations for the two new order parameters $A$ and $C$ (see Section~S-IV.C for details):
\begin{eqnarray}
    A &=& e^{\Omega_{\operatorname{MF}}} 
    \Braket{\frac{\partial}{\partial h}
    \left[\sum_{x\in \operatorname{Crt}(E_{\operatorname{MF}}[h])}e^{xz+\mu\bar{\mathcal{I}}(x)} x \right]
    }_{h,z} - t, \nonumber \\
    C &=& e^{\Omega_{\operatorname{MF}}} 
        \Braket{\frac{\partial^2 Z_{\operatorname{MF}} }{\partial h^2}}_{h,z}.
    \label{eq:ACselfcons}
\end{eqnarray}

Furthermore, we show in Section~S-IV.C that the cavity derived self-consistent equations for the order parameters $q$ in \eqref{eq:qselfcons}, $t$ in \eqref{eq:tselfcons} and $A$ and $C$ in \eqref{eq:ACselfcons} are collectively equivalent to the $4$ equations obtained from extremizing the grand potential $\Omega$ in \eqref{eq:grandpotgencavity} with respect to $q,t,A$ and $C$. This extremization yields the highly compact self-consistent equations
\begin{eqnarray}
\left[\begin{matrix}
        q & A \\
        A & C
    \end{matrix}\right] &=& 
    -\left[\begin{matrix}
        2 \partial_C \Omega_{\operatorname{MF}}   
        & 
        \partial_t \Omega_{\operatorname{MF}} \\
        \partial_t \Omega_{\operatorname{MF}}
        & 
        2 \partial_q \Omega_{\operatorname{MF}}
    \end{matrix}\right], \nonumber \\
t &=& \partial_A \Omega_{\operatorname{MF}}
                 - A. 
\label{eq:selfconscompact}
\end{eqnarray}

\subsection{Equivalence between the generalized cavity method and the supersymmetry broken replica method}

We now show the equivalence between the generalized cavity method and supersymmetry breaking in the replica method, identifying the cavity order parameters $A$ and $C$ in the generalized cavity method in \eqref{eq:orderparamqAC} with the SUSY breaking order parameters $A$ and $C$ in the replica method in \eqref{eq:complexity-replica}, \eqref{eq:Px-replica}, and \eqref{eq:self-consistent-replica}.  

First we note that the distribution of spins $P(x)$ in the generalized cavity method in \eqref{eq:Px_cavity} is entirely equivalent to the distribution of spins derived via the SUSY broken replica method in \eqref{eq:Px-replica} for any values of the order parameters $A$ and $C$ (as well as $q$ and $t$).  This can be seen by applying the Kac-Rice formula to \eqref{eq:Px_cavity}, and directly performing the integrals over $h$ and then $z$ (see Section~S-IV.F for details). 

Second, we note that the formula for the grand potential $\Omega$ derived by the generalized cavity method in \eqref{eq:grandpotgencavity} is entirely equivalent to that obtained by the replica method in \eqref{eq:complexity-replica}.  
However, the generalized cavity method now provides a simple interpretation of each of the terms in \eqref{eq:complexity-replica}.  
In particular, first part $-\frac{1}{2}(Cq + A^2) - A t$ of \eqref{eq:complexity-replica} is equivalent to the first term in \eqref{eq:grandpotgencavity}, and is simply the susceptibility of the grand potential to a change in variance, derived in \eqref{eq:cavitysuscvar}. Its origin lies in the mismatch between the size of the cavity system ($N-1$) and its connectivity variance ($\frac{1}{N}$). The remaining term in \eqref{eq:complexity-replica} is equivalent to the remaining term $-\Omega_{\operatorname{MF}}$ in \eqref{eq:grandpotgencavity}, and is simply the grand potential of an ensemble of {\it single} spins defined in \eqref{eq:grandpotmf} and \eqref{eq:partfuncmf}.  
This equivalence can be seen by applying the Kac-Rice formula to the mean field partition function in \eqref{eq:partfuncmf}, and performing the integrals over $h$ and $z$ in \eqref{eq:grandpotmf}. This calculation yields the final term in \eqref{eq:complexity-replica} (see Section~S-IV.F for details).  

Thus we conclude that the expressions for $\Omega$ in the generalized cavity method in \eqref{eq:grandpotgencavity} and the replica method in \eqref{eq:complexity-replica} are equivalent.  However, the generalized cavity method provides the important intuition, embodied in \eqref{eq:grandpotgencavity}, that the grand potential density $\Omega$ of the full $N$ dimensional system is simply the mean field grand potential $\Omega_{\operatorname{MF}}$ of an ensemble of random $1$ dimensional systems in \eqref{eq:grandpotmf}, plus a correction given in \eqref{eq:cavitysuscvar} due to the extreme reactivity of the landscape to changes in connectivity variance. 

Finally, given the equivalence of the grand potentials derived via the generalized cavity and replica methods, as well as the demonstration in the previous subsection that the self-consistent equations for the order parameters in the generalized cavity method can be obtained by extremizing the grand potential, we can conclude that these self-consistent equations are the same in both methods. To further corroborate this conclusion, we provide a direct proof in Section~S-IV.F that the self-consistent equations derived via the generalized cavity method for $A$ and $C$ in \eqref{eq:ACselfcons} are equivalent to those derived via the replica method in \eqref{eq:self-consistent-replica}.  Moreover the equations for $q$ and $t$ in \eqref{eq:qselfcons} and \eqref{eq:tselfcons} respectively, are manifestly equivalent to those derived via the replica method in \eqref{eq:self-consistent-replica} given the equivalence of the distributions $P(x)$ in \eqref{eq:Px_cavity} and \eqref{eq:Px-replica}.  

\subsection{Supersymmetry breaking order parameters in terms of landscape susceptibility}

The equivalence of the generalized cavity method and the replica method at the level of the grand potential $\Omega$, self-consistent equations for order parameters, and distribution of spins $P(x)$, thus identifies the parameters $A$ and $C$ in the generalized cavity method with the supersymmetry breaking order parameters of the replica method. We now provide a further geometric interpretation of these order parameters in terms of the susceptibility of the grand potential $\Omega$ to changes in the energy landscape.  In particular, consider adding two extra perturbative terms to the original energy landscape $E_{\operatorname{tot}}(\mathbf x)$ in \eqref{eq:Etot} to obtain
\begin{equation}
    E_{\operatorname{tot}}'(\mathbf x) = E_{\operatorname{tot}}(\mathbf x) - \frac{1}{2} a_0|\mathbf{x}|^2 + \sqrt{2s_0}\mathbf{g}\cdot \mathbf{x}.
    \label{eq:Epert}
\end{equation}
Here $\mathbf{g}$ is a zero mean random Gaussian vector with identity covariance.  Given the structure of the single site energy function $E_I(x)$ in \eqref{eq:EsingleOPO}, the first perturbation in \eqref{eq:Epert} corresponds in the CIM to changing the laser gain from $a$ to $a+a_0$.  The second perturbation corresponds to applying a random Gaussian field on the landscape with a variance on each component of $2s_0$. We can then consider computing the grand potential density $\Omega(a_0, s_0)$ (we suppress the dependence on $\beta=0$ and $\mu$ here) in \eqref{eq:grandpotdef} by replacing $E_{\operatorname{tot}}$ in \eqref{eq:Etot} with $E_{\operatorname{tot}}'$ in \eqref{eq:Epert} and further averaging over the random field $\mathbf{g}$.  We show in Section~S-IV.D that the supersymmetry breaking order parameters $A$ and $C$ are very simply related to the (connectivity and field averaged) susceptibility of the grand potential $\Omega$ with respect to the perturbation strengths $a_0$ and $s_0$ respectively:
\begin{equation}
A  = \left.\frac{\partial \Omega}{\partial a_0}\right|_{a_0=0,s_0=0} \qquad 
C  = \left.\frac{\partial \Omega}{\partial s_0}\right|_{a_0=0,s_0=0}.
\label{eq:ACsusc}
\end{equation}
This result {\it directly} connects the supersymmetry breaking order parameters to the extreme reactivity of the landscape to two specific small perturbations of the energy function.  In particular, $A$ and $C$ are nonzero if and only if the potential $\Omega$, and therefore the landscape complexity $\Sigma$, are sensitive to these perturbations.  

The expressions for $A$ and $C$ in \eqref{eq:ACsusc} also have a counterpart in the mean-field theory of a single spin. 
Consider adding the same two perturbations to the mean-field energy function $E_{\operatorname{MF}}[h]$ in \eqref{eq:E_MF}, obtaining the perturbed energy function  
\begin{equation}
E'_{\mathrm{MF}}[h,a_0,s_0](x) = E_{\mathrm{MF}}[h](x) - \frac{1}{2}{a_0}x^2 + \sqrt{2s_0}gx,
\end{equation}
where $g$ is now a zero mean unit variance random Gaussian scalar field. We can then consider computing the mean field grand potential $\Omega_{\operatorname{MF}}(a_0,s_0)$ obtained by replacing $E_{\mathrm{MF}}[h](x)$ with $E'_{\mathrm{MF}}[h,a_0,s_0](x)$ in the defining formula for $\Omega_{\operatorname{MF}}$ in \eqref{eq:grandpotmf}, and also further averaging over $g$.  Then we show in Section~S-IV.D that the supersymmetry breaking order parameters $A$ and $C$ are also very simply related to the susceptibility of the mean-field grand potential $\Omega_{\operatorname{MF}}$ with respect to the perturbation strengths $a_0$ and $s_0$ respectively:
\begin{equation}
A  = \left.\frac{\partial \Omega_{\operatorname{MF}}}{\partial a_0}\right|_{a_0=0,s_0=0} \qquad 
C  = \left.\frac{\partial \Omega_{\operatorname{MF}}}{\partial s_0}\right|_{a_0=0,s_0=0}.
\label{eq:ACsuscMF}
\end{equation} 
This result provides an additional way to interpret the SUSY breaking order parameters $A$ and $C$ within the mean-field theory, and exhibits an appealing correspondence to \eqref{eq:ACsusc}. 

\subsection{Structural stability of critical points implies preserved supersymmetry}

We further connect the Hessian eigenspectrum to SUSY breaking by showing in Section~S-IV.E that if the typical Hessian eigenspectrum of critical points has a gap away from $0$, then the SUSY is preserved, and $A=C=0$.  We do this by working at large but finite $\gamma$ and directly calculating $A$ and $C$ through \eqref{eq:orderparamqAC}, and averaging over $\mathbf{x}^{/0}$ (or equivalently $\mathbf{x}$) with respect to a distribution with partition function given by $Z$ in \eqref{eq:partitionfunc}, with $\delta$-functions softened to Gaussians via \eqref{eq:softgauss}.  The key idea is that this distribution concentrates in the vicinity of critical points, and if the Hessian has a gap at typical critical points, one can perform a change of variables from $\mathbf{x}$ to $\nabla E(\mathbf x)$ since there is a one-to-one map between these quantities in the neighborhood of any critical point with a gapped Hessian eigenspectrum. Direct calculation of the integral over gradients in the vicinity of a critical point then reveals that $A=C=0$.  

Thus if typical critical points are structurally stable (i.e. with gapped Hessian eigenspectra) SUSY is preserved.  The contrapositive of this statement then tells us that SUSY breaking implies a vanishing gap in the Hessian eigenspectrum, and therefore structural instability in typical critical points. If exponentially many critical points have such structural instability, then SUSY will be broken. This analytic calculation provides further justification for why the generalized cavity method (and not just the naive cavity method) is necessary when exponentially many critical points are structurally unstable.

\subsection{Convexity of the mean-field energy landscape implies preserved supersymmetry} 

 Suppose the mean-field energy function $E_{\operatorname{MF}}[h]$ in \eqref{eq:E_MF} is strictly convex. This happens only if $a+t < 0$. We prove in Section~S-IV.E that under this assumption of convexity, the self-consistent equations for the order parameters in \eqref{eq:ACselfcons} admit SUSY preserving solutions with $A=C=0$. We will find below in Section~\ref{sec:numericaltests} that these SUSY preserving solutions correctly predict the distribution of spins and Hessian eigenspectra when $a+t < 0$.  In contrast, we will also see that when $a+t>0$, and therefore the mean-field energy function $E_{\operatorname{MF}}[h]$ is nonconvex, we must use SUSY breaking solutions to correctly predict the distribution of spins and Hessian eigenspectra.  Thus just as SUSY breaking implies a highly reactive landscape in the full $N$ dimensional system, as evidenced by the susceptibility formulas in \eqref{eq:ACsusc}, at the level of the mean field $1$ dimensional system, SUSY breaking is closely related to nonconvexity in the mean field energy landscape. 

We further show in Section~S-IV.E that when the mean field energy function is convex, SUSY preserving solutions also exhibit vanishing complexity $\Sigma$.  Thus convexity of the mean field landscape implies simplicity of the full landscape under SUSY.  Thus in summary, the predictions our replica and generalized cavity theories are that for $a+t<0$, SUSY is preserved and landscape complexity is $0$, and when $a+t>0$, SUSY is broken.

\section{Numerical tests of supersymmetry breaking for typical critical points and minima}

\label{sec:numericaltests}

\subsection{A supersymmetry breaking phase transition in the properties of typical critical points}
\label{sec:typical-crt}

\begin{figure*}[!htbp]
    \includegraphics[width=7in]{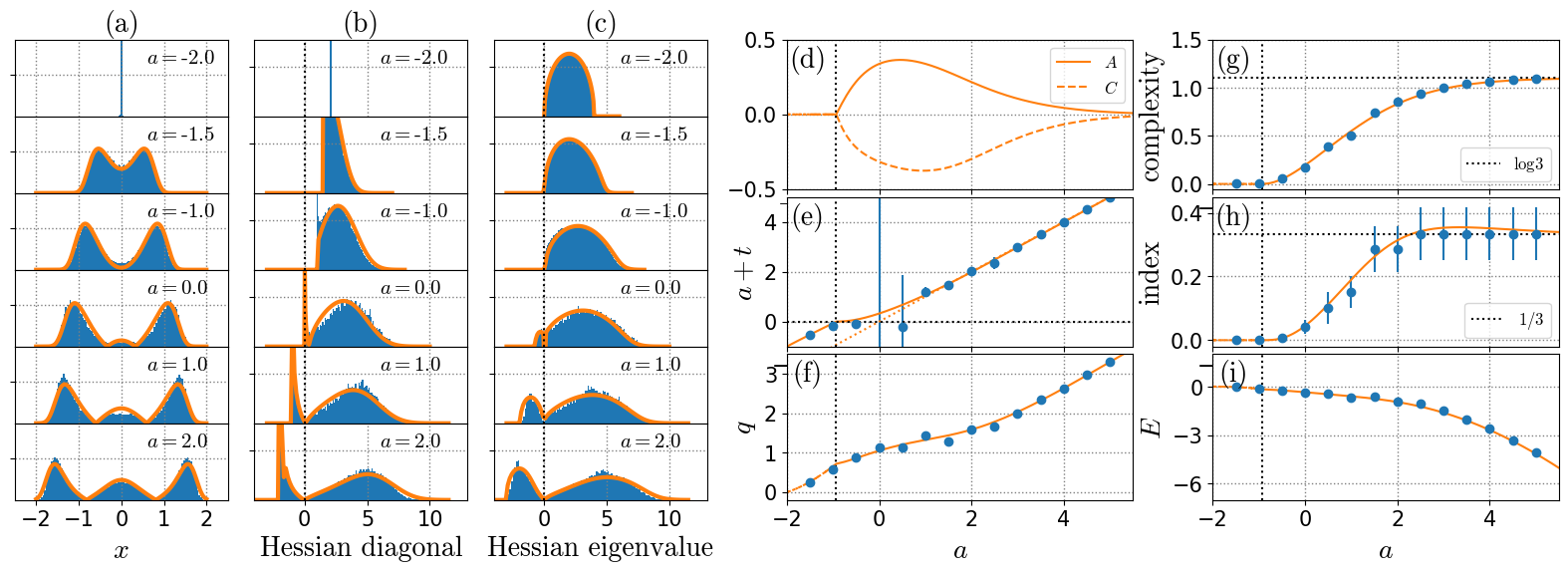} 
    \caption{{\bf A supersymmetry breaking phase transition in the properties of typical critical points.} All panels indicate theoretical results (orange curves) and experimental results (blue histograms and markers). Details of experimental results are in \ref{sec:num-process} and details of theoretical solutions are in \ref{sec:selfconssol}. (a) The distribution of spins or OPO amplitudes as laser gain $a$ increases. (b) The distribution of Hessian diagonal elements $H^I(\mathbf x)_{ii}$. (c) The Hessian eigenspectrum $\rho_H(\lambda)$. The Hessian eigenspectra in (c) can be understood intuitively as the outcome of Dyson's Brownian motion starting from the initial condition of diagonal elements in (b).  (d,e,f) the order parameters $A$ and $C$, $a+t$, and $q$ respectively.  A supersymmetry breaking phase transition is observed at the theoretically predicted point $a=a_t\sim-0.93$ (vertical dotted line). (g,h,i). The complexity $\Sigma$, the intensive index $r$, and the intensive energy $E$, respectively, of typical critical points.} 
    \label{fig:typical-crt}
\end{figure*} 

Here we test our theoretical predictions for the structure of typical critical points derived from the grand potential in \eqref{eq:complexity-replica}, or equivalently \eqref{eq:grandpotgencavity} with $\mu=0$.  
We directly sample critical points of all indices in many finite size SK models. Across this ensemble of critical points, the distribution of intensive index $r$ and energy $E$ peak sharply at their respective most likely values.  
Focusing on these typical critical points (see \ref{sec:num-process} for details of the numerical sampling of critical points), we can measure the distribution of spins $P(x)$, the distribution diagonal elements
$H^I(\mathbf x)_{ii}$ in \eqref{eq:diagonalHessian}, 
and the distribution of Hessian eigenvalues $\rho_H(\lambda)$, 
shown as blue histograms in Fig.~\ref{fig:typical-crt}-abc respectively.  
We can further compare these observables to the theoretical predictions for $P(x)$ in \eqref{eq:Px-replica} or \eqref{eq:Px_cavitySUSY}, the distribution of $H^I(\mathbf x)_{ii}$ derived from $P(x)$, and the Hessian eigenspectrum $\rho_H(\lambda)$ derived from \eqref{eq:inversionformula} and \eqref{eq:resolventeqn}.  
We obtain an excellent match between theory and experiment for a range of laser gain $a$ (compare orange curves and  blue histograms in Fig.~\ref{fig:typical-crt}-abc). 

A key feature of these results is that, as the laser gain $a$ is increased, the distribution of OPO amplitudes in Fig.~\ref{fig:typical-crt}A bifurcates into a bimodal then trimodal distribution with an increasing density of uncommitted spins with values near the origin. Correspondingly, the distribution of diagonal Hessian eigenvalues in Fig.~\ref{fig:typical-crt}-b exhibits an increasing density of negative values originating from these uncommitted spins, which then corresponds to an increasing density of negative Hessian eigenvalues in Fig.~\ref{fig:typical-crt}-c via Dyson's Brownian motion in \eqref{eq:dysonmotion}, starting from the initial condition in Fig.~\ref{fig:typical-crt}-b.   

We further compute the order parameters $q$, $t$, $A$ and $C$ arising from solutions of \eqref{eq:self-consistent-replica}, or equivalently \eqref{eq:ACselfcons} or \eqref{eq:selfconscompact} (see  \ref{sec:selfconssol} for numerical details of solving these self-consistent equations). Fig.~\ref{fig:typical-crt}-d shows the evolution of $A$ and $C$ with increasing laser gain $a$, indicating a supersymmetry breaking phase transition at $a=a_t\sim -0.93$, when $A$ and $C$ first acquire nonzero values. Fig.~\ref{fig:typical-crt}-e shows the evolution of $a+t$, which, both in theory and experiment,  transitions from negative to positive also at $a=a_t\sim -0.93$.  Recall that the mean field energy function $E_{\operatorname{MF}}[h](x)$ in \eqref{eq:E_MF} and \eqref{eq:EsingleOPO} is convex if and only if $a+t<0$.  Thus together Fig.~\ref{fig:typical-crt}-d,e confirm our theoretical prediction that SUSY is broken precisely when the mean-field energy function becomes non-convex. 

Finally, Fig.~\ref{fig:typical-crt}-f,g,h,i demonstrates an excellent match between theory (orange curves) and experiments (blue dots) for the order parameter $q$, the complexity $\Sigma$ (derived from \eqref{eq:legendre}), the intensive index $r$ (derived from \eqref{eq:rformula}), and the intensive energy $E$ (derived from \eqref{eq:eformula}), respectively.  In particular, the complexity in Fig.~\ref{fig:typical-crt}-g becomes nonzero at the same transition $a=a_t\sim -0.93$ when SUSY is broken and the mean field energy function becomes nonconvex. 

Quite remarkably, the intensive index $r$ and complexity $\Sigma$ is exactly zero for $a<a_t$ (Fig.~\ref{fig:typical-crt}-g,h). This means that most critical points have a vanishing intensive index, and the number of critical points is sub-exponential. Therefore we can expect that the energy landscape is relatively flat and not so rugged at such low laser gain $a$.  In contrast, at very large $a$, the complexity approaches $\log 3$ and the intensive index $r$ approaches $1/3$, as expected from the discussion in Sec \ref{subsubsection:largegain}, which suggests the existence of $3^N$ critical points at large $a$ located near the points $\{-\sqrt{a},0,\sqrt{a}\}^N$, in which a typical critical point has $1/3$ of its spins uncommitted near $0$, contributing to a typical index of $r=1/3$.  

In summary, our combined theory and experiment uncovers a phase transition between a supersymmetric phase (when $a<a_t\sim -0.93$), where the intensive index and complexity of typical critical points is $0$, the number of critical points is subexponential in $N$, and the mean-field energy function is convex, and a supersymmetry broken phase (when $a>a_t\sim -0.93$) where the intensive index and complexity of critical points is finite, there are exponentially many structurally unstable critical points, and the mean-field energy function is non-convex.

\subsection{A supersymmetry breaking phase transition in the properties of typical minima}
\label{sec:typical-min}

We next test our theoretical predictions for the structure of typical minima, derived from the grand potential in \eqref{eq:complexity-replica}, or equivalently \eqref{eq:grandpotgencavity} with $\mu\rightarrow -\infty$.  The theoretical calculations are entirely parallel to those of the previous subsection, with the sole replacement of $\mu=0$ with $\mu \rightarrow -\infty$, and the experimental results are also parallel with the sampling restricted to minima as opposed to saddle points of any index.  We further compute the binned energy of all minima found, and focus on the typical minima with the most likely binned value of energy (see \ref{sec:num-process} for details).  As we will see below, the intensive energy of typical minima can be strictly higher than the energy of the global minimum especially at large laser gain. 

\begin{figure*}[!htbp]
\includegraphics[width=7in]{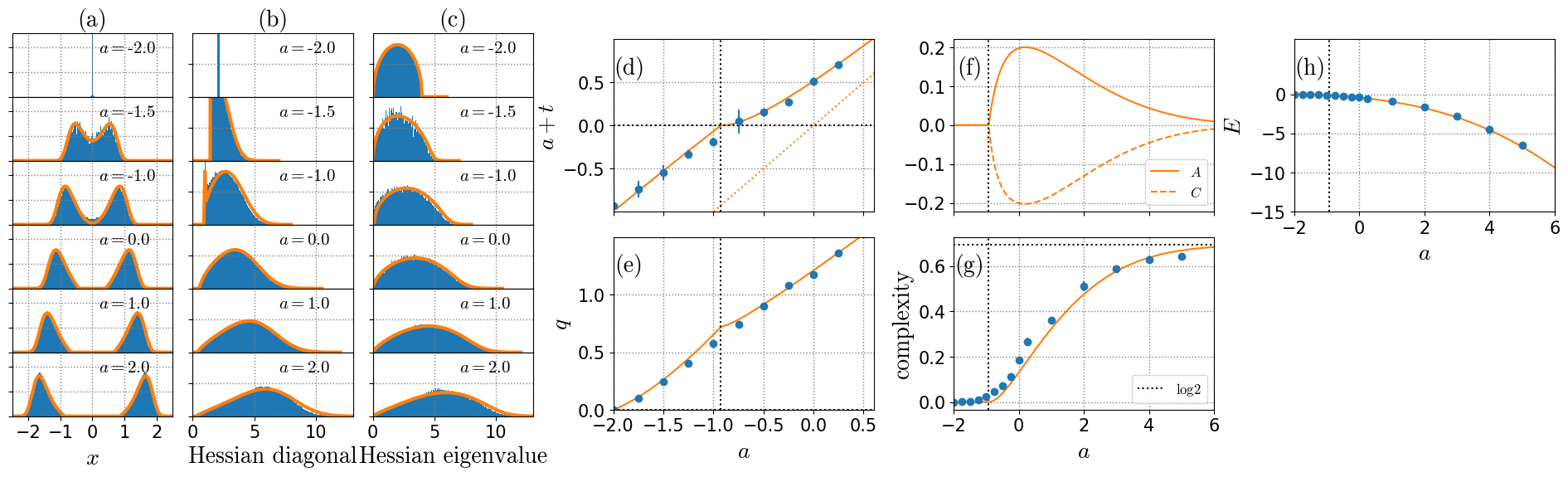} 
    \caption{{\bf A supersymmetry breaking phase transition in the properties of typical minima.} All panels indicate theoretical results (orange curves) and experimental results (blue histograms and markers). Details of experimental results are in \ref{sec:num-process} and details of theoretical solutions are in \ref{sec:selfconssol}. (a) The distribution of spins or OPO amplitudes as laser gain $a$ increases. (b) The distribution of Hessian diagonal elements $H^I(\mathbf x)_{ii}$. (c) The Hessian eigenspectrum $\rho_H(\lambda)$. The Hessian eigenspectra in (c) can be understood intuitively as the outcome of Dyson's Brownian motion starting from the initial condition of diagonal elements in (b).  (d,e,f) the order parameters $a+t$, $q$ and $A$ and $C$, respectively.  A supersymmetry breaking phase transition is observed at the theoretically predicted point $a=a_t\sim-0.93$ (vertical dotted line). (g,h). The complexity $\Sigma$ and the intensive energy $E$, respectively, of typical minima.} 
    \label{fig:typical-min}
\end{figure*}

Fig.\ref{fig:typical-min}-a,b,c demonstrates an excellent match between theory and experiment for the distribution of spins $P(x)$, 
the distribution diagonal elements
$H^I(\mathbf x)_{ii}$ in \eqref{eq:diagonalHessian}, 
and the distribution of Hessian eigenvalues $\rho_H(\lambda)$, respectively.  Interestingly, the distribution of spins $P(x)$ at large laser gain exhibits exactly $0$ density for a range of $x$ values around $x=0$ (see e.g. the cases of $a=0,1,2$ in Fig.~\ref{fig:typical-min}-a). This vanishing density can be understood through the cavity method as a simple consequence of the structure of the mean-field energy function $E_{\operatorname{MF}}[h](x)$ in \eqref{eq:E_MF} and \eqref{eq:EsingleOPO}, and its associated mean-field index function $\bar{\mathcal{I}}(x)$ defined in \eqref{eq:mean-field-index}, which is simply the index of $E_{\operatorname{MF}}$ evaluated at $x$. $\bar{\mathcal{I}}(x)$ plays a role in determining $P(x)$ through \eqref{eq:Px_cavity}, and when $\mu \rightarrow -\infty$, this equation indicates that $P(x)$ must vanish whenever $\bar{\mathcal{I}}(x)>0$, or equivalently, $P(x)$ must vanish over any range of $x$ where $E_{\operatorname{MF}}[h](x)$ has a negative Hessian. For the particular double-well form of $E_{\operatorname{MF}}[h](x)$ in \eqref{eq:E_MF} and \eqref{eq:EsingleOPO} with $a+t>0$, the theory implies $P(x)$ must vanish exactly when $|x| \leq \sqrt{(a+t)/3}$.  Remarkably, this striking prediction of vanishing density in $P(x)$ for typical minima at large $a$ when $E_{\operatorname{MF}}[h](x)$ is nonconvex, is verified in experiments (vanishing of blue histograms in Fig.\ref{fig:typical-min}-a).       

Furthermore, Fig.~\ref{fig:typical-min}-d and Fig.~\ref{fig:typical-min}-e show a match between theory and experiment for the order parameters $a+t$ and $q$ respectively, while Fig.~\ref{fig:typical-min}-f shows the evolution of $A$ and $C$ with $a$.  Fig.~\ref{fig:typical-min}-f indicates a supersymmetry breaking phase transition \cite{crisantiComplexityMeanfieldSpinglass2005} at $a=a_t\sim -0.93$, when $A$ and $C$ first become nonzero as laser gain $a$ increases.  This is exactly the same transition value at which supersymmetry breaking occurs for typical critical points. Indeed, this phase transition shares several similar properties with that of typical critical points. At this transition, $a+t$ first becomes positive as $a$ increases (Fig.~\ref{fig:typical-min}-d), which means the mean-field energy function $E_{\operatorname{MF}}[h](x)$ first becomes non-convex. This non-convexity of $E_{\operatorname{MF}}[h](x)$ then generates an increasingly large region of vanishing density in $P(x)$ for typical minima around $x=0$, as $a$ increases beyond $a_t$ (Fig.~\ref{fig:typical-min}-a), as discussed above. 

Finally, the complexity $\Sigma$ first becomes nonzero just above $a=a_t$ (Fig.\ref{fig:typical-min}-g). In the supersymmetric phase with $a\leq a_t$ and $A=C=0$, the complexity $\Sigma$ of typical minima is $0$, indicating the number of minima in the energy landscape is sub-exponential in $N$. On the other hand, there are exponentially many minima in the supersymmetry broken phase. As $a$ becomes large, we expect $2^N$ minima, and indeed the complexity converges to $\log 2$ in the large-$a$ limit. Finally, Fig.~\ref{fig:typical-min}-h depicts the evolution of the intensive energy of typical minima with increasing $a$, again indicating an excellent match between theory and experiment.

\begin{figure*}[!htbp]
    \includegraphics[width=7in]{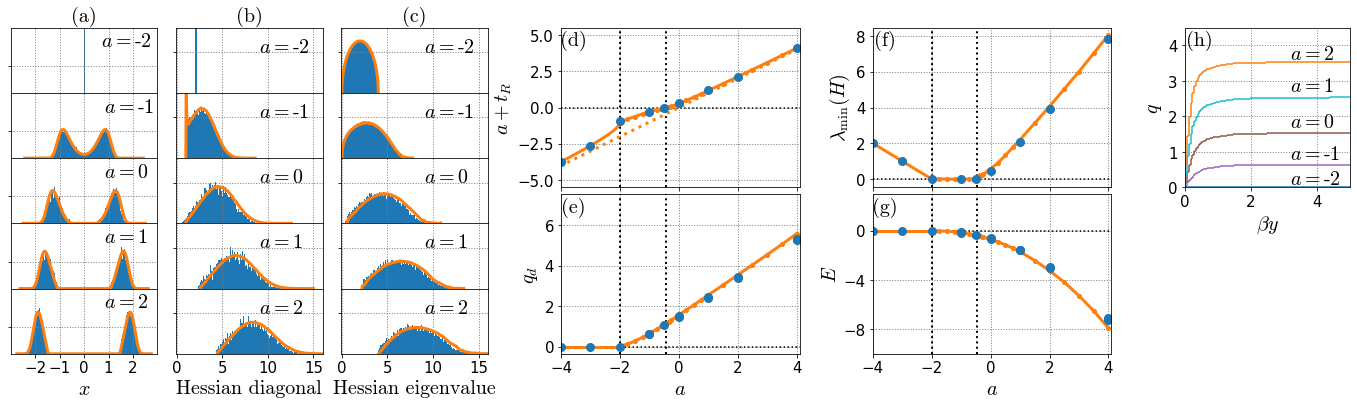}
    \caption{{\bf Replica symmetry breaking and rigidity phase transitions in the properties of global minima.} 
    All panels indicate theoretical results (orange curves) and experimental results (blue histograms and markers). Details of experimental results are in \ref{sec:num-process} and details of theoretical solutions are in \ref{sec:selfconssol}. (a) The distribution of spins or OPO amplitudes as laser gain $a$ increases. (b) The distribution of Hessian diagonal elements $H^I(\mathbf x)_{ii}$. (c) The Hessian eigenspectrum $\rho_H(\lambda)$. The Hessian eigenspectra in (c) can be understood intuitively as the outcome of Dyson's Brownian motion starting from the initial condition of diagonal elements in (b).  (d,e,f) the order parameters $a+t$, $q_d$ and $\lambda_\text{min}(H)$, respectively. At $a=a_r=-2$ there is a replica symmetry breaking phase transition where $q_d$ first acquires a nonzero value as $a$ increases (see panel e). At $a=a_g\sim-0.45$, there is a rigidity phase transition in the global minimum when $a+t$ first becomes positive (panel d) and the minimum Hessian eigenvalue transitions from $0$ to positive (panel f). (g) The intensive energy $E$ of global minima. (h) The overlap function $q(y)$ transitions from flat for $a<-2$ to continuously increasing for $a>-2$ indicating a replica symmetric to full replica symmetry breaking transition at $a=a_r=-2$. The two phase transitions in replica symmetry breaking at $a=a_r=-2$ and rigidity at $a=a_g\sim-0.45$ are shown as dotted vertical lines in panels d,e,f, and g.}
    \label{fig:global-min}
\end{figure*}

\section{Full replica symmetry breaking and rigidity phase transitions in global minima}
\label{sec:fullrsb}

We next move on from typical critical points and typical minima to the properties of global energy minima.  We define global minima as those with the lowest intensive energy. For a single sample, different global minima with the same intensive energy in the large $N$ limit could have different {\it extensive} energies with subleading $o(N)$ differences. Note that this definition allows the landscape to have multiple global minima. We find that at large laser gain $a$, global minima have lower intensive energies than local minima, and to describe such low energy global minima, we must break replica symmetry, just like in the SK model.  In contrast for local minima, as described above, replica symmetric solutions, albeit with broken SUSY, sufficed to match numerical experiments.     

\subsection{A full replica symmetry breaking theory of global minima}
\label{sec:global-min}
 
We performed a full replica symmetry breaking calculation (see Section~S-II G and S-III for details) for global minima, which yields the following formula for the grand potential in the low-temperature $\beta \rightarrow \infty$ limit:
\begin{eqnarray}
    \Omega(\infty, -\infty) 
    = \lim_{\beta\to\infty}&&\frac{1}{2}q_dt + \frac{\beta}{4}(q_d^2 - \int^1_0 dy q^2(y))\nonumber\\
    && - f(0,0),
    \label{eq:grandpotfRSB}
\end{eqnarray}
where the function $f(y,h)$ obeys the Parisi differential equation
\begin{equation}
   \frac{\partial}{\partial y}f(y, h)=-\frac{1}{2}\frac{dq}{dy}\left[\frac{\partial^{2} f}{\partial h^{2}}+\beta y\left(\frac{\partial f}{\partial h}\right)^{2}\right],
\end{equation}
with the boundary condition
\begin{equation}
    f(1, h)=\beta^{-1}\log\left[
    \sum_{x\in\operatorname{Crt}_0(E_{MF}(\cdot, h))} e^{-\beta E_{MF}(x, h)} \right].
    \label{eq:f_susy}
\end{equation}
This expression has order parameters $t, q_d$, and $q(y)$, where $q(y)$ is a non-decreasing non-negative function defined in $y\in[0,1]$. 
The values of these order parameters are chosen to extremize the grand potential in \eqref{eq:grandpotfRSB}. 
The order parameter $t$ reflects, as above in \eqref{eq:orderparamt}, the trace of the susceptibility matrix of the system to a small external field, but this time while the system is in a global minimum. 
$q_d$ refelcts the self-overlap (i.e., the average of $\frac{1}{N}\sum_i x_i^2$). 
The function $q(y)$ is called the overlap function, whose functional inverse represents the cumulative probability density of the overlap $\frac{1}{N}\sum_i x_i^1 x_i^2$ of two different randomly sampled global minima $\mathbf{x}^1$ and $\mathbf{x}^2$.

Furthermore, the distribution $P(x)$ can be obtained from a solution to the following differential equation for a propagator $P(y,h)$ \cite{sommersDistributionFrozenFields1984}
\begin{equation}
    \frac{\partial P}{\partial y}  = \frac{1}{2} \frac{dq}{dy} \left( \frac{\partial^2 P}{\partial h^2} - 2\beta y\frac{\partial f}{\partial h} \frac{\partial P}{\partial h} \right), 
    \label{eq:DiffEqForPmain}
\end{equation}
with a boundary condition at $y=0$ given by 
\begin{equation}
P(0,h) = (2\pi q(0))^{-1} \exp\left(-\frac{h^2}{2q(0)}\right).
\end{equation}
The distribution of spins $P(x)$ in global minima can then be written in terms of the propagator evaluated at $y=1$ and the value of $h = E_I(x) - t x$ which solves the extremization condition for the mean-field energy function $E_{\operatorname{MF}}[h]$ (see Section~S-II G details):
\begin{equation}
    P(x) = |\partial^2 E_I(x) - t| P(1,\partial E_I(x) - t x).
    \label{eq:globalminima}
\end{equation}
Finally, with $P(x)$ in hand, we can calculate the distribution of Hessian eigenvalues $\rho_H(\lambda)$ as above, using \eqref{eq:inversionformula} and \eqref{eq:resolventeqn}.

\subsection{Numerical tests of full replica symmetry breaking for global minima}

To test our theory for global minima, we numerically solved the extremization conditions for the grand potential in \eqref{eq:grandpotfRSB} by approximating the overlap function $q(y)$ to be a sum of $37$ step functions. This corresponds to a 37-step replica symmetry breaking solution, approximating full replica symmetry breaking. We exploited numerical techniques addressed in \cite{schmidtReplicaSymmetryBreaking2008} to find the order parameters satisfying the extremization conditions. 

In Fig.~\ref{fig:global-min}, we compare our theoretical predictions with numerical experiments on finite-size systems. For the numerical experiments, we found the lowest energy minimum among many sampled minima for each sample of $J$ (see \ref{sec:num-process} for details of the sampling). Fig.~\ref{fig:global-min}-a,b,c demonstrates an excellent match between theory and experiment for the distribution of spins $P(x)$, 
the distribution of diagonal elements
$H^I(\mathbf x)_{ii}$ in \eqref{eq:diagonalHessian}, 
and the distribution of Hessian eigenvalues $\rho_H(\lambda)$, respectively.  Fig.~\ref{fig:global-min}-d,e,g shows a match between theory and experiment for the order parameters $a+t$, $q_d$ and the intensive energy respectively. Finally, Fig.~\ref{fig:global-min}-h shows the overlap function $q(y)$. 

Fig.~\ref{fig:global-min} implies the existence of two phase transitions as $a$ increases. First, a phase transition occurs at $a=\lambda_{\min}(J)=-2$. This is the point where the origin $\mathbf{x}=0$ first bifurcates, and the landscape starts to be non-convex.  Indeed, Fig.~\ref{fig:global-min}-e shows that the self-overlap $q_d$ starts to have a finite value at $a=-2$, which implies that the global minimum is no longer at the origin. Moreover, the overlap function $q(y)$ function undergoes a transition from a vanishing flat function for $a<-2$, indicating replica symmetry, to a continuously increasing function for $a>-2$, indicating full replica symmetry breaking (FRSB) (Fig.~\ref{fig:global-min}-h). 

Within this FRSB regime, we find another phase transition similar to the case of the typical minima, which is again characterized by the sign of the $a+t$, which now goes from negative to positive at the critical point $a=a_g\sim -0.45$ (Fig.~\ref{fig:global-min}-d). The global minimum is supersymmetric (see Section~S-IIG for details) so this transition is not characterized by spontaneous SUSY breaking as in the case of typical minima.  Instead, it is characterized by the minimum eigenvalue of the Hessian. Fig.~\ref{fig:global-min}-c,f clearly shows that this minimum eigenvalue is close to vanishing for $a_r<a<a_g$. This indicates that even the global minima, like all typical minima, are marginally stable with soft or flat directions corresponding to vanishingly small Hessian eigenvalues. On the other hand, for $a>a_g$, the global minimum undergoes a rigidity phase transition in which the soft, flat directions disappear because the Hessian eigenspectrum is gapped away from $0$ (Fig.~\ref{fig:global-min}-c,f). Thus the global minimum is rigid or stable to small perturbations. In contrast, typical local minima remain soft and flat for all values of $a$ considered, since the Hessian eigenspectrum always reaches to $0$ (Fig.~\ref{fig:typical-min}-c).

Note that while the global minima are marginally stable when $-2<a<a_g$, they can still be described without SUSY-breaking order parameters $A$ and $C$, in contrast to the case of the typical local minima. This phenomenon can happen when the complexity of global minima is zero. In this case, even if states are highly reactive to perturbations of the energy landscape, the change in the number of the global minima, in response to small changes in the energy landscape, is still sub-exponential. Thus the susceptibility of the grand potential and complexity to such perturbations is $0$, which implies via \eqref{eq:ACsusc} that the SUSY breaking order parameters obey $A=C=0$. 

\section{The phase diagram of geometric landscape annealing}
\label{sec:phase-diagram}
\begin{figure}[!htbp]
    \includegraphics[width=\linewidth]{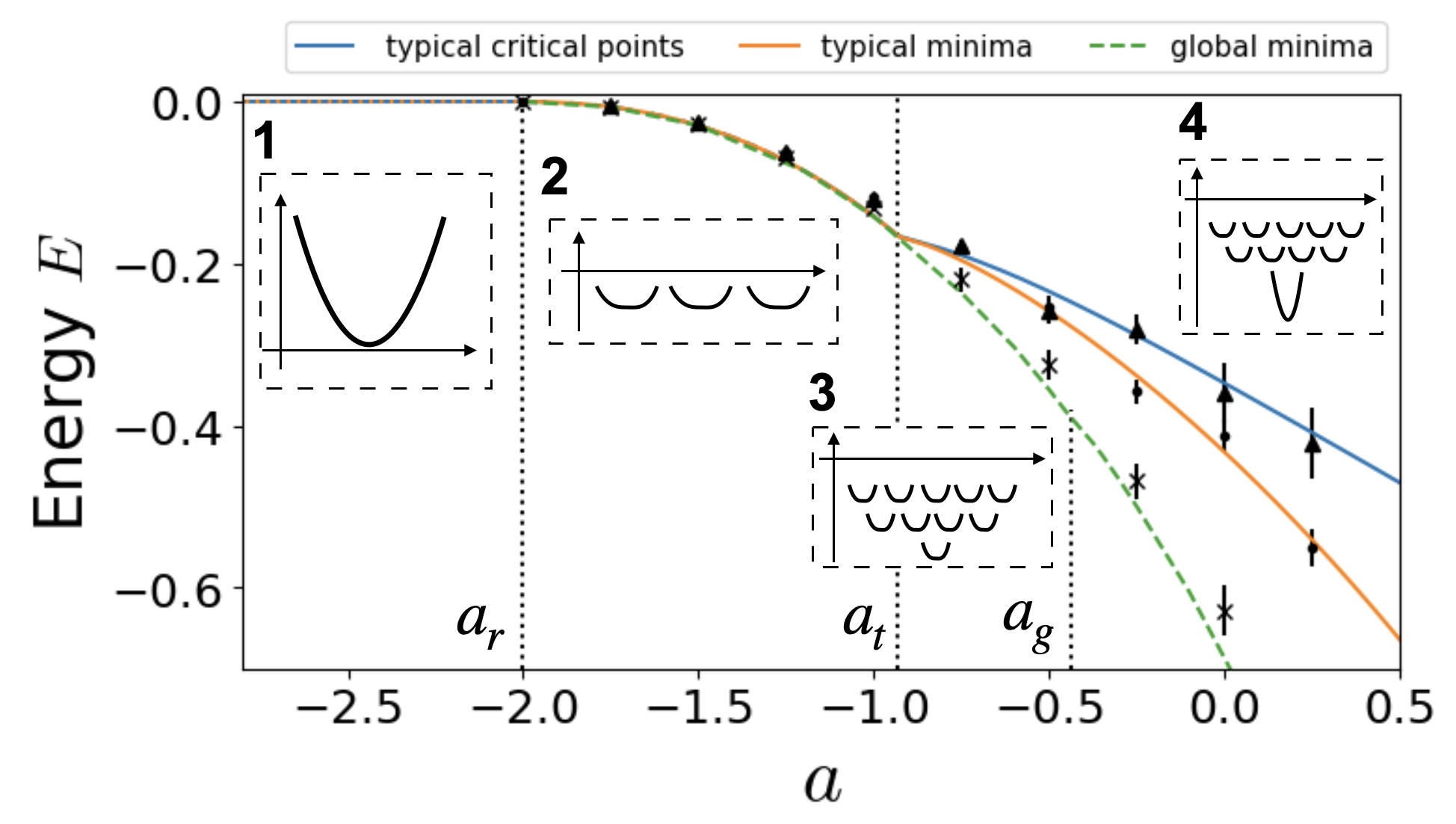} 
    \caption{{\bf The phase diagram of the energy landscape.} The curves are theoretically predicted energy of the typical critical points (the top blue line), typical minima (the middle orange line), and global minima (the green dashed line). The markers are the numerically obtained energies for typical critical points (the triangles), typical minima (the circles), and global minima (the crosses), respectively. We observe three phase transitions at $a=a_r,a_t$, and $a_g$. The insets are sketches of the energy landscape in the 4 different phases.}
    \label{fig:phase-diagram}
\end{figure}

We can now put together a global view of the geometry of the evolving energy landscape as the laser gain $a$ is annealed. Overall,  the energy landscape experiences three important phase transitions: \circled{1} the replica symmetry breaking transition for global minima at $a=a_r=-2$; \circled{2} the SUSY-breaking transition for typical minima and typical critical points at $a=a_t\sim -0.93$; and \circled{3} the rigidity phase transition for global minima at $a=a_g\sim -0.45$. The entire phase diagram is shown in Fig.~\ref{fig:phase-diagram}. 

When $a<a_r=-2$, the landscape is convex, and the single global minimum occurs at the origin $\mathbf{x}=0$. Then, at the first phase transition at $a=a_r$, the origin bifurcates, and just above $a=a_r$ many minima start to appear. Fig.~\ref{fig:phase-diagram} shows that the energy of global minima, typical minima, and typical critical points are all essentially equivalent for $a_r<a<a_t$. This means that the majority of critical points are minima, along with associated saddles of finite or at most subleading $o(N)$ index and energy barrier heights.  Moreover, typical minima are also almost global minima. 
Finally, due to zero complexity of typical critical points and minima in the range $a<a_t$ (Fig.~\ref{fig:typical-crt}-g and Fig.~\ref{fig:typical-min}-g respectively), the total number of critical points is sub-exponential within this phase, and hence so is the number of minima.  Thus SUSY is preserved due to zero complexity in the range $a_r<a<a_t$, despite the fact that typical critical points, minima, and global minima have a Hessian eigenspectrum that extends to $0$ in this range (Fig.~\ref{fig:typical-crt}-c, Fig.~\ref{fig:typical-min}-c, and Fig.~\ref{fig:global-min}-c). Thus overall, in the SUSY phase $a_r<a<a_t$, the non-convex energy landscape is relatively flat, with all sub-exponentially many critical points having essentially the same intensive energies, and all having soft or flat directions with near zero Hessian eigenvalues.  

At $a=a_t$, both typical critical points and minima experience SUSY breaking, due to the proliferation of exponentially many critical points and minima with nonzero complexity (Fig.~\ref{fig:typical-crt}-g and Fig.~\ref{fig:typical-min}-g respectively) in conjunction with their marginal stability (Fig.~\ref{fig:typical-crt}-c and Fig.~\ref{fig:typical-min}-c). Moreover, an intensive energy gap starts to appear between typical minima and global minima. Hence we expect that finding the lowest CIM energy state starts to get difficult at $a>a_t$ due to the exponential number of higher energy typical minima. 

Finally, while global minima are marginally stable until $a=a_g$, with many soft or flat modes, they become fully rigid for $a>a_g$ due to a Hessian eigenspectrum gapped away from $0$ (Fig.~\ref{fig:global-min}-c).

\begin{figure}[!htbp]  \includegraphics[width=\linewidth]{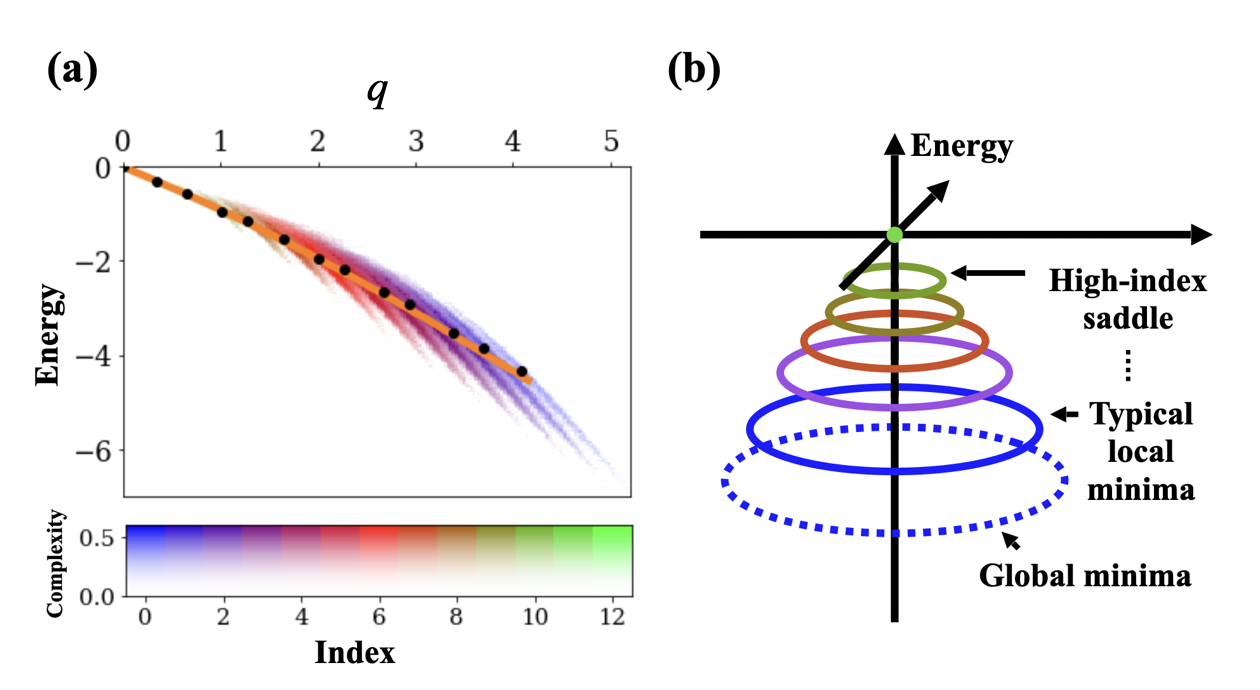} 
    \caption{{\bf A schematic view of the energy landscape at large laser gain.} (a) A two-dimensional heat map of the experimentally derived distribution of intensive energy and squared radius $q$ of critical points with $a=4, N=12$, derived from the sampled points depicted in Fig.~\ref{fig:typical-crt}. The color gradient denotes the index of critical points, while the opacity illustrates the complexity. The black points are the experimentally derived most likely location of critical points for each index $I=0,\cdots,12$. The orange curve is a theoretically predicted relationship between the energy and squared radius $q$ of critical points obtained by solving \eqref{eq:self-consistent-replica} for varying $\mu$ in \eqref{eq:complexity-replica} and plotting $q(\mu)$ versus $E(\mu)$ (given in \eqref{eq:eformula}).   (b) A schematic depiction of the energy landscape at large laser gain, consisting of concentric shells of increasing radius and decreasing index and energy.}
    \label{fig:heatmap-crt}
\end{figure}

In addition to a global view of how typical critical points, minima and global minima evolve as a function of laser gain $a$, as depicted in Fig.~\ref{fig:phase-diagram}, we can also obtain a global view of the energies and locations of critical points of {\it all} indices at a fixed laser gain $a$.  Fig.~\ref{fig:heatmap-crt} depicts this global view for both theory and experiment at large $a$ (in this case $a=4$), which is the important case when the CIM energy function approximates well the Ising energy function of interest (see \eqref{eq:CIM_Ising}). 

In particular, Fig.~\ref{fig:heatmap-crt}a shows a heat map of the numerically estimated complexity of critical points in a finite size ($N=12$) system, as a joint function of their energy and squared radius $q = \frac{1}{N} \sum_i x_i^2$ and colored by their index. This complexity heatmap shows a clear correlation between index, radius and energy, with lower index critical points occurring at lower energy and larger radius.  Moreover, for each index we plotted the most likely location in the energy-squared radius plane (i.e. the location where the complexity of critical points of that index is maximized) as black dots in Fig.~\ref{fig:heatmap-crt}-a. 

We then compared the locations of these black dots with theoretical predictions derived from our SUSY-breaking theory of critical points.  In particular, we continuously varied the chemical potential $\mu$ in the expression for the grand potential in \eqref{eq:complexity-replica} (or equivalently \eqref{eq:grandpotgencavity}) and solved the self-consistent equations for the order parameters in \eqref{eq:self-consistent-replica} (or equivalently \eqref{eq:ACselfcons} and \eqref{eq:selfconscompact}) as a function of $\mu$, as well as computed the intensive energy as function of $\mu$ via \eqref{eq:eformula}. Altogether, this yields the average squared radius $q(\mu)$ and intensive energy $E(\mu)$ over a weighted averaged of critical points controlled by $\mu$, as in the partition function in \eqref{eq:grandpotdef} with $\beta=0$.  As $\mu \rightarrow -\infty$, this weighted averaged is dominated by index $0$ critical points, or minima. As $\mu$ increases, the weighted average is dominated by higher index critical points.  Thus a theoretical prediction is that the curve $E(\mu)$ versus $q(\mu)$ as $\mu$ varies from $-\infty$ to $+\infty$ should provide information about the most likely location of saddle points of increasing index in the $E$-$q$ plane, thereby going through all the black points in Fig.~\ref{fig:heatmap-crt}-a. Remarkably, this prediction is confirmed in Fig.~\ref{fig:heatmap-crt}-a: the orange curve is a plot of the theoretically derived curve $E(\mu)$ versus $q(\mu)$, and it does indeed go through all the black points, which indicate the experimentally derived most likely locations in the $E$-$q$ plane for critical points of each index.  

A schematic view of the energy landscape which is justified by Fig.~\ref{fig:heatmap-crt}-a is shown in Fig.~\ref{fig:heatmap-crt}-b.  Schematically, at large $a$, the CIM energy landscape exhibits a highly rough structure with concentric shells of critical points of increasingly lower index occurring at increasing lower energy and increasingly larger radius.  In particular, the global minimum occurs at the largest radius and lowest energy.  But just above this in energy and at a smaller radius, there is a wall of exponentially many typical local minima that stand as a potential barrier.  Thus despite the fact that at large $a$, the CIM energy landscape has the nice property that it mimics the Ising energy landscape of interest (see \eqref{eq:CIM_Ising}), direct optimization at large $a$ starting from the origin poses a difficult problem, as energy minimization must traverse successively lower index saddles and minima at lower energy and larger radius, that may prevent reaching the deepest global minima at the largest radius.

\section{The relationship between annealing performance and energy landscape geometry}

\label{sec:relationship}

We next discuss the relationship between the phase transitions in the energy landscape geometry discussed above and the performance of geometric landscape annealing.  This analysis also reveals an optimal annealing schedule to arrive at a low value of the Ising energy.  Indeed it was this annealing schedule that we used to attain good performance in Fig.~\ref{fig:dyson-brownian} in Section~\ref{sec:performance}.

\begin{figure}[!htbp]
    \includegraphics[width=\linewidth]{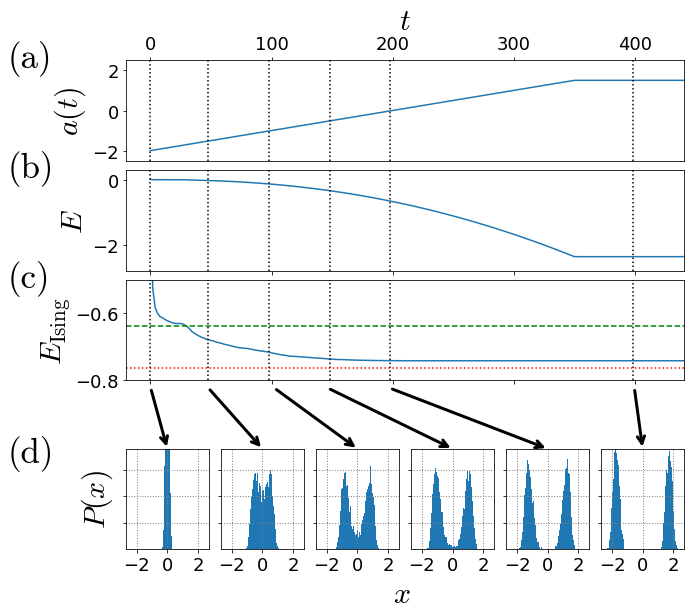} 
    \caption{{\bf An example of annealing trajectories} (a) The annealing schedule given by \eqref{eq:schedule} with $a_{\max}=1.5$ and $\tau=10^2$. (b) the trajectory of soft-spin network's energy $E$. (c) The trajectory of Ising energy $E_{\text{Ising}}$ of the corresponding spin configuration. The green dashed line represents the Ising energy obtained by the spectral method $E_{\mathrm{Ising}}=-\frac{2}{\pi}$, and the red dotted line the ground state energy in the large-$N$ limit ($\sim -0.74$). (d) The local variables' distributions $P(x)$. The plots represent the snapshots of the distribution at time $t=0, 50, 100, 150, 200, 400$, which are shown as vertical dotted lines above. }
    \label{fig:performance-1exp}
\end{figure}

We simulated the annealing processes with various annealing schedules with a system size of $N=10^4$.
The schedules $a(t)$ are chosen as 
\begin{equation}
    a(t) = \min\left(\frac{t}{\tau} + a(0), a_{\max}\right),
    \label{eq:schedule}
\end{equation}
i.e., $a(t)$ linearly increases from $a(0)$ with slope of $\tau^{-1}$ until it saturates at $a=a_{\max}$ (see top panel of Fig.~\ref{fig:performance-1exp}). We set $a(0) $ to be the smallest eigenvalue of $J$, i.e. approximately $a_r = -2$, because the state $\mathbf x$ is always trivially at the origin for $a$ smaller than the eigenvalue. The initial state is chosen as a random Gaussian vector with independent components each drawn from a Gaussian distribution with zero mean and standard deviation $0.1$. We verified that the annealing performance is not influenced by the choice of the standard deviation unless it is {\it much} smaller than $O(1)$. In that case, the initial state is very close to the origin and takes a long time to escape the saddle point at the origin.  When it does, it aligns with the principal eigenvector of $J$, which is the most negative curvature direction around the saddle at the origin. Therefore initializing very close to the origin is almost equivalent to the case of initializing the state along the principal eigenvector of $J$, a possibility which we will discuss further below.

In Fig.~\ref{fig:performance-1exp}, we show the trajectory of an annealing process with $a_{\max} = 1.5$ and $\tau = 100$ for the CIM energy (second panel), Ising energy (third panel) and several snapshots of the distribution of OPO amplitudes $P(x)$ (fourth panel).  Since the initial state is a random vector around the origin, both the CIM energy $E$ and the Ising energy $E_{\mathrm{Ising}}$ are close to zero \footnote{The Ising energy of random spin states is almost zero for the following reason. Flipping some of the Ising spins, we can make all spins $+1$. By flipping the sign of $J_{ij}$ properly at the same time, we can keep the Ising energy unchanged. This new $J$-matrix follows GOE as well. The Ising energy is given by the sum of all the elements of the new $J$-matrix. By the law of large numbers, this quantity converges to its mean $0$.}. As the laser gain $a(t)$ increases, both energies decrease monotonically and the distribution $P(x)$ gradually transforms from an unimodal shape around the origin to a bimodal shape. During this transformation, small OPO amplitudes $x_i$ around the origin are driven to either large positive or negative values, causing sign flips of $x_i$ that lower $E_{\mathrm{Ising}}$. However, once the distribution $P(x)$ gets completely separated into positive and negative parts at $t\sim 200$, fewer amplitudes $x_i$ can flip their signs, and the Ising energy freezes.

\begin{figure*}
\includegraphics[width=\linewidth]{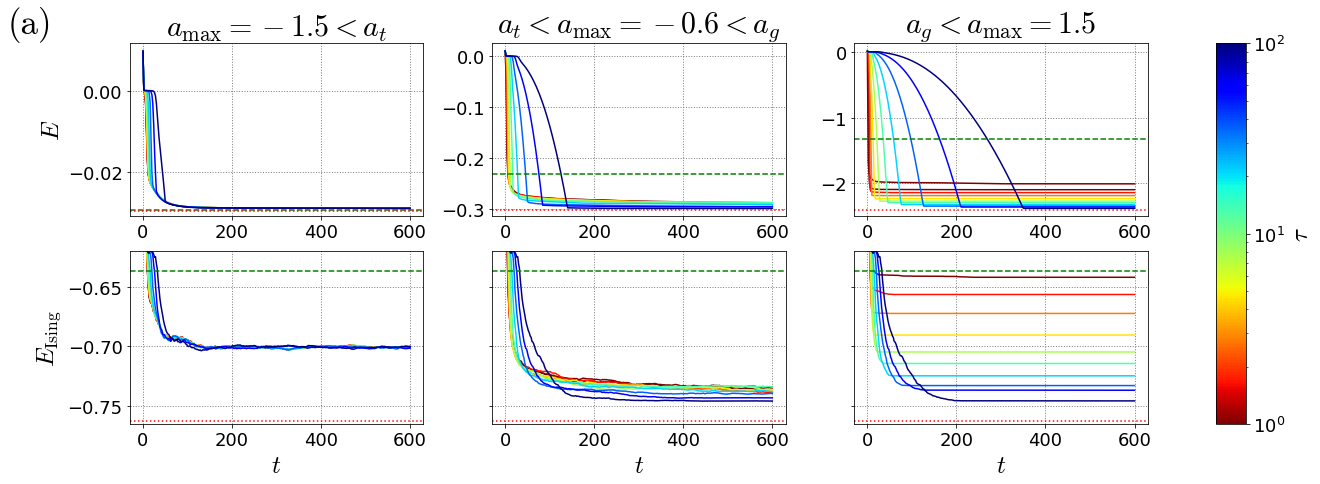}
    \includegraphics[width=\linewidth]{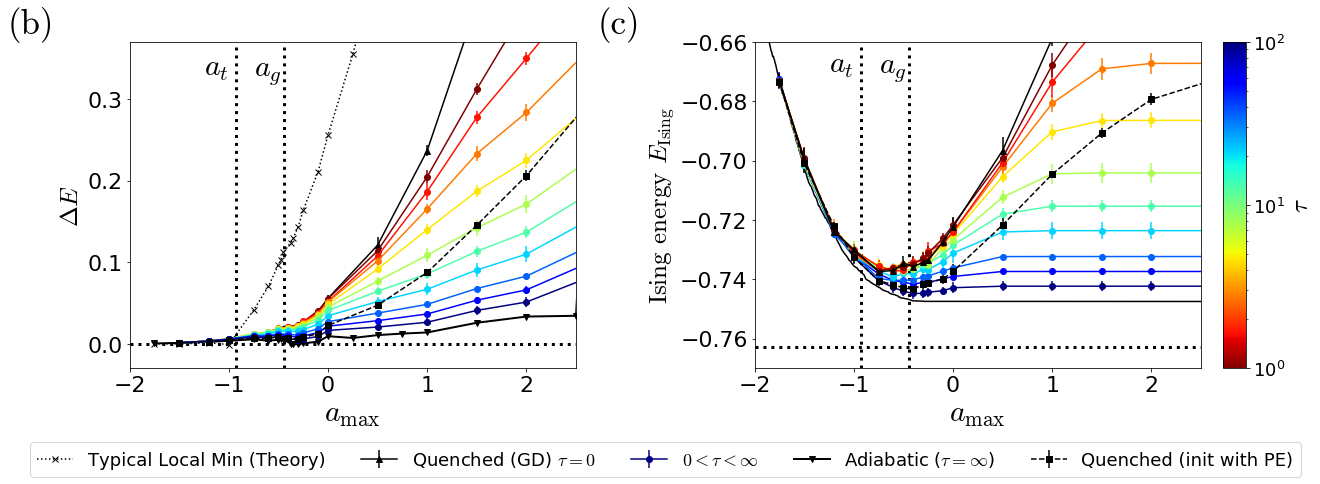} 
    \caption{{\bf The performance of geometric landscape annealing and its dependence on annealing schedules.}
    We simulated geometric landscape annealing with a two-parameter family of annealing schedules in \eqref{eq:schedule} paramterized by the final laser gain $a_{\max}$ and time constant $\tau$. (a) Trajectories of the CIM energy $E$ (top) and corresponding Ising energy $E_{\mathrm{Ising}}$ (bottom) in the three different phases for $a_{\max}$. The color bar indicates the annealing time constant $\tau$. In the three top plots of CIM energy, the green dashed and red dotted lines represent the energy of typical local minima and the global minima, respectively. In the bottom plots of Ising energy, the green dashed and red dotted horizontal lines represent $E_{\mathrm{sp}}=-2/\pi$ and $E_{\mathrm{SK}}\sim-0.763$, respectively.
    (b) The solid colored lines (with color indicating annealing time constant according to the color bar)  indicate the final achieved CIM energy $E$ for each $a_{\max}$ at $t=6\times 10^2$ {\it minus} the theoretically calculated CIM ground state energy for the same value of $a_{\max}$.  The dotted horizontal line of $\Delta E=0$ represents the baseline ground state energy. The very top dotted black line with x markers indicates the theoretically calculated higher CIM energy of typical local minima (again, minus the energy of the corresponding global minima). (c) The final achieved Ising energy at $t=6\times 10^2$ as a function of annealing time constant $\tau$ indicated by color and final gain $a_{\max}$ on the horizontal axis. The horizontal dotted line is the ground state Ising energy $E_{\text{SK}}\approx -0.763$. In (b) and (c), the solid colored lines are the annealing processes with mean and standard deviation computed across $5$ different initializations and connectivities $J$. The black line with triangle markers above these colored lines is the case of $\tau=0$, which corresponds to rapid quench from $a=a_r$ to $a=a_{\max}$. The black line with inverted triangle markers below all the colored lines is the trajectory of the energy obtained by integrating the adiabatic differential equation \eqref{eq:adiabatic_evol}, which essentially corresponds to $\tau=\infty$. The black dashed line with square markers in the midst of the colored lines represents the energy trajectory of gradient descent dynamics initialized along the principal eigenvector of $J$ when $a$ is fixed at $a_{\max}$.}
    \label{fig:performance}
\end{figure*}

To understand the dependency of the final CIM and Ising energies on the annealing schedule, we simulated the annealing processes with various $a_{\max}$ and $\tau$. Fig.~\ref{fig:performance}-a shows the CIM and Ising energy trajectories for $3$ different $a_{\max}$ and $10$ different $\tau$. Each row and color corresponds to a certain value of $a_{\max}$ and $\tau$, respectively. Fig.~\ref{fig:performance}-b,c shows the {\it final} achieved CIM and Ising energies respectively at $t=6\times 10^2$, averaged over $5$ different realizations of the random initial state and the connectivity $J$. In particular, Fig.~\ref{fig:performance}-b shows for each $a_{\max}$, the difference $\Delta E$ between the final CIM energy achieved by annealing to $a_{\max}$ and the corresponding global minimum energy at $a_{\max}$, while Fig.~\ref{fig:performance}-c shows the corresponding Ising energy $E_{\text{Ising}}$ of the Ising sign pattern of the CIM state found by annealing. The colored solid lines represent the final achieved energy for different $a_{\max}$ on the x-axis and different colors for different $\tau$.
The dotted black curve with triangle markers above these colored solid lines corresponds to $\tau=0$, i.e. a rapid quench or gradient descent from a random initial state at fixed $a_{\max}$.
On the other hand, the solid black curve with inverted triangle markers below all the colored solid lines corresponds to $\tau=\infty$, or the slowest possible annealing process obtained by integrating the following adiabatic evolution
\begin{equation}
    \frac{d\mathbf{x}}{da} = H^{-1}\mathbf{x},
    \label{eq:adiabatic_evol}
\end{equation}
with Hessian $H$ given in (\ref{eq:Hessian}).
Note that this equation can be obtained by differentiating the stationary condition $\frac{dE_{\operatorname{tot}}(\mathbf{x})}{dx_i} =0$ with respect to $a$.
For comparison, Fig.~\ref{fig:performance}-b,c also shows the results of gradient descent starting from the principal eigenvector (the black dashed line with rectangle markers) \footnote{We set the norm of the initialized principal eigenvector to match that of a Gaussian random initialization.}. This corresponds to the limit of an extremely small standard deviation of the initial random Gaussian state. In Fig.~\ref{fig:performance}-b, we also show for reference the CIM energy of typical minima (the highest black dotted line with ``x" markers).

In the following subsections, we discuss the major features observed in Fig.~\ref{fig:performance} when the annealing process terminates in different phases of the energy landscape geometry revealed in previous sections.  In particular, we discuss in succession \circled{1} the small-gain supersymmetric phase where $a_r < a_{\max}<a_t$ (the left column of Fig.~\ref{fig:performance}-a); \circled{2} the intermediate gain supersymmetry breaking phase where $a_t<a_{\max}<a_g$ (the middle column of Fig.~\ref{fig:performance}-a); and \circled{3} large-gain rigid global minimum phase of $a_{\max} > a_g$ (the right column in Fig.~\ref{fig:performance}-a). Note that when the gain is smaller than $a_r=-2$, the energy landscape is convex and the CIM state is confined to the origin. 

\subsubsection*{The small-gain supersymmetric phase \texorpdfstring{$a_r < a_{\max}<a_t$}{a_r< a_max < a_t}}
 As is shown in Fig.~\ref{fig:performance}-b, all the final achieved CIM energies are very close to the $\Delta E = 0$ horizontal line, regardless of annealing time constant $\tau$. This is because the majority of minima in the supersymmetric phase are almost global minima (Fig.~\ref{fig:phase-diagram}). The final achieved Ising energy is also almost independent of the annealing schedule, as is shown in Fig.~\ref{fig:performance}-c. However, unlike the final achieved CIM energy in Fig.~\ref{fig:performance}-b, the final achieved Ising energy in Fig.~\ref{fig:performance}-c decreases rapidly with increasing $a_{\max}$ in the range $a_r < a_{\max} < a_t$. This decrease occurs because the distribution of OPO amplitudes $P(x)$ has a finite density at the origin, and so ramping up the laser gain allows some of these small amplitude spins to flip their signs, thereby lowering the achieved Ising energy.
 
\subsubsection*{The intermediate-gain SUSY breaking phase \texorpdfstring{$a_t<a_{\max}<a_g$}{a_t<a_max<a_g}}
Once $a$ exceeds $a_t$, the complexity of minima becomes positive (Fig.~\ref{fig:typical-min}-g), and the energy of typical minima becomes strictly larger than that of global minima (Fig.~\ref{fig:phase-diagram}). This means there are exponentially many local minima above the global minima in the energy landscape. Thus, if we rapidly increase the laser gain $a$ and relax the system from a high-energy state near the origin, the resultant trajectory is likely to be trapped by a high-energy local minimum. This effect makes the final achieved CIM energy of rapid annealing higher than that of slow annealing (Fig.~\ref{fig:performance}-a middle-top and Fig.~\ref{fig:performance}-b). Moreover, since the CIM energy is correlated with the Ising energy, rapid annealing also yields higher Ising energy than slow annealing (Fig.~\ref{fig:performance}-a middle-bottom and Fig.~\ref{fig:performance}-c). However, since the complexity of minima is still relatively low in the range $a_t<a_{\max}<a_g$, the increment of final achieved CIM and Ising energies with annealing speed is relatively small. 

\begin{figure}[!htbp]
    \includegraphics[width=\linewidth]{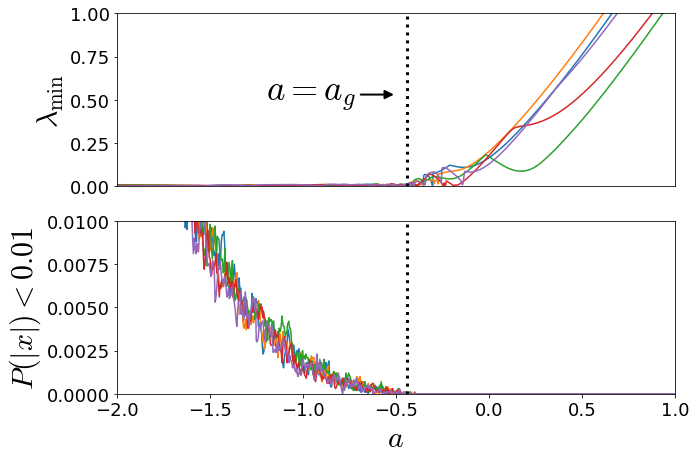} 
    \caption{{\bf Slow annealing trajectories exhibit the same phase transition as that of CIM global minima.} We simulated the geometric landscape annealing process by integrating \eqref{eq:graddescent} with 5 different samples of $J$ with system size $N=10^4$.
    The annealing schedule is given by (\ref{eq:schedule}) with $\tau=10^2$ and $a_{\max}=4$. Each color represents a trajectory of a single instance. (top panel) The minimum eigenvalue of the Hessian along the trajectory.  We observe a similar rigidity phase transition as that of global minima in Fig.~\ref{fig:global-min}-f wherein the minimal eigenvalue of the Hessian transitions from $0$ to nonzero values precisely at $a=a_g$. 
    (bottom panel) The fraction of the number of OPOs with small amplitudes $|x_i|<0.01$ starts to vanish at $a=a_g$, similar to how the distribution of amplitudes $P(x=0)$ for global minima vanishes at $a=a_g$.  These two observations provide further evidence that the slow annealing process can find the near-global CIM energy minima at least around $a=a_g$.}
    \label{fig:annealing}
\end{figure}

On the other hand, another effect decreases the Ising energy with increasing $a_{\max}$ at all annealing speeds in the range $a_t<a_{\max}<a_g$ (Fig.~\ref{fig:performance}-c). 
As seen in the top middle panel of Fig.~\ref{fig:performance}-a and Fig.~\ref{fig:performance}-b, the final CIM energy achieved by annealing is still lower than that of typical minima, and is rather closer to that of global minima.  Therefore the states obtained by annealing are likely to have features of the global minima rather than the typical minima.
Indeed, the slowest annealing process has energy very close to that of global minima.  This observation is bolstered by the numerical results shown in Fig.~\ref{fig:annealing}, which indicates that the trajectories of the slowest possible annealing processes have the features of the global minima's rigidity phase transition, i.e., the trajectory experiences the transition from marginal stability to full stability, and the localization of $P(x)$ exactly at $a=a_g$. Since $P(x)$ has a finite density around the origin below $a_g$, the signs of some OPO amplitudes can still be flipped by ramping up the gain up to $a=a_g$. This effect allows $E_{\text{Ising}}$ to decrease further in the range $a_t<a_{\max}<a_g$, especially for slow annealing, as seen in Fig.~\ref{fig:performance}-c. 

\subsubsection*{The large-gain rigid global minima phase \texorpdfstring{$a_{\max} > a_g$}{a_max>a_g}}
First, we focus on the black bottom lines in Fig.~\ref{fig:performance}-b,c representing the trajectory of the slowest possible annealing process. As we discussed above, this trajectory experiences a rigidity phase transition similar to that of global minima, and the Hessian of the energy landscape along this trajectory becomes gapped away from $0$ for $a>a_g$, just as it does for global minima (Fig.~\ref{fig:global-min}-f). This implies that the adiabatic evolution \eqref{eq:adiabatic_evol} is non-singular, and the time derivative of $x_i$ is upper bounded. Because the distribution $P(x)$ is localized and separated into two sets of $x_i$ with positive and negative signs, only a few numbers of $x_i$ can flip their signs by this bounded state evolution. Hence it is unlikely that many $x_i$ flip their signs, and therefore it is also unlikely that the Ising energy is lowered for $a>a_g$. Indeed the final achieved Ising energy for the slowest annealing process (lower solid black curve in Fig.~\ref{fig:performance}-c) is flat for $a_{\max} > a_g$. Thus, interestingly, as we ramp up the laser gain beyond $a_g$, even though the CIM energy landscape becomes more like the Ising landscape, the final achieved Ising energy via annealing cannot be lowered. In other words, the geometric landscape annealing process at slow annealing speeds is effectively terminated by the rigidity phase transition in global minima at $a=a_g$, well before the CIM energy landscape looks like the Ising energy landscape at large $a$ as in \eqref{eq:CIM_Ising}.

When the annealing speed is faster, the trajectory is more likely to be trapped by a higher-energy local minimum, leading to both higher final CIM and Ising energies.  This increase of final energies with increased annealing speeds becomes stronger as the final gain $a_{\max}$ increases because of both the complexity growth of typical minima with $a$ (Fig.~\ref{fig:typical-min}-g) and the growing energy gap between typical and global minima (Fig.~\ref{fig:phase-diagram}). Likely because of both of these landscape properties, the final achieved CIM and Ising energies are significantly larger under faster annealing compared to slower process at very large $a_{\max}>a_g$ (Fig.~\ref{fig:performance}-a right, and Fig.~\ref{fig:performance}-b,c).

\subsubsection*{The optimal annealing schedule terminates at the rigidity phase transition for global minima} 

In summary, out of the general space of annealing schedules in \eqref{eq:schedule}, the optimal schedule with lowest achieveable Ising energy is given by  $a_{\max}=a_g$ and large $\tau$. If the annealing speed is slow enough, a further increase of $a_{\max}$ does not further lower the Ising energy because the annealing trajectory tracks the evolution of a CIM global minimum found at $a=a_g$, which remains rigid for $a>a_g$. Also, more rapid annealing (smaller $\tau$) yields much less optimal, much higher Ising energies at larger $a_{\max}>a_g$.  However, at $a=a_g$, the dependence of the final achieved Ising energy on the speed of the annealing process is remarkably weak, ranging from $\sim-0.75$ to $\sim-0.73$ as $\tau$ ranges from $1$ to $100$.  Thus geometric landscape annealing in this case is surprisingly robust to annealing speed, provided annealing is optimally terminated at the rigidity phase transition for global minima.

Note that we also discovered that the suboptimal higher Ising energies found by fast annealing can be mitigated by initializing the state along the principal eigenvector of the connectivity $J$. The performance of gradient descent at fixed $a_{\max}$ with this initialization is shown as a dashed back line with square markers in Fig.~\ref{fig:performance}-b,c. The final Ising energy of this process also achieves the lowest value around $a=a_g$, and the Ising energy Ising difference from the slowest annealing process is $<0.01$.
Since the slow annealing process can reach a global minimum of the CIM energy at $a=a_g$, the best achievable Ising energy can be simply characterized as the Ising energy of the CIM global minimum {\it specifically} when $a=a_g$. Note however, that as $a_{\max}$ increases beyond $a_g$, even at extremely slow annealing, the CIM energy found by annealing is {\it no longer} equivalent to the CIM energy of the global minimum, as reflected by the detachment of the lower black solid line from the lowest horizontal dotted line of $\Delta E=0$ in Fig.~\ref{fig:performance}-b for $a \gg a_g$. This means geometric landscape annealing cannot find a CIM global energy minimum for $a\gg a_g$. Since the CIM energy global minimum at very large $a$ is also an Ising energy global minimum, according to \eqref{eq:CIM_Ising}, this means the annealing process cannot find the exact Ising ground state either, as reflected by the gap in  Fig.~\ref{fig:performance}-c, between the solid black bottom curve and the dotted horizontal line of $E_{SK}\sim-0.763$. As discussed above, the adiabatic evolution for $a>a_g$ is continuous due to the non-degeneracy of Hessian along the trajectory. This implies that the near-global CIM energy minimum at $a=a_g$ that originated from  the bifurcation at the origin, is itself not continuously connected to global CIM energy minima at very large $a$. 
This type of discontinuity around phase transitions has been known as a major challenge for annealing processes such as simulated annealing and quantum annealing.


\section{Discussion}
\label{sec:conclusion}

In an effort to develop a theoretical understanding of how a physical computing device, the coherent Ising machine, solves discrete combinatorial optimization problems by embedding them in annealed, nonlinear analog dynamics, we engaged in an extensive study of the geometry of the energy landscape of this system and how it evolves as the laser gain is annealed, when the system is attempting to find the ground state of the SK spin glass.  We were able to quantitatively describe the geometry of the landscape at all laser gains in terms of the number of critical points, and their locations (distance from the origin), energies, indices, and Hessian eigenspectra.  We found at large laser gain, when the CIM energy function mimics the Ising energy function, the CIM energy landscape exhibits a complex hierarchical concentric shell structure in which saddle points of successively lower index and energy are located at successively larger radii (Fig. \ref{fig:heatmap-crt}).  This complex landscape presents a challenge to dissipative gradient descent dynamics, which at a fixed large laser gain cannot come close to either the CIM or Ising energy global minimum (top solid quenched $\tau=0$ black line with triangle markers in Fig. \ref{fig:performance}-bc).  

However, annealing the laser gain takes the CIM landscape through a sequence of phase transitions, each one introducing successive complexity.  For $a<a_r$ the landscape is convex with a single global minimum at the origin.  Then for $a_r < a < a_t$, there are many (though subexponential in $N$) critical points. The intensive energies of all critical points are close to those of both typical and global minima.  Hessian eigenspectra of all critical points extend to zero, indicating extensively many soft modes. This represents a highly flat landscape with many minima with similar energies tightly concentrated around a specific value, separated by saddle points whose energy barrier heights relative to minima, and whose indices, both scale sublinearly in $N$. This situation is described by supersymmetric solutions of both the replica and cavity methods.  Then for $a_t < a < a_g$, supersymmetry for typical critical points and minima is broken; there are exponentially many of them with Hessian eigenspectra extending to $0$. Furthermore, the intensive energies of typical critical points, typical minima, and global minima start to separate, indicating the beginnings of a rugged landscape (Fig.~\ref{fig:phase-diagram}).  The global minimum still has extensively many soft modes.  Finally, for $a > a_g$, the global minimum undergoes a rigidity phase transition and all its soft modes disappear.  Moreover, our cavity method for deriving these results yields conceptual insight into the meaning of SUSY breaking and the resultant order parameters, in terms of the extreme reactivity of the landscape to specific external perturbations, originating from exponentially many critical points with extensively many soft modes. 

This detailed analysis of the landscape not only provides conceptual insights into why geometric landscape annealing works, through annealing the laser gain of the CIM, but also suggests an optimal annealing schedule.  Basically, the Ising energy along a slow CIM annealing trajectory continuously decreases as $a$ increases until $a$ hits $a_g$.  At this point, the CIM annealing trajectory, whose energy has been following that of the CIM global minimum (bottom solid adiabatic $\tau=\infty$ black line with inverted triangle markers in Fig.~\ref{fig:performance}-b), becomes trapped in a rigid minimum that it cannot escape with further annealing.  Thus no further sign flips can occur and the Ising energy is fixed; there is no advantage to terminating the annealing process at any $a_{\text{max}}>a_g$.  In fact there is a disadvantage: if one terminates annealing at some $a_{\text{max}}>a_g$, the results can depend strongly on the annealing speed $\tau$ (i.e. substantial height variation of the colored lines for $a_{\text{max}} \gg a_g$ in Fig.~\ref{fig:performance}-bc). However, if one terminates at $a_{\text{max}} = a_g$ the final achieved CIM or Ising energies do not depend strongly on annealing speed (i.e. very little height variation of the colored lines for $a_{\text{max}} = a_g$ in Fig.~\ref{fig:performance}-bc). This robustness to annealing speed is a consequence of the landscape geometry: for $a$ up to $a_g$ the intensive energy gap between typical local minima and global minima is not so large (Fig.~\ref{fig:phase-diagram}), so if faster annealing results in trapping by local minima, such trapping cannot lead to substantially higher CIM energy.  

All of this landscape analysis together points to an optimal and robust annealing schedule: simply anneal $a$ to $a_{\text{max}} = a_g$ when the global minima of the CIM energy landscape become rigid. The slower the annealing the better, but excessive slowness is not required due to the robustness of the final energies to annealing speed. Indeed, this landscape-derived annealing schedule allowed us to find SK spin configurations with energies in the large $N$ limit close to within about $1\%$ of the true ground state energy (Fig.\ref{fig:ising-energy-by-cim}).  This final mismatch between the Ising energy found by CIM annealing and the actual Ising energy means that the global minimum of the CIM energy landscape at $a=a_g$ (or at least the state, with energy close to that of the global minimum, found by annealing) is not continuously connected to the global minimum of the CIM energy landscape at $a \gg a_g$, when the CIM energy landscape approximates well the Ising energy landscape. One possible scenario is an energy level crossing between two far apart local minima between $a_g$ and large $a$, which leads to a different state becoming the global minimum at large $a$ than the state that is a global minimum at $a=a_g$.  

Overall, this extensive analysis of energy landscape geometry and its relation to annealing dynamics opens the door to several interesting directions.  First, while we have focused on the SK spin glass problem, there exist many other ensembles of random optimization problems that can be efficiently mapped to Ising energy minimization, including for example, partitioning, covering, packing, matching, clique finding, graph coloring, minimum spanning trees, and the traveling salesman problem~\cite{lucasIsingFormulationsMany2014}.  Each of these ensembles of random problems could exhibit different geometric properties under landscape annealing, and analyzing the relationship between the evolution of landscape geometry, optimal annealing schedules, and annealing performance in these different ensembles could shed light on different universality classes of possible scenarios.

Second, we have considered gradient descent dynamics on an evolving energy landscape.  One could also analyze non-gradient descent dynamics, either due to the addition of non-conservative feedback \cite{leleuDestabilizationLocalMinima2019} or the addition of asymmetric parts to the connectivity matrix \cite{sternDynamicsRandomNeural2014}.  Such additional non-gradient dynamics can typically induce chaos, and destabilize the least stable minima.   Just as the Kac-Rice formula can be used to count critical points of an energy landscape, as we have done here, it can also be used to count fixed points in non-gradient dynamical systems, for example in neural network dynamics~\cite{Wainrib2013-of} or ecological dynamics~\cite{
biroliMarginallyStableEquilibria2018, 
buninEcologicalCommunitiesLotkaVolterra2017, 
ipsenKacRiceFixed2018}. 
Such a Kac-Rice analysis of the CIM dynamics with an asymmetric connectivity component, which can be implemented physically in the CIM hardware, may provide an intriguing window into whether and how chaos might aid optimization.

Third, our analysis methods may also be useful for exploring the potential utility of non-degenerate OPO dynamics~\cite{Roy2021} for evading obstacles in the CIM energy landscape. Such non-degenerate OPO dynamics can be modelled as a set of coupled oscillatory phase variables, akin to a network of Kuramoto oscillators \cite{Acebron2005-vi}, which have also been employed in physical computing devices to solve Ising energy minimization problems \cite{Wu2011-ev,Albertsson2021-yw, wang2019oim}. The geometric landscape annealing considered here can be thought of as gradually interpolating between soft-spin variables to strongly bistable binary variables while keeping the Ising connectivity fixed. On the other hand, the flexibility of physical OPO devices also opens the door to more general {\it dynamics} annealing strategies that interpolate between non-degenerate oscillatory phase-like dynamics, and degenerate soft-spin dynamics or strongly bistable binary  dynamics~\cite{Roy2021}.  Exploring and analyzing the utility of this broader class of annealing strategies in solving diverse optimization problems constitutes an interesting direction for future research.  

Fourth, and perhaps most interestingly, OPO networks can be constructed in ways that interpolate between classical and quantum operating regimes,  as a function, {\it e.g.}, of linear decoherence rates relative to coherent nonlinear dynamical rates~\cite{Jankowski2023}). Our work in the classical setting here provides a foundation for exploring how novel emergent information dynamics in the classical-quantum crossover~\cite{Yanagimoto2023} may impact optimization performance.  Indeed a key open question is how does open dissipative quantum dynamics negotiate high dimensional spaces riddled with saddle points and local minima as in Fig. \ref{fig:heatmap-crt}?  Is there some balance between coherent quantum evolution and environment-induced dissipation that can aid in optimization through energy minimization? Perhaps an interesting place to start is small systems of $N=4$ coupled OPOs whose open dissipative quantum dynamics can both be tractably simulated on classical computers \cite{InuiYamamoto2020}, as well as physically implemented in circuit QED \cite{BlaisGrimsmoGirvinWallraff2021} or nanophotonic \cite{Jankowski2023} devices.  An interesting question is to map out the computational phase diagram of such problems, parameterized by $4$ by $4$ connectivity matrices, and determine the boundaries between two computational phases in which the classical CIM either succeeds or fails. Then one could explore how the quantum CIM behaves differently in each of these phases. 

Along these quantum lines, recent work has examined how quantum or other physical effects in open dissipative physical systems modify their classical dissipative dynamics, yielding optimization benefits.  For example, in a multimode cavity QED system whose classical dynamics mimics a Hopfield associative memory \cite{Hopfield1982-fx}, the natural cavity dynamics yield the steepest energy descent dynamics that enhance both the capacity and robustness of memories relative to that of the classical Hopfield model \cite{Marsh2021-mz}.  Also, when the same cavity QED system implements an SK spin glass with spin 1/2 particles, simulations of the system reveal that the open dissipative quantum dynamics drive the coupled spins to enter highly entangled quantum states, which in turn allow the system to evade semiclassical energy barriers, thereby arriving at lower energy states more quickly relative to the more semi-classical dissipative dynamics \cite{Marsh2023-wy}.  It would be interesting to explore whether analogous effects related to optimization benefits arise in quantum versions of the CIM.

In summary, the solution of combinatorial optimization algorithms using novel physical computing hardware is a rich and emerging field.  Our initial theoretical analysis of the coherent Ising machine in the classical limit reveals a rich theory with diverse connections across physics and mathematics, spanning spin glasses, the replica method, the cavity method, supersymmetry breaking, Dyson's Brownian motion, random matrix theory, and the statistical mechanics of random landscapes.  Moreover, analysis combining these topics yields geometric insights into the nature of optimal annealing schedules and the computational power of geometric landscape annealing in optimization.  Future directions of theory suggest the potential for usefully connecting to even more diverse topics, including Kuramoto networks, chaos, and open dissipative quantum dynamics.  Given the recent emerging interest in diverse physical computing devices, spanning spintronic \cite{grimaldi2023evaluating}, memristor \cite{english2022ising}, photonic/optical \cite{MohseniMcMahonByrnes2022, pierangeli2020adiabatic}, and CMOS substrates \cite{wang2019new}, for solving diverse NP-hard combinatorial optimization problems, we hope our initial theoretical analysis may inspire much future work aimed at understanding general approaches for how annealed nonlinear analog dynamical systems can aid in solving discrete optimization problems, thereby merging the primarily analog worlds of physics with the primarily discrete worlds of computer science.      
\begin{center}
    {\bf Acknowledgement}
\end{center}

This work has been supported by the National Science Foundation under award CCF-1918549. The authors wish to thank NTT Research for their support. Atsushi Yamamura is supported by the Masason Foundation. Surya Ganguli acknowledges funding from NSF
CAREER award \#1845166. The authors thank Timothée Leleu, Evan Laksono, Daniel Wennberg, Niharika Gunturu, Edwin Ng, and Ryotatsu Yanagimoto for helpful discussions.



\newpage
\renewcommand{\thesection}{Appendix}
\section{Details of comparison of theory and numerical experiments in Fig.~\ref{fig:typical-crt},\ref{fig:typical-min},\ref{fig:global-min}}

\subsection{Numerical exploration of critical points}
\label{sec:num-process}
Here we explain how we numerically sampled typical critical points in Fig.~\ref{fig:typical-crt}, typical minima in Fig.~\ref{fig:typical-min}, and global minima in Fig.~\ref{fig:global-min} for different realizations of the connectivity $J$. 

In Fig.~\ref{fig:typical-crt}
we sampled critical points via Newton's method with many different initializations.
Note that Newton's method converges not only to local minima but also to critical points of any index\cite{dauphin2014identifying}, and hence works as an efficient sampler of all critical points.
Table~I shows the experimental parameters, including the system size $N$, chosen for each value of $a$.
The chosen system size $N$ decreases as $a$ grows.
This is because the number of critical points increases exponentially, and hence it is difficult to sample all the critical points with large $N$ in the large-$a$ regime.
The initial states are sampled from a centered gaussian distribution for $a\leq 1$, while for $a > 1$, we initialize at every point of $\{-\sqrt{a},0,\sqrt{a}\}^N$.
After the deduplication of the sampled critical points, we estimated the most frequent values of the energy and the index. 
In order to select specifically the most typical critical points, we focused only on sampled critical points with the most frequent values of both the energy and index. 
The most frequent index is plotted in the figure with the error bar of length $1/N$ reflecting the minimum discretization of the fractional index $r$ at finite $N$. 
To find the most likely energy, we discretized energy into $N_{\operatorname{bins}}$ bins and included only critical points whose energy is in the most likely bin.
From this restricted set of the most typical critical points, we can directly compute order parameters $q$ in \eqref{eq:orderparamq}, and $t$ in \eqref{eq:orderparamt} for each typical critical point. 
Then, we computed the average values of $q$ and $t$ over all the typical critical points for each instance. 
For each sampled instance $J$, we compute the means of order parameters across all the typical critical points.
In the plots of $q$ and $t$ in Fig.~\ref{fig:typical-crt}, the error bars represent the standard deviation of the mean over the different instances of $J$. 
We sampled more realizations of $J$ at smaller $a$ to compensate for the fact that there are fewer critical points at smaller $a$.
To compute the distribution of spins $P(x)$, we computed the empirical histogram of individual spin values across the ensemble of typical critical points with the most frequent energy and index.
The Hessian eigen-spectrum was computed similarly as ensembles of the eigenvalues of those sampled critical points.

In Fig.~\ref{fig:typical-min}
we minimize the energy function by the Newton-CG method from many randomly initialized points to sample minima. Table~II shows the number of samples $N_{\operatorname{sample}}$ as well as the other experimental parameters, such as the system size $N$ chosen for each value of $a$. After the deduplication of the sampled minima, we focus on the typical minima, defined as the minima with the most likely energy in a bin among $N_{\operatorname{bins}}$ energy bins (just as we did for critical points above).
We computed the order parameters $q,t$ from this set of minima. The error bars in the figure represent the standard deviation across instances of $J$ of the values over all typical minima for each instance.
The complexity is calculated for $a>-1.5$ as $N^{-1}\log (\mathcal{N}_{\max}/\delta E)$, where $\mathcal{N}_{\max}$ is the number of minima in the most likely energy bin and $\delta E$ is the bin width. 
For $a\leq -1.5$, the numbers of minima we obtained are not enough to estimate the density $\mathcal{N}_{\max}/\delta E$, and hence we instead estimated the complexity as the logarithm of the total number of minima, divided by $N$.
We compute the ensemble of $P(x)$ (Fig.~\ref{fig:typical-min}-a) and the Hessian spectrum (Fig.~\ref{fig:typical-min}-b) across all minima in the most likely energy bin. 
The system size $N$ is selected according to the value of $a$, from $N=15$ to $N=10^3$ (Table~II).
Since the complexity $\Sigma$ is smaller for smaller $a$, we used a larger $N$ for smaller $a$ to be able to more accurately estimate the smaller complexity.  

In Fig.~\ref{fig:global-min} we sampled minima in the same manner above using the Newton-CG method with $N_{\operatorname{sample}}$ different initialized points and chose the lowest energy state from each of $N_J$ instances. The values of $N$, $N_{\operatorname{sample}}$, and $N_J$ for each value of $a$ are displayed in Table~III. The order parameters and the distributions are computed as the ensemble of those sampled lowest energy states.

\subsection{Solving self-consistent equations for order parameters}
\label{sec:selfconssol}
In the following, we describe our approach to solving the self-consistent equations \eqref{eq:self-consistent-replica}, for each value of the gain parameter $a$. Our solution is obtained by iteratively updating the order parameters until they satisfy the self-consistent relations. To simplify the iterative process, we fix a value of $a_{\operatorname{eff}}=a+t$ rather than fixing the gain parameter $a$.

With a fixed value of $a_{\operatorname{eff}}$, we proceed to iteratively update $q, A$, and $Cq$ with the following equations starting from  $q= 1.0$, $A=0.5$, and $Cq = 2.0$.
\begin{eqnarray}
\left\{
  \begin{aligned}
  & q \leftarrow \braket{x^2}\\
  & A \leftarrow \frac{\braket{xh(x)}}{2q} - \frac{1}{2}\Braket{\frac{1}{3x^2 - a_{\operatorname{eff}}}}\\
  & Cq \leftarrow -1 + q^{-1}\Braket{h^2(x)} - 2q^{-1}A\Braket{xh(x)} +A^2.
  \end{aligned}\right.
  \nonumber\\
\end{eqnarray}

We perform these updates for a total of 300 iterations, after which the variables converge within an error margin of $10^{-6}$. Subsequently, we compute $t$ using the following equation: $t=\Braket{\frac{1}{3x^2 - a_{\operatorname{eff}}}}$. Finally, we determine the value of $a$ by calculating $a = a_{\operatorname{eff}}-t$.

\begin{widetext}
\begin{minipage}{\linewidth}
\centering
\begin{tabular}{l|l|cccccccccccccccc}
\toprule
    parameters & description & values\\
    \hline
    $a$& gain parameter&-1.5& -1.0&-0.5& 0.0& 0.5& 1.0& 1.5& 2.0& 2.5 & 3.0& 3.5 & 4.0 & 4.5 & 5.0 \\
     $N$& system size&400& 400&120&48& 20&20&14&14&12&12&12&12&12&12\\
     $N_{\operatorname{sample}}$&\begin{tabular}{l} number of sampled\\minima for \\each instance\end{tabular}
&1e3&2e3&1e5&3e5&4e5&1.6e6&$3^{14}$&$3^{14}$&$3^{12}$&$3^{12}$&$3^{12}$&$3^{12}$&$3^{12}$&$3^{12}$\\
    $N_{J}$ & \begin{tabular}{l}number of sampled\\instances \end{tabular} &20& 20& 20& 10&10&5&5&5&5&5&5&5&5&5\\
$N_{\operatorname{bins}}$&\begin{tabular}{l}number of bins for\\energy histogram\end{tabular}  &100&100&100&100&100&100&100&100&100&100&100&100&100&100\\
\hline
\end{tabular}
\\
\vspace{2mm}
TABLE I {\bf Experimental parameters for sampling typical critical points.}
\vspace{5mm}
\label{table:typical-crt}
\end{minipage}
\begin{minipage}{\linewidth}
\centering
\begin{tabular}{l|l|cccccccccccccccc}
\toprule
    parameters & description & values\\
    \hline
    $a$& gain parameter & -2.0& -1.75& -1.5& -1.25& -1.0& -0.75& -0.5& -0.25& 0.0& 0.25& 1.0& 2.0& 3.0& 4.0 & 5.0 \\
     $N$& system size & 1000 & 1000& 1000& 1000& 400& 200& 150& 100& 50& 30& 30& 20& 20 & 15 &20 \\
     $N_{J}$ & \begin{tabular}{l}number of sampled\\instances\end{tabular}  &20& 20& 20& 20& 20& 20&20& 20& 20&20& 20& 20&20& 20& 20\\
    $N_{\operatorname{sample}}$&\begin{tabular}{l}number of sampled\\minima for \\each instance\end{tabular}&7e1& 7e3& 7e3& 7e3& -7e3& 7e5& 1.4e6& 2.8e6& 2.8e6& 2.8e6& 2.8e6& 7e6& 7e6& 7e6& 7e6\\
$N_{\operatorname{bins}}$&\begin{tabular}{l}number of bins for\\energy histogram
  \end{tabular}  &10&10&10&20&25&25&50&100&100&50&25&500&500&500&500\\
\hline
\end{tabular}
\\
\vspace{2mm}
TABLE II {\bf Experimental parameters for sampling typical minima}
\vspace{5mm}
\label{table:typical-min}
\end{minipage}

\begin{minipage}{\linewidth}
\centering
\begin{tabular}{l|l|cccccccccc}
\toprule
    parameter & description & values\\
    \hline 
    $a$&gain 
    parameter&-4.0& -3.0& -2.0& -1.25& -1.0& -0.5& 0.0& 1.0& 2.0& 4.0 \\
     $N$&system size&1000& 1000& 1000& 1000& 800& 200& 100& 40& 40& 20 \\
     $N_J$&\begin{tabular}{l}number of sampled\\instances\end{tabular} &20& 20& 20& 20& 20& 20& 20& 200& 200 & 200\\
    $N_{\operatorname{sample}}$&\begin{tabular}{l}number of sampled\\minima for \\each instance\end{tabular}& 7e1& 7e1& 7e1& 7e3& 2.8e6& 1.4e6& 2.8e6& 7e6& 7e6 & 7e6\\
\hline
\end{tabular}
\\
\vspace{2mm}
TABLE III {\bf Experimental parameters for sampling global minima}
\vspace{5mm}
\label{table:global-min}
\end{minipage}
\end{widetext}
\widetext
\clearpage
\begin{center}
\textbf{\large Supplemental Materials:\\ Geometric landscape annealing as an optimization principle
underlying the coherent Ising machine}
\end{center}
\setcounter{equation}{0}
\setcounter{section}{0}
\setcounter{subsection}{0}
\renewcommand{\thesection}{S-\Roman{section}}
\renewcommand{\theequation}{S-\arabic{equation}}
\renewcommand{\thefigure}{S-\arabic{figure}}

\section{Overview of supplementary material}

In the supplementary material, we provide detailed derivations of the properties of typical critical points, typical local minima, and the global minimum for the energy landscape of the coherent Ising machine (CIM) as a function of the laser pump parameter $a$ in the case of a Sherrington-Kirkpatrick (SK) type random connectivity.  In particular, the properties of critical points we derive include their complexity, relevant order parameters, the distribution of spins, and the eigenvalue distributions of their Hessians. Moreover, starting from the Kac-Rice formula, we provide two different derivations for each of these properties, one based on the replica method, and the other based on the cavity method, and we demonstrate their equivalence.    

Research on calculating the complexity of critical points in random high-dimensional energy landscapes has a long history. Indeed such calculations have been performed for the Thouless-Anderson-Palmer(TAP) free energy landscape of a few spin-glass models such as the Sherrington-Kirkpatrick (SK) model \cite{brayMetastableStatesSpin1980,brayMetastableStatesSolvable1981}, Ising p-spin model \cite{crisantiComplexityMeanfieldSpinglass2005}, and the p-spin spherical model \cite{crisantiComplexitySphericalMathsfp2003}. In these calculations, the authors disregard the modulus of determinant in the Kac-Rice formula, and simply average over the determinant itself. Disregarding this modulus results in the calculation of the Euler characteristic (i.e. the sum of the number of critical points with even index minus the number with odd index), not the total number of critical points. However, it turned out that the results of those calculations happen to be equivalent to the complexity of minima for several models. For example, disregarding the modulus of the determinant can be justified for the SK model as follows \cite{aspelmeierComplexityIsingSpin2004}. The TAP free energy landscape for the SK model has only local minima and index-one saddles, which occur in equal numbers and appear as pairs except for the trivial minimum at the origin in the space of magnetizations. The Hessian eigenspectra of these critical points consist of a continuous positive band and an isolated eigenvalue fluctuating around the origin, whose sign controls the index. By ignoring the small fluctuation of the isolated eigenvalue, one can count all the critical points without the negative weight for index-one saddles, which yields the complexity of minima plus index-one saddles. Since the number of index-one saddles and minima differ just by one, the result is equivalent to the complexity of minima.

As opposed to the SK model's TAP free energy, the CIM energy landscape has many high-index saddles, and the Hessian eigenspectra of its critical points generically do not have an isolated eigenvalue near the origin, but rather have a continuous density approaching the origin. Therefore we cannot ignore the modulus of the determinant in the Kac-Rice formula. In this supplemental material, we will show how to handle this issue.

In section \ref{supsec:replica}, we start from the Kac-Rice formula and develop a replica based average formula for the grand potential that includes both supersymmetry (SUSY) breaking order parameters and replica symmetry breaking order parameters.  We further derive self-consistent equations for these order parameters, as well as formulas for the distributions of spins and Hessian eigenvalues associated with various critical points of the energy landscape. We discuss the cases of typical critical points and typical minima (for which we break SUSY but not replica symmetry) and global minima (for which we preserve SUSY but break replica symmetry).  These formulas are all successfully tested against numerical experiments as described in the main paper. In section ~\ref{supsec:freeenergy} we calculate the free energy of the CIM with SK connectivity under full replica symmetry breaking. Furthermore, as a demonstration of the internal consistency of our calculations, we show that this free energy calculation fully agrees with the complexity calculation of the grand potential of the global minimum in the low-temperature limit in section \ref{supsec:replica}. Next, in section \ref{supsec:generalcavity} we provide a second derivation of the geometry of critical points in the CIM energy landscape using a generalized cavity method, and demonstrate that the results are equivalent to those in section \ref{supsec:replica}.  However, this derivation has the advantage that it is more transparent. First, it provides a new geometric interpretation of the supersymmetry breaking order parameters in terms of the susceptibility of the grand potential to small changes in the CIM energy function.  Second, it provides new simple formulae that show that counting critical points in the high dimensional CIM energy landscape with SK connectivity can be reduced to counting critical points in an equivalent ensemble of low dimensional random mean field energy landscapes over a {\it single} spin.  Thus as usual, the cavity derivation in section \ref{supsec:generalcavity} provides further insights into the replica derivation in section \ref{supsec:replica}.    

\section{Replica calculation of the complexity of critical points}
\label{supsec:replica}
In this section, we derive the complexity, distribution of spins, and Hessian eigenspectra  of critical points for the high dimensional energy landscape given by 
\begin{equation}
    E(\mathbf{x}) = \sum_{i=1}^{N}E_I(x_i) - \frac{1}{2} \sum_{i,j=1}^{N} J_{ij}x_i x_j.
\end{equation}
Here $\mathbf{x} \in \mathbb{R}^N$ represents a configuration of $N$ soft-spins, the $N$ by $N$ interaction matrix $J$ is randomly sampled from the Gaussian Orthogonal Ensemble (GOE), and $E_I(x)$ is the common single-site energy function of every spin. For the CIM, we have $E_I(x) = \frac{1}{4}x^4 - \frac{a}{2}x^2$ where $a$ is the laser pump parameter. 
Every critical point by definition satisfies the stationary condition
\begin{equation}
    \sum_j J_{ij} x_j = \partial E_I(x_i),
\end{equation}
where $\partial E_I(x) = \frac{d}{dx} E_I(x)$. Also the Hessian of the energy landscape at any point in spin configuration space $\mathbf{x}$ is given by the $N$ by $N$ matrix with entries
\begin{equation}
    H(\mathbf{x})_{ij} = \delta_{ij} \partial^2 E_I(x_i) - J_{ij}. 
\label{eq:hessdef}
\end{equation}

The structure of this section is the following. First, in section \ref{supsubsec:replkacrice} we derive the quenched-level complexity of critical points with a fixed index and an energy level. From this general formula, we derive the complexities of the typical critical points, typical minima and the global minima. These formulas are successfully compared against numerical experiments in our main paper.

\subsection{A replicated Kac-Rice formula for the grand potential}
\label{supsubsec:replkacrice}
The complexity of critical points at energy level $Ne$ and index $Nr$ is defined as
\begin{equation}
    \Sigma(e, r|J) = \frac{1}{N}\log\sum_{\alpha\in \operatorname{Crt}(E)}
    \delta(e - N^{-1}E(\mathbf{x^{\alpha}}))\delta(r - N^{-1}\mathcal{I}(\mathbf{x^\alpha})).
    \label{eq:complexitydef}
\end{equation}
Here $\operatorname{Crt}(E)$ denotes the set of critical points of the energy landscape $E(\mathbf{x})$, $\mathbf{x}^\alpha$ represents the spin-configuration at critical point $\alpha$, and $\mathcal{I}(\mathbf{x}^\alpha)$ is the index of the critical point at $\mathbf{x}^\alpha$, which is by definition the number of negative eigenvalues of the Hessian $H(\mathbf{x}^\alpha)$ in  \eqref{eq:hessdef}.

Instead of calculating the complexity directly, we first calculate the grand potential $\Omega(\beta,\mu|J)$ defined as
\begin{equation}
    \exp(-N\beta\Omega(\beta, \mu|J)) = \sum_{\alpha\in \operatorname{Crt}(E)}
        e^{-\beta E(x^\alpha) + \mu \mathcal{I}(x^\alpha)}.
    \label{eqs:grandpotdef}
\end{equation}
This grand potential is the Legendre transform of the complexity with the effective inverse  temperature $\beta$ acting as the Legendre dual of the energy density $e$, and the chemical potential $\mu$ acting as the Legendre dual of the intensive index $r$. This Legendre dual relationship can be seen as follows:
\begin{eqnarray}
    \beta\Omega(\beta, \mu|J) &&= -N^{-1}\log \sum_{\alpha\in \operatorname{Crt}(E)}
        e^{-\beta E(x^\alpha) + \mu \mathcal{I}(x^a)} \nonumber\\
    &&= -N^{-1}\log \int de dr e^{N\Sigma(e,r|J)}
        e^{-N\beta e + N\mu r} \nonumber\\
    &&= \mathrm{inf}_{e,r} \left[ \beta e - \mu r - \Sigma(e,r|J)\right].
    \label{eq:grandpotdef1}
\end{eqnarray}
As a consequence, the complexity can be obtained from the grand potential via the inverse Legendre transform:
\begin{equation}
    \Sigma(e,r|J) = \mathrm{inf}_{\beta,\mu} \left[ \beta e -\mu r  - \beta\Omega(\beta, \mu|J)\right].
    \label{eq:invlegendre}
\end{equation}

Now we expect that both the grand potential $\beta\Omega(\beta, \mu|J)$ and the complexity $\Sigma(e,r|J)$ should be self-averaging, i.e. their fluctuations across realizations of $J$ should concentrate tightly about their mean over $J$.  We therefore begin by computing the quenched average of the grand potential, exploiting the replica method:
\begin{equation}
    -\beta\Omega(\beta,\mu) := -\beta\Braket{\Omega(\beta,\mu|J)}_J = N^{-1} \Braket{\log(Z)}_J = N^{-1}\lim_{n\to 0} n^{-1}\log \Braket{Z^n}_J,
\label{eq:potav}    
\end{equation}
where $\braket{\cdot}_J$ represents an ensemble average over the realizations of the connectivity matrix $J$, and
\begin{eqnarray}
    Z
    &:=& \sum_{\alpha\in \operatorname{Crt}(E)}e^{- \beta E(\mathbf{x}^{\alpha}) 
        + \mu \mathcal{I}(\mathbf{x^\alpha})}\nonumber \\
    &=& \int \prod_{i}dx_i\ \left[\prod_i\delta(\partial_i E(\mathbf{x}))\right] |\det[H(x)]|
        e^{-\beta E(\mathbf{x}) + \mu \mathcal{I}(\mathbf{x})} \nonumber \\
    &=& \int \prod_{i}dx_i \int^{i\infty}_{-i\infty} \prod_{i} \frac{du_i}{2\pi i}\ \exp\left[\sum_i u_i \partial_i E(\mathbf{x}) -\beta E(\mathbf{x})\right] |\det[H(\mathbf{x})]|
    e^{ \mu \mathcal{I}(\mathbf{x})}.
    \label{eq:Z}
\end{eqnarray}
Here $\partial_i := \frac{\partial}{\partial x_i}$ and $H(\mathbf{x})$ is the Hessian defined in  \eqref{eq:hessdef}. In the second line of  \eqref{eq:Z} we have used the Kac-Rice formula \cite{adler2007random} to replace the sum over all critical points with an integral over all spin-configurations. Also in the third line of  \eqref{eq:Z} we have used a standard integral representation of the $\delta$ function, thereby introducing new integration variables $u_i$ that are dual to the gradient conditions $\partial_i E(\mathbf{x})=0$. The replicated partition function $Z^n$ can then be written as an integral over replicated variables $x^a_i$ and $u^a_i$ for $a=1,\dots,n$ via
\begin{equation}
    Z^n = \int \prod_{i,a}dx^a_i \int^{i\infty}_{-i\infty} \prod_{i,a} \frac{du^a_i}{2\pi i}\ \exp\left[\sum_i u^a_i \partial_i E(\mathbf{x}^a) -\beta E(\mathbf{x}^a)\right] |\det[H(x^a)]|
    e^{ \mu \mathcal{I}(\mathbf{x}^a)}.
    \label{eq:Zn}
\end{equation}
Then inserting this expression for $Z^n$ into  \eqref{eq:potav}, performing a change of variables in $J_{ij}$ \cite{brayMetastableStatesSpin1980},  and introducing new auxiliary variables via Hubbard-Stratonovich transformations \cite{annibaleRoleBecchiRouet2003}, we obtain (see Appendix \ref{app:hubbard} for details):
\begin{eqnarray}
    -\beta\Omega(\beta, \mu)
    &=&\lim_{n\to 0}\frac{1}{n} 
            \operatorname{Ext}_\Theta \left(\Omega'_0(\Theta) 
            + N^{-1}\log \left[\int \prod_{i,a}dx_i^a \int^{i\infty}_{-i\infty} \prod_{i,a} \frac{du_i^a}{2\pi i} e^{\sum_i S'(\Theta, \mathbf{x}_i, \mathbf{u}_i)}
          \Braket{\prod_a\lvert\det H(\mathbf{x}^a)\rvert e^{\mu\mathcal{I}(\mathbf{x}^a)}}_J\right] \right)\nonumber\\
    \label{eq:Omega_before_detH}
\end{eqnarray}
Here $\Theta$ is a tuple of several auxiliary variables, 
\begin{equation}
    \Theta = (\{q^{ab}\}, \{w^{ab}\}, \{\lambda^{ab}\}),
\end{equation}
$\operatorname{Ext}_{\Theta}$ means extremization with respect all variables $\Theta$, and 
$\Omega'_0(\Theta)$ and $S'(\Theta, x, u)$ are given by
\begin{eqnarray}
    \Omega'_0(\Theta) &=& \sum_{a,b}
    \left[-\frac{1}{2}w^{ab}w^{ba} - \beta w^{ab}q^{ab}
        - \frac{\beta^2}{4}(q^{ab})^2 - \lambda^{ab} q^{ab}\right] \nonumber\\
    S'(\Theta, x, u) &=&
    \sum_a (-\beta E_I(x^a) + u^a \partial E_I(x^a)) 
    + \sum_{ab} \left[w^{ab}x^a u^b + \frac{1}{2} q^{ab}u^a u^b + \lambda^{ab} x^a x^b\right].
    \label{eq:action}
\end{eqnarray}

Now in \eqref{eq:Omega_before_detH}, we must still average, with respect to the connectivity matrix $J$, a term involving 
products of the modulus of the determinant of Hessian, $\lvert\det H(\mathbf{x}^a)\rvert$ and 
the chemical potential factor $e^{\mu\mathcal{I}(\mathbf{x}^a)}$. For simplicity, we neglect correlations between replicas and those between the determinants and the chemical potential factors in the average over $J$, thereby
assuming that the average of the product factorizes into the product of the averages.
\begin{eqnarray}
    \Braket{\prod_a\lvert\det H(\mathbf{x}^a)\rvert
e^{\mu\mathcal{I}(\mathbf{x}^a)}}_J \approx 
\prod_a\Braket{\lvert\det H(\mathbf{x}^a)\rvert}_J\Braket{ e^{\mu\mathcal{I}(\mathbf{x}^a)}}_J.
\end{eqnarray}
In the following, we will evaluate these averages $\Braket{\lvert\det H(\mathbf{x})\rvert}_J$ and $\Braket{ e^{\mu\mathcal{I}(\mathbf{x})}}_J$ for a given vector $\mathbf{x}$, and substitute them into \eqref{eq:Omega_before_detH}. We exploit the formula \cite{cavagnaIndexDistributionRandom2000}
\begin{equation}
    \mathcal{I}(\mathbf{x}) =  \lim_{\epsilon\to 0}\frac{1}{2\pi i}
        \left( \log\det[H(\mathbf{x}) - i\epsilon]
            - \log\det[H(\mathbf{x}) + i\epsilon]\right).
    \label{eq:index}
\end{equation}
This formula can be derived via Cauchy's residue theorem with a contour integral along a loop $C$ surrounding all the negative eigenvalues.
\begin{eqnarray}
    \mathcal{I}(\mathbf{x}) &=& \frac{1}{2\pi i}\int_C \sum_i\frac{1}{z -\lambda_i} dz\nonumber\\
    &=& \lim_{\epsilon\to 0}\frac{1}{2\pi x i}\left(\int_{s_l+i\epsilon}^{i\epsilon} \sum_i\frac{1}{z -\lambda_i} dz - \int_{s_l-i\epsilon}^{-i\epsilon} \sum_i\frac{1}{z -\lambda_i} dz\right)\nonumber\\
    &=&  \lim_{\epsilon\to 0}\frac{1}{2\pi i}\sum_i\left( (\log(\lambda_i - i\epsilon) - \log(\lambda_i - s_l - i\epsilon)) -  (\log(\lambda_i + i\epsilon) - \log(\lambda_i + s_l - i\epsilon))\right)\nonumber\\
    &=&  \lim_{\epsilon\to 0}\frac{1}{2\pi i}\sum_i\left( \log(\lambda_i - i\epsilon) -  \log(\lambda_i + i\epsilon)\right) \nonumber\\
    &=& \text{(RHS of \eqref{eq:index})}.
\end{eqnarray}
Here $s_L < \min_i \lambda_i$, and we choose the branch cut of the complex logarithm function such that $\log(\lambda_i - s_l \pm i\epsilon)$ are on opposite sides of the branch cut. Using this formula, we see that \cite{fyodorovCountingStationaryPoints2005}
\begin{eqnarray}
    e^{\mu\mathcal{I}(\mathbf{x})}
    &=&\lim_{\epsilon\to 0}
        \left[\det[H(\mathbf{x}) - i\epsilon]\right]^{\mu/ (2\pi i)}
            \left[\det[H(\mathbf{x}) + i\epsilon]\right]^{-\mu/ (2\pi i)}.  \nonumber\\
    |\det[H(\mathbf{x})]|
    &=&\det(H(\mathbf{x})) e^{\xi\pi i \mathcal{I}(\mathbf{x})}\nonumber\\
    &=&\lim_{\epsilon\to 0}
        \det[H(\mathbf{x})]\left[\det[H(\mathbf{x}) - i\epsilon]\right]^{\xi/2}
            \left[\det[H(\mathbf{x}) + i\epsilon]\right]^{-\xi/2}.  
\end{eqnarray}
Here $\xi$ can be either $+1$ or $-1$, and the choice of $\xi$ will be judiciously made later to simplify expressions. 

Next, we further neglect correlations between determinants of $H(x)\pm i\epsilon$ and $H(x)$.
\begin{eqnarray}
    \Braket{e^{\mu\mathcal{I}(\mathbf{x})}}_J
    &\approx&\lim_{\epsilon\to 0}
        \Braket{\det[H(\mathbf{x}) - i\epsilon]}_{J}^{\mu/ (2\pi i)}
            \Braket{\det[H(\mathbf{x}) + i\epsilon]}_J^{-\mu/ (2\pi i)}.  \nonumber\\
    \Braket{|\det[H(\mathbf{x})]|}_J
    &\approx&\lim_{\epsilon\to 0}
        \Braket{\det[H(\mathbf{x})]}_J\Braket{\det[H(\mathbf{x}) - i\epsilon]}_J^{\xi/2}
            \Braket{\det[H(\mathbf{x}) + i\epsilon}_J^{-\xi/2}.  
\end{eqnarray}
We can compute the average of $\det[H(\mathbf{x}) - zI]$ with respect to $J$
for a general $z\in\mathbb{C}$ (eventually we will focus on the specific values $z=0\pm i\epsilon$) as follows (see Appendix \ref{app:rmthess} for a detailed derivation):
\begin{equation}
    \Braket{\det[H(\mathbf{x}) - zI]}_J =\int^{i\infty}_{-i\infty} dt
    \exp\left(\frac{N}{2}t^2 +\sum_i \log(\partial^2 E_I(x_i)- z - t)\right).
    \label{eq:det_H_integral}
\end{equation}
This integral can be performed via a saddle point approximation.  The saddle point condition, obtained by extremizing the exponent, yields
\begin{equation}
    t(z) = N^{-1}\sum_i\frac{1}{\partial^2 E_I(x_i)- z - t(z)}.
    \label{eq:saddletz}
\end{equation}
The saddle point approximation to the integral in  \eqref{eq:det_H_integral} then yields
\begin{equation}
    \log\Braket{\det[H(\mathbf{x}) - zI]}_J = \frac{N}{2}t^2(z) +\sum_i \log(\partial^2 E_I(x_i)- z - t(z)),
    \label{eq:logdet_H-z}
\end{equation}
where $t(z)$ obeys the saddle point condition in  \eqref{eq:saddletz}. 
Note that $t(z^*) = t^*(z)$ and therefore for $z=0$, we can have two relevant extrema $t(\pm i0) = t_R \pm t_I i$ if $t_I\neq 0$. At these extrema, we can evaluate $\Braket{e^{\mu\mathcal{I}(\mathbf{x})}}_J$ and $\Braket{\lvert\det[H(\mathbf{x})]\rvert }_J$ as follows.

\subsubsection{Average of the chemical potential factor}
\begin{eqnarray}
    \log\Braket{e^{\mu\mathcal{I}(\mathbf{x})}}_J
    &\approx&\frac{\mu}{2\pi i}\lim_{\epsilon\to 0}\left[\log\Braket{\left[\det[H(\mathbf{x}) - i\epsilon]\right]}_J- \log\Braket{\left[\det[H(\mathbf{x}) + i\epsilon]\right]}_J\right]\nonumber\\
    &=& \frac{\mu}{2\pi i}\lim_{\epsilon\to 0}\left[\frac{N}{2}(t^2(i\epsilon) -t^2(-i\epsilon)) +\sum_i \log(\partial^2 E_I(x_i)- i\epsilon - t(i\epsilon)) - \log(\partial^2 E_I(x_i) +i\epsilon - t(-i\epsilon)) \right]\nonumber\\
    &=&\mu\lim_{\epsilon\to 0}\sum_{i}
             \left(\frac{t_R t_I}{\pi}+\frac{1}{2\pi i}\left(\log\left[\partial^2 E_I(x_i)-i\epsilon-t(i\epsilon)\right] - \log\left[\partial^2 E_I(x_i)+i\epsilon-t(-i\epsilon)\right]\right)\right)\nonumber\\
    &=&\mu\sum_i\overline{\mathcal{I}}(x_i),
    \label{eq:logchempot}
\end{eqnarray}
where
\begin{equation}
    \overline{\mathcal{I}}(x) = \frac{t_R t_I}{\pi}-\frac{1}{\pi}\tan^{-1}\left(\frac{t_I}{\partial^2 E_I(x)-t_R} \right) + \lim_{\epsilon\to0}\Theta\left(\frac{\operatorname{Im}(t(i\epsilon))+\epsilon}{\partial^2 E_I(x)-\operatorname{Re}(t(i\epsilon))}\right).
    \label{eqs:mean-field-index}
\end{equation}
Here $\Theta$ is the Heaviside step function.  We note, as described near the end of subsection \ref{supsubsec:mfindex}, that when $t_I=0$, the mean-field index function $\overline{\mathcal{I}}(x)$ reduces to a simple indicator function given in  \eqref{eq:mean-field-index-t_I0}, and this indicator function is simply the index of a simple scalar mean field energy function given in  \eqref{eq:emf}.  We furthermore note, as described near the end of \ref{supsubsec:eigdisthess}, that the simplification $t_I=0$ can made precisely when the eigenvalue density of the Hessian at the origin vanishes.

\subsubsection{Average of the modulus of the determinant}
For the average of the modulus of the determinant, we obtain: 
\begin{eqnarray}
    &&\log\Braket{\lvert\det[H(\mathbf{x})]\rvert}_J\nonumber\\
    &\approx& \lim_{\epsilon\to 0}\left[\log\Braket{\det[H(\mathbf{x})]}_J+\frac{\xi}{2}\log\Braket{\left[\det[H(\mathbf{x}) - i\epsilon]\right]}_J-\frac{\xi}{2} \log\Braket{\left[\det[H(\mathbf{x}) + i\epsilon]\right]}_J\right]\nonumber\\
    &=& \frac{N}{2}(t_R \pm t_I i)^2 +\sum_i \log(\partial^2 E_I(x_i) - t_R -\mp t_Ii) \nonumber\\
    &&+ \lim_{\epsilon\to 0}\frac{\xi}{2}\left[\frac{N}{2}(t^2(i\epsilon) -t^2(-i\epsilon)) +\sum_i \log(\partial^2 E_I(x_i)- i\epsilon - t(i\epsilon)) - \log(\partial^2 E_I(x_i) +i\epsilon - t(-i\epsilon)) \right]\nonumber\\
    &=&\lim_{\epsilon\to 0}\sum_{i} 
            \frac{1}{2}(t_R^2 - t_I^2) + (\xi\pm 1)t_R t_I i
        +\frac{\xi}{2}\log\frac{\partial^2 E_I(x_i) - i\epsilon-t(i\epsilon)}{\partial^2 E_I(x_i)+i\epsilon-t(-i\epsilon)} +\log\left[\partial^2 E_I(x_i)-t_R \mp t_I i \right].
    \label{eq:ave_logdet_H}
\end{eqnarray}
Here the choice of signs corresponds to the choice of the extremum. By choosing the sign of $\xi$ wisely, we can see that the quantity above is independent of the choice of the extremum.
\begin{eqnarray}
    &&\text{(the right hand side of \eqref{eq:ave_logdet_H})}\nonumber\\
    &=&\lim_{\epsilon\to 0}\sum_{i} 
            \frac{1}{2}(t_R^2 - t_I^2) + (\mp 1\pm 1)t_R t_I i
        \mp\frac{1}{2}\log\frac{\partial^2 E_I(x_i) - i\epsilon-t(i\epsilon)}{\partial^2 E_I(x_i)+i\epsilon-t(-i\epsilon)} +\log\left[\partial^2 E_I(x_i)-t_R \mp t_I i \right]\nonumber\\
        &=& \sum_{i} \frac{1}{2}(t_R^2 - t_I^2)
        +\log\sqrt{(\partial^2 E_I(x_i)-t_R)^2 + (t_I)^2}.
\end{eqnarray}
Now we obtain the formula for the average of the modulus of the determinant
\begin{equation}
     \log\Braket{\lvert\det[H(\mathbf{x})]\rvert}_J=\sum_{i} \frac{1}{2}(t_R^2 - t_I^2)
        +\log\sqrt{(\partial^2 E_I(x_i)-t_R)^2 + (t_I)^2}.
\label{eq:logmoddet}
\end{equation}
In total, we get
\begin{eqnarray}
    \log\Braket{\lvert\det[H(\mathbf{x})]\rvert e^{\mu\mathcal{I}(\mathbf{x})}}_J=\sum_{i} \left(
            \frac{1}{2}(t_R^2 - t_I^2)
        +\log\sqrt{(\partial^2 E_I(x_i)-t_R)^2 + (t_I)^2}+ \mu \overline{\mathcal{I}}(x_i)\right).
    \label{eq:chemhessav}
\end{eqnarray}

Note that this equation is for a given fixed vector $\mathbf{x}$. When we evaluate $\Braket{\lvert\det[H(\mathbf{x})]\rvert e^{\mu\mathcal{I}(\mathbf{x})}}_J$ in \eqref{eq:Omega_before_detH}, we first substitute the equation before applying the saddle point method,
\begin{eqnarray}
    \Braket{\lvert\det[H(\mathbf{x})]\rvert e^{\mu\mathcal{I}(\mathbf{x})}}_J = \lim_{\epsilon\to 0}\int dt_0 dt_+ dt_- \prod_i\exp\left[\frac{1}{2}\left(t_0^2+ \left(\frac{\mu}{2\pi i}+\frac{\xi}{2}\right)(t^2_+-t^2_-)\right)\right]\nonumber\\
    \times (\partial^2 E_I(x_i) - t_0)\left(\frac{\partial^2 E_I(x_i) - i\epsilon - t_+}{\partial^2 E_I(x_i) + i\epsilon - t_-}\right)^{\frac{\mu}{2\pi i}+\frac{\xi}{2}}
    \label{eq:detH-e_mu-before-saddle-point-method}
\end{eqnarray}
and then later we apply the saddle point method, in the same manner we discussed above.
\subsubsection{Remarks on the index, the Hessian determinant, and decoupling across spin variables}
\label{supsubsec:locality}

Note, interestingly, that before averaging over the connectivity matrix $J$, 
both $e^{\mu\mathcal{I}(\mathbf{x})}$ 
and $\lvert\det[H(\mathbf{x})]\rvert$ 
are quite complex coupled functions of all the spin variables $x_i$, depending on many higher order products of spin variables. 
However, after averaging over $J$, we obtain simple expressions 
for both $\log\Braket{e^{\mu\mathcal{I}(\mathbf{x})}}_J$
in  \eqref{eq:logchempot} 
and $\log\Braket{\lvert\det[H(\mathbf{x})]\rvert}_J$ in  \eqref{eq:logmoddet} that involve a sum of terms each depending on a {\it single} spin variable. Thus these expressions achieve a decoupling across the site index $i$, provided of course we have a solution to the saddle point condition $t(z)$ in  \eqref{eq:saddletz} evaluated at $t(z = \pm i\epsilon) = t_R \pm i t_I$. 

This solution $t(z)$ depends on all the $x_i$ in a nonlocal manner of course. Thus all of the non-local dependence on the spins $x_i$ of the averaged chemical potential factor and the averaged modulus of the Hessian determinant are isolated to the variable $t(z = \pm i\epsilon) = t_R \pm i t_I$. We will see below that $t(z)$ is nothing other than the resolvent of the Hessian $H(\mathbf{x})$.  Thus the resolvent of the Hessian, evaluated in the complex plane just above the origin, plays a key role in determining both the average chemical potential and modulus of the Hessian determinant.

With this resolvent in hand, we can think of $\overline{\mathcal{I}}(x_i)$ used in  \eqref{eq:logchempot} and defined in  \eqref{eqs:mean-field-index} as an effective contribution of each individual spin $x_i$ to the overall ensemble averaged index. More precisely, assuming the average of the exponential is well approximated by the exponential of the average, we could rewrite  \eqref{eq:logchempot} suggestively as
\begin{equation}
    \Braket{\mathcal{I}(\mathbf{x})}_J =  \sum_i\overline{\mathcal{I}}(x_i).
\end{equation}
This formula represents an interesting decomposition of the connectivity averaged index of the random matrix $H(\mathbf{x})$ defined in  \eqref{eq:hessdef} into a simple sum of mean-field contributions from each spin $x_i$. Each of these site specific contributions to the overall average index is determined by the mean field index function $\overline{\mathcal{I}}(x_i)$ in \eqref{eqs:mean-field-index}.

Similarly in  \eqref{eq:logmoddet} we can think of each term in the right hand side as the logarithm of an effective mean-field eigenvalue magnitude, and the sum of these site specific mean field terms equals the logarithm of the ensemble averaged modulus of the Hessian determinant.  More precisely, we could rewrite  \eqref{eq:logmoddet} suggestively as 
\begin{equation}
    \Braket{\lvert\det[H(\mathbf{x})]\rvert}_J\nonumber = \prod_i \lvert\lambda_{eff}(x_i) \rvert,
\end{equation}
where
\begin{equation}
    \lvert\lambda_{eff}(x) \rvert = e^{\frac{1}{2}(t_R^2 - t_I^2)
        +\log\sqrt{(\partial^2 E_I(x)-t_R)^2 + (t_I)^2}}.
\end{equation}
We will next exploit this decoupling across sites $i$ of both the ensemble averaged index and the ensemble averaged modulus of the Hessian determinant to complete our derivation of a mean-field replica theoretic formula for the grand potential. 

\subsubsection{A replica expression for the grand potential}
We now return to  \eqref{eq:Omega_before_detH} and note that the only obstruction to decoupling the integral in  \eqref{eq:Omega_before_detH} across the site index $i$ involves the connectivity averaged product of Hessian determinants and chemical potential factors.  However,  \eqref{eq:detH-e_mu-before-saddle-point-method} reveals that for a single replica, 
the average $\Braket{\lvert\det[H(\mathbf{x})]\rvert e^{\mu\mathcal{I}(\mathbf{x})}}_J$ 
has an expression that decouples, or factorizes across the sites.  Furthermore, as previously mentioned, we 
neglect correlations between replicas in this average over $J$, i.e. we assume that 
$\Braket{\prod_a\lvert\det H(\mathbf{x}^a)\rvert
e^{\mu\mathcal{I}(\mathbf{x}^a)}}_J \approx 
\prod_a\Braket{\lvert\det H(\mathbf{x}^a)\rvert e^{\mu\mathcal{I}(\mathbf{x}^a)}}_J$.  
Then under this assumption, by substituting  \eqref{eq:detH-e_mu-before-saddle-point-method} into  \eqref{eq:Omega_before_detH}, we achieve a complete decoupling of sites, and can express the grand potential in terms of an integral over $n$ replicated {\it single} spins $x^a$:
\begin{eqnarray}
    -\beta\Omega(\beta, \mu)
    &=&\lim_{n\to 0}\frac{1}{n} 
    \operatorname{Ext}\left(\Omega'_0
            + N^{-1}\lim_{\epsilon\to 0}\log\int \prod_a[dt^a_0 dt^a_+ dt^a_-]\left( \int \prod_a [dx^a du^a] e^{S+T}\right)^N\right)
\end{eqnarray}
with
\begin{eqnarray}
    \Omega'_0 &=& \sum_{a,b}
    \left[ -\frac{1}{2}w^{ab}w^{ba} - \beta w^{ab}q^{ab}
        - \frac{\beta^2}{4}(q^{ab})^2 - \lambda^{ab} q^{ab}\right] \nonumber\\
    S &=&
    \sum_a (-\beta E_I(x^a) + u^a \partial E_I(x^a)) 
    + \sum_{ab} \left[w^{ab}x^a u^b + \frac{1}{2} q^{ab}u^a u^b + \lambda^{ab} x^a x^b\right]\nonumber\\
    T &=&\sum_a\frac{1}{2}\left((t_0^a)^2+ \left(\frac{\mu}{2\pi i}+\frac{\xi}{2}\right)((t^a_+)^2-(t^a_-)^2)\right)+\log\left[(\partial^2 E_I(x^a) - t^a_0)\left(\frac{\partial^2 E_I(x^a) - i\epsilon - t^a_+}{\partial^2 E_I(x^a) + i\epsilon - t^a_-}\right)^{\frac{\mu}{2\pi i}+\frac{\xi}{2}}\right]
\end{eqnarray}
Next, we apply the saddle point method on $t_0^a,t_+^a$, and $t_-^a$. The saddle point equations give the following self-consistent equations similar to \eqref{eq:saddletz}.
\begin{eqnarray}
    t^a_0 &=& \Braket{\frac{1}{\partial^2 E_I(x^a) - t^a_0}}_{S,T}\nonumber\\
    t^a_\pm &=& \Braket{\frac{1}{\partial^2 E_I(x^a) \mp i\epsilon - t^a_\pm}}_{S,T},
\end{eqnarray}
where the average $\Braket{\cdot}_{S,T}$ is defined as
\begin{equation}
    \Braket{X}_{S,T} = \frac{\int \prod_a [dx^a du^a] Xe^{S+T}}{\int \prod_a [dx^a du^a] e^{S+T}}.
\end{equation}
Following the argument in the previous subsection, we define $t_R$ and $t_I$ as $t_R \pm t_Ii = t_\pm^a$, and we obtain
\begin{eqnarray}
    -\beta\Omega(\beta, \mu)
    &=&\lim_{n\to 0}\frac{1}{n} 
            \operatorname{Ext}\left(\Omega_0
            + \log \int \prod_a [dx^a du^a w(x^a)] e^{S+ \mu\sum_a\overline{\mathcal{I}}(x^a)}\right)
    \label{eq:Omega}
\end{eqnarray}
with
\begin{eqnarray}
    w(x) &:=& \sqrt{(\partial^2 E_I(x)-t^a_R)^2 + (t^a_I)^2} \nonumber\\
    \Omega_0 &=& \sum_{a,b}
    \left[ -\frac{1}{2}w^{ab}w^{ba} - \beta w^{ab}q^{ab}
        - \frac{\beta^2}{4}(q^{ab})^2 - \lambda^{ab} q^{ab} + \frac{1}{2}((t^a_R)^2 - (t^a_I)^2)\right] \nonumber\\
    S &=&
    \sum_a (-\beta E_I(x^a) + u^a \partial E_I(x^a)) 
    + \sum_{ab} \left[w^{ab}x^a u^b + \frac{1}{2} q^{ab}u^a u^b + \lambda^{ab} x^a x^b\right].
    \label{eq:Omega_S_replica_w_lambda}
\end{eqnarray}

In order to further simplify this expression, we introduce new variables $A^{ab}$ and $C^{ab}(=C^{ba})$ and, as is done for supersymmetry-breaking complexity calculations \cite{mullerMarginalStatesMeanfield2006}, we perform the change of variables 
\begin{eqnarray}
    \begin{cases}
        w^{ab} = -t^a_R \delta_{ab} - \beta q_{ab} - A^{ab} \nonumber \\
        \lambda^{ab} = \frac{\beta}{2} t^a_R \delta_{ab} + \frac{\beta^2}{2} q^{ab}  +\frac{\beta}{2} (A^{ab} + A^{ba}) + \frac{1}{2}C^{ab}.
    \end{cases}
\end{eqnarray}
In terms of the new variables $A^{ab}$ and $C^{ab}$, we obtain (see Appendix \ref{app:susytrans} for a detailed derivation),
\begin{eqnarray}
    \Omega_0 &=& \sum_{a,b}
    \left[ -\frac{1}{2}A^{ab}A^{ba} - A^{aa}t^a_R\delta_{ab}
        - \frac{\beta}{2}q^{aa} t^a_R\delta_{ab} - \frac{\beta}{4}(q^{ab})^2 - \beta A^{ab} q^{ab} - \frac{1}{2} C^{ab}q^{ab} - \frac{1}{2}(t^a_I)^2\right] \nonumber\\
    S &=&
    \sum_{ab}\left[\frac{1}{2}q^{ab}u^a u^b + A^{ab} u^a x^b + \frac{1}{2} C^{ab} x^a x^b\right] + \sum_{a} \left[ -u^a h^a + \beta x^a h^a -\beta (E_I(x^a) -\frac{t^a_R}{2}(x^a)^2) \right],
    \label{eq:Omega_S_replica}
\end{eqnarray}
where $h^a = \partial E_I(x^a) - t^a_R x^a$. 

In summary, we have a derived the formula for the grand potential of critical points in the form of \eqref{eq:Omega} with $\Omega_0$ and $S$ given by \eqref{eq:Omega_S_replica}, where order parameters $A^{ab}, C^{ab}, q^{ab}, t^a_R+t^a_Ii$ are obtained by extremizing \eqref{eq:Omega}. This extremization results in the stationary conditions
\begin{eqnarray}
    A^{ab} &=& \braket{x^au^b}_S-t_R\delta_{ab}\nonumber\\
    C^{ab} &=& \braket{u^au^b}_S -\beta [A^{ab} + t_R/2 \delta_{ab}]\nonumber\\
    q^{ab} &=& \Braket{x^ax^b}_S\nonumber\\
    t^a_R+t^a_Ii &=& \Braket{\frac{1}{\partial^2 E(x^a)-t^a_R-t^a_Ii}}_S.
    \label{eq:fullstatcond}
\end{eqnarray}
Here the average $\Braket{\cdot}_S$ is defined as
\begin{equation}
    \Braket{X}_S = \frac{\int \prod_a [dx^a du^a w(x^a)] Xe^{S+ \mu\sum_a\overline{\mathcal{I}}(x^a)}}{\int \prod_a [dx^a du^a w(x^a)] e^{S+ \mu\sum_a\overline{\mathcal{I}}(x^a)}}.
\end{equation}
Note that the effective action $S$ defined in \eqref{eq:Omega_S_replica} is symmetric under permutations of replicas and therefore the self-consistent equation for $t^a$ \eqref{eq:fullstatcond} is symmetric as well. This implies that $t^a_R$ and $t^a_I$ are independent of the replica index $a$.

Finally, inserting the order parameters which solve the stationary conditions  \eqref{eq:fullstatcond} into the formula for the grand potential in \eqref{eq:Omega}, and then performing the inverse Legendre transform in  \eqref{eq:invlegendre}, allows us to find connectivity averaged complexity $\Sigma(e,r)$ of critical points as a function of their intensive energy $e$ and intensive index $r$. We will perform this procedure below after making different ansatzes about the structure of the solutions to  \eqref{eq:fullstatcond} for typical critical points, typical minima, and global minima. However first, in the next two subsections, we describe other properties of critical points that we can also derive, beyond the complexity, namely the distribution of spins and the eigenvalue distribution of the Hessian. 

\subsection{The distribution of spins in ensembles of critical points}
We derive the distribution $P(x)$ of spins $\{x_i\}_{i=1,2,\cdots,N}$ across critical points.  To define such a distribution, we must first work with a distribution over critical points. We consider a distribution over critical points such that each critical point $\alpha$ with spin configuration $\mathbf{x}^\alpha$ is weighted by the same Boltzmann factor as that used in the definition of the grand potential in  \eqref{eqs:grandpotdef}, namely the factor $e^{-\beta E(x^\alpha) + \mu \mathcal{I}(x^\alpha)}$.  With this choice, the connectivity averaged distribution of spins $P(x)$ across this ensemble of critical points is defined as
\begin{equation}
    P(x)= \Braket{Z^{-1}\sum_{\alpha\in \operatorname{Crt}(E)} e^{-\beta E_I(\mathbf{x}^\alpha) + \mu\mathcal{I}(\mathbf{x}^\alpha)}  \left(\frac{1}{N}\sum_{i=1}^{N} \delta(x^\alpha_i-x)\right)}_J, 
    \label{eqs:defpx}
\end{equation}
where $Z = \sum_{\alpha} e^{-\beta E(x^\alpha) + \mu \mathcal{I}(x^\alpha)}$. This distribution $P(x)$ depends on the effective inverse temperature $\beta$ and the chemical potential $\mu$.  

If on the other hand we wish to compute the distribution of spins in a different of ensemble of critical points, namely one with a fixed intensive energy $e$ and a fixed intensive index $r$, i.e. those critical points contributing to the complexity $\Sigma(e,r)$ in  \eqref{eq:complexitydef}, then we can do so via the inverse Legendre transform in  \eqref{eq:invlegendre}.  In particular, for a given pair $e$ and $r$, we find the corresponding Legendre dual $\beta$ and $\mu$ that achieve the extremum in  \eqref{eq:invlegendre}, and insert this $\beta$ and $\mu$ into  \eqref{eqs:defpx}. The resulting distribution $P(x)$ then reflects the distribution of spins across all critical points of the given intensive energy $e$ and index $r$, since the sum over all critical points in  \eqref{eqs:defpx} is dominated by critical points with values of the energy $e$ and index $r$ that are Legendre dual to the inverse temperature $\beta$ and chemical potential $\mu$. 

Below we will consider a few special cases corresponding to simple choices for $\beta$ and $\mu$.  For example, the case of a typical critical point, chosen from a ``whitened" or uniform distribution over all possible critical points of any index and energy, corresponds to the choice $\beta=\mu=0$.  The case of a typical minimum, chosen from a uniform distribution over all possible local minima regardless of their energy, corresponds to the case $\beta=0$ and $\mu \rightarrow -\infty$.  This latter choice for $\mu$ focuses the sum over critical points in  \eqref{eqs:defpx} to a sum over index $0$ critical points, or local minima.  Finally, for the cases of the lowest energy critical points corresponding to global minima, we can take $\beta \rightarrow \infty$ and $\mu=0$.

Now, in order to compute $P(x)$ in  \eqref{eqs:defpx} for arbitrary $\beta$ and $\mu$, we first define an arbitrary function $O(x_i)$ of $x_i$, and calculate its expectation value
\begin{equation}
   \Braket{O} := \Braket{Z^{-1}\sum_{\alpha\in \operatorname{Crt}(E)} e^{-\beta E(\mathbf{x}^\alpha) + \mu\mathcal{I}(\mathbf{x}^\alpha)} \left(N^{-1}\sum_iO(x_i)\right)}_J.
\end{equation}
We will then show that  
\begin{equation}
    \braket{O} = \int dx P(x) O(x),
\end{equation}
where $P(x)$ is derived below as \eqref{eq:Px_susy_breaking}. Since these two expressions for $\braket{O}$ are equivalent for any arbitrary observable $O(x)$, we then conclude that $P(x)$ in \eqref{eq:Px_susy_breaking} is a formula for the distribution of local spins defined in \eqref{eqs:defpx}. To implement this strategy, we calculate the expectation value $\braket{O}$ as a derivative of a grand potential
\begin{equation}
    \Braket{O} := N^{-1} \left. \frac{d}{ds}\right|_{s=0} \Braket{\log \sum_{\alpha \in\operatorname{Crt}(E)} e^{-\beta E(\mathbf{x}^\alpha) + \mu\mathcal{I}(\mathbf{x}^\alpha) + s \sum_i O(x_i)}}_J.
\end{equation}
In a sequence of steps that are very similar to those carried out in Subsection \ref{supsubsec:replkacrice}, we can compute $\Braket{O}$ via the replica method, obtaining
\begin{eqnarray}
    \braket{O} &=& \lim_{n\to 0}\frac{1}{n}  \left.\frac{d}{ds}\right|_{s=0}
    \left(\Omega_0
    + \log \int \prod_a [dx^a du^a w(x^a)] e^{S+ \mu\sum_a\overline{\mathcal{I}}(x^a)+ s \sum_a O(x^a)} \right) \nonumber\\
    &=& \lim_{n\to 0}\frac{1}{n}  \left.\frac{\partial}{\partial s}\right|_{s=0}
    \log \int \prod_a [dx^a du^a w(x^a)] e^{S+ \mu\sum_a\overline{\mathcal{I}}(x^a)+ s \sum_a O(x^a)}\nonumber\\
    &=& \lim_{n\to 0}\frac{1}{n}  
    \sum_b \int \prod_a [dx^a du^a w(x^a)]\frac{e^{S+ \mu\sum_a\overline{\mathcal{I}}(x^a)}}
    {\int \prod_a [dx^a du^a w(x^a)] e^{S+ \mu\sum_a\overline{\mathcal{I}}(x^a)}} O(x^b)
\end{eqnarray}
Therefore the distribution is
\begin{equation}
    P(x) \propto \int \prod_a [dx^a du^a w(x^a)] \delta(x-x^b) e^{S+ \mu\sum_a\overline{\mathcal{I}}(x^a)},
    \label{eq:Px_susy_breaking}
\end{equation}
where $b$ is an arbitrarily chosen replica index.  Later, we will derive the detailed form of this distribution for typical critical points, typical minima and global minima.

\subsection{The eigenvalue distribution of the Hessian}
\label{supsubsec:eigdisthess}
Let $\lambda_i^\alpha$ denote the $i$'th eigenvalue of the Hessian $H(\mathbf{x}^\alpha)$ at a critical point $\alpha$ with spin configuration $\mathbf{x}^\alpha$. Similar to the way we defined the connectivity averaged distribution of spins $P(x)$ in  \eqref{eqs:defpx} in an ensemble of critical points characterized by an inverse temperature $\beta$ and a chemical potential $\mu$, we can also define the connectivity averaged Hessian eigenvalue distribution $\rho(\lambda)$ over the same ensemble of critical points:
\begin{equation}
    \rho(\lambda) = 
    \Braket{Z^{-1}\sum_{\alpha\in \operatorname{Crt}(E)} e^{-\beta E_I(\mathbf{x}^\alpha) + \mu\mathcal{I}(\mathbf{x}^\alpha)}  \left(\frac{1}{N}\sum_{i=1}^{N} \delta(\lambda - \lambda^\alpha_i)\right)}_J, 
    \label{eq:defrholam}
\end{equation}
where again $Z = \sum_{\alpha} e^{-\beta E(x^\alpha) + \mu \mathcal{I}(x^\alpha)}$.
In this subsection, we derive a formula for this Hessian eigenvalue distribution.

We first recall some basic facts about eigenvalue distributions and resolvents. First, consider any individual symmetric matrix, for example the Hessian $H(\mathbf{x})$ at any spin configuration $\mathbf{x}$ with eigenvalues $\lambda_i$ for $i=1,\dots,N$. For any such symmetric matrix,  its eigenvalue distribution $\rho(\lambda) = \frac{1}{N}\sum_i \delta(\lambda - \lambda_i)$ can be obtained from the resolvent $R(z)$ (also known as the Stieltjes transform of the eigenvalue spectrum) where $R(z)$ is defined as 
\begin{equation}
    R(z) := \frac{1}{N} \operatorname{Tr}[H(\mathbf{x})-zI]^{-1} = \int d\lambda \frac{\rho(\lambda)}{\lambda - z}.
    \label{eq:defrz}
\end{equation}
In particular, given the resolvent, or Stieltjes transform $R(z)$, one can recover the eigenvalue spectrum $\rho(\lambda)$ via the Stieltjes inversion formula 
\begin{equation}
    \rho(\lambda)=\lim _{\varepsilon \rightarrow 0^{+}} \frac{R(\lambda-i \varepsilon)-R(\lambda+i \varepsilon)}{2 \pi i}.
    \label{eq:Stieltjes-Perron}
\end{equation}
Note that the resolvent obeys the relation $R(z^*)= R^*(z)$.  Therefore the eigenvalue density $\rho(\lambda)$ in  \eqref{eq:defrz} is simply proportional to the imaginary part of the resolvent near the real axis.  
Moreover, $R(z)$ itself can be written as
\begin{equation}
    R(z) = -\frac{1}{N}\left.\frac{d}{d\hat z}\right|_{\hat z=z}\frac{\det[H(\mathbf{x})-\hat zI]}{\det[H(\mathbf{x})-zI]}.
    \label{eq:resdet}
\end{equation}
This formula is suitable for averaging over the connectivity $J$. Thus instead of directly averaging the eigenvalue spectrum over both the ensemble of critical points $\alpha$ at a fixed inverse temperature $\beta$ and chemical potential $\mu$, and then averaging over the connectivity $J$, as in  \eqref{eq:defrholam}, we instead calculate the average of   $\log\det[H(\mathbf{x}^\alpha)-\hat zI] - \log\det[H(\mathbf{x}^\alpha)-zI]$ over the same ensemble, and then obtain an ensemble averaged resolvent through  \eqref{eq:resdet} and an ensemble averaged eigenvalue spectrum through the inversion formula in  \eqref{eq:Stieltjes-Perron}. 

Following this strategy, the average of $\log\det[H(\mathbf{x}^\alpha)-\hat zI] - \log\det[H(\mathbf{x}^\alpha)-zI]$ is given by 
\begin{eqnarray}
    &&\Braket{\log\det[H(\mathbf{x}^\alpha)-\hat zI] - \log\det[H(\mathbf{x}^\alpha)-zI]}_{\beta,\mu}\nonumber\\
    &=& N^{-1} \left. \frac{d}{ds}\right|_{s=0} \log \sum_{\alpha \in\operatorname{Crt}(E)} \exp\left(-\beta E_I(\mathbf{x}^\alpha) + \mu\mathcal{I}(\mathbf{x}^\alpha) + s (\log\det[H(\mathbf{x}^\alpha)-\hat zI] - \log\det[H(\mathbf{x}^\alpha)-zI])\right).
\end{eqnarray}
Here $\Braket{\cdot}_{\beta,\mu}$ represents the weighted average over all critical points $\mathbf{x}^\alpha$ with the Boltzmann weights $\exp\left(-\beta E_I(\mathbf{x}^\alpha) + \mu\mathcal{I}(\mathbf{x}^\alpha)\right)$.
The modified free energy parameterized by $s$ can be calculated using the replica method in a manner entirely parallel to the calculation in subsection \ref{supsubsec:replkacrice}, and also by exploiting the formula for the average of the determinant in  \eqref{eq:logdet_H-z}. In the end, we obtain
\begin{eqnarray}
    &&\Braket{\log \sum_{\alpha \in\operatorname{Crt}(E)} \exp\left(-\beta E_I(\mathbf{x}^\alpha) + \mu\mathcal{I}(\mathbf{x}^\alpha) + s (\log\det[H(\mathbf{x}^\alpha)-\hat zI] - \log\det[H(\mathbf{x}^\alpha)-zI])\right)}_J\nonumber\\
    &=& \lim_{n\to 0}\frac{1}{n} 
     \operatorname{Ext}\left(\Omega_0
            + \log \int \prod_a [dx^a du^a w(x^a)] \exp\left(S+ \mu\sum_a\overline{\mathcal{I}}(x^a)\right.\right.\nonumber\\
            &&\ \ \ \ \left. \left.+ s \left[\frac{N}{2}(t^a(\hat z))^2 + \log(\partial E_I(x^a) -\hat z-t^a(\hat z)) -\frac{N}{2}(t^a(z))^2 - \log(\partial E_I(x^a) -z-t^a(z))\right]\right)\right),
\end{eqnarray}
where the stationary condition gives
\begin{equation}
    t^a(z) = \Braket{\frac{1}{\partial^2 E(x^a)-z-t^a(z)}}_S,
    \label{eq:Pastur}
\end{equation}
where $\Omega_0$ and $S$ are defined in \eqref{eq:Omega_S_replica}.
By taking the derivative with respect to $s$, we obtain 
\begin{equation}
    \Braket{\log\det[H(\mathbf{x}^\alpha)-\hat zI] - \log\det[H(\mathbf{x}^\alpha)-zI]}_{\beta,\mu} = \frac{N}{2}(t(\hat z))^2 - \frac{N}{2}(t(z))^2 +N \Braket{\log\frac{\partial E_I(x^a) -\hat z-t(\hat z)}{\partial E_I(x^a) -z-t(z)}}_S.
\end{equation}
Here we exploit the fact that $t^a$ is independent of the replica index $a$, and write it as $t$ without its replica index.
By assuming that the quantity above is self-averaging, the resolvent is
\begin{eqnarray}
    R(z) &=& -N^{-1}\left.\frac{d}{d\hat z}\right|_{\hat z=z}\exp\left(\frac{N}{2}(t(\hat z))^2 - \frac{N}{2}(t(z))^2 +N \Braket{\log\frac{\partial E_I(x^a) -\hat z-t(\hat z)}{\partial E_I(x^a) -z-t(z)}}_S\right)\nonumber\\
    &=&t(z)t'(z) - \Braket{[\partial E_I(x^a) -z-t(z))]^{-1} (1+t'(z))}_S\nonumber\\
    &=& t(z).
\end{eqnarray}
Therefore the function $t(z)$ defined in subsection \ref{supsubsec:replkacrice} is nothing other than the resolvent of the Hessian. Moreover, through  \eqref{eq:Stieltjes-Perron}, the eigenvalue density of the Hessian is simply proportional to the imaginary part of $t(z)$ near the real axis. Thus in the definition $t(z=0 \pm i \epsilon) = t_R \pm i t_I$, used in \eqref{eq:fullstatcond}, the eigenvalue density of the Hessian at the origin $\lambda=0$ is simply proportional to $t_I$. Finally, we note that \eqref{eq:Pastur} is an analog of Pastur's self-consistent equation \cite{pasturSpectrumRandomMatrices1972} for the resolvent of random matrices of the form in  \eqref{eq:hessdef}.  

\subsection{The full replica symmetry breaking solution for the grand potential}
In order to derive the full replica symmetry breaking solution for the grand potential, we transform the previous integral expression in  \eqref{eq:Omega} to the following differential form
\begin{equation}
    -\beta\Omega(\beta, \mu)= \lim_{n\to 0}\frac{1}{n} 
            \left(\Omega_0(\Theta, t_R, t_I) 
            + \log\left[ \exp\left(\frac{1}{2} \Delta \right)\prod_a \sum_{x\in \operatorname{Crt}(E_{MF}(\cdot, h^a))} \tilde w(x) e^{-\beta E_{MF}(x, h^a) + \mu\overline{\mathcal{I}}(x) + z^a x }\right]_{\mathbf{h}=\mathbf{z}=0}\right),
\end{equation}
with the mean-field energy $E_{MF}(x,h) := E_I(x) - \frac{t_R^a}{2} x^2 - xh$, the prefactor of the exponential
\begin{eqnarray}
    \tilde w(x) &:=& \sqrt{(\partial^2 E_I(x)-t_R)^2 + t_I^2} / |\partial^2 E_I(x)-t_R |,
\end{eqnarray}
and the following $2n\times 2n$ symmetric second-derivative operator $\Delta$
\begin{equation}
    \Delta = \left(\begin{matrix}
        q^{ab}\partial_{h^a} \partial_{h^b} & A^{ab} \partial_{h^a} \partial_{z^b} \\
        A^{ab} \partial_{z^a} \partial_{h^b} & C^{ab} \partial_{z^a} \partial_{z^b}\\
    \end{matrix}\right).
\end{equation}
Assuming a Parisi-style hierarchical ansatz for the matrices $q, A, C$, and the taking the continuum limit in the division of replica indices into hierarchical groups, we can replace the matrices with functions which represent the distribution of  non-diagonal entries and scalars representing the diagonal entries \cite{mezardSpinGlassTheory1986}:
\begin{equation}
    q^{ab} \to q(y), q_d,\ A^{ab}\to A(y), A_d,\ C^{ab}\to C(y), C_d.
\end{equation}
These functions above are defined in $[0,1]$.
In this limit, $\Omega$ can be written as
\begin{eqnarray}
    -\beta\Omega(\beta,\mu) &=& -\frac{1}{2} A_d^2 - A^dt_R - \frac{\beta}{2}q_dt_R - \frac{\beta^2}{4}q_d^2 - \beta A_dq_d - \frac{1}{2} C_dq_d - \frac{1}{2}t_I^2  \nonumber \\
    &&\ \ \ + \int^1_0 dy \left(\frac{1}{2}A^2(y) + \frac{\beta^2}{4}q^2(y) +\beta A(y)q(y) + \frac{1}{2} C(y)q(y)\right)\nonumber\\
    &&\ \ \ + \beta\int dh dz K(h,z, M(0)) \phi(0,h,z),
    \label{eq:omegafullrsb}
\end{eqnarray}
where $K(h,z,M)$ is the Gaussian function
\begin{equation}
    K(h,z,M) := \frac{1}{2\pi\sqrt{\det M} }\exp\left(-\frac{1}{2}
    \left(\begin{matrix}
        h \\ z
    \end{matrix}\right)^T M^{-1}
    \left(\begin{matrix}
        h \\ z
    \end{matrix}\right)
\right),
\end{equation}
the matrix $M(y)$ is defined as
\begin{equation}
    M(y) := \left(\begin{matrix}
        q(y) & A(y) \\
        A(y) & C(y)
    \end{matrix}\right),
\end{equation}
and the function $\phi(y,h,z)$ obeys the following Parisi's differential equation
\begin{equation}
   \frac{\partial}{\partial y}\phi(y, h, z)=-\frac{\dot{q}}{2}\left[\frac{\partial^{2} \phi}{\partial h^{2}}+\beta y\left(\frac{\partial \phi}{\partial h}\right)^{2}\right]-\dot{A}\left[\frac{\partial^{2} \phi}{\partial h \partial z}+\beta y \frac{\partial \phi}{\partial h} \frac{\partial \phi}{\partial z}\right]
    -\frac{\dot{C}}{2}\left[\frac{\partial^{2} \phi}{\partial z^{2}}+\beta y\left(\frac{\partial \phi}{\partial z}\right)^{2}\right],
    \label{eq:phi_parisi_pde}
\end{equation}
with the boundary condition
\begin{equation}
    \phi(1, h, z)=\beta^{-1}\log\left[\int dh' dz' K(h-h', z-z', M_d - M(1))
    \sum_{x\in \operatorname{Crt}(E_{MF}(\cdot, h'))} \tilde w(x) e^{-\beta E_{MF}(x, h') + \mu\overline{\mathcal{I}}(x) + z' x } \right],
\end{equation}
where 
\begin{equation}
    M_d = \left(\begin{matrix}
        q_d & A_d \\
        A_d & C_d
    \end{matrix}\right).
\end{equation}
In the following subsections, we derive specific cases associated with typical critical points,  typical minima, and global minima from this general formula. Before going into these cases, we briefly discuss the form of the mean-field index function $\overline{\mathcal{I}}(x)$ for minima.

\subsection{The mean-field index function \texorpdfstring{$\overline{\mathcal{I}}(x)$}{} for the case of minima}
\label{supsubsec:mfindex}

The relationship between the typical intensive index $r$ that dominates the sum over critical points in  \eqref{eqs:grandpotdef} and the chemical potential $\mu$ can be obtained via the value of $\mu$ which achieves the extremum in  \eqref{eq:invlegendre} at a fixed $r$. This extremum condition yields the following relation between $r$ and $\mu$:
\begin{eqnarray}
    r &=& -\frac{\partial\beta\Omega(\beta,\mu)}{\partial \mu} =\lim_{n\to 0}n^{-1}\sum_a\braket{\overline{\mathcal{I}}(x^a)}_{S}.
    \label{eq:r}
\end{eqnarray}
In the second equality, we have exploited the expression in  \eqref{eq:Omega} for $\Omega$. 
If we consider the limit of infinitely large negative chemical potential $\mu\to-\infty$, the integral in \eqref{eq:Omega} over the region with $\overline{\mathcal{I}}(x^a) > 0$ goes to zero, so the above relation implies that $r=0$ as $\mu \rightarrow -\infty$. Therefore the limit $\mu\to-\infty$ corresponds to the case of minima.

Now, assuming $t_I\neq0$, the set of solutions for $x$ of $\overline{\mathcal{I}}(x)=0$ in  \eqref{eqs:mean-field-index} consists of a discrete set. This in turn implies that the distribution of local spins $P(x)$ in  \eqref{eq:Px_susy_breaking} also consists of a discrete set of probability masses. This result is clearly inconsistent with the numerically measured distribution of spins in minima. Therefore we assume that mean field solutions that describe minima obey $t_I = 0$. Note by the discussion at the end of subsection \ref{supsubsec:eigdisthess}, the assumption of $t_I=0$ is equivalent to the assumption that the eigenvalue density of the Hessian at $\lambda=0$ vanishes (due to the Stieltjes inversion formula in \eqref{eq:Stieltjes-Perron}). Interestingly, the vanishing of Hessian eigenspectrum at $0$ for minima can then be thought of as a prediction of our theory.  This prediction is confirmed through numerical experiments in the main paper.  

Finally, with the assumption that $t_I=0$, the mean field index function $\overline{\mathcal{I}}(x)$ in  \eqref{eqs:mean-field-index} simplifies to an indicator function
$
    \overline{\mathcal{I}}(x) \to \Theta\left(\pm(\partial^2 E_I(x)-t_R)\right),
$
where the sign inside the Heaviside step function $\Theta$ depends on the sign of $\operatorname{Im}(t(i\epsilon))+\epsilon$. If this is positive, the distribution of local spins $P(x)$ in typical minima should vanish in the region of $\partial^2 E_I(x)-t_R < 0$. This in disagreement with the numerical observations shown in the main paper, and thus we choose the minus sign here, i.e., when $t_I=0$  \eqref{eqs:mean-field-index} reduces to
\begin{equation}
    \overline{\mathcal{I}}(x) = \Theta\left(-(\partial^2 E_I(x)-t_R)\right).
    \label{eq:mean-field-index-t_I0}
\end{equation}
Interestingly, this indicator function expression for the mean field index can be interpreted in terms of the index of a simple, scalar mean field energy function  
\begin{equation}
    E_{MF}(x) = E_I(x) - \frac{1}{2} t_R x^2.
    \label{eq:emf}
\end{equation}
The Hessian of this scalar mean field energy is simply $H(x) = \partial^2 E_I(x)-t_R$.  Thus the index of $E_{MF}(x)$ is $1$ if $H(x) < 0$ and $0$ if $H(x) \geq 0$. However this is exactly what $\overline{\mathcal{I}}(x)$ in  \eqref{eq:mean-field-index-t_I0} computes.  Thus in summary, whenever the eigenvalue density of the Hessian in an ensemble of critical points vanishes at the origin, or equivalently whenever $t_I=0$, the mean-field index function $\overline{\mathcal{I}}(x)$ is simply the index of the mean-field energy function $E_{MF}(x)$ defined in  \eqref{eq:emf}.  In particular, we have shown for typical minima that $t_I=0$ is the only solution for $t_I$ that is consistent with numerics.

\subsection{A supersymmetry broken but replica symmetric ansatz for typical critical points and typical minima}
The sum over critical points in  \eqref{eqs:grandpotdef} becomes a ``white" sum over all critical points when the inverse temperature $\beta=0$  and the chemical potential $\mu=0$.  In this setting, the grand potential is dominated by, and contains information about, typical critical points, regardless of their energy or index.  Similarly, when $\beta=0$ and $\mu \rightarrow -\infty$, the sum in  \eqref{eqs:grandpotdef} is dominated by typical minima.  To access the properties of typical critical points, and typical minima, we therefore focus on computing the grand potential at $\beta=0$. We additionally assume that all the matrices $q^{ab}, A^{ab}, C^{ab}$ in  \eqref{eq:Omega_S_replica} are diagonal with equal elements on the diagonal (i.e. replica symmetry).  Under this assumption, the different replicas are decoupled, and this in essence corresponds also to an annealed approximation. We let the scalars $q, A, C$ denote the identical diagonal elements of these three matrices respectively. However, we do not assume that $A$ and $C$ are zero.  Indeed nonzero values of $A$ and $C$ are related to supersymmetry breaking \cite{crisantiComplexitySherringtonKirkpatrickModel2003}. Substituting this supersymmetry broken but replica symmetric annealed ansatz into \eqref{eq:Omega_S_replica}, we obtain 
\begin{eqnarray}
    &&-\beta \Omega(0, \mu)\nonumber\\
    &=& -\frac{1}{2} (Cq + A^2) - At_R - \frac{1}{2}t_I^2 
        + \log \int^{i\infty}_{-i\infty} \frac{du}{2\pi i}\int dx\ w(x)\exp\left(
            \frac{1}{2} 
            \left(\begin{matrix}
                u \\ x
            \end{matrix}\right)^T 
            \left(\begin{matrix}
                q & A \\ A & C
            \end{matrix}\right)
            \left(\begin{matrix}
                u  \\ x
            \end{matrix}\right)
            - uh + \mu \overline{\mathcal{I}}(x)
        \right) \nonumber \\
     &=& -\frac{1}{2} (Cq + A^2) - At_R - \frac{1}{2}t_I^2 
        +\log\left[
            (2\pi q)^{-\frac{1}{2}} \int dx\  w(x)
           \exp\left( -\frac{1}{2q}h(x)^2 + \frac{A}{q} xh(x) + \frac{1}{2} \frac{qC - A^2}{q}x^2 + \mu\overline{\mathcal{I}}(x)\right)
        \right] \nonumber\\
    &=& -\frac{1}{2} (Cq + A^2) - At_R - \frac{1}{2}t_I^2 
    +\log\left[
        (2\pi q)^{-\frac{1}{2}} \int dh \sum_{x\in\operatorname{Crt}(E_{MF}[h])} \tilde w(x)
        \exp\left( -\frac{1}{2q}h^2 + \frac{A}{q} xh + \frac{1}{2} \frac{qC - A^2}{q}x^2 + \mu\overline{\mathcal{I}}(x)\right)
    \right], \nonumber \\
    \label{eq:Omega-typical-crt}
\end{eqnarray}
where $h(x) = \partial E_I(x) - t_R x$, $E_{MF}[h](x) := E_I(x) - \frac{t_R^a}{2} x^2 - xh$ and $\tilde w(x) = \sqrt{(\partial^2 E_I(x)-t_R)^2 + t_I^2}/ |\partial^2 E_I(x)-t_R|$. For the last equality, we changed the integration variable from $x$ to $h(x)$.

Similarly, under our supersymmetry broken but replica symmetric ansatz, the distribution of spins $P(x)$ in  \eqref{eq:Px_susy_breaking} simplifies to 
\begin{equation}
    P(x) \propto w(x)
    \exp\left( -\frac{1}{2q}h(x)^2 + \frac{A}{q} xh(x) + \frac{1}{2} \frac{qC - A^2}{q}x^2 + \mu\overline{\mathcal{I}}(x)\right).
    \label{eq:Px_typical}
\end{equation}
The order parameters $q, A,C, t_R$ and $t_I$ satisfy the following self-consistent stationary equations
\begin{eqnarray}
\left\{
  \begin{aligned}
  & q = \braket{x^2}\\
  & A = \frac{\braket{xh(x)}}{2q} - t_R/2\\
  & C = -q^{-1} + q^{-2}\Braket{h^2(x)} - 2q^{-2}A\Braket{xh(x)} +q^{-1}A^2\\
  & t_R+t_Ii= \Braket{\frac{1}{\partial^2 E_I(x) - (t_R+t_Ii)}}
  \end{aligned}\right.,
  \label{eqs:self-consistent-replica}
\end{eqnarray}
where $\braket{\cdot}$ denotes an average with respect to the mean field distribution of spins $P(x)$ in  \eqref{eq:Px_typical}. These expressions above for the grand potential, the distribution of spins, and the self-consistent equations for the order parameters will be derived again by the cavity method below. This alternate derivation will yield insights into the geometric interpretation of supersymmetry breaking.

We can solve these self-consistent equations through fixed point iteration on the order parameters. In the case of typical critical points ($\mu=0$) and typical local minima ($\mu \rightarrow -\infty$) for the single site CIM energy function $E_I(x) = x^4/4 - ax^2/2$, we observe a phase transition at $a\sim -0.95$. Below this value of the laser pump parameter, we find solutions with $A=C=0$, reflecting unbroken supersymmetry. Moreover, in this regime, the complexity is zero. On the otherhand, above this value of laser pump power, we find solutions where $A,C$ are non-zero, corresponding to broken supersymmetry, and the complexity is strictly positive.

Lastly, we drive an explicit formula for the case of typical local minima by taking the limit $\mu \to -\infty$. In this case, we need to calculate $\overline{\mathcal{I}}(x)$ up to the order of $\mu^{-1}$. As we discussed above in subsection \ref{supsubsec:mfindex}, $t_I$ should converge to zero and therefore we can expand $t_I$ in powers of $1/\mu$ as 
\begin{equation}
    t_I = \mu^{-1} t^1_I + O(\mu^{-2}),
\end{equation}
where $t^1_I$ is a coefficient of the first-order term. With this coefficient, we can expand $\overline{\mathcal{I}}(x)$ of \eqref{eqs:mean-field-index} as
\begin{equation}
    \overline{\mathcal{I}}(x) = \Theta\left(-(\partial^2 E_I(x)-t_R)\right)  + \frac{t^1_I}{\mu\pi}\left(t_R -  \frac{1}{\partial E_I(x)^2 - t_R}\right) + O(\mu^{-2}).
    \label{eq:mean_index_expansion}
\end{equation}
By substituting this into the grand potential $\Omega$ and the distribution $P(x)$, we get
\begin{eqnarray}
    -\beta \Omega(0, -\infty)
     = -\frac{1}{2} (Cq + A^2) - At_R
        &+&\log\left[
            (2\pi q)^{-\frac{1}{2}} \int dh \sum_{x\in\operatorname{Crt}_0(E_{MF}[h])}
            \exp\left( -\frac{1}{2q}h^2 + \frac{A}{q} xh \right. \right. \nonumber \\
            &&\left. \left. + \frac{1}{2} \frac{qC - A^2}{q}x^2 + \frac{t^1_I}{\pi}\left(t_R -  \frac{1}{\partial E_I(x)^2 - t_R}\right)\right)
        \right]
         \label{eq:Omega-typical-min}
\end{eqnarray}
\begin{equation}
    P(x)|_{\mu\to-\infty} \propto \begin{cases}
        w(x)
    \exp\left( -\frac{1}{2q}h(x)^2 + \frac{A}{q} xh(x) + \frac{1}{2} \frac{qC - A^2}{q}x^2  -\frac{t^1_I}{\pi} \frac{1}{\partial E_I(x)^2 - t_R}\right) & (\text{if } x\in \operatorname{Crt_0}(E_{\operatorname{MF}}[h]))\\
    0 & (\text{if } x\in \operatorname{Crt_1}(E_{\operatorname{MF}}[h]) )
    \end{cases}.
\end{equation}
Here $\operatorname{Crt_n}(\cdot)$ represents the set of critical points of index $n$, and recall that $w(x) = |\partial^2 E_I(x)-t_R|$. 

Lastly we derive the complexity of the typical local minima from the grand potential. Recall that the complexity can be obtained by the Legendre transform of the grand potential
\begin{equation}
    \Sigma = -\beta\Omega - \mu r,
\end{equation}
where $r = -\beta\frac{\partial }{\partial \mu}\Omega = \int \bar{\mathcal{I}}(x) P(x) dx$. Under the limit of $\mu\to-\infty$, we know that $r\to 0$, and therefore we need to evaluate $r\mu$ carefully to compute the complexity. To understand the asymptotic behavior of $r$, we we perform Taylor expansion of $\int \bar{\mathcal{I}}(x) P(x) dx$ with respect to $\mu^{-1}$. We have already obtained the expansion of $\bar{\mathcal{I}}(x)$ in \eqref{eq:mean_index_expansion}. On the other hand, as we can see from \eqref{eq:Px_typical},  $\left.\frac{\partial P(x)}{\partial \mu^{-1}}\right|_{\mu^{-1}=0} = 0$. Therefore
\begin{eqnarray}
    \lim_{\mu\to-\infty}\mu r &=& \lim_{\mu\to-\infty}\int \left(\mu\Theta(-\partial^2 E_I(x)+t_R) + \frac{t^1_I}{\pi}\left(t_R -  \frac{1}{\partial E_I(x)^2 - t_R}\right)\right) P(x)|_{\mu=-\infty} dx\nonumber\\
    &=&\int  \frac{t^1_I}{\pi}\left(t_R -  \frac{1}{\partial E_I(x)^2 - t_R}\right) P(x)|_{\mu=-\infty}\nonumber\\
    &=& 0.
\end{eqnarray}
Here for the second equality, we exploit the fact that $P(x)|_{\mu=-\infty}$ vanishes when $-\partial^2 E_I(x)+t_R < 0$, and the second equality follows from the self-consistent equation of $t_R$. This result implies that the complexity of typical local minima is equivalent to the grand potential, i.e. 
\begin{equation}
    \Sigma(r=0) = -\beta\Omega(0,-\infty).
\end{equation}

\subsection{A supersymmetric but full replica symmetry breaking ansatz for global minima}
\label{supsubsec:gpotgmin}
For a few spin-glass models, it is known that metastable states with low free energy can often be described by supersymmetric solutions in which  $A^{ab}=C^{ab}=0$ \cite{crisantiQuenchedComputationDependence2004,crisantiComplexityMeanfieldSpinglass2005}. Here we assume this supersymmetric ansatz for the calculation of the complexity of global minima.
As we discussed above in subsection \ref{supsubsec:mfindex}, for minima the chemical potential taken should be chosen as $\mu \rightarrow -\infty$, the mean field index function simplifies to 
$\overline{\mathcal{I}}(x) = \Theta(-\partial^2 E_I(x)+t_R)$, 
and the order parameter $t_I$ should be $0$ (see discussion around  \eqref{eq:mean-field-index-t_I0}).  Substituting these simplifying choices into  \eqref{eq:omegafullrsb} while assuming the full replica symmetry breaking structure of $q^{ab}$, the grand potential $\Omega(\beta,-\infty)$ becomes
\begin{eqnarray}
 \beta\Omega(\beta,-\infty) 
 &=& \frac{\beta}{2}q_dt_R + \frac{\beta^2}{4}(q_d^2 - \int^1_0 dy q^2(y)) - \beta\int \frac{dh}{2\pi}\exp\left(-\frac{h^2}{2}\right)f(0,\sqrt{q(0)}h),
\end{eqnarray}
where the function $f(y,h)$ obeys the following Parisi's differetial equation
\begin{equation}
   \frac{\partial}{\partial y}f(y, h)=-\frac{\dot{q}}{2}\left[\frac{\partial^{2} f}{\partial h^{2}}+\beta y\left(\frac{\partial f}{\partial h}\right)^{2}\right]  ,
\end{equation}
with the boundary condition
\begin{equation}
    f(1, h)=\beta^{-1}\log\left[\int \frac{dh'}{2\pi\sqrt{q_d - q(1)} }\exp\left(-\frac{(h-h')^2}{2(q_d-q(1))}
    \right) 
    \sum_{x\in \operatorname{Crt_0}(E_{MF}(\cdot, h'))} e^{-\beta E_{MF}(x, h')} \right].
    \label{eq:f1_susy}
\end{equation}

The order parameters $q(x)$ and $q_d$ can be determined by the argument in the next subsection. In the case of the CIM single site energy function $E_I(x) = x^4/4 - ax^2/2$, we observe the replica symmetry $q(x)=0$ for $a<-2$, while in $a>2$, $q(x)$ is increasing, i.e, full replica symmetry breaking occurs.
\subsubsection{determination of \texorpdfstring{$q(x)$}{q(x)} and  \texorpdfstring{$q_d$}{qd}}
The function $q(y)$ and the constant $q_d$ are determined by associated self-consistent stationary conditions. The stationary condition for  $q(y)$, is given by
\begin{equation}
    q(y) = \int dh M^2(y,h) P(y,h),
\end{equation}
where $M:= \frac{\partial f}{\partial h}$, which can be computed by solving the following differential equation
\begin{equation}
    \frac{\partial M}{\partial y}  = -\frac{1}{2} \frac{dq}{dy} \left( \frac{\partial^2 M}{\partial h^2} + 2\beta yM \frac{\partial M}{\partial h} \right),\text{  } M(1,h) = \braket{x^2}_{MF}
\end{equation}
and $P(y, h)$ is a propagator that satisfies the following differential equation.\cite{sommersDistributionFrozenFields1984}
\begin{equation}
    \frac{\partial P}{\partial y}  = \frac{1}{2} \frac{dq}{dy} \left( \frac{\partial^2 P}{\partial h^2} - 2\beta yM \frac{\partial P}{\partial h} \right), \text{  } P(0,h) = (2\pi q(0))^{-1} \exp\left(-\frac{h^2}{2q(0)}\right).
    \label{eq:DiffEqForP}
\end{equation}
Here $\braket{\cdot}_{MF}$ represents the thermal average with mean-field energy $E_{MF}$.

Therefore for $y=0$ we obtain
\begin{equation}
    q(0) =  \int dh \exp\left(-\frac{h^2}{2}\right) M^2(0, \sqrt{q(0)}h).
\end{equation}
This indicates $q(0)=0$. On the other hand, at the other edge $y=1$, we see
\begin{equation}
    q(1) = \int dh\braket{x}^2_{MF} P(1, h).
\end{equation}
We can also calculate the stationary condition for $q_d$, i.e. the diagonal part of $n$ by $n$ replica overlap matrix $Q$ as follows:
\begin{equation}
    q_d = 2 \beta^{-1} \frac{\partial}{\partial (q_d-q(1))} f(0,0) = \int dh \braket{x^2}_{MF} P(1, h).
\end{equation}
In the low temperature limit $\beta \to \infty$, $q(1)$ converges to $q_d$.
Hence, for the global minima, i.e. $\beta\to\infty$, \eqref{eq:f1_susy} can be re-written as
\begin{equation}
    f(1, h)=\beta^{-1}\log\left[
    \sum_{x\in\operatorname{argmin}(E_{MF}(\cdot, h))} e^{-\beta E_{MF}(x, h')} \right].
    \label{eq:f1_susy_beta_limit}
\end{equation}
We will show below in section \ref{supsec:freeenergy} that this result is indeed equivalent to the one derived from the free energy of the CIM in the zero temperature limit. This internal self-consistency check validates our ansatz of supersymmetry for global minima. 

\subsubsection{The distribution of spins in global minima}
In the full replica symmetry breaking scheme, the expectation value of arbitrary function $O(x)$ is
\begin{equation}
    \braket{O} = \beta\left.\frac{\partial}{\partial s}\right|_{s=0} f(0, 0, s)
\end{equation}
where $f(y,h,s)$ obeys the PDE \eqref{eq:ParisiDiffEq} and the boundary condition
\begin{eqnarray}
    f(1, h, s) &=& \beta^{-1}\log \left[
        \frac{1}{\sqrt{2\pi(q_d-q(1))}}
        \int dh' \exp\left(-\frac{(h-h')^2}{2(q_d-q(1))}\right)
        \sum_{x\in\operatorname{argmin}(E_{MF}(\cdot, h'))} e^{-\beta E_{MF}(x, h') + sO(x)}
    \right]\nonumber\\
    &=& \beta^{-1}\log \left[
        \sum_{x\in\operatorname{argmin}(E_{MF}(\cdot, h))} e^{-\beta E_{MF}(x, h) + sO(x)}
    \right].
\end{eqnarray}
For the second equality, we use the fact that $q_d-q(0)$ goes to zero under the low-temperature limit.
Since the derivative of $f(0,0,s)$ can be expressed with the propagator as
\begin{equation}
    \frac{\partial}{\partial s}f(0, 0, s) = \int dh P(1, h) \frac{\partial}{\partial s}f(1, h, s),
\end{equation}
we obtain
\begin{equation}
    \Braket{O} = \int dh P(1,h) \frac{\sum_{x\in\operatorname{argmin}(E_{MF}(\cdot, h))} O(x)}{|\{x\in\operatorname{argmin}(E_{MF}(\cdot, h))\}|} = \int dx |\partial^2 E_I(x) - t_R| P(1,\partial E_I(x) - t_R x) O(x),
\end{equation}
where for the second equality, we use the fact that $|\{x\in\operatorname{argmin}(E_{MF}(\cdot, h))\}|=1$ is true almost everywhere in $\mathbb{R}$, and we change the variable of integration from $h$ to $x$.
This equality indicates that the distribution of $x_i$ of global minima is given by
\begin{equation}
    P(x) = |\partial^2 E_I(x) - t_R| P(1,\partial E_I(x) - t_R x).
\end{equation}

\section{Replica calculation of CIM free energy}
\label{supsec:freeenergy}
In this section, we calculate the CIM free energy with the replica method \cite{mezardSpinGlassTheory1986}, and derive the properties of the global minima by taking the low-temperature limit $\beta\to\infty$. We will find that the obtained result coincides with the supersymmetric grand potential computed in subsection \ref{supsubsec:gpotgmin}, after taking the limit $\beta\to\infty, \mu\to-\infty$. We discuss an intuitive reason for why they should coincide later in this section.

As usual, we assume that the free energy is self-averaging, i.e. in the large N limit, the free energy associated with a single realization of the connectivity $J$ tightly concentrates about the mean of the free energy over the connectivity $J$.  Thus we compute the connectivity averaged free energy 
\begin{equation}
    F = \lim_{N \rightarrow \infty} \braket{F_J}_J
\end{equation}
To do this, we will exploit the replica trick, 
\begin{equation}
    F = -\frac{1}{\beta} \lim_{n\rightarrow 0} \frac{\log\braket{Z^n}_J}{n},
\end{equation}
where $Z=\int \prod_{i} dx_i e^{-\beta E(\mathbf{x})}$.

The average of $Z^n$ can be calculated by introducing replicas and  Hubbard-Stratonovich transforms as follows:
\begin{eqnarray}
\braket{Z^n}_J
&=& \int \mathcal{D}J \int \prod_{i,a} dx_i^a \exp\left[\beta\sum_{a}\left(\sum_{i,j}\left(\frac{1}{2} J_{ij} x_i^a x_j^a \right) -\beta \sum_i E_I(x^a_i)\right)\right] \nonumber \\
&=& \int \prod_{i,a} dx_i^a \exp\left[\frac{\beta^2}{2N}\sum_{a,b}\sum_{i<j}\left(x_i^a x_i^b x_j^a x_j^b \right) - \beta \sum_{a,i} E_I(x^a_i)\right] \nonumber \\
&=& \int \prod_{i,a} dx_i^a \exp\left[\frac{\beta^2}{4N}\sum_{a,b}\left(\left(\sum_i x_i^a x_i^b\right)^2\right) - \frac{\beta^2}{4N} \sum_i\left(\sum_a (x_i^a)^2\right)^2 - \beta \sum_{a,i}E_I(x^a_i)\right] \nonumber \\
&=& \int \prod_{i,a} dx_i^a \exp \left[-\frac{\beta^2}{4N} \sum_i \left(\sum_a \left(x_i^a\right)^2\right)^2 - \beta \sum_{a,i}E_I(x^a_i)\right] \int \prod_{a\leq b} dQ_{ab} \exp \left(-\sum_{a,b} \left( \frac{N}{4} \beta^2  Q_{ab}^2 - \frac{\beta^2}{2}\sum_i x_i^a x_i^b Q_{ab}\right)\right) \nonumber \\
&=& \int \prod_{a\leq b} dQ_{ab} \left(\prod_{a} dx^a \exp \left[-\frac{\beta^2}{4N} \left(\sum_a \left(x^a\right)^2\right)^2 - \beta \sum_{a}E_I(x^a)\right] -\sum_{a,b} \left( \frac{N}{4} \beta^2  Q_{ab}^2 - \frac{\beta^2}{2} x^a x^b Q_{ab}\right)\right)^N \nonumber \\
&=& \int \prod_{a\leq b} dQ_{ab}\exp \left(N S[Q]\right).
\end{eqnarray}
For the last equality, we defined a effective action $S[Q] := \log\left( \int \prod_a dx^a e^{\Tilde{S}[\{x^a\},Q] + O(N^{-1})}\right)$ with
\begin{equation}
    \Tilde{S}[\{x^a\}, Q] = -\beta \sum_a E_I(x^a) - \sum_{a,b} \left( \frac{1}{4} \beta^2  Q_{ab}^2 - \frac{\beta^2}{2} x^a x^b Q_{ab}\right).
\end{equation}
Since $N$ goes to infinity, the integration over $Q_{ab}$ is dominated by $S[Q]$ around its extremum.

\subsection{The full replica symmetry breaking solution}
Assuming that that replica symmetry is broken an infinite number of times, the matrix $Q_{ab}$ consists of several blocks of size $n\geq m_1 \geq m_2 \geq \cdots \geq 1$, and the value inside the block of size $m_i$ is $q_i$.
In this case, the non-diagonal part of the matrix $Q_{ab}$ can be fully characterized by a function $q(y)$ defined as follows:
\begin{equation}
q(y) = q_i \ \ (\text{for }m_{i+1}< y \leq m_i)
\end{equation}
On top of that, we assume that the diagonal part is constant;
\begin{equation}
Q_{aa} =: q_d \ \text{for } \forall a\in[n].
\end{equation}
By taking the limit $n\to 0$, the inequality above is flipped; 
\begin{equation}
0\leq m_1 \leq m_2 \leq \cdots \leq 1
\end{equation}
and the function $q(y)$ becomes a continuous increasing function on $[0,1]$. With this function $q(y)$ and a constant $q_d$, we can express the free energy as follows:
\begin{equation}
F = -\frac{\beta^2}{4} \left[\int^{1}_{0}q^2(y) dy - q_d^2\right] 
- \beta\int dz \exp\left(-\frac{1}{2}z^2\right) f(0, \sqrt{q(0)}z),
\label{eq:FRSB_free_energy}
\end{equation}
where the function $f(y, h)$ obeys Parisi's differential equation
\begin{equation}
    \frac{\partial}{\partial y}f(y, h)=-\frac{\dot{q}}{2}\left[\frac{\partial^{2} f}{\partial h^{2}}+\beta y\left(\frac{\partial f}{\partial h}\right)^{2}\right],
\label{eq:ParisiDiffEq}
\end{equation}
with a boundary condition at $y=1$ given by
\begin{equation}
f(1, h) = \beta^{-1}\log \left(\int dx \exp(-\beta E_{MF}(x, h))\right),\ \ \left(E_{MF}(x,h):= E_I(x) - xh - \frac{1}{2}\Delta a x^2\right),
\end{equation}
and where $\Delta a := \beta(q_d - q(1))$. $f(1,h)$ can be interpreted as the free energy density under a mean-field approximation, with an 'Onsager reaction' term proportional to $\Delta a$.

\subsection{The free energy of global minima}
In the low-temperature limit $\beta\to \infty$, $\Delta a$ can be evaluated as
\begin{equation}
    \Delta a = \int dh \beta(\braket{x^2}_{MF} - \braket{x}^2_{MF})P(1, h) = \int dh \frac{1}{\partial^2 E_{MF}(x_*(h))} P(1, h),
    \label{eq:Delta_a}
\end{equation}
where $x_*(h) := \operatorname{argmin}_{x\in\mathbb{R}}E_{MF}(x,h)$.
With this $\Delta a$, the free energy with $\beta\to\infty$ is obtained as follows:
\begin{equation}
    F = \lim_{\beta\to \infty}\left[- \frac{\beta}{2}q_d \Delta a -\frac{\beta^2}{4} \left[\int^{1}_{0}q^2(y) dy - (q_d-\Delta a)^2\right] - \beta\int dz \exp\left(-\frac{1}{2}z^2\right) f(0, \sqrt{q(0)}z) \right],
\end{equation}
where $f(x,h)$ obeys \eqref{eq:ParisiDiffEq} and the boundary condition
\begin{equation}
    f(1, h) = \beta^{-1}\log \sum_{x\in\operatorname{argmin}(E_{MF}(\cdot,h))} \exp(-\beta E_{MF}(x, h)).
\end{equation}
Since $t_R$ in the previous section and $\Delta a$ obey the same self-consistent equation, by the re-definition $q_d -\Delta a\to q_d$, we can see that the free energy $F$ and the grand potential $\Omega$ have the exactly same expression.

This coincidence can be intuitively explained as follows. The Gibbs distribution $e^{-\beta E(\mathbf{x})}$ concentrates around the global minima with large $\beta$ and we see that
\begin{equation}
    F = \sum_{\mathbf{x}\in \operatorname{Crt}_0(E)} e^{-\beta E(\mathbf{x})} = \lim_{\mu\to -\infty} -r\mu - \beta \Omega(\beta,\mu),
\end{equation}
where the normalized index $r$ is given by
\begin{eqnarray}
    r &=& -\frac{\partial\beta\Omega(\beta,\mu)}{\partial \mu} = \Braket{\bar I(\mathbf{x}^a)}_S
\end{eqnarray}
Here $a$ is an arbitrary replica index. To evaluate $r\mu$ in the limit of $\mu\to-\infty$, we compute the term of order $O(\mu^{-1})$. This term can be calculated from \eqref{eq:mean_index_expansion}
\begin{equation}
    \frac{\partial r}{\partial \mu^{-1}} = \frac{t^1_I}{\pi}\left(t_R - \Braket{ \frac{1}{\partial E_I(x)^2 - t_R}}_S\right) = 0.
\end{equation}
Therefore 
\begin{equation}
    F = \lim_{\mu\to -\infty} -r\mu - \beta \Omega(\beta,\mu) = - \beta \Omega(\beta,\mu).
\end{equation}

\subsection{The replica symmetric solution for the free energy}
Finally, we calculate the replica symmetric solution of the free energy and discuss when the symmetry breaks.
Under the assumption of replica symmetry, the function $q(y)$ is constant. Considering the fact that $q(0) = 0$, we can say $q(y)= 0$.
In this case, the r.h.s. of the partial differential equation \eqref{eq:ParisiDiffEq} vanishes, and the free energy is
\begin{eqnarray}
    F= \frac{\beta^2}{4} q_d^2 -\log\left[\int dx e^{-\beta E_{MF}(x,0)}\right].
\end{eqnarray}
Under the low-temperature limit $\beta\to\infty$, $q_d\to q(1) = 0$, which means  $\Braket{x^2}_{MF} = 0$.
Hence the mean-field potential should have a minimum at the origin, i.e., $a+\Delta a \leq 0$. By solving the self-consistent equation for $\Delta a$, we see
\begin{equation}
    \Delta a = \frac{1}{2}(-a\pm\sqrt{a^2-4}).
\label{eq:rbeta}
\end{equation}
The sign in the equation above can be chosen properly by analyzing the stability of the solution. The Hessian in replica space around the extrema can be obtained as follows:
\begin{eqnarray}
    \frac{\partial^2 S[Q]}{\partial Q_{(a,b)} \partial Q_{(c,d)}}
   &=& \begin{cases}
  \frac{1}{2} \beta^2-\frac{1}{4}\beta^4\left(\braket{(x^a)^4}_{\tilde S}- q_d^2 \right) =\frac{1}{2}\beta^2 - \frac{1}{2}\beta^4q_d^2    & (a=b=c=d) \\
    \beta^2  -\beta^4 q_d^2& ((a,b) = (c,d), a\neq b) \\
0 & (otherwise)
\end{cases}
\end{eqnarray}
In order to have non-negative eigenvalues, $1-\beta^2 q_d^2$ needs to be non-negative, i.e. $\Delta a \leq 1$. 
Then both conditions $a+\Delta a \leq 0$ and $\Delta a \leq 1$ hold only when we choose the minus sign for \eqref{eq:rbeta} and $a\leq -2$. Therefore we can conclude that replica symmetry breaking occurs at $a=-2$, which agrees with numerical observation (see our main paper).

\section{A geometric interpretation of supersymmetry breaking via a generalized cavity method}
\label{supsec:generalcavity}

\subsection{Overview of the generalized cavity method, marginal stability and landscape sensitivity}
The cavity method \cite{mezardSpinGlassTheory1986} generally involves analyzing a system with $N$ degrees of freedom by first removing or isolating $1$ degree of freedom and separating it from the remaining $N-1$ degrees of freedom. This latter system in which one degree of freedom is removed is called the cavity system.  Then this single degree of freedom is added back in and the properties of the full system are computed self-consistently by relating the order parameters of the cavity system, and its response or susceptibility to the addition of $1$ degree of freedom, to the order parameters of the full system (which themselves are close to or identical to that of the cavity system due to the sheer existence of a large $N$ thermodynamic limit).  Thus a key step in applying the cavity method lies in understanding how a large system with $N-1$ degrees of freedom responds to the addition of a single new degree of freedom. Often, in many mean field models, this response can be treated perturbatively, which makes the cavity method tractable. 

In our context of landscape analysis, where the quantities of interest are critical points of a given index or energy and their associated distribution of spins and Hessian eigenspectra, the cavity method is straightforward to apply when the Hessian eigenspectra of critical points are non-degenerate or stable, which by definition means that there are no Hessian eigenvalues close to zero.  In the case of minima, this means the energy landscape in the vicinity of the minimum has no flat modes and every minimum exhibits a stiff response to any small external perturbation, including that derived from adding a single degree of freedom.  One can see this by noting that if the cavity system spin configuration is located at a particular energy minimum, then the zero-temperature susceptibility matrix that translates external fields into changes in the location of the minimum is simply the inverse of the Hessian.  Therefore if the Hessian eigenvalues are bounded away from $0$, the susceptibility matrix has bounded eigenvalues, and therefore all small external fields yield similarly small changes in the location of each non-degenerate minimum.  Moreover, in such a situation where every minimum is non-degenerate, one can expect a one to one correspondence between minima in the $N-1$ dimensional cavity system before the addition of a spin, and minima in the full $N$ dimensional system after the addition of the spin, as the bounded response to a single spin is unlikely to either destroy or create new minima given the stiffness of the response of each individual minimum. Therefore the complexity of minima (and more generally of critical points if they are also non-degenerate) will not change under the addition of a single spin. Similarly, if all critical points are non-degenerate, the grand potential will also remain unchanged after the addition of a single spin.  Thus the setting of non-degenerate minima (and also more generally non-degenerate critical points of any index) is an especially simple setting in which to apply the cavity method for two key reasons: that upon the addition of a single spin, (1) individual critical points move a small amount, and (2) critical points are neither created or destroyed.   

The situation changes dramatically however if critical points are degenerate or marginally stable, so that there is no gap in the Hessian eigenvalue density away from $0$. This means that for minima for example, there are arbitrarily soft or flat modes associated with Hessian eigenvalues that are very close to zero. These modes have a large susceptibility to external perturbations.  In particular, one cannot treat the response of such a minimum to the addition of a single spin perturbatively.  These flat modes also mean that the addition of a single spin might create or destroy critical points or change the index of a critical point.  Thus degeneracy of critical points can in principle destroy the property of one-to-one correspondence between critical points before and after the addition of a single spin.  Therefore the grand potential and the complexity could be highly susceptible to external perturbations when critical points are degenerate.  In such a setting, a correct application of the cavity method, which takes into account this extreme sensitivity of the geometric structure of the energy landscape to the addition of a single degree of freedom, can be much more involved.

In this section, we introduce a generalized version of the cavity method that takes into account this sensitivity.  
We note that this issue of marginal stability has been previously discussed in the context of the TAP free energy landscape, especially for the SK model. In this model, it is known that the Hessian eigenspectrum of critical points in the large $N$ limit consists of a continuous positive band with a gap away from $0$, plus a single isolated point at the origin. The isolated eigenvalue makes the state marginally stable \cite{aspelmeierComplexityIsingSpin2004}. Rizzo \cite{rizzoTAPComplexityCavity2005} proposed a generalized cavity method, where an infinitesimally small regularization term is added to the energy function to move the isolated point away from the origin, which makes the state fully stable. This method works well for the SK model, but our CIM model has Hessian eigenspectra with many infinitesimally small eigenvalues, given that the eigenvalue density is nonzero all the way down to the origin.  Thus the method of  \cite{rizzoTAPComplexityCavity2005} cannot be applied to this case.  Cavagna et al. \cite{cavagnaCavityMethodSupersymmetrybreaking2005} also proposed another generalized cavity method for the SK model, which again cannot be applied in our case due to the multiplicity of small eigenvalues.  Moreover, neither of these works provided a geometric interpretation of the supersymmetry breaking order parameters $A$ and $C$ above. 

As we see below, our generalized cavity method can successfully deal with a multiplicity of small Hessian eigenvalues, can recover the grand potential obtained by the above supersymmetry breaking calculation, and can provide a clear geometric understanding of the meaning of the supersymmetry breaking order parameters $A$ and $C$ in terms of the susceptibility of the grand potential to infinitesimal external perturbations of the landscape.  We note that another previous geometric interpretation of supersymmetry breaking was provided for the SK model in terms of certain statistics of the single, isolated marginal mode \cite{mullerMarginalStatesMeanfield2006}. Of course such an interpretation does not apply to the CIM energy landscape with its more general high multiplicity of marginal modes. 

\subsection{A generalized cavity derivation of the grand potential}
We define the grand potential $\Omega(\mu)$ and rewrite it in terms of the Kac-Rice formula as follows: 
\begin{eqnarray}
    \exp(N\Omega(\mu))
    &=& \mathbb{E}_J \sum_{\alpha\in \operatorname{Crt}(E)}
        e^{\mu \mathcal{I}(\mathbf{x}^\alpha)} \nonumber\\
    &=& \mathbb{E}_J\left[\int \prod_{i=0}^{N-1} dx_i \prod_{i=0}^{N-1} \delta\left(\partial E_I(x_i) + \sum_{j=0}^{N-1} J_{ij} x_j\right) \left|\det H(\mathbf{x})\right| e^{\mu I(\mathbf{x})}\right].
    \label{eq:Omega-cavity}
\end{eqnarray}
Note that the grand potential $\Omega$ in this section is different from one in section \ref{supsec:replica} by a factor of $-\beta$, and moreover we work simply with $\beta=0$, using only the Lagrange dual chemical potential $\mu$ to select amongst critical points of different intensive index $r$. In particular $\mu=0$ corresponds to typical critical points regardless of index $r$ and $\mu \rightarrow -\infty$ corresponds to typical local minima of index $r=0$. We are considering a system with $N$ spins ($i=0,1,\cdots,N-1$) where $H(\mathbf{x})$ is an $N$ by $N$ Hessian defined in  \eqref{eq:hessdef}, and $\mathbb{E}_J$ represents an average over the $N\times N$ connectivity matrix $J$ sampled from Gaussian Orthogonal Ensemble (GOE). 

Anticipating that soft or flat modes of the energy landscape will play a critical role in determining its geometry and its response to perturbations, we first relax the hard $\delta$ function constraint on the gradient and replace it with a soft Gaussian with an effective inverse temperature $\beta$, obtaining
\begin{eqnarray}
    \exp(N\Omega_\beta)
    &=& \mathbb{E}_J\left[\int \prod_{i=0}^{N-1} dx_i \left(\frac{\beta}{\pi}\right)^{N/2}\exp\left(-\beta \sum_{i=0}^{N-1} \left(\partial E_I(x_i) + \sum_{j=0}^{N-1} J_{ij} x_j\right)^2\right) \left|\det H(\mathbf{x})\right| e^{\mu I(\mathbf{x})}\right].
    \label{eq:Omega-cavity-finite-beta}
\end{eqnarray}
Eventually we will take the low temperature $\beta\to\infty$ limit to recover the original grand potential of interest. However, this intermediate finite temperature Gaussian representation of the $\delta$ function allows our analysis to carefully track the structure of the landscape in the vicinity of critical points, including the putative soft modes.  

Along the lines of the cavity method, we will split the $N$ degrees of freedom in the spin configuration $\mathbf{x}$ into the first spin $x_0$ and the remaining $N-1$ spins $\mathbf{x}^{/0}:= (x_1, x_2, \cdots, x_{N-1})$ which form the cavity system. With this split, the joint distribution of $x_0$ and the gradient of the energy with respect to $x^0$ will play a key role in the cavity analysis. We denote this joint distribution by $P_\beta(x,y)$ where $x=x_0$ and $y = \nabla_0 E(\mathbf{x})$, and it is given by 
\begin{eqnarray}
    &&P_\beta(x,y)\nonumber\\
    &=& \exp(-N\Omega_\beta)\mathbb{E}_J\left[\int \prod_{i=0}^{N-1} dx_i \left(\frac{\beta}{\pi}\right)^{N/2}\exp\left(-\beta \sum_{i=0}^{N-1} \left(\partial E_I(x_i) + \sum_{j=0}^{N-1} J_{ij} x_j\right)^2\right) \left|\det H(\mathbf{x})\right| e^{\mu I(\mathbf{x})} )\delta(x_0-x)\delta(\nabla_0 E(\mathbf{x})-y))\right].\nonumber\\
    \label{eq:P(x,y)-cavity}
\end{eqnarray}
Note that as $\beta \rightarrow \infty$ this distribution concentrates onto the zero gradient $y=0$ as expected, but at finite $\beta$ this distribution allows us to explore regions of nonzero gradient in the vicinity of critical points. 

Next, following the cavity method, we attempt to write the grand potential $\Omega_\beta$ of the system of size $N$ in terms of a product of the grand potential of the cavity system of size $N-1$, a Boltzmann factor involving only $x_0$, and an interaction term that couples $x_0$ to simple order parameters associated with the cavity system $\mathbf{x}^{/0}$.  We first consider the gradient on the cavity spins $i=1,\dots,N-1$:
\begin{equation}
    \nabla_i E(\mathrm{x}) = \partial E_I(x_i) + \sum_{i=1}^{N-1} J_{ij} x_j + J_{i0} x_0, \qquad \text{for } i=1,\dots,N.
\end{equation}
Here we have split the gradient into a contribution from the cavity system   $\mathbf{x}^{/0}$ alone (first two terms) and a contribution from $x_0$ alone (third term).  In the absence of $x_0$, the first two terms constitute the exact gradient of the cavity system.  Adding $x_0$ to this cavity system can then be thought of as exerting an external field $J_{i0} x_0$ on each spin $i=1,\dots,N$ in the cavity system.  

This point of view suggests the following factorization of the grand potential of the full system of $N$ spins:
\begin{eqnarray}
    \exp(N\Omega_\beta) = \mathbb{E}_{J^{/0}}\left[\int \prod_{i=0}^{N-1} dx_i  \mathbb{E}_{\mathbf{J}_0}\left[\omega(x_0 \mathbf{J}_0, \mathbf{x}^{/0})   \left(\frac{\beta}{\pi}\right)^{1/2}\exp\left(-\beta \left(\sum_{i=1}^{N-1} J_{0i} x_i + \partial E_I(x_0) \right)^2\right) 
    \frac{\left|\det H(\mathbf{x})\right|}{\left|\det H(\mathbf{x}^{/0})\right|}
    e^{\mu (\mathcal{I}(\mathbf{x}) - \mathcal{I}(\mathbf{x}^{/0}))}\right]\right].\nonumber\\
    \label{eq:P_x_cavity}
\end{eqnarray}
Here $H(\mathbf{x}^{/0})$ is the $N-1 \times N-1$ sub-matrix of $H(\mathbf{x})$ and is simply the Hessian of the cavity system in the absence of $x_0$.  $\mathbb{E}_{\mathbf{J}_0}$ represents an average over the vector $\mathbf{J}_0 := (J_{01}, J_{02},\cdots, J_{0(N-1)})$ which couples $x_0$ to the cavity system $\mathbf{x}^{/0}$. Also $\mathbb{E}_{J^{/0}}$ represents an average over $J^{/0}$ which is the $N-1\times N-1$ sub-matrix of $J$ corresponding to the connectivity matrix of the cavity. Finally $\omega(\mathbf{s}, \mathbf{x}^{/0})$ is the grand potential density of the cavity system $\mathbf{x}^{/0}$ in the absence of $x^0$ but in the presence of an external field $\mathbf{s}$ that tilts the gradient:
\begin{equation}
    \omega(\mathbf{s}, \mathbf{x}^{/0}) := \left(\frac{\beta}{\pi}\right)^{(N-1)/2}\exp\left(-\beta \sum_{i=1}^{N-1} \left(\partial E_I(x_i) + \sum_{j=1}^{N-1} J_{ij} x_j +s_i\right)^2\right) \left|\det H(\mathbf{x}^{/0})\right| e^{\mu \mathcal{I}(\mathbf{x}^{/0})}.
    \label{eq:grandpotdens}
\end{equation}
Thus in  \eqref{eq:P_x_cavity} we have begun to express the grand potential of the system of $N$ spins $\mathbf{x}$ recursively in terms of a grand potential density involving $N-1$ spins $\mathbf{x}^{/0}$, in the presence of an external field $\mathbf{s} = x_0 \mathbf{J}_0$. In addition to this term, the integrand of  \eqref{eq:P_x_cavity} also contains a second Gaussian factor constraining the gradient of spin $x_0$.  However, to account for the discrepancy between the grand potential density of $N-1$ spins, and the grand potential density of $N$ spins,  \eqref{eq:P_x_cavity} also contains two important factors: (1) a ratio of the modulus of the determinant of the full system $\left|\det H(\mathbf{x})\right|$ to the modulus of the determinant of the cavity system $\left|\det H(\mathbf{x}^{/0})\right|$, and (2) a chemical potential term involving the change in index upon adding $x^0$ to the cavity system $\mathbf{x}^{/0}$, i.e. $\mathcal{I}(\mathbf{x}) - \mathcal{I}(\mathbf{x}^{/0})$. These latter two terms on the surface seem to couple $x^0$ to $\mathbf{x}^{/0}$ in an intricate and complex manner, thereby making the application of the cavity method potentially difficult.  However, we can simplify these two terms to reveal that in both cases, $x^0$ actually couples to the cavity system in $\mathbf{x}^{/0}$ only through a simple interaction between $x^0$ and a certain order parameter associated with the susceptibility of the cavity system.  

In the following we simplify \eqref{eq:P_x_cavity} by first simplifying the ratio of determinants and the chemical potential term, and then simplifying the grand potential density $\omega(x_0\mathbf{J}_0, \mathbf{x}^{/0})$ in the presence of an external field.  After this we perform an average over the spin-cavity interaction connectivity $\mathbf{J}_0$.

\subsubsection{Spin-cavity interaction through a determinantal ratio}

First we simplify the ratio of determinants in \eqref{eq:P_x_cavity}. We make use of the Schur complement formula, which expresses the determinant of a block matrix in terms of determinants of (functions of) its sub-blocks. Associated with the split of  $\mathbf{x}$ into a single spin $x^0$ and the cavity system $\mathbf{x}^{/0}$, the full Hessian $H(\mathbf{x})$ has a spin-cavity block structure
\begin{equation}
    H(\mathbf{x}) = \left(\begin{matrix} \partial^2 E_I(x_0) & \mathbf{J}_0^T \\ \mathbf{J}_0 & H(\mathbf{x}^{/0}) \end{matrix}\right).
    \label{eq:hessblock}
\end{equation}
Applying the Schur complement formula to $H(\mathbf{x})$ using this split into 4 blocks, we obtain 
\begin{equation}
    \left| \frac{\det H(\mathbf{x})}{\det H(\mathbf{x}^{/0})} \right| = \left|\partial^2 E_I(x_0) - \mathbf{J}_0^T H(\mathbf{x}^{/0})^{-1} \mathbf{J}_0 \right|.
\end{equation}
We note this is a random quantity due to the randomness in the spin-cavity connectivity vector $\mathbf{J}_0$.  
However, at large $N$ we expect this quantity to concentrate about its mean over $\mathbf{J}_0$, given by
\begin{eqnarray}
    \left| \frac{\det H(\mathbf{x})}{\det H(\mathbf{x}^{/0})} \right|
    &=&  |\partial^2 E_I(x_0) - N^{-1} \operatorname{Tr}H^{-1}(\mathbf{x}^{/0})|\nonumber\\
    &=& |\partial^2 E_I(x_0) - N^{-1} \operatorname{Tr}H^{-1}(\mathbf{x})|.
\end{eqnarray}
Furthermore in this last equality, we assume that the addition of a single column and row to go from $H(\mathbf{x}^{/0})$ to $H(\mathbf{x})$ in  \eqref{eq:hessblock} does not change the the overall trace to leading order in $\frac{1}{N}$, i.e. we assume that  $\frac{1}{N} \operatorname{Tr}H^{-1}(\mathbf{x}) = \frac{1}{N} \operatorname{Tr}H^{-1}(\mathbf{x}^{/0})$.

Thus in the determinantal ratio, the single spin $x_0$ interacts with the cavity system $\mathbf{x}^{/0}$ only through the simple order parameter $t_R$, defined as
\begin{equation}
    t_R := \frac{1}{N} \operatorname{Tr}H^{-1}(\mathbf{x}^{/0}).
\end{equation}
Recall that near stable non-degenerate minima, the inverse Hessian is simply the zero temperature susceptibility matrix that transforms small external fields to changes in the location of the minimum. Thus in this setting $t_R$ is simply the average of the diagonal of the susceptibility matrix of the cavity system.  Or more generally, $t_R$ is simply the resolvent $R(z)$ of the Hessian in  \eqref{eq:defrz} evaluated at $z=0+i \epsilon$. This quantity satisfies the following self-consistent equation, which can be derived using methods similar to that in subsection \ref{supsubsec:eigdisthess}, yielding 
\begin{equation}
    t_R = \int dx dy P_\beta(x,y) \frac{1}{\partial^2 E_I(x) - t_R}.
    \label{eq:selfconstr}
\end{equation}
This is an analog of the Pastur formula \cite{pasturSpectrumRandomMatrices1972}. 
We note that in this entire derivation of the generalized cavity method, we assume that this self-consistent equation has a real-valued solution. This is equivalent to the assumption that the eigenvalue density of the Hessian vanishes at the origin and the resolvent near the origin is real (see discussion at the end of subsection \ref{supsubsec:eigdisthess}).  This assumption is justified by numerics in the main paper for typical critical points and typical local minima, to which we will largely apply the generalized cavity method. 

In summary, we have the substantial simplification
\begin{equation}
     \left| \frac{\det H(\mathbf{x})}{\det H(\mathbf{x}^{/0})} \right| = \left| \partial^2 E_I(x_0) - t_R \right|,
     \label{eq:detratiotR}
\end{equation}
where $t_R$ satisfies  \eqref{eq:selfconstr}.  Thus as promised, in the determinantal ratio, $x_0$ interacts with the cavity system $\mathbf{x}^{/0}$ only through a simple order parameter, namely $t_R$.

\subsubsection{Spin-cavity interaction through a difference of indices}
To simplify the difference of indices $\mathcal{I}(\mathbf{x}) - \mathcal{I}(\mathbf{x}^{/0})$, we exploit the relation between the index and the determinant of the Hessian in \eqref{eq:index} to obtain 
\begin{eqnarray}
    \mathcal{I}(\mathbf{x}) - \mathcal{I}(\mathbf{x}^{/0}) &=& \lim_{\epsilon\to 0}\frac{1}{2\pi i}
        \left( \log\frac{\det[H(\mathbf{x}) - i\epsilon]}{\det[H(\mathbf{x}^{/0}) - i\epsilon]}
            - \log\frac{\det[H(\mathbf{x}) + i\epsilon]}{\det[H(\mathbf{x}^{/0}) + i\epsilon]}\right) \nonumber\\
    &=& \lim_{\epsilon\to 0}\frac{1}{2\pi i}
        \left( \log\left(\partial^2 E_I(x_0) - i\epsilon- N^{-1} \operatorname{Tr}[H(\mathbf{x}) -i\epsilon]^{-1}
        \right) - \log\left(\partial^2 E_I(x_0) + i\epsilon - N^{-1} \operatorname{Tr}[H(\mathbf{x}) +i\epsilon]^{-1}\right)\right)\nonumber\\
    &=& \lim_{\epsilon\to 0}\frac{1}{2\pi i}
        \left( \log\left(\partial^2 E_I(x_0) - i\epsilon- t_R
        \right) - \log\left(\partial^2 E_I(x_0) + i\epsilon - t_R \right)\right) \nonumber\\
    &=& \Theta\left(-(\partial^2 E_I(x_0)-t_R)\right) \nonumber\\
    &=& \overline{\mathcal{I}}(x_0).
    \label{eq:cavderivmfindex}
\end{eqnarray}
Here in the third line we have assumed that  $\lim_{\epsilon\to0}N^{-1}\operatorname{Tr}[H(
(\mathbf{x}))-i\epsilon]^{-1} = t_R$. 

Note that this calculation recovers none other than the mean field index function $\overline{\mathcal{I}}(x_0)$ calculated via the replica method in  \eqref{eqs:mean-field-index} in the special case in which $t_I=0$, so that $\overline{\mathcal{I}}(x)$ reduces to a simple indicator function derived both here and in  \eqref{eq:mean-field-index-t_I0}. Moreover, this indicator function is simply the index of a simple scalar mean field energy function given in  \eqref{eq:emf}.  We furthermore note, as described near the end of \ref{supsubsec:eigdisthess}, that the special case $t_I=0$ can chosen precisely when the eigenvalue density of the Hessian at the origin vanishes.
 
We note that  \eqref{eq:cavderivmfindex} suggests a striking regularity in random matrices of the form in  \eqref{eq:hessdef}. In particular, suppose that the distribution of the components of the cavity system $\mathbf{x}^{/0}$ is such that the eigenvalue density of its Hessian  $H(\mathbf{x}^{/0})$ vanishes at the origin. Suppose one then adds a new spin fixed at a value $x_0$ with a given random connectivity $\mathbf{J}_0$ to create a larger Hessian $H(\mathbf{x})$ as in the block matrix in  \eqref{eq:hessblock}.  Then  \eqref{eq:cavderivmfindex} implies that, with very high probability over the random choice of $\mathbf{J}_0$, the index of $H(\mathbf{x})$ will be one larger than the index of $H(\mathbf{x}^{/0})$ if and only if $\partial^2 E_I(x_0) < t_R$, and otherwise, the index will remain unchanged upon the addition of $x_0$.  Thus basically the curvature $\partial^2 E_I(x_0)$ has to be below the threshold $t_R$ in order to create an additional negative eigenvalue, and if this is not the case, then no additional negative eigenvalue will be created or lost. In a more complex situation where the eigenvalue density of $H(\mathbf{x}^{/0})$ is nonzero at the origin (i.e. proportional to $t_I$), then the typical change in the index is governed by the more complex mean field index function $\overline{\mathcal{I}}(x_0)$ in  \eqref{eqs:mean-field-index}.

Finally, we note that the sequence of steps in  \eqref{eq:cavderivmfindex} suggests a substantial simplification in the nature of the interaction between the new spin $x^0$ and the cavity system $\mathbf{x}^{/0}$. In particular, in the interaction term $\mathcal{I}(\mathbf{x}) - \mathcal{I}(\mathbf{x}^{/0})$, $x^0$ on the surface appears to be coupled to $\mathbf{x}^{/0}$ in a highly intricate manner. However  \eqref{eq:cavderivmfindex} reveals that in the large $N$ limit (and when the Hessian eigenvalue density at the origin is $0$), $x_0$ couples to the cavity system $\mathbf{x}^{/0}$ only through the simple order parameter $t_R$, just as it did for the ratio of determinants in  \eqref{eq:detratiotR}.

\subsubsection{Spin-cavity interaction due to the grand potential density in an external field}

We next simplify the grand potential density $\omega(\mathbf{s},\mathbf{x}^{/0})$ (see \eqref{eq:grandpotdens}) of the cavity system in an external field $\mathbf{s}$ where the external field $\mathbf{s}=x_0 \mathbf{J}_0$ is due to the addition of a new spin $x_0$ and its random connectivity $\mathbf{J}_0$ to the cavity system. Our goal is to write this grand potential density in this external field as the product of the grand potential density of the cavity system $\mathbf{x}^{/0}$ in the {\it absence} of an external field times an interaction term between the new spin $x^0$ and the cavity system $\mathbf{x}^{/0}$. This can be done straightforwardly by factoring out terms that specifically include $x_0$:
\begin{eqnarray}
    &&\omega(x_0 \mathbf{J}_0,\mathbf{x^{/0}})\nonumber\\
    &=&  \left(\frac{\beta}{\pi}\right)^{(N-1)/2}\exp\left(-\beta \sum_{i=1}^{N-1} \left(\partial E_I(x_i) + \sum_{j=1}^{N} J_{ij} x_j + x_0 J_{0i}\right)^2\right) \left|\det H(\mathbf{x}^{/0})\right| e^{\mu \mathcal{I}(\mathbf{x}^{/0})}\nonumber\\
    &=& \omega(0,\mathbf{x^{/0}})\exp\left(-2x_0\beta \sum_{i=1}^{N-1}\nabla_i E(\mathbf{x}^{/0}) J_{0i} -\beta x_0^2\sum_{i=1}^{N-1} J_{0i}^2\right)\nonumber\\
    &=& \omega(0,\mathbf{x^{/0}})\exp\left(-2x_0\beta \sum_{i=1}^{N-1}\nabla_i E(\mathbf{x}^{/0}) J_{0i} -\beta x_0^2\right),
    \label{eq:grandpotdenssimp}
\end{eqnarray}
where $\nabla_i E(\mathbf{x}^{/0})= \partial E_I(x_i) + \sum_{j=1}^{N} J_{ij} x_j$. 

\subsubsection{Averaging over the connectivity between the new spin and the cavity}

Now inserting into  \eqref{eq:P_x_cavity} the simplified formulas for the interaction between the new spin $x^0$ and the cavity $\mathbf{x}^{/0}$ coming from the determinantal ratio in  \eqref{eq:detratiotR}, the difference of indices in  \eqref{eq:cavderivmfindex}, and the grand potential density in an external field in  \eqref{eq:grandpotdenssimp}, we obtain the following expression for the grand potential of the full system $\Omega_\beta$:
\begin{eqnarray}
    &&\exp(N\Omega_\beta)\nonumber\\
    = &&\mathbb{E}_{J^{/0}}\left[\int \prod_{i=1}^{N-1} dx_i \omega(0,\mathbf{x}^{/0})\mathbb{E}_{\mathbf{J}_0}\left[\exp\left(- 2x_0\beta \sum_{i=0}^{N-1}\nabla_i E(\mathbf{x}^{/0}) J_{0i}-\beta x_0^2\right) \left(\frac{\beta}{\pi}\right)^{1/2}\exp\left(-\beta \left(\sum_{i=1}^{N-1} J_{0i} x_i + \partial E_I(x_0) \right)^2\right)\right]\right]\nonumber\\
    &&\ \ \ \times \left| \partial^2 E_I(x_0) - t_R \right| e^{\mu\bar{\mathcal{I}}(x_0)}.
\end{eqnarray}
The advantage of this expression is that the coupling terms between the new spin $x^0$ and the cavity system $\mathbf{x}^{/0}$ are made simple and manifest. Of course this coupling depends on the coupling vector $\mathbf{J}_0$ between the spin and the cavity, which we now average over. 

Inside the inner bracket $[\cdot]$, the entire dependence on the coupling vector $\mathbf{J}_0$ arises only through two scalar variables:
\begin{eqnarray}
    \bar h&:=&-\sum_{i=1}^{N-1} J_{0i} x_i \nonumber\\
    \bar z&:=&-2\beta\sum_{i=1}^{N-1}\nabla_i E(\mathbf{x}^{/0}) J_{0i}.\nonumber
\end{eqnarray} 
The scalars $\bar h$ and $\bar z$ measure the alignment of the coupling vector $\mathbf{J}_0$ to the cavity state $\mathbf{x}^{/0}$ and the gradient of the cavity state $\nabla E(\mathbf{x}^{/0})$ respectively.  Due to the randomness in $\mathbf{J}_0$, these two scalars are jointly Gaussian distributed with zero mean and covariance given by 
\begin{eqnarray}
    &&\Braket{\bar h^2}_{\mathbf{J}_0} = N^{-1}|\mathbf{x}^{/0}|^2\nonumber\\
    &&\Braket{\bar h\bar z}_{\mathbf{J}_0} = 2N^{-1}\beta \nabla E(\mathbf{x}^{/0})\cdot  \mathbf{x}^{/0}\nonumber\\
    &&\Braket{\bar z^2}_{\mathbf{J}_0} = 4N^{-1}\beta^2 |\nabla E(\mathbf{x}^{/0})|^2.
\end{eqnarray}
Thus the average over the $N-1$ dimensional spin-cavity coupling vector $\mathbf{J}_0$ can be simplified to a Gaussian average over the two correlated scalars which we denote by $\Braket{\cdot}_{\bar h,\bar z}$. This simplification yields
\begin{eqnarray}
    \exp(N\Omega_\beta)=\mathbb{E}_{J^{/0}}\left[\int \prod_{i=0}^{N-1} dx_i  \omega(0,\mathbf{x}^{/0}) \left(\frac{\beta}{\pi}\right)^{1/2}\Braket{\exp\left( x_0\bar z-\beta \left(\partial E_I(x_0) -\bar h \right)^2-\beta x_0^2\right) }_{\bar h,\bar z}
    \left| \partial^2 E_I(x_0) - t_R \right| e^{\mu\bar{\mathcal{I}}(x_0)}\right].\nonumber\\
\end{eqnarray}
Furthermore, we perform a change of variables of integration $\bar z\to z$ and $\bar h \to h$ according to the formula in \eqref{eq:h-z-formula} with $t=t_R$ and $s=2\beta$. This change of variables is helpful, first because we can eliminate the term $-\beta x_0^2$, and secondly, because we can make the term in the absolute value ($\partial^2 E_I(x_0)-t_R$) be the derivative of the term $\partial E_I(x_0)-t_Rx_0 - h$ in the Gaussian expression. This derivative relationship will enable us to apply the inverse of the Kac-Rice formula below to further simplify our expression for the grand potential. This change of variables achieves both the objectives, yielding
\begin{eqnarray}
    \exp(N\Omega_\beta)=\mathbb{E}_{J^{/0}}\left[\int \prod_{i=0}^{N-1} dx_i  \omega(0,\mathbf{x}^{/0}) \left(\frac{\beta}{\pi}\right)^{1/2}\Braket{\exp\left(x_0 z-\beta \left(\partial E_I(x_0) - t_R x_0 - h \right)^2\right) }_{h, z}
    \left| \partial^2 E_I(x_0) - t_R \right| e^{\mu\bar{\mathcal{I}}(x_0)}\right],\nonumber\\
\end{eqnarray}
where the covariance of new variables $z$ and $h$ is now given by
\begin{eqnarray}
    &&\Braket{h^2} = N^{-1}|\mathbf{x}^{/0}|^2\nonumber\\
    &&\Braket{hz} = 2N^{-1}\beta \nabla E(\mathbf{x}^{/0})\cdot \mathbf{x}^{/0} - t_R\nonumber\\
    &&\Braket{z^2} = 4N^{-1}\beta^2 |\nabla E(\mathbf{x}^{/0})|^2 - 2\beta.
\end{eqnarray}
We can further simplify the equation above by changing the variable of integration from $x_0$ to $y=\partial E_I(x_0) - t_R x_0 - h$. In each region $U_i\subset\mathbb{R}$ ($i=1,2,\cdots,n$) where $y=f(x_0) := \partial E_I(x_0) - t_R x_0 - h$ is monotonically increasing or decreasing, the restricted function $f|_{U_i}:U_i\to\mathbb{R}$ is injective, and we can change the the integration variable to obtain
\begin{eqnarray}
    &&\int_{U_i} dx_0\left(\frac{\beta}{\pi}\right)^{1/2}\Braket{\exp\left(x_0 z-\beta \left(\partial E_I(x_0) - t_R x_0 - h \right)^2\right) }_{h, z}
    \left| \partial^2 E_I(x_0) - t_R \right| e^{\mu\bar{\mathcal{I}}(x_0)} \nonumber\\
    &=& \int_{f(U_i)} dy \left(\frac{\beta}{\pi}\right)^{1/2}e^{-\beta y^2}e^{f|_{U_i}^{-1}(y)z+\mu\bar{\mathcal{I}}(f|_{U_i}^{-1}(y))}.
\end{eqnarray}
Note that the Jacobian of this change of variable is given by $\frac{\partial y}{\partial x_0} = \partial E_I - t_R x$.
Assuming that $f(x_0)$ does not have any flat region with finite measure, the integral over the entire space $\mathbb{R}$ can be obtained by summing up the integral above over region $U_i$, and 
\begin{eqnarray}
    &&\int_\mathbb{R} dx_0\left(\frac{\beta}{\pi}\right)^{1/2}\Braket{\exp\left(x_0 z-\beta \left(\partial E_I(x_0) - t_R x_0 - h \right)^2\right) }_{h, z}
    \left| \partial^2 E_I(x_0) - t_R \right| e^{\mu\bar{\mathcal{I}}(x_0)} \nonumber\\
    &=& \sum_{i= 1,\cdots,n}\int_{f(U_i)} dy \left(\frac{\beta}{\pi}\right)^{1/2}e^{-\beta y^2}e^{f|_{U_i}^{-1}(y)z+\mu\bar{\mathcal{I}}(f|_{U_i}^{-1(y)})}\nonumber\\
    &=& \int_{\mathbb{R}} dy \left(\frac{\beta}{\pi}\right)^{1/2}e^{-\beta y^2}\sum_{x\in X(y)}e^{xz+\mu\bar{\mathcal{I}}(x)},
\end{eqnarray}
where $X(y) = \{f|_{U_i}^{-1}(y)\}_{1\leq i\leq n}$. By noticing that $X(y)$ is the set of points $x$ that satisfies $\partial E_I(x) - t_R x - (y+h) = 0$, which is the stationary condition of a 1-dimensional mean-field energy function 
\begin{equation}
    E_{\operatorname{MF}}[h+y](x) := E_I(x) - \frac{t_R}{2}x^2 - (h+y)x,
\end{equation}
we can say that
\begin{equation}
    X(y) = \operatorname{Crt}(E_{\operatorname{MF}}[y+h])
\end{equation}
where $\operatorname{Crt}(E_{\operatorname{MF}}[y+h])$ represents the set of critical points of the 1-dimensional mean-field energy function $E_{\operatorname{MF}}[y+h](x)$. Therefore,
\begin{eqnarray}
    \exp(N\Omega_\beta)=\mathbb{E}_{J^{/0}}\left[\int \prod_{i=1}^{N-1} dx_i \omega(0,\mathbf{x}^{/0}) \Braket{\int dy \left(\frac{\beta}{\pi}\right)^{1/2}e^{-\beta y^2}\sum_{x\in \operatorname{Crt}(E_{\operatorname{MF}}[y+h])}e^{xz+\mu\bar{\mathcal{I}}(x)}}_{h,z}\right],
    \label{eq:Omega-before-qAC}
\end{eqnarray}
This sequence of steps starting from an integral over $x^0$ of the determinant of a term that is the derivative of a quantity appearing inside the Gaussian representation of a delta function, to finally summing over the critical points of a $1$ dimensional mean field energy function, is basically tantamount to inverting the Kac-Rice formula.

\subsubsection{Introduction of three more cavity order parameters q, A, C}
In \eqref{eq:Omega-before-qAC}, the Gaussian average $\Braket{\cdot}_{h,z}$ has a covariance matrix that depends on the cavity system $\mathbf{x}^{/0}$ only through the following three quantities, which we define to be $q(\mathbf{x}^{/0}),A(\mathbf{x}^{/0})$, and  $C(\mathbf{x}^{/0})$:
\begin{equation}
    \begin{cases}
    q(\mathbf{x}^{/0}) = (N-1)^{-1}|\mathbf{x}^{/0}|^2 \approx \Braket{h^2}\nonumber\\
    A(\mathbf{x}^{/0}) = 2(N-1)^{-1}\beta \nabla E(\mathbf{x}^{/0})\cdot\mathbf{x}^{/0}-t_R \approx \Braket{hz}\nonumber\\
    C(\mathbf{x}^{/0}) = 4(N-1)^{-1}\beta^2 \lvert\nabla E(\mathbf{x}^{/0})\rvert^2-2\beta \approx \Braket{z^2}
    \end{cases}
\end{equation}
We assume that $q(\mathbf{x}^{/0}), A(\mathbf{x}^{/0}), C(\mathbf{x}^{/0})$ concentrate around their expectation values $q,A,C$ under the cavity grand potential density $\omega(0,\mathbf{x}^{/0})$. Under this assumption, the average $\Braket{\cdot}_{h,z}$ can be taken outside of the integral over $\mathbf{x}^{/0}$, obtaining
\begin{eqnarray}
    \exp(N\Omega_\beta)=\exp\left((N-1)\tilde\Omega_\beta\right) \Braket{\int dy \left(\frac{\beta}{\pi}\right)^{1/2}e^{-\beta y^2}\sum_{x\in \operatorname{Crt}(E_{\operatorname{MF}}[y+h])}e^{xz+\mu\bar{\mathcal{I}}(x)} }_{h,z},
    \label{eq:expNomegabeta}
\end{eqnarray}
where the covariance of $h$ and $z$ is given by $\left(\begin{matrix}q&A \\A&C\end{matrix}\right)$ and $\tilde\Omega_\beta$ is defined as
\begin{eqnarray}
    \exp((N-1)\tilde\Omega_\beta)) := \mathbb{E}_{J^{/0}}\left[\int \prod_{i=1}^{N-1} dx_i \omega(0,\mathbf{x^{/0}})\right].
    \label{eq:tildeOmega_beta(qA'C)}
\end{eqnarray}
Now taking the logarithm of both sides of  \eqref{eq:expNomegabeta} and then subtracting $(N-1)\Omega_\beta$ from both sides yields a formula for the grand potential of the full system:
\begin{equation}
    \Omega_{\beta} = (N-1)(\tilde\Omega_\beta - \Omega_\beta) + \log\Braket{\int dy \left(\frac{\beta}{\pi}\right)^{1/2}e^{-\beta y^2}\sum_{x\in \operatorname{Crt}(E_{\operatorname{MF}}[y+h])}e^{xz+\mu\bar{\mathcal{I}}(x)} }_{h,z}.
    \label{eq:omegabcavint}
\end{equation}

\subsubsection{Analyzing a cavity system with a mismatch between system size and connectivity variance}

We next need to evaluate the difference $\tilde \Omega_\beta - \Omega_\beta$. 
The reason this difference is nonzero is because of a mismatch in the system size of the cavity and the variance of its connectivity. In particular, in the definition of the original grand potential $\Omega_\beta$ in Eq \eqref{eq:Omega-cavity-finite-beta}, the full system size is $N$ and the variance of each connection is $\frac{1}{N}$. 
However, in the definition of $\tilde\Omega_\beta$ in  \eqref{eq:tildeOmega_beta(qA'C)}, the cavity system size is $N-1$, but its connectivity variance is still $\frac{1}{N}$. 
In contrast for the same system size $N-1$ the original $\Omega_\beta$ should have a connectivity variance of $\frac{1}{N-1}$. 
To account for this mismatch between system size and variance, we fix the system size to be $N-1$ and we parameterize the variance as $\sigma N^{-1}$ where $\sigma=1+(N-1)^{-1}$ for $\Omega_\beta$ (yielding the correct variance $\frac{1}{N-1}$) and $\sigma=1$ for $\tilde\Omega_\beta$ (yielding a mismatched variance $\frac{1}{N}$).  
We consider a generalized grand potential as a function of the (potentially mismatched) variance parameter $\sigma$ and we denote it as $\Omega_\beta(\sigma)$. The derivative of this function with respect to $\sigma$ is equivalent to $-(N-1)(\tilde \Omega_\beta - \Omega_\beta)$ in the large-$N$ limit, as we can see in the following equation.
\begin{equation}
    \lim_{N\to \infty} (N-1)(\tilde \Omega_\beta - \Omega_\beta) = -\lim_{N\to \infty}\frac{\Omega_\beta(1+(N-1)^{-1}) - \Omega_\beta(1)}{(N-1)^{-1}} =- \left.\frac{d\Omega_{\beta}(\sigma)}{d\sigma}
     \right|_{\sigma=1}.
     \label{eq:mismatchderiv}
\end{equation}
The derivative can be calculated as follows.
\begin{eqnarray}
    &&\frac{d\Omega_{\beta}(\sigma)}{d\sigma}\nonumber\\ 
    &=& (N-1)^{-1}\frac{d}{d\sigma}\log\Braket{\int \prod^{N-1}_{i=1}dx_i \left(\frac{\beta}{\pi}\right)^{(N-1)/2}\exp\left(-\beta \sum_{i=1}^{N-1} \left(\partial E_I(x_i) + \sum_{j=1}^{N} J_{ij} x_j\right)^2\right) \left|\det H(\mathbf{x})\right| e^{\mu \mathcal{I}(\mathbf{x})}}_{J}\nonumber\\
    &=& (N-1)^{-1}N^{-1}e^{-(N-1)\Omega_\beta}\nonumber\\
    &&\ \ \ \ \times\frac{1}{2}\sum_{0<i<j\leq n}\Braket{\frac{\partial^2}{\partial J_{ij}^2}\int \prod^{N-1}_{i=1}dx_i \left(\frac{\beta}{\pi}\right)^{(N-1)/2}\exp\left(-\beta \sum_{i=1}^{N-1} \left(\partial E_I(x_i) + \sum_{j=1}^{N} J_{ij} x_j\right)^2\right) \left|\det H(\mathbf{x})\right| e^{\mu \mathcal{I}(\mathbf{x})}}_{J}\nonumber\\
\end{eqnarray}
Note that we used the following formula for any centered Gaussian random vector $\mathbf{g}$ with a covariance matrix $M$:
\begin{eqnarray}
    \frac{\partial}{\partial M_{ij}}\mathbb{E}X(\mathbf{g}) = \frac{1}{2}\mathbb{E}\left[\frac{\partial^2}{\partial g_i\partial g_j}X(\mathbf{g})\right].
    \label{eq:gaussderivcov}
\end{eqnarray}

Now, we need to evaluate the second derivative with respect to  $J_{ij}$.
There are three terms; the second derivative of the determinant, the second derivative of Gaussian, and the product of two first derivatives of the Gaussian and of determinant. The last term represents the correlation between the Gaussian and the determinant of Hessian. We can safely neglect this correlation in the large $N$ limit (this can be confirmed by successfully confronting our theory with numerical simulations as is done in the main paper).  Therefore, we will evaluate just the first two terms.

The derivative of the determinant can be calculated as follows. The first derivative is
\begin{equation}
    \frac{\partial}{\partial J_{ij}}|\det H| = 2|\det H| H^{-1}_{ij}.
\end{equation}
Therefore,
\begin{eqnarray}
    N^{-2}\sum_{0<i<j\leq n}\left[\frac{\partial^2}{\partial J_{ij}^2}|\det H|\right] / |\det H|&=& 2N^{-2}\sum_{0<i<j\leq n}\left[\frac{\partial}{\partial J_{ij}}|\det H| H^{-1}_{ij} \right]/ |\det H|\nonumber\\
    &=& 2N^{-2}\sum_{0<i<j\leq n} \left(2\left(H^{-1}_{ij}\right)^2 - H^{-1}_{ii}H^{-1}_{jj} - \left(H^{-1}_{ij}\right)^2\right)\nonumber\\
    &=& N^{-2}\sum_{0<i\neq j\leq n} \left(\left(H^{-1}_{ij}\right)^2 - H^{-1}_{ii}H^{-1}_{jj}\right)\nonumber\\
    &=& N^{-2}\sum_{0<i, j\leq n} \left(\left(H^{-1}_{ij}\right)^2 - H^{-1}_{ii}H^{-1}_{jj}\right)\nonumber\\
    &=& N^{-2}\operatorname{Tr}H^{-2} - (N^{-1}\operatorname{Tr}H^{-1})^2\nonumber\\
    &=& -t_R^2.
\end{eqnarray}
For the last equality, we assume that $N^{-2}\operatorname{Tr}H^{-2}$ vanishes under the large-$N$ limit.
Next, the second derivative of the Gaussian part is obtained as follows.
\begin{eqnarray}
    &&N^{-2}\left[\sum_{0<i<j\leq n}\frac{\partial^2}{\partial J_{ij}^2} \exp\left(-\beta \sum_{i=1}^{N-1} \left(\partial E_I(x_i) + \sum_{j=1}^{N} J_{ij} x_j\right)^2\right)\right] / \exp\left(-\beta \sum_{i=1}^{N-1} \left(\partial E_I(x_i) + \sum_{j=1}^{N} J_{ij} x_j\right)^2\right)
    \nonumber\\
    &=& N^{-2}\sum_{0<i<j\leq n}\left(-2\beta\nabla_i E(\mathbf{x}) x_j -2\beta\nabla_j E(\mathbf{x}) x_i\right)^2 -2\beta (x_i^2+x_j^2)\nonumber\\
    &=& N^{-2}\sum_{0<i<j\leq n}8\beta^2 (\nabla_i E(\mathbf{x}))^2 x^2_j + 8\beta^2 \nabla_i E(\mathbf{x})x_i \nabla_j E(\mathbf{x}) x_j -2\beta (x_i^2+x_j^2)\nonumber\\
    &=& N^{-2}\sum_{0<i\neq j\leq n}4\beta^2 (\nabla_i E(\mathbf{x}))^2 x^2_j + 4\beta^2 \nabla_i E(\mathbf{x})x_i \nabla_j E(\mathbf{x}) x_j -\beta (x_i^2+x_j^2)\nonumber\\
    &=& N^{-2}\sum_{0<i,j\leq n}4\beta^2 (\nabla_i E(\mathbf{x}))^2 x^2_j + 4\beta^2 \nabla_i E(\mathbf{x})x_i \nabla_j E(\mathbf{x}) x_j -\beta (x_i^2+x_j^2)\nonumber\\
    &=& qC + (A+t_R)^2\nonumber\\
\end{eqnarray}
For the second to the last equality, we ignored the terms with $i=j$, since the number of terms is $N$, which is much smaller than $N^{2}$.

Thus in total, we have
\begin{eqnarray}
    \left.\frac{d\Omega_\beta(\sigma)}{d\sigma}\right|_{\sigma=1} &=& \frac{1}{2}qC + \frac{1}{2}A^2 +A t_R.
\end{eqnarray}
Inserting this result into  \eqref{eq:mismatchderiv} and then in turn into  \eqref{eq:omegabcavint} we obtain 
\begin{eqnarray}
    \Omega_\beta &=& -\frac{1}{2}(Cq + A^2) - A t_R +\log\Braket{\int dy \left(\frac{\beta}{\pi}\right)^{1/2}e^{-\beta y^2}\sum_{x\in \operatorname{Crt}(E_{\operatorname{MF}}[y+h])}e^{xz+\mu\bar{\mathcal{I}}(x_0)} }_{h,z}.
    \label{eq:Omega_beta}
\end{eqnarray}
Finally, by taking the limit $\beta\to\infty$, we obtain our first main result of the generalized cavity derivation:
\begin{eqnarray}
    \Omega &=& -\frac{1}{2}(Cq + A^2) - A t_R+\log \Braket{\sum_{x\in\mathrm{Crt}(E_{\operatorname{MF}}[h])}e^{xz+\mu\bar{\mathcal{I}}(x)}}_{ h,z}.
    \label{eq:Omega-cavity2}
\end{eqnarray}
This result yields a simple formula for the connectivity averaged grand potential originally defined in \eqref{eq:Omega-cavity}.  There it involved summing over critical points, with a chemical potential factor weight, in an ensemble of $N$ dimensional energy landscapes.  Here we have reduced it to summing over critical points in an ensemble of $1$ dimensional mean field energy landscapes, with both the chemical potential factor weight, and an additional random weight of the form $e^{xz}$.  This mean field expression for the grand potential depends on $4$ order parameters $A$, $C$, $q$ and $t_R$. We will next show how to derive self-consistent equations for these order parameters.  Eventually we will show that our generalized cavity method yields identical results to that of the replica method for both the potential and the self-consistent equations for the order parameters.

\subsection{Deriving self-consistent cavity equations for the order parameters}
As is usually done in the cavity method, we assume that the joint distribution of $\mathbf{x}$ and $\nabla E(\mathbf{x})$ can be factorized into a product of $N$ copies of the independent distribution $P_\beta(x,y)$, defined in \eqref{eq:P(x,y)-cavity}. Under this assumption any self-averaging function of $\mathbf{x}$ and $\nabla E(\mathbf{x})$ that tightly concentrates about it's mean can be well approximated as an average over the mean field distribution $P_\beta(x,y)$.  For example, for any sufficiently well-behaved function $F(x,y)$ we assume that in the large $N$ limit we have the convergence 
\begin{equation}
    \frac{1}{N} \sum_{i=1}^N F(x_i, \nabla_i E(\mathbf{x})) \rightarrow
    \int dxdy P_\beta(x,y) F(x,y).
\end{equation}
Under this assumption, the functions $q(\mathbf{x}),A(\mathbf{x}),C(\mathbf{x})$ will concentrate about their means as follows: 
\begin{equation}
    \begin{cases}
    q(\mathbf{x}) = N^{-1}|\mathbf{x}|^2 \to \int dxdy P_\beta(x,y)x^2\\
    A(\mathbf{x}) = 2N^{-1}\beta \nabla E(\mathbf{x})\cdot\mathbf{x}-t_R \to 2\beta\int dxdy P_\beta(x,y)xy- t_R\\
    C(\mathbf{x}) = 4N^{-1}\beta^2 \lvert\nabla E(\mathbf{x})\rvert^2-2\beta \to 4\beta^2\int dxdy P_\beta(x,y)y^2 - 2\beta.
    \label{eq:self-consistent-1}
    \end{cases}
\end{equation}
The distribution $P_\beta(x,y)$ can be obtained in the same manner as we discussed above, by noticing that $\nabla_0 E(\mathbf{x}) = \partial E_I(x_0) + \sum_i J_{0i} x_i = \partial E_I(x_0) -\bar h \to \partial E_I(x_0) - t_Rx_0 -h$. (Recall that we replace the Gaussian variable $\bar h$ with $h$ by applying \eqref{eq:h-z-formula}. Then we obtain 
\begin{eqnarray}
    P_\beta(x,y)&=& Z_\beta^{-1}\Braket{\int dy_* \left(\frac{\beta}{\pi}\right)^{1/2}e^{-\beta y_*^2}\sum_{x_*\in \operatorname{Crt}(E_{\operatorname{MF}}[y_*+h])}e^{x_*z+\mu\bar{\mathcal{I}}(x_*)} \delta(x_*-x)\delta(\partial E_I(x_*)-t_Rx_* -h - y)}_{h,z},\nonumber\\
    &=& Z_\beta^{-1}\Braket{\int dy_* \left(\frac{\beta}{\pi}\right)^{1/2}e^{-\beta y_*^2}\sum_{x_*\in \operatorname{Crt}(E_{\operatorname{MF}}[y_*+h])}e^{x_*z+\mu\bar{\mathcal{I}}(x_*)} \delta(x_*-x)\delta(y_* - y)}_{h,z},
    \label{eq:P-beta-xy}
\end{eqnarray}
where
\begin{equation}
    Z_\beta = \Braket{\int dy \left(\frac{\beta}{\pi}\right)^{1/2}e^{-\beta y^2}\sum_{x\in \operatorname{Crt}(E_{\operatorname{MF}}[y+h])}e^{xz+\mu\bar{\mathcal{I}}(x)} }_{h,z}.
\end{equation}
Here the second equality holds, since $x_*$ is a critical point of $E_{\operatorname{MF}}[h+y_*](x) := E_I(x) - \frac{t_R}{2}x^2 - (h+y_*)x$, and thus $y_* = \partial E_I(x_*)-t_Rx_* -h$.
Thus substituting \eqref{eq:P-beta-xy} into  \eqref{eq:self-consistent-1} leads to self-consistent equations for the order parameters:
\begin{eqnarray}
    q = Z_\beta^{-1}\Braket{\int dy \left(\frac{\beta}{\pi}\right)^{1/2}e^{-\beta y^2}\sum_{x\in \operatorname{Crt}(E_{\operatorname{MF}}[y+h])}e^{xz+\mu\bar{\mathcal{I}}(x)} x^2}_{h,z}
\end{eqnarray}
\begin{eqnarray}
    A&=& 2\beta Z_\beta^{-1}\Braket{\int dy \left(\frac{\beta}{\pi}\right)^{1/2}e^{-\beta y^2}\sum_{x\in \operatorname{Crt}(E_{\operatorname{MF}}[y+h])}e^{xz+\mu\bar{\mathcal{I}}(x)} xy}_{h,z}-t_R\nonumber\\
    &=& -Z_\beta^{-1}\Braket{\int dy \left(\frac{\beta}{\pi}\right)^{1/2}\left[\frac{d}{dy}e^{-\beta y^2}\right]\sum_{x\in \operatorname{Crt}(E_{\operatorname{MF}}[y+h])}e^{xz+\mu\bar{\mathcal{I}}(x)} x}_{h,z}-t_R\nonumber\\
    &=& Z_\beta^{-1}\Braket{\int dy \left(\frac{\beta}{\pi}\right)^{1/2}e^{-\beta y^2}\frac{d}{dh}\sum_{x\in \operatorname{Crt}(E_{\operatorname{MF}}[y+h])}e^{xz+\mu\bar{\mathcal{I}}(x)} x}_{h,z}-t_R\nonumber\\
\end{eqnarray}
\begin{eqnarray}
    C &=& 4\beta^2Z_\beta^{-1}\Braket{\int dy \left(\frac{\beta}{\pi}\right)^{1/2}e^{-\beta y^2}\sum_{x\in \operatorname{Crt}(E_{\operatorname{MF}}[y+h])}e^{xz+\mu\bar{\mathcal{I}}(x)} y^2}_{h,z}-2\beta\nonumber\\
    &=& Z_\beta^{-1}\Braket{\int dy \left(\frac{\beta}{\pi}\right)^{1/2}\left[\frac{d^2}{dy^2}e^{-\beta y^2}\right]\sum_{x\in \operatorname{Crt}(E_{\operatorname{MF}}[y+h])}e^{xz+\mu\bar{\mathcal{I}}(x)} }_{h,z}\nonumber\\
    &=& Z_\beta^{-1}\Braket{\int dy \left(\frac{\beta}{\pi}\right)^{1/2}e^{-\beta y^2}\frac{d^2}{dh^2}\sum_{x\in \operatorname{Crt}(E_{\operatorname{MF}}[y+h])}e^{xz+\mu\bar{\mathcal{I}}(x)} }_{h,z}.
\end{eqnarray}

Lastly, by taking the low temperature limit $\beta\to\infty$, we obtain self-consistent equations for the order parameters associated specifically with critical points where the gradient vanishes exactly:
\begin{equation}
    \label{eq:cavityselfconsistent}
    \begin{cases}
    q = Z^{-1}\Braket{\sum_{x\in \operatorname{Crt}(E_{\operatorname{MF}}[h])}e^{xz+\mu\bar{\mathcal{I}}(x)} x^2}_{h,z}\\
    A = Z^{-1}\Braket{\frac{d}{dh}\sum_{x\in \operatorname{Crt}(E_{\operatorname{MF}}[h])}e^{xz+\mu\bar{\mathcal{I}}(x)} x}_{h,z} - t_R\\
    C = Z^{-1}\Braket{\frac{d^2}{dh^2}\sum_{x\in \operatorname{Crt}(E_{\operatorname{MF}}[h])}e^{xz+\mu\bar{\mathcal{I}}(x)}}_{h,z},
    \end{cases}
\end{equation}
with $Z = \Braket{\sum_{x\in \operatorname{Crt}(E_{\operatorname{MF}}[h])}e^{xz+\mu\bar{\mathcal{I}}(x)}}_{h,z}$. Note the function $\sum_{x\in \operatorname{Crt}(E_{\operatorname{MF}}[h])}e^{xz+\mu\bar{\mathcal{I}}(x)}$ can be non-differentiable with respect to $h$ for some values of $h$, but we can define the expectation value of the derivatives through integration by parts. Similarly, the self-consistent equation for $t_R$ in  \eqref{eq:selfconstr}  can be represented as follows by substituting the expression for  $P_{\beta}(x,y)$ and taking the limit $\beta\to\infty$
\begin{eqnarray}
\label{eq:tRselfconsistentcavity}
    t_R &=&
    \lim_{\beta\to\infty}\int dx dy P_{\beta}(x,y) \frac{1}{\partial^2 E_I(x)-t_R}\nonumber\\
    &=& \int dx Z^{-1} \Braket{\sum_{x^*\in \operatorname{Crt}(E_{\operatorname{MF}}[h])}e^{x^*z+\mu\bar{\mathcal{I}}(x^*)}\delta(x-x^*) \frac{1}{\partial^2 E_I(x)-t_R}}_{h,z}\nonumber\\
    &=& Z^{-1} \Braket{\sum_{x\in \operatorname{Crt}(E_{\operatorname{MF}}[h])}e^{xz+\mu\bar{\mathcal{I}}(x)}\frac{1}{\partial^2 E_I(x)-t_R}}_{h,z}\nonumber\\
    &=& Z^{-1} \Braket{\sum_{x\in \operatorname{Crt}(E_{\operatorname{MF}}[h])}e^{xz+\mu\bar{\mathcal{I}}(x)}\frac{dx}{dh}}_{h,z}.
\end{eqnarray}
In the last step we have used linear response theory, namely the response $\frac{dx}{dh}$ of a critical point $x$ of the mean field energy function to an external field $h$ is given by the inverse of the Hessian of the energy function at $x$. 

These self-consistent equations can also be given a variational characterization. Indeed, they are equivalent to the equations obtained by extremizing the right hand side of \eqref{eq:Omega-cavity2} with respect to $q,A,C,t_R$ i.e.,
\begin{equation}
    \Omega = \underset{(q,A,C,t_R)}{\mathrm{ext}} -\frac{1}{2}(Cq + A^2) - A t_R+\log \Braket{\sum_{x\in\mathrm{Crt}(E_{\operatorname{MF}}[h])}e^{xz+\mu\bar{\mathcal{I}}(x)}}_{ h,z}.
\end{equation}
This equivalence can be seen by explicitly calculating the derivatives of the right hand side of \eqref{eq:Omega-cavity2} as follows: 
\begin{eqnarray}
    \label{eq:selfconsistentint}
    \frac{\partial\Omega(q,A,C,t_R)}{dq} &=& -\frac{C}{2} + \frac{1}{2}Z^{-1}\Braket{\frac{d^2}{dh^2}\sum_{x\in \operatorname{Crt}(E_{\operatorname{MF}}[h])}e^{xz+\mu\bar{\mathcal{I}}(x)}}_{h,z}  = 0\nonumber\\
    \frac{\partial\Omega(q,A,C,t_R)}{dA} &=& -A - t_R+ Z^{-1}\Braket{\frac{d^2}{dhdz}\sum_{x\in \operatorname{Crt}(E_{\operatorname{MF}}[h])}e^{xz+\mu\bar{\mathcal{I}}(x)}}_{h,z}  = 0\nonumber\\
    \frac{\partial\Omega(q,A,C,t_R)}{dC} &=& -\frac{q}{2} + \frac{1}{2}Z^{-1}\Braket{\frac{d^2}{dz^2}\sum_{x\in \operatorname{Crt}(E_{\operatorname{MF}}[h])}e^{xz+\mu\bar{\mathcal{I}}(x)}}_{h,z}  = 0\nonumber\\
    \frac{\partial\Omega(q,A,C,t_R)}{dt_R} &=& -A + Z^{-1}\frac{\partial}{\partial t_R}\Braket{\sum_{x\in \operatorname{Crt}(E_{\operatorname{MF}}[h])}e^{xz+\mu\bar{\mathcal{I}}(x)}}_{h,z} = 0.
\end{eqnarray}
Here we have used \eqref{eq:gaussderivcov} to convert derivatives of Gaussian averages of a function over $h$ and $z$ with respect to their covariance parameters $q$, $A$, and $C$, into Gaussian averages of derivatives of the same function with respect to $h$ and $z$. Now note that the first $3$ equations here are straightforwardly equivalent to the first $3$ self-consistent equations in \eqref{eq:cavityselfconsistent}. However, demonstrating the equivalence for the last equation for $t_R$ is a bit more involved, since the function inside the bracket is not differentiable with respect to $t_R$, for some values of $h$. To overcome this issue, we first transform the average $\Braket{\cdot}_{h,z}$ term as follows
\begin{eqnarray}
    \Braket{\sum_{x\in \operatorname{Crt}(E_{\operatorname{MF}}[h])}e^{xz+\mu\bar{\mathcal{I}}(x)}}_{h,z} &=&
    \int dh dz \exp\left(K(h,z)\right)\sum_{x\in \operatorname{Crt}(E_{\operatorname{MF}}[h])}e^{xz+\mu\bar{\mathcal{I}}(x)}\nonumber\\
    &=& \int dx dz w(x)\exp\left(K(\partial E_I(x)-t_R x ,z)\right)e^{xz+\mu\bar{\mathcal{I}}(x)}
\end{eqnarray}
where $K(h,z) = -\frac{1}{2} \left(\begin{matrix}h\\ z\end{matrix}\right)^T \left(\begin{matrix}q&A\\A&C \end{matrix}\right)^{-1}\left(\begin{matrix}h\\ z\end{matrix}\right)$ and $w(x) = |\partial^2 E_I(x)-t_R|$. This second step changes the integration variable from $h$ to $x$ using the relation $h = \partial E_I(x)-t_R x$ that holds at critical points of the mean field energy function. Thus we have, 
\begin{eqnarray}
    Z^{-1}\frac{\partial}{\partial t_R} \Braket{\sum_{x\in \operatorname{Crt}(E_{\operatorname{MF}}[h])}e^{xz+\mu\bar{\mathcal{I}}(x)}}_{h,z} 
    &=& Z^{-1}\int dx dz w(x)\frac{\partial}{\partial t_R}\exp\left(K(\partial E_I(x)-t_R x ,z)\right)e^{xz+\mu\bar{\mathcal{I}}(x)} \nonumber\\
    &&- Z^{-1}\int dx dz (\partial^2 E_I(x)-t_R)^{-1}w(x) \exp\left(K(\partial E_I(x)-t_R x ,z)\right)e^{xz+\mu\bar{\mathcal{I}}(x)}\nonumber\\
    &=& -Z^{-1}\int dx dz w(x) x\left.\frac{\partial}{\partial h}\right|_{h=\partial E_I(x)-t_R x}\left[\exp\left(K(h ,z)\right)\right]e^{xz+\mu\bar{\mathcal{I}}(x)} - t_R\nonumber\\
    &=&Z^{-1}\Braket{\frac{d}{dh}\sum_{x\in\mathrm{Crt}(E_{\operatorname{MF}}[h])}e^{xz+\mu\bar{\mathcal{I}}(x)}x}_{h,z} - t_R\nonumber\\
    &=& A.
\end{eqnarray}
Notice that in the second step we exploited the third line of \eqref{eq:tRselfconsistentcavity} (after changing the integration variable from $h$ to $x$) to convert the second term after the first equality to the second term after the second equality. We then arrive at an equation equivalent to the second equation in \eqref{eq:selfconsistentint}.  Thus taken together, the $4$ extremal conditions for the grand potential $\Omega$ in \eqref{eq:Omega-cavity2} are equivalent to the $4$ self-consistent cavity equations in \eqref{eq:cavityselfconsistent} and \eqref{eq:tRselfconsistentcavity}.

\subsection{A geometric interpretation of the supersymmetry breaking order parameters \texorpdfstring{$A, C$}{A,C}}
In this subsection, we show that the order parameters $A$ and $C$ can be interpreted as certain susceptibilities of the grand potential $\Omega$. We consider the following perturbation of the original energy function.
\begin{equation}
    E'(\mathbf{x}) = E(\mathbf{x}) - \frac{a}{2}|\mathbf{x}|^2 + \sqrt{2s}\mathbf{g}\cdot \mathbf{x},
\end{equation}
where $a$ is a constant for the quadratic term, and the other term $\mathbf{g}\cdot \mathbf{x}$ represents the coupling with the external random field $\mathbf{g}$, a centered Gaussian vector with unit variance. $s$ is the scalar coupling constant. The corresponding mean-field energy with the same perturbations is defined as
\begin{equation}
    E'_{\mathrm{MF}}[h,a,\sqrt{2s}g](x) = E_{\mathrm{MF}}[h](x) - \frac {a}{2}x^2 + \sqrt{2s}gx
    \left(= E_{I}(x) - \frac {t_R + a}{2}x^2 + \left(\sqrt{2s}g - h\right)x\right).
\end{equation}

In this subsection, we will show that $A,C$ are susceptibilities of the grand potential with respect to the quadratic perturbation and the random linear coupling, respectively. 
We first consider the case where we only have the quadratic perturbation ($a>0$ and $s=0$). The grand potential $\Omega(a)$ of the perturbed energy function  should be given by
\begin{equation}
    \Omega(a) = -\frac{1}{2}(C(a)q(a) + A^2(a)) - A(a) t_R(a)+\log \Braket{\sum_{x\in\mathrm{Crt}(E'_{\operatorname{MF}}[h,a,0])}e^{xz+\mu\bar{\mathcal{I}}(x)}}_{ h,z}.
\end{equation}
The order parameters $q(a),A(a),C(a),t_R(a)$ satisfy the self-consistent equations, which implies that they satisfy the stationary condition for $\Omega(a)$. Now we consider the derivative of $\Omega(a)$ with respect to $a$. 
\begin{eqnarray}
\label{eq:A-is-derivative-by-a}
    \left.\frac{d\Omega}{da}\right|_{a=0}
    &=& \left.\frac{d A}{d a}\right|_{a=0}\left.\frac{\partial \Omega}{\partial A} + \frac{d C}{d a}\right|_{a=0}\left.\frac{\partial \Omega}{\partial C} + \frac{dq}{d a}\right|_{a=0}\left.\frac{\partial \Omega}{\partial q} + \frac{d t_R}{d a}\right|_{a=0}\left.\frac{\partial \Omega}{\partial t_R} +\frac{\partial \Omega}{\partial a}\right|_{a=0}\nonumber\\ 
    &=& \left.\frac{\partial \Omega}{\partial a}\right|_{a=0}\nonumber\\ 
    &=& \left.\frac{\partial}{\partial a}\right|_{a=0}\log \Braket{\sum_{x\in\mathrm{Crt}(E'_{\operatorname{MF}}[h,a,0])}e^{xz+\mu\bar{\mathcal{I}}(x)}}_{ h,z}\nonumber\\ 
    &=&Z^{-1}\left.\frac{\partial}{\partial a}\right|_{a=0}\Braket{\sum_{x\in\mathrm{Crt}(E'_{\operatorname{MF}}[h,a,0])}e^{xz+\mu\bar{\mathcal{I}}(x)}}_{h,z}\nonumber\\
    &=& Z^{-1}\frac{\partial}{\partial t_R}\Braket{\sum_{x\in\mathrm{Crt}(E_{\operatorname{MF}}[h])}e^{xz+\mu\bar{\mathcal{I}}(x)}}_{h,z}\nonumber\\
    &=& A.
\end{eqnarray}
Here for the second equality, we exploit the fact that $\Omega$ should satisfy the stationary condition with respect to order parameters $A, C, q$, and $t_R$.
This equality shows that $A$ can be interpreted as a susceptibility of the grand potential with respect to the strength of the quadratic perturbation.

Similarly, we consider the case only with the coupling with the random external field (i.e., $a=0$, $s>0$). We define the averaged grand potential as follows.
\begin{eqnarray}
    \exp(N\Omega(s))
    &=& \mathbb{E}_g\mathbb{E}_J\left[\int \prod_{i=0}^{N-1} dx_i \prod_{i=0}^{N-1} \delta\left(\partial E_I(x_i) + \sum_{j=0}^{N-1} J_{ij} x_j + \sqrt{2s}g_i\right) \left|\det H(\mathbf{x})\right| e^{\mu I(\mathbf{x})}\right],
\end{eqnarray}
where $\mathbb{E}_g$ represents the average over the random external field $\mathbf{g}$. By calculating the average over $\mathbf{g}$, we get
\begin{eqnarray}
    \exp(N\Omega(s))
    &=& \mathbb{E}_J\left[\int \prod_{i=0}^{N-1} dx_i \left(\frac{1}{4s\pi}\right)^{N/2}\exp\left(-\frac{1}{4s} \sum_{i=0}^{N-1} \left(\partial E_I(x_i) + \sum_{j=0}^{N-1} J_{ij} x_j\right)^2\right) \left|\det H(\mathbf{x})\right| e^{\mu I(\mathbf{x})}\right]\nonumber\\
    &=& \exp(N\Omega_{\frac{1}{4s}}).
\end{eqnarray}
Therefore the derivative with respect to $s$ is given by
\begin{eqnarray}
    \left.\frac{d\Omega}{ds}\right|_{s=0} = \left.2\frac{d\Omega_\beta}{d\frac{1}{2}\beta^{-1}}\right|_{\beta=\infty} = Z_\beta^{-1}\left.\Braket{\int dy \left(\frac{\beta}{\pi}\right)^{1/2}e^{-\beta y^2}\frac{d^2}{dy^2}\sum_{x\in \operatorname{Crt}(E'_{\operatorname{MF}}[y+h])}e^{xz+\mu\bar{\mathcal{I}}(x)} }_{h,z}\right|_{\beta=\infty} = C.
    \end{eqnarray}
Thus, $C$ can be interpreted as the susceptibility of the grand potential with respect to the coupling with the random external field. (or equivalently, the susceptibility with respect to the effective temperature $\beta^{-1}$.)

The argument above clearly shows that the supersymmetric order parameter $A$ and $C$ can be interpreted as the derivatives of the grand potential $\Omega(a,s)$. However, it is not difficult to see that they can be understood also as the derivatives of the {\it mean-field} grand potential $\Omega_{\mathrm{MF}}(a,s):= \log\Braket{\sum_{x\in\mathrm{Crt}(E'_{\operatorname{MF}}[h,a,\sqrt{2s}g])}e^{xz+\mu\bar{\mathcal{I}}(x)}}_{ h,z,g}$.
Indeed, for the order parameter $A$, \eqref{eq:A-is-derivative-by-a} implies $A = \frac{\partial}{\partial a}|_{a=0}\Omega_{\mathrm{MF}}$. Similarly, the following argument shows $C = \frac{\partial}{\partial s}|_{s=0}\Omega_{\mathrm{MF}}$. Notice that the quantity inside the braket $\sum_{x\in\mathrm{Crt}(E'_{\operatorname{MF}}[h,0,\sqrt{2s}g])}e^{xz+\mu\bar{\mathcal{I}}(x)}$ is a function of $h+\sqrt{2s}g$, which is a centered random Gaussian variable. Its variance and covariance with $z$ can be easily obtained as follows.
\begin{equation}
    \begin{cases}
        \Braket{(h+\sqrt{2s}g)^2} = \Braket{h^2} + \Braket{2sg^2} = q + 2s\\
        \Braket{(h+\sqrt{2s}g)z} = \Braket{hz} + \Braket{\sqrt{2s}gz} = A.
    \end{cases}
\end{equation}
Hence the Gaussian average over $h, z$, and $g$ is equivalent to the average over $h+\sqrt{2s}g$ and z with the covariance matrix 
$\left(\begin{matrix}
    q+2s& A \\ A&C
\end{matrix}\right)$.
Therefore the derivative by $s$ is equivalent to the derivative by $\frac{q}{2}$.
Thus
\begin{eqnarray}
    \left.\frac{\partial \Omega_{\mathrm{MF}}}{\partial s}\right|_{s=0}
    &=& \left.\frac{\partial }{\partial s}\right|_{s=0}\Braket{\sum_{x\in\mathrm{Crt}(E'_{\operatorname{MF}}[h,0,\sqrt{2s}g])}e^{xz+\mu\bar{\mathcal{I}}(x)}}_{h,z,g}\nonumber\\
    &=&2\frac{d}{dq}\Braket{\sum_{x\in\mathrm{Crt}(E_{\operatorname{MF}}[h])}e^{xz+\mu\bar{\mathcal{I}}(x)}}_{h,z}\nonumber\\
    &=& \Braket{\frac{d^2}{dh^2}\sum_{x\in\mathrm{Crt}(E_{\operatorname{MF}}[h])}e^{xz+\mu\bar{\mathcal{I}}(x)}}_{h,z}
    = C.
\end{eqnarray}

\subsection{Non-degeneracy of critical points implies supersymmetry and structural stability
}
In this subsection, we show that if typical critical points have eigenvalues bounded away from 0, then its supersymmetry breaking order parameters are vanishing $C = A= 0$, which implies the structural stability of the  typical critical points (since $A$ and $C$ are susceptibilities of the grand potential to certain perturbations of energy landscape.)

Recall that the order parameter $A,C$ are defined as follows with a finite $\beta$,
\begin{equation}
    \begin{cases}
    A(\mathbf{x}) = 2N^{-1}\beta \nabla E(\mathbf{x})\cdot\mathbf{x}-t_R \\
    C(\mathbf{x}) = 4N^{-1}\beta^2 \lvert\nabla E(\mathbf{x})\rvert^2-2\beta.
    \end{cases}
\end{equation}
If $\beta$ is large enough, we anticipate that the order parameters of a typical critical point can be obtained by averaging $A(\mathbf{x}),C(\mathbf{x})$ over a small neighborhood $U$ around the critical point, i.e.,
\begin{eqnarray}
    A + t_R &=& \int_U \prod_{i=0}^{N-1} dx_i 2N^{-1}\beta [\nabla E(\mathbf{x})\cdot\mathbf{x}]\exp\left(-\beta|\nabla E(\mathbf{x})|^2\right) \left|\det H(\mathbf{x})\right| e^{\mu I(\mathbf{x})}/\int_U \prod_{i=0}^{N-1} dx_i \exp\left(-\beta|\nabla E(\mathbf{x})|^2\right) \left|\det H(\mathbf{x})\right| e^{\mu I(\mathbf{x})}\nonumber\\
    &=& \int_U \prod_{i=0}^{N-1} dx_i 2N^{-1}\beta \nabla E(\mathbf{x})\cdot\mathbf{x}\exp\left(-\beta|\nabla E(\mathbf{x})|^2\right) \left|\det H(\mathbf{x})\right|/\int_U \prod_{i=0}^{N-1} dx_i \exp\left(-\beta|\nabla E(\mathbf{x})|^2\right) \left|\det H(\mathbf{x})\right|.
\end{eqnarray}
Here we exploit the fact that $\mathcal{I}(\mathbf{x})$ is constant in the small neighborhood because the eigenvalues are bounded away from zero. Similarly
\begin{equation}
    C+2\beta= \int_{U} \prod_{i=0}^{N-1} dx_i 4N^{-1}\beta^2 |\nabla E(\mathbf{x})|^2\exp\left(-\beta|\nabla E(\mathbf{x})|^2\right)\left|\det H(\mathbf{x})\right| /\int_{U} \prod_{i=0}^{N-1} dx_i \exp\left(-\beta|\nabla E(\mathbf{x})|^2\right) \left|\det H(\mathbf{x})\right|.
\end{equation}

By the assumption, we can apply the inverse function theorem to $y_i :=\nabla_i E(\mathbf{x})$ around the neighborhood of the critical point. The Jacobian is given by the Hessian $\frac{dy_i}{dx_j} = H_{ij}$, and we approximately see that $x_i = H^{-1}_{ij}y_j$. Therefore by changing the coordinate variables from $\mathbf{x}$ to $\mathbf{y}$,
\begin{eqnarray}
    A+t_R &=& 2N^{-1}\beta \int_U \prod_{i=0}^{N-1} dy_i \mathbf{y}^T H^{-1}\mathbf{y}\exp\left(-\beta |\mathbf{y}|^2\right) /\int_U \prod_{i=0}^{N-1} dy_i \exp\left(-\beta |\mathbf{y}|^2\right) \nonumber\\
    &\approx& 2N^{-1}\beta \int_{\mathbb{R}^N} \prod_{i=0}^{N-1} dy_i \mathbf{y}^T H^{-1}\mathbf{y}\exp\left(-\beta |\mathbf{y}|^2\right) /\int_{\mathbb{R}^N} \prod_{i=0}^{N-1} dy_i \exp\left(-\beta|\mathbf{y}|^2\right) \nonumber\\
    &=& N^{-1}\operatorname{Tr}H^{-1} = t_R,
\end{eqnarray}
Thus $A=0$. Similarly, we can see that $C=0$ as follows.
\begin{eqnarray}
    C+2\beta &=& 4N^{-1}\beta^2 \int_U \prod_{i=0}^{N-1} dy_i |\mathbf{y}|^2\exp\left(-\beta |\mathbf{y}|^2\right) /\int_U \prod_{i=0}^{N-1} dy_i \exp\left(-\beta |\mathbf{y}|^2\right) \nonumber\\
    &\approx& 2N^{-1}\beta^2 \int_{\mathbb{R}^N} \prod_{i=0}^{N-1} dy_i |\mathbf{y}|^2\exp\left(-\beta |\mathbf{y}|^2\right) /\int_{\mathbb{R}^N} \prod_{i=0}^{N-1} dy_i \exp\left(-\beta |\mathbf{y}|^2\right) \nonumber\\
    &=& 2\beta.
\end{eqnarray}

The contrapositive of this result then immediately tells us that if supersymmetry breaking order parameters are nonzero, then typical critical points are marginally unstable (with no gap in the eigenvalue density away from $0$). Earlier results additionally imply such marginally unstable critical points are also structurally unstable to small perturbations in the energy landscape.  

\subsubsection*{Convexity of the mean-field energy landscape implies vanishing complexity and supersymmetry}

In the following discussion, we address two statements on the convexity of the mean-field energy landscape and supersymmetry of the typical critical points.
First, we show that the strict convexity of the mean-field energy landscape $E_{\mathrm{MF}}[0](x)$ is a sufficient condition for the self-consistent equations to have a supersymmetric fixed point with $A=C=0$.
We also show that this supersymmetric solution implies vanishing complexity.
Furthermore, while the convexity might not be a necessary condition for existence of supersymmetric solutions, we prove that it is the case when the mean-field energy is given by the quartic function $E_{\mathrm{MF}}[0](x)=x^4/4 - a_{\operatorname{eff}}x^2/2$.
Suppose that the supersymmetry-breaking order parameters vanish $A=C=0$. In this setting, the Gaussian random variable $z$ is always zero, and the average $\Braket{\cdot}_{h,z}$ are only over $h$ with variance of $q$, which we denote $\Braket{\cdot}_{h}$. Therefore the self-consistent equations for $A,C$ are reduced to
\begin{equation}
    \begin{cases}
    A = Z^{-1}\Braket{\frac{d}{dh}\sum_{x\in \operatorname{Crt}(E_{\operatorname{MF}}[h])}x}_{h}-t_R\\
    C = Z^{-1}\Braket{\frac{d^2}{dh^2}\sum_{x\in \operatorname{Crt}(E_{\operatorname{MF}}[h])}1}_{h}
    \end{cases}
    \label{eq:self-consistent-A=C=0}
\end{equation}
It is easy to see that these self-consistent equations hold when the mean-field energy function without external field $E_{MF}[0](x)$ is strictly convex and has a monotonically increasing gradient. 
Indeed,  $\operatorname{Crt}(E_{\operatorname{MF}}[h])$ contains only single element for any $h\in\mathbb{R}$, and therefore
\begin{eqnarray}
    A  &=& Z^{-1}\Braket{\sum_{x\in \operatorname{Crt}(E_{\operatorname{MF}}[h])}\frac{d}{dh} x}_{h}-t_R = t_R-t_R = 0
\end{eqnarray}
\begin{eqnarray}
    C &=& Z^{-1}\Braket{\frac{d^2}{dh^2} \sum_{x\in \operatorname{Crt}(E_{\operatorname{MF}}[h])}1}_{h} = 0.
\end{eqnarray}
Hence the convexity of the mean-field energy implies supersymmetry and, and therefore also implies structural stability of typical critical points.

Moreover, in this case, the grand potential (and therefore the complexity) of typical critical points vanishes:
\begin{eqnarray}
        \Omega&=& -\frac{1}{2}(Cq + A^2) - A t+\log \Braket{\sum_{x\in\mathrm{Crt}(E_{\operatorname{MF}}[h])}e^{xz}}_{ h,z} \nonumber\\
         &=& \log \Braket{\sum_{x\in\mathrm{Crt}(E_{\operatorname{MF}}[h])}1}_{h} = 0.
    \end{eqnarray}
Next, we focus on the case of quartic energy function $E_{\mathrm{MF}}[0](x)=x^4/4 - a_{\operatorname{eff}}x^2/2$, and will show that the strong convexity of $E_{\mathrm{MF}}[0](x)$, i.e., $a_{\operatorname{eff}}\leq 0$ is a necessary condition for the existence of supersymmetric solutions, i.e., if $a_{\operatorname{eff}}>0$, the self-consistent equations do not have any supersymmetric solution.
Suppose $a_{\operatorname{eff}}>0$ and a supersymmetric solution exists, with which \eqref{eq:self-consistent-A=C=0} holds.
The second equation of \eqref{eq:self-consistent-A=C=0} can be calculated as follows for the quartic energy.
\begin{eqnarray}
    C &=& Z^{-1}\int^{\infty}_{-\infty} dh \frac{1}{\sqrt{2\pi q}} \exp\left(-\frac{h^2}{2q}\right) \left(\frac{d^2}{dh^2}\sum_{x\in \operatorname{Crt}(E_{\operatorname{MF}}[h])}1\right)\nonumber\\
    &=& Z^{-1}\int^{\infty}_{-\infty} dh \frac{1}{\sqrt{2\pi q}} \exp\left(-\frac{h^2}{2q}\right) \left(\frac{d}{dh}\delta\left(-\tilde{h}\right)-\delta\left(\tilde{h}\right)\right)\nonumber\\
&=& Z^{-1}\int^{\infty}_{-\infty} dh \left(-\frac{h}{q}\right)\frac{1}{\sqrt{2\pi q}} \exp\left(-\frac{h^2}{2q}\right) \left(\delta\left(-\tilde{h}\right)-\delta\left(\tilde{h}\right)\right)\nonumber\\
&=& \frac{2\tilde{h}}{\sqrt{2\pi q^3}} \exp\left(-\frac{\tilde{h}^2}{2q}\right),
\end{eqnarray}
where $\tilde{h} = \frac{2a_{\operatorname{eff}}^{2/3}}{3\sqrt{3}}$.
This is strictly positive and hence contradicts our assumption of supersymmetry $A=C=0$.
Therefore SUSY solution does not exist.

\subsection{Equivalence between the cavity method and the replica method}
We here present another set of expressions of the grand potential $\Omega$, the distribution $P(
x)$ of local variables and the self-consistent equations for order parameters, that is more suitable for numerical computations. They turn out to be equivalent to those derived by the replica calculation (under the assumption $t_I = 0$.) 

We exploit the formula \eqref{eq:integrate-z} to integrate out the Gaussian variable $z$ and to obtain the new expression. It is easy to see that the grand potential \eqref{eq:Omega-cavity2} is expressed as
\begin{equation}
    \Omega = -\frac{1}{2}(Cq + A^2) - A t_R+\log\frac{1}{\sqrt{2\pi q}}\int dh \sum_{x\in\mathrm{Crt}(E_{\operatorname{MF}}[h])}\exp\left( -\frac{1}{2q}h^2 + \frac{A}{q} xh + \frac{1}{2} \frac{qC - A^2}{q}x^2 + \mu\overline{\mathcal{I}}(x)\right)
\end{equation}
This is equivalent to \eqref{eq:Omega-typical-crt}. Similarly, the distribution $P(x)$ can be obtained by marginalizing \eqref{eq:P-beta-xy}, taking the limit $\beta\to\infty$, and applying \eqref{eq:integrate-z}
\begin{eqnarray}
    P(x) &=& Z^{-1}\int dh \sum_{x_*\in\mathrm{Crt}(E_{\operatorname{MF}}[h])}\exp\left( -\frac{1}{2q}h^2 + \frac{A}{q} xh + \frac{1}{2} \frac{qC - A^2}{q}x^2 + \mu\overline{\mathcal{I}}(x) \right)\delta(x-x_*)\nonumber\\
    &=& Z^{-1}w(x)\exp\left( -\frac{1}{2q}h^2(x) + \frac{A}{q} xh(x) + \frac{1}{2} \frac{qC - A^2}{q}x^2 + \mu\overline{\mathcal{I}}(x) \right),
\end{eqnarray}
where $h(x) = \partial E_I(x) - t_R x$ and $w(x) = |\frac{\partial{h}}{\partial x}| = |\partial^2 E_I(x) - t_R|$ is a Jacobian in the change of variables from $h$ to $x$. This is equivalent to \eqref{eq:Px_typical}. With this distribution $P(x)$, it is easy to see that the self-consistent equations for $q$ and $t_R$ are given by
\begin{eqnarray}
    q &=& \int dx P(x) x^2 \nonumber\\
    t_R &=& \int dx P(x) (\partial^2 E_I(x) - t_R)^{-1}.
\end{eqnarray}
The reduction of the self-consistent equations for $A$ and $C$ can be done as follows. Recall that the braket $\Braket{\cdot}_{h,z}$ is an integration over $h$ and $z$ with Gaussian weight $\exp(K(h,z))$ where  $K(h,z) = -\frac{1}{2} \left(\begin{matrix}h\\ z\end{matrix}\right)^T \left(\begin{matrix}q&A\\A&C \end{matrix}\right)^{-1}\left(\begin{matrix}h\\ z\end{matrix}\right)$.

\begin{eqnarray}
    A &=& Z^{-1}\int dh dz \exp\left(K(h,z)\right)\frac{\partial}{\partial h}\sum_{x\in\mathrm{Crt}(E_{\operatorname{MF}}[h])}  e^{xz+\mu\bar{\mathcal{I}}(x)}x - t_R\nonumber\\
    &=& -Z^{-1}\int dh dz \frac{\partial}{\partial h}\exp\left(K(h,z)\right)\sum_{x\in\mathrm{Crt}(E_{\operatorname{MF}}[h])}  e^{xz+\mu\bar{\mathcal{I}}(x)}x - t_R\nonumber\\
    &=& Z^{-1}\int dh dz(qC-A^2)^{-1}\left(Ch-Az\right) \exp\left(K(h,z)\right)\sum_{x\in\mathrm{Crt}(E_{\operatorname{MF}}[h])}  e^{xz+\mu\bar{\mathcal{I}}(x)}x - t_R\nonumber\\
    &=& Z^{-1}\int dh dz \left(q^{-1}h+\frac{A}{q} \frac{\partial}{\partial z}\right)\left[\exp\left(K(h,z)\right)\right]\sum_{x\in\mathrm{Crt}(E_{\operatorname{MF}}[h])}  e^{xz+\mu\bar{\mathcal{I}}(x)}x - t_R\nonumber\\
     &=& q^{-1}\int dx P(x) h(x)x -\frac{A}{q} Z^{-1}\int dx dz\exp\left(K(h,z)\right)\frac{\partial}{\partial z}\sum_{x\in\mathrm{Crt}(E_{\operatorname{MF}}[h])}  e^{xz+\mu\bar{\mathcal{I}}(x)}x - t_R\nonumber\\
     &=& q^{-1}\int dx P(x) h(x)x - A - t_R\nonumber\\
\end{eqnarray}
Therefore
\begin{equation}
    A = \frac{1}{2q}\int dx P(x) h(x)x -\frac{t_R}{2}.
\end{equation}
\begin{eqnarray}
    C &=& Z^{-1}\int dh dz \exp\left(K(h,z)\right)\frac{\partial^2}{\partial h^2}\sum_{x\in\mathrm{Crt}(E_{\operatorname{MF}}[h])}  e^{xz+\mu\bar{\mathcal{I}}(x)}\nonumber\\
    &=& Z^{-1}\int dh dz \frac{\partial^2}{\partial h^2}\left[ \exp\left(K(h,z)\right)\right]\sum_{x\in\mathrm{Crt}(E_{\operatorname{MF}}[h])}  e^{xz+\mu\bar{\mathcal{I}}(x)}\nonumber\\
    &=& Z^{-1}\int dh dz \left(q^{-1}h+\frac{A}{q} \frac{\partial}{\partial z}\right)^2\left[ \exp\left(K(h,z)\right)\right]\sum_{x\in\mathrm{Crt}(E_{\operatorname{MF}}[h])}  e^{xz+\mu\bar{\mathcal{I}}(x)}-q^{-1}\nonumber\\
    &=& Z^{-1}\int dh dz  \exp\left(K(h,z)\right)\sum_{x\in\mathrm{Crt}(E_{\operatorname{MF}}[h])}  \left(q^{-1}h-\frac{A}{q} x\right)^2 e^{xz+\mu\bar{\mathcal{I}}(x)} -q^{-1}\nonumber\\
    &=& \int dx P(x)\left(q^{-1}h(x)-\frac{A}{q}x\right)^2 - q^{-1}\nonumber\\
\end{eqnarray}
These self-consistent equations are equivalent to those obtained by the replica method in \eqref{eqs:self-consistent-replica}.

\setcounter{section}{0}
 \renewcommand{\thesection}{S-App.\Roman{section}}
 \section{Simplifying the replicated Kac-Rice formula: derivation of \ref{eq:Omega_before_detH}}
 \label{app:hubbard}
 In the integrand of \eqref{eq:Zn}, terms including $J$-matrix entries are
 \begin{eqnarray}
    &&\int \prod_{i<j} dJ_{ij} \prod_{i<j} \exp\left(-\frac{N}{2}J_{ij}^2\right) 
    \exp\left(\sum_{a,i,j} -u^a_i J_{ij} x^a_j + \frac{\beta}{2} x^a_i J_{ij} x^a_j\right) \prod_a\left[|\det H(\mathbf{x}^a)| e^{\mu\mathcal{I}(\mathbf{x}^a)}\right] \nonumber\\
    &=& \int \prod_{i<j} dJ_{ij} \prod_{i<j} \exp\left(-\frac{N}{2}(J_{ij} - N^{-1}\sum_a (- u^a_i x^a_j - u^a_j x^a_i + \beta x^a_i x^a_j))^2 \right)\nonumber \\
    &&\times \exp\left(\frac{1}{2N}(\sum_a - u^a_i x^a_j - u^a_j x^a_i + \beta x^a_i x^a_j)^2\right)
     \prod_a\left[|\det H(\mathbf{x}^a)| e^{\mu\mathcal{I}(\mathbf{x}^a)}\right]
     \label{eq:a-1}
 \end{eqnarray}
Now we perform the translations of variables $J_{ij} \to J_{ij} - N^{-1}\sum_a(- u^a_i x^a_j - u^a_j x^a_i + \beta x^a_i x^a_j)$. These translations are so small that the effect on the determinant of the hessian and the index is negligible. Then,
\begin{eqnarray}
    \text{(r.h.s of \eqref{eq:a-1})} 
    &=&\exp\left(\frac{1}{2N}\sum_{i< j}\left(\sum_a -( u^a_i x^a_j + u^a_j x^a_i) + \beta x^a_i x^a_j\right)^2 \right) \Braket{\prod_a\lvert\det H(\mathbf{x}^a)\rvert e^{\mu\mathcal{I}(\mathbf{x}^a)}}_J
\end{eqnarray}

The term in the exponential function can be further reduced. First of all, since the sum of terms with $i=j$ is the order of $O(1)$, we can add it without changing the leading term.
\begin{eqnarray}
    \frac{1}{2N}\sum_{i< j}\left(\sum_a -( u^a_i x^a_j + u^a_j x^a_i) + \beta x^a_i x^a_j\right)^2 = \frac{1}{4N}\sum_{i,j}\left(\sum_a -( u^a_i x^a_j + u^a_j x^a_i) + \beta x^a_i x^a_j\right)^2 + O(1)
\end{eqnarray}
The right hand side can be transformed as follows.
\begin{eqnarray}
    &&\frac{1}{4N}\sum_{i,j}\left(\sum_a -( u^a_i x^a_j + u^a_j x^a_i) + \beta x^a_i x^a_j\right)^2 \nonumber\\
    &=& \frac{1}{2N}\sum_{i,j}\sum_{a,b} \left[ u^a_i u^b_i x^a_j x^b_j + u^a_i u^b_j x^a_j x^b_i + \frac{\beta^2}{2} x^a_i x^b_i x^a_j x^b_j- 2\beta u^a_i x^b_i x^a_j x^b_j \right] \nonumber\\
    &=& \frac{1}{2N}\sum_{a,b}\left(\left(\sum_i x^a_i x^b_i\right)\left(\sum_i u^a_i u^b_i\right) + \left(\sum_i x^a_i u^b_i\right)\left(\sum_i x^b_i u^a_i\right) + \frac{\beta^2}{2}\left(\sum_i x^a_i x^b_i\right)^2 - 2\beta \left(\sum_i x^a_i x^b_i\right)\left(\sum_i x^a_i u^b_i\right) \right) \nonumber\\
    &=& \frac{1}{2N}\sum_{a,b}\left(\left(\sum_i x^a_i x^b_i\right)\left(\sum_i u^a_i u^b_i\right) + \left(\sum_i\beta x^a_i x^b_i - x^a_i u^b_i\right)\left(\sum_i\beta x^b_i x^a_i - x^b_i u^a_i\right) - \frac{\beta^2}{2}\left(\sum_i x^a_i x^b_i\right)^2 \right)
\end{eqnarray}
To linearize the second term, we exploit Hubbard-Stratonovich transform with new auxiliary variables $w^{ab}$.
\begin{eqnarray}
    &&\text{(r.h.s of \eqref{eq:a-1})} \nonumber\\
    &=& \int \prod_{a,b}dw^{ab} \prod_{a,b}\exp\left(-\frac{N}{2}w^{ab}w^{ba} - w^{ab}\left(\sum_i\beta x^a_i x^b_i - x^a_i u^b_i\right) \right) \nonumber \\
    &&\ \ \ \times\exp\left(\frac{1}{2N}\left(\left(\sum_i x^a_i x^b_i\right)\left(\sum_i u^a_i u^b_i\right) - \frac{\beta^2}{2}\left(\sum_i x^a_i x^b_i\right)^2 \right)\right)\Braket{\prod_a\lvert\det H(\mathbf{x}^a)\rvert e^{\mu\mathcal{I}(\mathbf{x}^a)}}_J
    \label{eq:a-2}
\end{eqnarray}
Next we insert the following delta function and integrate it with respect to $q_{ab}$.
\begin{equation}
    \delta(Nq^{ab}-\sum_i x^a_i x^b_i) = \int^{i\infty}_{-i\infty} \frac{d\lambda^{ab}}{2\pi i} \exp\left( -N\lambda^{ab}q^{ab} + \lambda^{ab} \sum_i x^a_i x^b_i\right)
\end{equation}
Then we get
\begin{eqnarray}
    &&\text{(r.h.s. of \eqref{eq:a-1})}\nonumber\\
    &=& \int \prod_{(a,b)} dq^{ab} \frac{d\lambda^{ab}}{2\pi i}\prod_{a,b} dw^{ab}
    \prod_{a,b}\exp\left(-\frac{N}{2}w^{ab} w^{ba} -N\beta w^{ab}q^{ab} -\frac{N\beta^2}{4}(q^{ab})^2 - N\lambda^{ab} q^{ab}\right) \nonumber \\
    &&\ \ \ \times\prod_i\exp\left(w^{ab}x_i^a u_i^b +\frac{1}{2}q^{ab}u_i^au^b_i +\lambda^{ab}x_i^a x^b_i\right)\Braket{\prod_a\lvert\det H(\mathbf{x}^a)\rvert e^{\mu\mathcal{I}(\mathbf{x}^a)}}_J
\end{eqnarray}
We recover \eqref{eq:Omega_before_detH}. 

\section{Random matrix theory of the Hessian: derivation of \ref{eq:det_H_integral}}
\label{app:rmthess}
The determinant can be expanded by Grassmann variables $\psi_i,\bar\psi_i$ as follows
\begin{equation}
    \det[H(\mathbf{x}) - zI] =\int\prod_{i} d\bar\psi_i d\psi_i \exp\left(\sum_{i,j} \bar\psi_i (H_{ij}(\mathbf{x}) -z)\psi_j\right),
\end{equation}
where $\psi_i,\bar\psi_i$ satisfy the anti-commutation relation $\{\psi_i,\psi_j\}=\{\bar\psi_i,\bar\psi_j\}=\delta_{ij}$ and $\{\psi_i,\bar\psi_j\}=0$. We are going to calculate its average over $J$. Here we first integrate out $J$-matrix, next introduce auxiliary variables by the Hubbard-Stratonovich transformation, and then integrate over the Grassmann variables.
\begin{eqnarray}
    && \Braket{\det H(\mathbf{x}) - zI}_J \nonumber\\
    &=&
     \int \prod_{(i,j)} dJ_{ij}\exp\left(-\frac{N}{2}J_{ij}^2\right) \int \frmint
    \exp\left[
        \sum_{i,j} \bar\psi_i \left(H_{ij}(\mathbf{x})
            - z \delta_{ij}\right)\psi_j\right] \nonumber\\
    &=&
    \int \frmint \int \prod_{(i,j)} dJ_{ij} \exp\left(-\frac{N}{2}J_{ij}^2\right) \prod_{i,j}\exp\left( J_{ij}\bar\psi_i \psi_j \right)\exp\left( \sum_{i} \bar\psi_i \left(\partial^2 E_I(x_i^a)
        - z \right)\psi_i\right) \nonumber\\
    &=&
    \int \frmint\exp\left(\frac{1}{2N}\sum_{(i,j)} \left(\bar\psi_i \psi_j + \bar\psi_j \psi_i\right)^2 +
    \sum_{i} \bar\psi_i \left(\partial^2 E_I(x^a_i)
        - z \right)\psi_i\right)\nonumber\\
    \label{eq:ap2-1}
    \end{eqnarray}
The first term in the exponential function can be reduced to
\begin{eqnarray}
&&\frac{1}{2N}\sum_{(i,j)} \left(\bar\psi_i \psi_j + \bar\psi_j \psi_i\right)^2= -\frac{1}{2N}\sum_{i,j} \bar\psi_i \psi_i \bar\psi_j \psi_j= -\frac{1}{2N} \left(\sum_{i}\bar\psi_i \psi_i\right)^2.\nonumber
\end{eqnarray}
We linearize this term by introducing a new auxiliary variable $t$
\begin{eqnarray}
    \text{(r.h.s. of \eqref{eq:ap2-1})}&=&\int \frmint\int dt
    \exp\left(\frac{N}{2}t^2 - t\left(\sum_i \bar\psi_i \psi_i\right)\right)\exp\left(\sum_{i} \bar\psi_i \left(\partial^2 E_I(x^a_i) - z \right)\psi_i\right)  \nonumber\\
    &=& \int^{i\infty}_{-i\infty} dt
    \exp\left(\frac{N}{2}t^2 +\sum_i \log(\partial^2 E_I(x^a_i)- z - t)\right)
\end{eqnarray}
This integral is dominated by the stationary point of the exponent. The stationary condition is given by
\begin{equation}
    t = N^{-1}\sum_i\frac{1}{\partial^2 E_I(x^a_i)- z - t}.
\end{equation}
This is equivalent to \eqref{eq:det_H_integral}.

\section{derivation of  \ref{eq:Omega_S_replica}}
\label{app:susytrans}
The expression for $\Omega$ can be derived just by substituting the expressions for $w^{ab}$ and $\lambda^{ab}$. We here discuss the derivation of the expression for $S$. In  \eqref{eq:Omega_S_replica_w_lambda}, the terms with $u^a$ are $\frac{1}{2}q^{ab}u^{a}u^b + w^{ab}u^{a}x^b + u^a \partial E_I(x^a)$.
Since this is quadratic, we can perform the Gaussian integral over $u^a$.
\begin{eqnarray}
    &&\int^{i\infty}_{-i\infty} \prod_a du^a\exp\left(\sum_{a,b} \frac{1}{2}q^{ab}u^{a}u^b + w^{ab}u^{a}x^b + u^a \partial E_I(x^a)\right)\nonumber\\
    &=&
    \sqrt{\frac{\pi^n}{\det q}}\exp\left(\sum_{a,b}-\frac{1}{2} [q^{-1}]^{ab} \left(h^a - \sum_c  A^{ac}x^c - \beta q^{ac} x^c\right) \left(h^b - \sum_c  A^{bc}x^c - \beta q^{bc} x^c\right)\right) \nonumber\\
    &=& \sqrt{\frac{\pi^n}{\det q}}\exp\left(\sum_{a,b}-\frac{1}{2}[q^{-1}]^{ab} \left(h^a - \sum_c  A^{ac}x^c \right) \left(h^b - \sum_c  A^{bc}x^c\right) \right.\nonumber\\
    &&\ \ \ \ \  \left. + \beta \sum_{a} x^a \left(h^a - \sum_c  A^{ac}x^c \right) - \frac{\beta^2}{2} \sum_{a,b} x^a q^{ab} x^b\right),
\end{eqnarray}
where $h^a := \partial E_I(x^a) - t^a x^a$.
Again we can introduce new auxiliary variables $u^a$ and get
\begin{eqnarray}
    = &&\int^{i\infty}_{-i\infty} \prod_a du^a\exp\left(\sum_{a,b}\frac{1}{2}q^{ab}u^a u^b +\sum_{a} u^a \left(h^a - \sum_c  A^{ac}x^c \right) + \beta \sum_{a} x^a \left(h^a - \sum_c  A^{ac}x^c \right)- \frac{\beta^2}{2} \sum_{a,b} x^a q^{ab} x^b\right)\nonumber\\
\end{eqnarray}

Substituting this and $\lambda^{ab} =\frac{\beta}{2} t_R \delta_{ab} + \frac{\beta^2}{2} q^{ab}  +\frac{\beta}{2} (A^{ab} + A^{ba}) + \frac{1}{2}C^{ab}$ to $S$, we see
\begin{eqnarray}
    S &=& \sum_{a,b}\frac{1}{2}q^{ab}u^a u^b -\sum_{a} u^a \left(h^a - \sum_c  A^{ac}x^c \right) + \beta \sum_{a} x^a \left(h^a - \sum_c  A^{ac}x^c \right)- \frac{\beta^2}{2} \sum_{a,b} x^a q^{ab} x^b\nonumber\\
    &&+ \sum_{a,b} \left(\frac{\beta}{2} t_R \delta_{ab} + \frac{\beta^2}{2} q^{ab}  +\beta A^{ab} + \frac{1}{2}C^{ab}\right)x^a x^b + \sum_a -\beta E_I(x^a) \nonumber\\
    &=& \sum_{ab}\left[\frac{1}{2}q^{ab}u^a u^b + A^{ab} u^a x^b + \frac{1}{2} C^{ab} x^a x^b\right]+ \sum_{a} \left[ -u^a h^a + \beta x^a h^a -\beta (E_I(x^a) -\frac{t_R}{2}(x^a)^2) \right].
\end{eqnarray}

\section{A formula for a 2D Gaussian vector}
\label{ap:h-z-formula}
Let $\bar h,\bar z$ be a pair of centered Gaussian random variables with covariance $\left(\begin{matrix}q& A \\A& C\end{matrix}\right)$ and $h,z$ be another pair with covariance $\left(\begin{matrix}q& A-t \\A-t& C-s\end{matrix}\right)$. The following equality holds
\begin{eqnarray}
    \Braket{\exp(f(\bar h) + x\bar z)}_{\bar h,\bar z} =  \Braket{\exp(f(h+tx) + xz+\frac{s}{2}x^2)}_{h,z},
    \label{eq:h-z-formula}
\end{eqnarray}
where $\Braket{\cdot}$ represents the expectation value,  $f(\bar h)$ is an arbitrary function of $\bar h$, and $x$ is a constant.

This can be shown as follows. By explicitly integrating over $z$,
\begin{eqnarray}
    \Braket{\exp(f(\bar h) + x\bar z)}_{\bar h,\bar z} = \frac{1}{\sqrt{2\pi q}}\int d\bar h \exp\left(f(\bar h) - \frac{\bar h^2}{2q} + q^{-1}A\bar hx + \frac{1}{2q}(qC-A^2)x^2\right).
    \label{eq:integrate-z}
\end{eqnarray}
We can apply this result also for the right hand side of \eqref{eq:h-z-formula},
\begin{eqnarray}
    \Braket{\exp(f(h+t) + xz)}_{h,z} &=& \frac{1}{\sqrt{2\pi q}}\int dh \exp\left(f(h+tx) - \frac{h^2}{2q} + q^{-1}(A-t)hx + \frac{1}{2q}(q(C-s)-(A-t)^2)x^2\right)\nonumber\\
    &=&  \frac{1}{\sqrt{2\pi q}}\int dh \exp\left(f(h+tx) - \frac{(h+tx)^2}{2q} + q^{-1}A(h+tx)x + \frac{1}{2q}(qC-A^2)x^2 - \frac{s}{2}x^2\right)\nonumber\\
    &=& \Braket{\exp(f(\bar h) + x\bar z)}_{\bar h,\bar z} \exp\left(-\frac{s}{2}x^2\right)\nonumber.\\
\end{eqnarray}
Multiplying both sides of this last equation by 
$\exp\left(\frac{s}{2}x^2\right)$ yields the desired result in \eqref{eq:h-z-formula}. 

\bibliography{main}

\begin{thebibliography}{101}%
\makeatletter
\providecommand \@ifxundefined [1]{%
 \@ifx{#1\undefined}
}%
\providecommand \@ifnum [1]{%
 \ifnum #1\expandafter \@firstoftwo
 \else \expandafter \@secondoftwo
 \fi
}%
\providecommand \@ifx [1]{%
 \ifx #1\expandafter \@firstoftwo
 \else \expandafter \@secondoftwo
 \fi
}%
\providecommand \natexlab [1]{#1}%
\providecommand \enquote  [1]{``#1''}%
\providecommand \bibnamefont  [1]{#1}%
\providecommand \bibfnamefont [1]{#1}%
\providecommand \citenamefont [1]{#1}%
\providecommand \href@noop [0]{\@secondoftwo}%
\providecommand \href [0]{\begingroup \@sanitize@url \@href}%
\providecommand \@href[1]{\@@startlink{#1}\@@href}%
\providecommand \@@href[1]{\endgroup#1\@@endlink}%
\providecommand \@sanitize@url [0]{\catcode `\\12\catcode `\$12\catcode `\&12\catcode `\#12\catcode `\^12\catcode `\_12\catcode `\%12\relax}%
\providecommand \@@startlink[1]{}%
\providecommand \@@endlink[0]{}%
\providecommand \url  [0]{\begingroup\@sanitize@url \@url }%
\providecommand \@url [1]{\endgroup\@href {#1}{\urlprefix }}%
\providecommand \urlprefix  [0]{URL }%
\providecommand \Eprint [0]{\href }%
\providecommand \doibase [0]{https://doi.org/}%
\providecommand \selectlanguage [0]{\@gobble}%
\providecommand \bibinfo  [0]{\@secondoftwo}%
\providecommand \bibfield  [0]{\@secondoftwo}%
\providecommand \translation [1]{[#1]}%
\providecommand \BibitemOpen [0]{}%
\providecommand \bibitemStop [0]{}%
\providecommand \bibitemNoStop [0]{.\EOS\space}%
\providecommand \EOS [0]{\spacefactor3000\relax}%
\providecommand \BibitemShut  [1]{\csname bibitem#1\endcsname}%
\let\auto@bib@innerbib\@empty
\bibitem [{\citenamefont {Papadimitriou}\ and\ \citenamefont {Steiglitz}(1998)}]{Papadimitriou1998-hj}%
  \BibitemOpen
  \bibfield  {author} {\bibinfo {author} {\bibfnamefont {C.~H.}\ \bibnamefont {Papadimitriou}}\ and\ \bibinfo {author} {\bibfnamefont {K.}~\bibnamefont {Steiglitz}},\ }\href@noop {} {\emph {\bibinfo {title} {Combinatorial Optimization: Algorithms and Complexity}}}\ (\bibinfo  {publisher} {Courier Corporation},\ \bibinfo {year} {1998})\BibitemShut {NoStop}%
\bibitem [{\citenamefont {Mohseni}\ \emph {et~al.}(2022)\citenamefont {Mohseni}, \citenamefont {McMahon},\ and\ \citenamefont {Byrnes}}]{MohseniMcMahonByrnes2022}%
  \BibitemOpen
  \bibfield  {author} {\bibinfo {author} {\bibfnamefont {N.}~\bibnamefont {Mohseni}}, \bibinfo {author} {\bibfnamefont {P.~L.}\ \bibnamefont {McMahon}},\ and\ \bibinfo {author} {\bibfnamefont {T.}~\bibnamefont {Byrnes}},\ }\bibfield  {title} {\bibinfo {title} {Ising machines as hardware solvers of combinatorial optimization problems},\ }\href {https://doi.org/10.1038/s42254-022-00440-8} {\bibfield  {journal} {\bibinfo  {journal} {Nature Reviews Physics}\ }\textbf {\bibinfo {volume} {4}},\ \bibinfo {pages} {369} (\bibinfo {year} {2022})}\BibitemShut {NoStop}%
\bibitem [{\citenamefont {Kalinin}\ and\ \citenamefont {Berloff}(2022)}]{kalinin2022computational}%
  \BibitemOpen
  \bibfield  {author} {\bibinfo {author} {\bibfnamefont {K.~P.}\ \bibnamefont {Kalinin}}\ and\ \bibinfo {author} {\bibfnamefont {N.~G.}\ \bibnamefont {Berloff}},\ }\bibfield  {title} {\bibinfo {title} {Computational complexity continuum within ising formulation of np problems},\ }\href {https://www.nature.com/articles/s42005-021-00792-0} {\bibfield  {journal} {\bibinfo  {journal} {Communications Physics}\ }\textbf {\bibinfo {volume} {5}},\ \bibinfo {pages} {20} (\bibinfo {year} {2022})}\BibitemShut {NoStop}%
\bibitem [{\citenamefont {Syed}\ and\ \citenamefont {Berloff}(2023)}]{syed2023physics}%
  \BibitemOpen
  \bibfield  {author} {\bibinfo {author} {\bibfnamefont {M.}~\bibnamefont {Syed}}\ and\ \bibinfo {author} {\bibfnamefont {N.~G.}\ \bibnamefont {Berloff}},\ }\bibfield  {title} {\bibinfo {title} {Physics-enhanced bifurcation optimisers: All you need is a canonical complex network},\ }\href {https://ieeexplore.ieee.org/stamp/stamp.jsp?tp=&arnumber=10012498} {\bibfield  {journal} {\bibinfo  {journal} {IEEE Journal of Selected Topics in Quantum Electronics}\ }\textbf {\bibinfo {volume} {29}},\ \bibinfo {pages} {1} (\bibinfo {year} {2023})}\BibitemShut {NoStop}%
\bibitem [{\citenamefont {Moln{\'a}r}\ \emph {et~al.}(2020)\citenamefont {Moln{\'a}r}, \citenamefont {Kharel}, \citenamefont {Hu},\ and\ \citenamefont {Toroczkai}}]{molnar2020accelerating}%
  \BibitemOpen
  \bibfield  {author} {\bibinfo {author} {\bibfnamefont {F.}~\bibnamefont {Moln{\'a}r}}, \bibinfo {author} {\bibfnamefont {S.~R.}\ \bibnamefont {Kharel}}, \bibinfo {author} {\bibfnamefont {X.~S.}\ \bibnamefont {Hu}},\ and\ \bibinfo {author} {\bibfnamefont {Z.}~\bibnamefont {Toroczkai}},\ }\bibfield  {title} {\bibinfo {title} {Accelerating a continuous-time analog sat solver using gpus},\ }\href {https://www.sciencedirect.com/science/article/pii/S0010465520302204} {\bibfield  {journal} {\bibinfo  {journal} {Computer Physics Communications}\ }\textbf {\bibinfo {volume} {256}},\ \bibinfo {pages} {107469} (\bibinfo {year} {2020})}\BibitemShut {NoStop}%
\bibitem [{\citenamefont {Dutta}\ \emph {et~al.}(2020)\citenamefont {Dutta}, \citenamefont {Khanna}, \citenamefont {Assoa}, \citenamefont {Paik}, \citenamefont {Schlom}, \citenamefont {Toroczkai}, \citenamefont {Raychowdhury},\ and\ \citenamefont {Datta}}]{dutta2020ising}%
  \BibitemOpen
  \bibfield  {author} {\bibinfo {author} {\bibfnamefont {S.}~\bibnamefont {Dutta}}, \bibinfo {author} {\bibfnamefont {A.}~\bibnamefont {Khanna}}, \bibinfo {author} {\bibfnamefont {A.~S.}\ \bibnamefont {Assoa}}, \bibinfo {author} {\bibfnamefont {H.}~\bibnamefont {Paik}}, \bibinfo {author} {\bibfnamefont {D.}~\bibnamefont {Schlom}}, \bibinfo {author} {\bibfnamefont {Z.}~\bibnamefont {Toroczkai}}, \bibinfo {author} {\bibfnamefont {A.}~\bibnamefont {Raychowdhury}},\ and\ \bibinfo {author} {\bibfnamefont {S.}~\bibnamefont {Datta}},\ }\bibfield  {title} {\bibinfo {title} {An ising hamiltonian solver using stochastic phase-transition nano-oscillators},\ }\href {https://doi.org/10.1038/s41928-021-00616-7} {\bibfield  {journal} {\bibinfo  {journal} {Nature Electron}\ }\textbf {\bibinfo {volume} {4}},\ \bibinfo {pages} {502–512} (\bibinfo {year} {2020})}\BibitemShut {NoStop}%
\bibitem [{\citenamefont {Honjo}\ \emph {et~al.}(2021)\citenamefont {Honjo}, \citenamefont {Sonobe}, \citenamefont {Inaba}, \citenamefont {Inagaki}, \citenamefont {Ikuta}, \citenamefont {Yamada}, \citenamefont {Kazama}, \citenamefont {Enbutsu}, \citenamefont {Umeki}, \citenamefont {Kasahara}, \citenamefont {ichi Kawarabayashi},\ and\ \citenamefont {Takesue}}]{Honjo2021}%
  \BibitemOpen
  \bibfield  {author} {\bibinfo {author} {\bibfnamefont {T.}~\bibnamefont {Honjo}}, \bibinfo {author} {\bibfnamefont {T.}~\bibnamefont {Sonobe}}, \bibinfo {author} {\bibfnamefont {K.}~\bibnamefont {Inaba}}, \bibinfo {author} {\bibfnamefont {T.}~\bibnamefont {Inagaki}}, \bibinfo {author} {\bibfnamefont {T.}~\bibnamefont {Ikuta}}, \bibinfo {author} {\bibfnamefont {Y.}~\bibnamefont {Yamada}}, \bibinfo {author} {\bibfnamefont {T.}~\bibnamefont {Kazama}}, \bibinfo {author} {\bibfnamefont {K.}~\bibnamefont {Enbutsu}}, \bibinfo {author} {\bibfnamefont {T.}~\bibnamefont {Umeki}}, \bibinfo {author} {\bibfnamefont {R.}~\bibnamefont {Kasahara}}, \bibinfo {author} {\bibfnamefont {K.}~\bibnamefont {ichi Kawarabayashi}},\ and\ \bibinfo {author} {\bibfnamefont {H.}~\bibnamefont {Takesue}},\ }\bibfield  {title} {\bibinfo {title} {100,000-spin coherent ising machine},\ }\href {https://doi.org/10.1126/sciadv.abh0952} {\bibfield  {journal} {\bibinfo  {journal} {Science Advances}\ }\textbf {\bibinfo {volume} {7}},\ \bibinfo
  {pages} {eabh0952} (\bibinfo {year} {2021})},\ \Eprint {https://arxiv.org/abs/https://www.science.org/doi/pdf/10.1126/sciadv.abh0952} {https://www.science.org/doi/pdf/10.1126/sciadv.abh0952} \BibitemShut {NoStop}%
\bibitem [{\citenamefont {Inagaki}\ \emph {et~al.}(2016)\citenamefont {Inagaki}, \citenamefont {Haribara}, \citenamefont {Igarashi}, \citenamefont {Sonobe}, \citenamefont {Tamate}, \citenamefont {Honjo}, \citenamefont {Marandi}, \citenamefont {McMahon}, \citenamefont {Umeki}, \citenamefont {Enbutsu}, \citenamefont {Tadanaga}, \citenamefont {Takenouchi}, \citenamefont {Aihara}, \citenamefont {Kawarabayashi}, \citenamefont {Inoue}, \citenamefont {Utsunomiya},\ and\ \citenamefont {Takesue}}]{inagakiCoherentIsingMachine2016}%
  \BibitemOpen
  \bibfield  {author} {\bibinfo {author} {\bibfnamefont {T.}~\bibnamefont {Inagaki}}, \bibinfo {author} {\bibfnamefont {Y.}~\bibnamefont {Haribara}}, \bibinfo {author} {\bibfnamefont {K.}~\bibnamefont {Igarashi}}, \bibinfo {author} {\bibfnamefont {T.}~\bibnamefont {Sonobe}}, \bibinfo {author} {\bibfnamefont {S.}~\bibnamefont {Tamate}}, \bibinfo {author} {\bibfnamefont {T.}~\bibnamefont {Honjo}}, \bibinfo {author} {\bibfnamefont {A.}~\bibnamefont {Marandi}}, \bibinfo {author} {\bibfnamefont {P.~L.}\ \bibnamefont {McMahon}}, \bibinfo {author} {\bibfnamefont {T.}~\bibnamefont {Umeki}}, \bibinfo {author} {\bibfnamefont {K.}~\bibnamefont {Enbutsu}}, \bibinfo {author} {\bibfnamefont {O.}~\bibnamefont {Tadanaga}}, \bibinfo {author} {\bibfnamefont {H.}~\bibnamefont {Takenouchi}}, \bibinfo {author} {\bibfnamefont {K.}~\bibnamefont {Aihara}}, \bibinfo {author} {\bibfnamefont {K.-i.}\ \bibnamefont {Kawarabayashi}}, \bibinfo {author} {\bibfnamefont {K.}~\bibnamefont {Inoue}}, \bibinfo {author} {\bibfnamefont
  {S.}~\bibnamefont {Utsunomiya}},\ and\ \bibinfo {author} {\bibfnamefont {H.}~\bibnamefont {Takesue}},\ }\bibfield  {title} {\bibinfo {title} {A coherent {{Ising}} machine for 2000-node optimization problems},\ }\href {https://doi.org/10.1126/science.aah4243} {\bibfield  {journal} {\bibinfo  {journal} {Science}\ }\textbf {\bibinfo {volume} {354}},\ \bibinfo {pages} {603} (\bibinfo {year} {2016})}\BibitemShut {NoStop}%
\bibitem [{\citenamefont {McMahon}\ \emph {et~al.}(2016)\citenamefont {McMahon}, \citenamefont {Marandi}, \citenamefont {Haribara}, \citenamefont {Hamerly}, \citenamefont {Langrock}, \citenamefont {Tamate}, \citenamefont {Inagaki}, \citenamefont {Takesue}, \citenamefont {Utsunomiya}, \citenamefont {Aihara}, \citenamefont {Byer}, \citenamefont {Fejer}, \citenamefont {Mabuchi},\ and\ \citenamefont {Yamamoto}}]{mcmahonFullyProgrammable100spin2016}%
  \BibitemOpen
  \bibfield  {author} {\bibinfo {author} {\bibfnamefont {P.~L.}\ \bibnamefont {McMahon}}, \bibinfo {author} {\bibfnamefont {A.}~\bibnamefont {Marandi}}, \bibinfo {author} {\bibfnamefont {Y.}~\bibnamefont {Haribara}}, \bibinfo {author} {\bibfnamefont {R.}~\bibnamefont {Hamerly}}, \bibinfo {author} {\bibfnamefont {C.}~\bibnamefont {Langrock}}, \bibinfo {author} {\bibfnamefont {S.}~\bibnamefont {Tamate}}, \bibinfo {author} {\bibfnamefont {T.}~\bibnamefont {Inagaki}}, \bibinfo {author} {\bibfnamefont {H.}~\bibnamefont {Takesue}}, \bibinfo {author} {\bibfnamefont {S.}~\bibnamefont {Utsunomiya}}, \bibinfo {author} {\bibfnamefont {K.}~\bibnamefont {Aihara}}, \bibinfo {author} {\bibfnamefont {R.~L.}\ \bibnamefont {Byer}}, \bibinfo {author} {\bibfnamefont {M.~M.}\ \bibnamefont {Fejer}}, \bibinfo {author} {\bibfnamefont {H.}~\bibnamefont {Mabuchi}},\ and\ \bibinfo {author} {\bibfnamefont {Y.}~\bibnamefont {Yamamoto}},\ }\bibfield  {title} {\bibinfo {title} {A fully programmable 100-spin coherent {{Ising}} machine with
  all-to-all connections},\ }\href {https://doi.org/10.1126/science.aah5178} {\bibfield  {journal} {\bibinfo  {journal} {Science}\ }\textbf {\bibinfo {volume} {354}},\ \bibinfo {pages} {614} (\bibinfo {year} {2016})}\BibitemShut {NoStop}%
\bibitem [{\citenamefont {Yamamoto}\ \emph {et~al.}(2020)\citenamefont {Yamamoto}, \citenamefont {Leleu}, \citenamefont {Ganguli},\ and\ \citenamefont {Mabuchi}}]{yamamotoCoherentIsingMachines2020}%
  \BibitemOpen
  \bibfield  {author} {\bibinfo {author} {\bibfnamefont {Y.}~\bibnamefont {Yamamoto}}, \bibinfo {author} {\bibfnamefont {T.}~\bibnamefont {Leleu}}, \bibinfo {author} {\bibfnamefont {S.}~\bibnamefont {Ganguli}},\ and\ \bibinfo {author} {\bibfnamefont {H.}~\bibnamefont {Mabuchi}},\ }\bibfield  {title} {\bibinfo {title} {Coherent {{Ising}} machines\textemdash{{Quantum}} optics and neural network {{Perspectives}}},\ }\href {https://doi.org/10.1063/5.0016140} {\bibfield  {journal} {\bibinfo  {journal} {Applied Physics Letters}\ }\textbf {\bibinfo {volume} {117}},\ \bibinfo {pages} {160501} (\bibinfo {year} {2020})}\BibitemShut {NoStop}%
\bibitem [{\citenamefont {Yamamoto}\ \emph {et~al.}(2017)\citenamefont {Yamamoto}, \citenamefont {Aihara}, \citenamefont {Leleu}, \citenamefont {Kawarabayashi}, \citenamefont {Kako}, \citenamefont {Fejer}, \citenamefont {Inoue},\ and\ \citenamefont {Takesue}}]{yamamotoCoherentIsingMachines2017}%
  \BibitemOpen
  \bibfield  {author} {\bibinfo {author} {\bibfnamefont {Y.}~\bibnamefont {Yamamoto}}, \bibinfo {author} {\bibfnamefont {K.}~\bibnamefont {Aihara}}, \bibinfo {author} {\bibfnamefont {T.}~\bibnamefont {Leleu}}, \bibinfo {author} {\bibfnamefont {K.-i.}\ \bibnamefont {Kawarabayashi}}, \bibinfo {author} {\bibfnamefont {S.}~\bibnamefont {Kako}}, \bibinfo {author} {\bibfnamefont {M.}~\bibnamefont {Fejer}}, \bibinfo {author} {\bibfnamefont {K.}~\bibnamefont {Inoue}},\ and\ \bibinfo {author} {\bibfnamefont {H.}~\bibnamefont {Takesue}},\ }\bibfield  {title} {\bibinfo {title} {Coherent {{Ising}} machines\textemdash optical neural networks operating at the quantum limit},\ }\href {https://doi.org/10.1038/s41534-017-0048-9} {\bibfield  {journal} {\bibinfo  {journal} {npj Quantum Information}\ }\textbf {\bibinfo {volume} {3}},\ \bibinfo {pages} {1} (\bibinfo {year} {2017})}\BibitemShut {NoStop}%
\bibitem [{\citenamefont {Wang}\ \emph {et~al.}(2013)\citenamefont {Wang}, \citenamefont {Marandi}, \citenamefont {Wen}, \citenamefont {Byer},\ and\ \citenamefont {Yamamoto}}]{wangCoherentIsingMachine2013}%
  \BibitemOpen
  \bibfield  {author} {\bibinfo {author} {\bibfnamefont {Z.}~\bibnamefont {Wang}}, \bibinfo {author} {\bibfnamefont {A.}~\bibnamefont {Marandi}}, \bibinfo {author} {\bibfnamefont {K.}~\bibnamefont {Wen}}, \bibinfo {author} {\bibfnamefont {R.~L.}\ \bibnamefont {Byer}},\ and\ \bibinfo {author} {\bibfnamefont {Y.}~\bibnamefont {Yamamoto}},\ }\bibfield  {title} {\bibinfo {title} {Coherent {{Ising}} machine based on degenerate optical parametric oscillators},\ }\href {https://doi.org/10.1103/PhysRevA.88.063853} {\bibfield  {journal} {\bibinfo  {journal} {Physical Review A}\ }\textbf {\bibinfo {volume} {88}},\ \bibinfo {pages} {063853} (\bibinfo {year} {2013})}\BibitemShut {NoStop}%
\bibitem [{\citenamefont {Yamamura}\ \emph {et~al.}(2017)\citenamefont {Yamamura}, \citenamefont {Aihara},\ and\ \citenamefont {Yamamoto}}]{yamamuraQuantumModelCoherent2017}%
  \BibitemOpen
  \bibfield  {author} {\bibinfo {author} {\bibfnamefont {A.}~\bibnamefont {Yamamura}}, \bibinfo {author} {\bibfnamefont {K.}~\bibnamefont {Aihara}},\ and\ \bibinfo {author} {\bibfnamefont {Y.}~\bibnamefont {Yamamoto}},\ }\bibfield  {title} {\bibinfo {title} {Quantum model for coherent {{Ising}} machines: {{Discrete}}-time measurement feedback formulation},\ }\href {https://doi.org/10.1103/PhysRevA.96.053834} {\bibfield  {journal} {\bibinfo  {journal} {Physical Review A}\ }\textbf {\bibinfo {volume} {96}},\ \bibinfo {pages} {053834} (\bibinfo {year} {2017})}\BibitemShut {NoStop}%
\bibitem [{\citenamefont {Strinati}\ \emph {et~al.}(2021)\citenamefont {Strinati}, \citenamefont {Pierangeli},\ and\ \citenamefont {Conti}}]{strinati2021all}%
  \BibitemOpen
  \bibfield  {author} {\bibinfo {author} {\bibfnamefont {M.~C.}\ \bibnamefont {Strinati}}, \bibinfo {author} {\bibfnamefont {D.}~\bibnamefont {Pierangeli}},\ and\ \bibinfo {author} {\bibfnamefont {C.}~\bibnamefont {Conti}},\ }\bibfield  {title} {\bibinfo {title} {All-optical scalable spatial coherent ising machine},\ }\href {https://journals.aps.org/prapplied/abstract/10.1103/PhysRevApplied.16.054022} {\bibfield  {journal} {\bibinfo  {journal} {Physical Review Applied}\ }\textbf {\bibinfo {volume} {16}},\ \bibinfo {pages} {054022} (\bibinfo {year} {2021})}\BibitemShut {NoStop}%
\bibitem [{\citenamefont {Barahona}(1982)}]{barahonaComputationalComplexityIsing1982}%
  \BibitemOpen
  \bibfield  {author} {\bibinfo {author} {\bibfnamefont {F.}~\bibnamefont {Barahona}},\ }\bibfield  {title} {\bibinfo {title} {On the computational complexity of {{Ising}} spin glass models},\ }\href {https://doi.org/10.1088/0305-4470/15/10/028} {\bibfield  {journal} {\bibinfo  {journal} {Journal of Physics A: Mathematical and General}\ }\textbf {\bibinfo {volume} {15}},\ \bibinfo {pages} {3241} (\bibinfo {year} {1982})}\BibitemShut {NoStop}%
\bibitem [{\citenamefont {Lucas}(2014)}]{lucasIsingFormulationsMany2014}%
  \BibitemOpen
  \bibfield  {author} {\bibinfo {author} {\bibfnamefont {A.}~\bibnamefont {Lucas}},\ }\bibfield  {title} {\bibinfo {title} {Ising formulations of many {{NP}} problems},\ }\bibfield  {journal} {\bibinfo  {journal} {Frontiers in Physics}\ }\textbf {\bibinfo {volume} {2}},\ \href {https://doi.org/10.3389/fphy.2014.00005} {10.3389/fphy.2014.00005} (\bibinfo {year} {2014})\BibitemShut {NoStop}%
\bibitem [{\citenamefont {Jankowski}\ \emph {et~al.}(2023)\citenamefont {Jankowski}, \citenamefont {Yanagimoto}, \citenamefont {Ng}, \citenamefont {Hamerly}, \citenamefont {McKenna}, \citenamefont {Mabuchi},\ and\ \citenamefont {Fejer}}]{Jankowski2023}%
  \BibitemOpen
  \bibfield  {author} {\bibinfo {author} {\bibfnamefont {M.}~\bibnamefont {Jankowski}}, \bibinfo {author} {\bibfnamefont {R.}~\bibnamefont {Yanagimoto}}, \bibinfo {author} {\bibfnamefont {E.}~\bibnamefont {Ng}}, \bibinfo {author} {\bibfnamefont {R.}~\bibnamefont {Hamerly}}, \bibinfo {author} {\bibfnamefont {T.~P.}\ \bibnamefont {McKenna}}, \bibinfo {author} {\bibfnamefont {H.}~\bibnamefont {Mabuchi}},\ and\ \bibinfo {author} {\bibfnamefont {M.~M.}\ \bibnamefont {Fejer}},\ }\bibfield  {title} {\bibinfo {title} {Ultrafast nonlinear photonics---from classical physics to non-gaussian quantum dynamics},\ }\href@noop {} {\bibfield  {journal} {\bibinfo  {journal} {Advances in Optics and Photonics}\ ,\ \bibinfo {pages} {to appear}} (\bibinfo {year} {2023})}\BibitemShut {NoStop}%
\bibitem [{\citenamefont {Yanagimoto}\ \emph {et~al.}(2023)\citenamefont {Yanagimoto}, \citenamefont {Nehra}, \citenamefont {Hamerly}, \citenamefont {Ng}, \citenamefont {Marandi},\ and\ \citenamefont {Mabuchi}}]{Yanagimoto2023}%
  \BibitemOpen
  \bibfield  {author} {\bibinfo {author} {\bibfnamefont {R.}~\bibnamefont {Yanagimoto}}, \bibinfo {author} {\bibfnamefont {R.}~\bibnamefont {Nehra}}, \bibinfo {author} {\bibfnamefont {R.}~\bibnamefont {Hamerly}}, \bibinfo {author} {\bibfnamefont {E.}~\bibnamefont {Ng}}, \bibinfo {author} {\bibfnamefont {A.}~\bibnamefont {Marandi}},\ and\ \bibinfo {author} {\bibfnamefont {H.}~\bibnamefont {Mabuchi}},\ }\bibfield  {title} {\bibinfo {title} {Quantum nondemolition measurements with optical parametric amplifiers for ultrafast universal quantum information processing},\ }\href {https://doi.org/10.1103/PRXQuantum.4.010333} {\bibfield  {journal} {\bibinfo  {journal} {PRX Quantum}\ }\textbf {\bibinfo {volume} {4}},\ \bibinfo {pages} {010333} (\bibinfo {year} {2023})}\BibitemShut {NoStop}%
\bibitem [{\citenamefont {Bilbro}\ \emph {et~al.}(1989)\citenamefont {Bilbro}, \citenamefont {Mann}, \citenamefont {Miller}, \citenamefont {Snyder}, \citenamefont {van~den Bout},\ and\ \citenamefont {White}}]{bilbroOptimizationMeanField1989}%
  \BibitemOpen
  \bibfield  {author} {\bibinfo {author} {\bibfnamefont {G.}~\bibnamefont {Bilbro}}, \bibinfo {author} {\bibfnamefont {R.}~\bibnamefont {Mann}}, \bibinfo {author} {\bibfnamefont {T.~K.}\ \bibnamefont {Miller}}, \bibinfo {author} {\bibfnamefont {W.~E.}\ \bibnamefont {Snyder}}, \bibinfo {author} {\bibfnamefont {D.~E.}\ \bibnamefont {van~den Bout}},\ and\ \bibinfo {author} {\bibfnamefont {M.}~\bibnamefont {White}},\ }\bibfield  {title} {\bibinfo {title} {Optimization by {{Mean Field Annealing}}},\ }in\ \href@noop {} {\emph {\bibinfo {booktitle} {Advances in {{Neural Information Processing Systems}} 1}}},\ \bibinfo {editor} {edited by\ \bibinfo {editor} {\bibfnamefont {D.~S.}\ \bibnamefont {Touretzky}}}\ (\bibinfo  {publisher} {{Morgan-Kaufmann}},\ \bibinfo {year} {1989})\ pp.\ \bibinfo {pages} {91--98}\BibitemShut {NoStop}%
\bibitem [{\citenamefont {Chaudhari}\ and\ \citenamefont {Soatto}(2017)}]{chaudhariEnergyLandscapeDeep2017}%
  \BibitemOpen
  \bibfield  {author} {\bibinfo {author} {\bibfnamefont {P.}~\bibnamefont {Chaudhari}}\ and\ \bibinfo {author} {\bibfnamefont {S.}~\bibnamefont {Soatto}},\ }\bibfield  {title} {\bibinfo {title} {On the energy landscape of deep networks},\ }\href@noop {} {\bibfield  {journal} {\bibinfo  {journal} {arXiv:1511.06485 [cs]}\ } (\bibinfo {year} {2017})},\ \Eprint {https://arxiv.org/abs/1511.06485} {arXiv:1511.06485 [cs]} \BibitemShut {NoStop}%
\bibitem [{\citenamefont {Fyodorov}\ and\ \citenamefont {Le~Doussal}(2014)}]{fyodorovTopologyTrivializationLarge2014}%
  \BibitemOpen
  \bibfield  {author} {\bibinfo {author} {\bibfnamefont {Y.~V.}\ \bibnamefont {Fyodorov}}\ and\ \bibinfo {author} {\bibfnamefont {P.}~\bibnamefont {Le~Doussal}},\ }\bibfield  {title} {\bibinfo {title} {Topology {{Trivialization}} and {{Large Deviations}} for the {{Minimum}} in the {{Simplest Random Optimization}}},\ }\href {https://doi.org/10.1007/s10955-013-0838-1} {\bibfield  {journal} {\bibinfo  {journal} {Journal of Statistical Physics}\ }\textbf {\bibinfo {volume} {154}},\ \bibinfo {pages} {466} (\bibinfo {year} {2014})}\BibitemShut {NoStop}%
\bibitem [{\citenamefont {Haribara}\ \emph {et~al.}(2017)\citenamefont {Haribara}, \citenamefont {Ishikawa}, \citenamefont {Utsunomiya}, \citenamefont {Aihara},\ and\ \citenamefont {Yamamoto}}]{haribaraPerformanceEvaluationCoherent2017}%
  \BibitemOpen
  \bibfield  {author} {\bibinfo {author} {\bibfnamefont {Y.}~\bibnamefont {Haribara}}, \bibinfo {author} {\bibfnamefont {H.}~\bibnamefont {Ishikawa}}, \bibinfo {author} {\bibfnamefont {S.}~\bibnamefont {Utsunomiya}}, \bibinfo {author} {\bibfnamefont {K.}~\bibnamefont {Aihara}},\ and\ \bibinfo {author} {\bibfnamefont {Y.}~\bibnamefont {Yamamoto}},\ }\bibfield  {title} {\bibinfo {title} {Performance evaluation of coherent {{Ising}} machines against classical neural networks},\ }\href {https://doi.org/10.1088/2058-9565/aa8190} {\bibfield  {journal} {\bibinfo  {journal} {Quantum Science and Technology}\ }\textbf {\bibinfo {volume} {2}},\ \bibinfo {pages} {044002} (\bibinfo {year} {2017})}\BibitemShut {NoStop}%
\bibitem [{\citenamefont {Hamerly}\ \emph {et~al.}(2019)\citenamefont {Hamerly}, \citenamefont {Inagaki}, \citenamefont {McMahon}, \citenamefont {Venturelli}, \citenamefont {Marandi}, \citenamefont {Onodera}, \citenamefont {Ng}, \citenamefont {Langrock}, \citenamefont {Inaba}, \citenamefont {Honjo}, \citenamefont {Enbutsu}, \citenamefont {Umeki}, \citenamefont {Kasahara}, \citenamefont {Utsunomiya}, \citenamefont {Kako}, \citenamefont {Kawarabayashi}, \citenamefont {Byer}, \citenamefont {Fejer}, \citenamefont {Mabuchi}, \citenamefont {Englund}, \citenamefont {Rieffel}, \citenamefont {Takesue},\ and\ \citenamefont {Yamamoto}}]{hamerlyExperimentalInvestigationPerformance2019}%
  \BibitemOpen
  \bibfield  {author} {\bibinfo {author} {\bibfnamefont {R.}~\bibnamefont {Hamerly}}, \bibinfo {author} {\bibfnamefont {T.}~\bibnamefont {Inagaki}}, \bibinfo {author} {\bibfnamefont {P.~L.}\ \bibnamefont {McMahon}}, \bibinfo {author} {\bibfnamefont {D.}~\bibnamefont {Venturelli}}, \bibinfo {author} {\bibfnamefont {A.}~\bibnamefont {Marandi}}, \bibinfo {author} {\bibfnamefont {T.}~\bibnamefont {Onodera}}, \bibinfo {author} {\bibfnamefont {E.}~\bibnamefont {Ng}}, \bibinfo {author} {\bibfnamefont {C.}~\bibnamefont {Langrock}}, \bibinfo {author} {\bibfnamefont {K.}~\bibnamefont {Inaba}}, \bibinfo {author} {\bibfnamefont {T.}~\bibnamefont {Honjo}}, \bibinfo {author} {\bibfnamefont {K.}~\bibnamefont {Enbutsu}}, \bibinfo {author} {\bibfnamefont {T.}~\bibnamefont {Umeki}}, \bibinfo {author} {\bibfnamefont {R.}~\bibnamefont {Kasahara}}, \bibinfo {author} {\bibfnamefont {S.}~\bibnamefont {Utsunomiya}}, \bibinfo {author} {\bibfnamefont {S.}~\bibnamefont {Kako}}, \bibinfo {author} {\bibfnamefont {K.-i.}\ \bibnamefont
  {Kawarabayashi}}, \bibinfo {author} {\bibfnamefont {R.~L.}\ \bibnamefont {Byer}}, \bibinfo {author} {\bibfnamefont {M.~M.}\ \bibnamefont {Fejer}}, \bibinfo {author} {\bibfnamefont {H.}~\bibnamefont {Mabuchi}}, \bibinfo {author} {\bibfnamefont {D.}~\bibnamefont {Englund}}, \bibinfo {author} {\bibfnamefont {E.}~\bibnamefont {Rieffel}}, \bibinfo {author} {\bibfnamefont {H.}~\bibnamefont {Takesue}},\ and\ \bibinfo {author} {\bibfnamefont {Y.}~\bibnamefont {Yamamoto}},\ }\bibfield  {title} {\bibinfo {title} {Experimental investigation of performance differences between coherent {{Ising}} machines and a quantum annealer},\ }\bibfield  {journal} {\bibinfo  {journal} {Science Advances}\ }\textbf {\bibinfo {volume} {5}},\ \href {https://doi.org/10.1126/sciadv.aau0823} {10.1126/sciadv.aau0823} (\bibinfo {year} {2019})\BibitemShut {NoStop}%
\bibitem [{\citenamefont {Vincent}\ \emph {et~al.}(1994)\citenamefont {Vincent}, \citenamefont {Mirko},\ and\ \citenamefont {{Jean-Paul}}}]{vincentSimulatedAnnealingProof1994}%
  \BibitemOpen
  \bibfield  {author} {\bibinfo {author} {\bibfnamefont {G.}~\bibnamefont {Vincent}}, \bibinfo {author} {\bibfnamefont {K.}~\bibnamefont {Mirko}},\ and\ \bibinfo {author} {\bibfnamefont {R.}~\bibnamefont {{Jean-Paul}}},\ }\bibfield  {title} {\bibinfo {title} {Simulated annealing: A proof of convergence},\ }\href {https://doi.org/10.1109/34.295910} {\bibfield  {journal} {\bibinfo  {journal} {IEEE Transactions on Pattern Analysis and Machine Intelligence}\ }\textbf {\bibinfo {volume} {16}},\ \bibinfo {pages} {652} (\bibinfo {year} {1994})}\BibitemShut {NoStop}%
\bibitem [{\citenamefont {Kadowaki}\ and\ \citenamefont {Nishimori}(1998)}]{kadowakiQuantumAnnealingTransverse1998}%
  \BibitemOpen
  \bibfield  {author} {\bibinfo {author} {\bibfnamefont {T.}~\bibnamefont {Kadowaki}}\ and\ \bibinfo {author} {\bibfnamefont {H.}~\bibnamefont {Nishimori}},\ }\bibfield  {title} {\bibinfo {title} {Quantum annealing in the transverse {{Ising}} model},\ }\href {https://doi.org/10.1103/PhysRevE.58.5355} {\bibfield  {journal} {\bibinfo  {journal} {Physical Review E}\ }\textbf {\bibinfo {volume} {58}},\ \bibinfo {pages} {5355} (\bibinfo {year} {1998})}\BibitemShut {NoStop}%
\bibitem [{\citenamefont {Leleu}\ \emph {et~al.}(2017)\citenamefont {Leleu}, \citenamefont {Yamamoto}, \citenamefont {Utsunomiya},\ and\ \citenamefont {Aihara}}]{leleuCombinatorialOptimizationUsing2017}%
  \BibitemOpen
  \bibfield  {author} {\bibinfo {author} {\bibfnamefont {T.}~\bibnamefont {Leleu}}, \bibinfo {author} {\bibfnamefont {Y.}~\bibnamefont {Yamamoto}}, \bibinfo {author} {\bibfnamefont {S.}~\bibnamefont {Utsunomiya}},\ and\ \bibinfo {author} {\bibfnamefont {K.}~\bibnamefont {Aihara}},\ }\bibfield  {title} {\bibinfo {title} {Combinatorial optimization using dynamical phase transitions in driven-dissipative systems},\ }\href {https://doi.org/10.1103/PhysRevE.95.022118} {\bibfield  {journal} {\bibinfo  {journal} {Physical Review E}\ }\textbf {\bibinfo {volume} {95}},\ \bibinfo {pages} {022118} (\bibinfo {year} {2017})}\BibitemShut {NoStop}%
\bibitem [{\citenamefont {Leleu}\ \emph {et~al.}(2019)\citenamefont {Leleu}, \citenamefont {Yamamoto}, \citenamefont {McMahon},\ and\ \citenamefont {Aihara}}]{leleuDestabilizationLocalMinima2019}%
  \BibitemOpen
  \bibfield  {author} {\bibinfo {author} {\bibfnamefont {T.}~\bibnamefont {Leleu}}, \bibinfo {author} {\bibfnamefont {Y.}~\bibnamefont {Yamamoto}}, \bibinfo {author} {\bibfnamefont {P.~L.}\ \bibnamefont {McMahon}},\ and\ \bibinfo {author} {\bibfnamefont {K.}~\bibnamefont {Aihara}},\ }\bibfield  {title} {\bibinfo {title} {Destabilization of {{Local Minima}} in {{Analog Spin Systems}} by {{Correction}} of {{Amplitude Heterogeneity}}},\ }\href {https://doi.org/10.1103/PhysRevLett.122.040607} {\bibfield  {journal} {\bibinfo  {journal} {Physical Review Letters}\ }\textbf {\bibinfo {volume} {122}},\ \bibinfo {pages} {040607} (\bibinfo {year} {2019})}\BibitemShut {NoStop}%
\bibitem [{\citenamefont {Mezard}\ \emph {et~al.}(1986)\citenamefont {Mezard}, \citenamefont {Parisi},\ and\ \citenamefont {Virasoro}}]{mezardSpinGlassTheory1986}%
  \BibitemOpen
  \bibfield  {author} {\bibinfo {author} {\bibfnamefont {M.}~\bibnamefont {Mezard}}, \bibinfo {author} {\bibfnamefont {G.}~\bibnamefont {Parisi}},\ and\ \bibinfo {author} {\bibfnamefont {M.}~\bibnamefont {Virasoro}},\ }\href {https://doi.org/10.1142/0271} {\emph {\bibinfo {title} {Spin {{Glass Theory}} and {{Beyond}}: {{An Introduction}} to the {{Replica Method}} and {{Its Applications}}}}},\ \bibinfo {series} {World {{Scientific Lecture Notes}} in {{Physics}}}, Vol.~\bibinfo {volume} {9}\ (\bibinfo  {publisher} {{WORLD SCIENTIFIC}},\ \bibinfo {year} {1986})\BibitemShut {NoStop}%
\bibitem [{\citenamefont {Bray}\ and\ \citenamefont {Moore}(1987)}]{brayChaoticNatureSpinGlass1987}%
  \BibitemOpen
  \bibfield  {author} {\bibinfo {author} {\bibfnamefont {A.~J.}\ \bibnamefont {Bray}}\ and\ \bibinfo {author} {\bibfnamefont {M.~A.}\ \bibnamefont {Moore}},\ }\bibfield  {title} {\bibinfo {title} {Chaotic {{Nature}} of the {{Spin}}-{{Glass Phase}}},\ }\href {https://doi.org/10.1103/PhysRevLett.58.57} {\bibfield  {journal} {\bibinfo  {journal} {Physical Review Letters}\ }\textbf {\bibinfo {volume} {58}},\ \bibinfo {pages} {57} (\bibinfo {year} {1987})}\BibitemShut {NoStop}%
\bibitem [{\citenamefont {Krzakala}\ and\ \citenamefont {Martin}(2002)}]{krzakalaChaoticTemperatureDependence2002}%
  \BibitemOpen
  \bibfield  {author} {\bibinfo {author} {\bibfnamefont {F.}~\bibnamefont {Krzakala}}\ and\ \bibinfo {author} {\bibfnamefont {O.}~\bibnamefont {Martin}},\ }\bibfield  {title} {\bibinfo {title} {Chaotic temperature dependence in a model of spin glasses},\ }\href {https://doi.org/10.1140/epjb/e2002-00221-y} {\bibfield  {journal} {\bibinfo  {journal} {The European Physical Journal B - Condensed Matter and Complex Systems}\ }\textbf {\bibinfo {volume} {28}},\ \bibinfo {pages} {199} (\bibinfo {year} {2002})}\BibitemShut {NoStop}%
\bibitem [{\citenamefont {Rizzo}\ and\ \citenamefont {Crisanti}(2003)}]{rizzoChaosTemperatureSherringtonKirkpatrick2003}%
  \BibitemOpen
  \bibfield  {author} {\bibinfo {author} {\bibfnamefont {T.}~\bibnamefont {Rizzo}}\ and\ \bibinfo {author} {\bibfnamefont {A.}~\bibnamefont {Crisanti}},\ }\bibfield  {title} {\bibinfo {title} {Chaos in {{Temperature}} in the {{Sherrington}}-{{Kirkpatrick Model}}},\ }\href {https://doi.org/10.1103/PhysRevLett.90.137201} {\bibfield  {journal} {\bibinfo  {journal} {Physical Review Letters}\ }\textbf {\bibinfo {volume} {90}},\ \bibinfo {pages} {137201} (\bibinfo {year} {2003})}\BibitemShut {NoStop}%
\bibitem [{\citenamefont {Krzakala}\ and\ \citenamefont {Zdeborov{\'a}}(2010)}]{krzakalaFollowingGibbsStates2010}%
  \BibitemOpen
  \bibfield  {author} {\bibinfo {author} {\bibfnamefont {F.}~\bibnamefont {Krzakala}}\ and\ \bibinfo {author} {\bibfnamefont {L.}~\bibnamefont {Zdeborov{\'a}}},\ }\bibfield  {title} {\bibinfo {title} {Following {{Gibbs}} states adiabatically \textemdash{{The}} energy landscape of mean-field glassy systems},\ }\href {https://doi.org/10.1209/0295-5075/90/66002} {\bibfield  {journal} {\bibinfo  {journal} {EPL (Europhysics Letters)}\ }\textbf {\bibinfo {volume} {90}},\ \bibinfo {pages} {66002} (\bibinfo {year} {2010})}\BibitemShut {NoStop}%
\bibitem [{\citenamefont {Laumann}\ \emph {et~al.}(2014)\citenamefont {Laumann}, \citenamefont {Pal},\ and\ \citenamefont {Scardicchio}}]{laumannManybodyMobilityEdge2014}%
  \BibitemOpen
  \bibfield  {author} {\bibinfo {author} {\bibfnamefont {C.~R.}\ \bibnamefont {Laumann}}, \bibinfo {author} {\bibfnamefont {A.}~\bibnamefont {Pal}},\ and\ \bibinfo {author} {\bibfnamefont {A.}~\bibnamefont {Scardicchio}},\ }\bibfield  {title} {\bibinfo {title} {Many-body mobility edge in a mean-field quantum spin glass},\ }\href {https://doi.org/10.1103/PhysRevLett.113.200405} {\bibfield  {journal} {\bibinfo  {journal} {Physical Review Letters}\ }\textbf {\bibinfo {volume} {113}},\ \bibinfo {pages} {200405} (\bibinfo {year} {2014})},\ \Eprint {https://arxiv.org/abs/1404.2276} {arXiv:1404.2276} \BibitemShut {NoStop}%
\bibitem [{\citenamefont {Foini}\ \emph {et~al.}(2010)\citenamefont {Foini}, \citenamefont {Semerjian},\ and\ \citenamefont {Zamponi}}]{foiniSolvableModelQuantum2010}%
  \BibitemOpen
  \bibfield  {author} {\bibinfo {author} {\bibfnamefont {L.}~\bibnamefont {Foini}}, \bibinfo {author} {\bibfnamefont {G.}~\bibnamefont {Semerjian}},\ and\ \bibinfo {author} {\bibfnamefont {F.}~\bibnamefont {Zamponi}},\ }\bibfield  {title} {\bibinfo {title} {Solvable {{Model}} of {{Quantum Random Optimization Problems}}},\ }\href {https://doi.org/10.1103/PhysRevLett.105.167204} {\bibfield  {journal} {\bibinfo  {journal} {Physical Review Letters}\ }\textbf {\bibinfo {volume} {105}},\ \bibinfo {pages} {167204} (\bibinfo {year} {2010})}\BibitemShut {NoStop}%
\bibitem [{\citenamefont {Altshuler}\ \emph {et~al.}(2010)\citenamefont {Altshuler}, \citenamefont {Krovi},\ and\ \citenamefont {Roland}}]{altshulerAndersonLocalizationMakes2010}%
  \BibitemOpen
  \bibfield  {author} {\bibinfo {author} {\bibfnamefont {B.}~\bibnamefont {Altshuler}}, \bibinfo {author} {\bibfnamefont {H.}~\bibnamefont {Krovi}},\ and\ \bibinfo {author} {\bibfnamefont {J.}~\bibnamefont {Roland}},\ }\bibfield  {title} {\bibinfo {title} {Anderson localization makes adiabatic quantum optimization fail},\ }\href {https://doi.org/10.1073/pnas.1002116107} {\bibfield  {journal} {\bibinfo  {journal} {Proceedings of the National Academy of Sciences}\ }\textbf {\bibinfo {volume} {107}},\ \bibinfo {pages} {12446} (\bibinfo {year} {2010})}\BibitemShut {NoStop}%
\bibitem [{\citenamefont {Bapst}\ \emph {et~al.}(2013)\citenamefont {Bapst}, \citenamefont {Foini}, \citenamefont {Krzakala}, \citenamefont {Semerjian},\ and\ \citenamefont {Zamponi}}]{bapstQuantumAdiabaticAlgorithm2013}%
  \BibitemOpen
  \bibfield  {author} {\bibinfo {author} {\bibfnamefont {V.}~\bibnamefont {Bapst}}, \bibinfo {author} {\bibfnamefont {L.}~\bibnamefont {Foini}}, \bibinfo {author} {\bibfnamefont {F.}~\bibnamefont {Krzakala}}, \bibinfo {author} {\bibfnamefont {G.}~\bibnamefont {Semerjian}},\ and\ \bibinfo {author} {\bibfnamefont {F.}~\bibnamefont {Zamponi}},\ }\bibfield  {title} {\bibinfo {title} {The quantum adiabatic algorithm applied to random optimization problems: {{The}} quantum spin glass perspective},\ }\href {https://doi.org/10.1016/j.physrep.2012.10.002} {\bibfield  {journal} {\bibinfo  {journal} {Physics Reports}\ }\textbf {\bibinfo {volume} {523}},\ \bibinfo {pages} {127} (\bibinfo {year} {2013})}\BibitemShut {NoStop}%
\bibitem [{\citenamefont {Roy}\ \emph {et~al.}(2021)\citenamefont {Roy}, \citenamefont {Jahani}, \citenamefont {Langrock}, \citenamefont {Fejer},\ and\ \citenamefont {Marandi}}]{Roy2021}%
  \BibitemOpen
  \bibfield  {author} {\bibinfo {author} {\bibfnamefont {A.}~\bibnamefont {Roy}}, \bibinfo {author} {\bibfnamefont {S.}~\bibnamefont {Jahani}}, \bibinfo {author} {\bibfnamefont {C.}~\bibnamefont {Langrock}}, \bibinfo {author} {\bibfnamefont {M.}~\bibnamefont {Fejer}},\ and\ \bibinfo {author} {\bibfnamefont {A.}~\bibnamefont {Marandi}},\ }\bibfield  {title} {\bibinfo {title} {Spectral phase transitions in optical parametric oscillators},\ }\href {https://doi.org/10.1038/s41467-021-21048-z} {\bibfield  {journal} {\bibinfo  {journal} {Nature Communications}\ }\textbf {\bibinfo {volume} {12}},\ \bibinfo {pages} {835} (\bibinfo {year} {2021})}\BibitemShut {NoStop}%
\bibitem [{\citenamefont {Bray}\ and\ \citenamefont {Moore}(1980)}]{brayMetastableStatesSpin1980}%
  \BibitemOpen
  \bibfield  {author} {\bibinfo {author} {\bibfnamefont {A.~J.}\ \bibnamefont {Bray}}\ and\ \bibinfo {author} {\bibfnamefont {M.~A.}\ \bibnamefont {Moore}},\ }\bibfield  {title} {\bibinfo {title} {Metastable states in spin glasses},\ }\href {https://doi.org/10.1088/0022-3719/13/19/002} {\bibfield  {journal} {\bibinfo  {journal} {Journal of Physics C: Solid State Physics}\ }\textbf {\bibinfo {volume} {13}},\ \bibinfo {pages} {L469} (\bibinfo {year} {1980})}\BibitemShut {NoStop}%
\bibitem [{\citenamefont {Bray}\ and\ \citenamefont {Moore}(1981)}]{brayMetastableStatesSolvable1981}%
  \BibitemOpen
  \bibfield  {author} {\bibinfo {author} {\bibfnamefont {A.~J.}\ \bibnamefont {Bray}}\ and\ \bibinfo {author} {\bibfnamefont {M.~A.}\ \bibnamefont {Moore}},\ }\bibfield  {title} {\bibinfo {title} {Metastable states in the solvable spin glass model},\ }\href {https://doi.org/10.1088/0305-4470/14/9/012} {\bibfield  {journal} {\bibinfo  {journal} {Journal of Physics A: Mathematical and General}\ }\textbf {\bibinfo {volume} {14}},\ \bibinfo {pages} {L377} (\bibinfo {year} {1981})}\BibitemShut {NoStop}%
\bibitem [{\citenamefont {Aspelmeier}\ \emph {et~al.}(2004)\citenamefont {Aspelmeier}, \citenamefont {Bray},\ and\ \citenamefont {Moore}}]{aspelmeierComplexityIsingSpin2004}%
  \BibitemOpen
  \bibfield  {author} {\bibinfo {author} {\bibfnamefont {T.}~\bibnamefont {Aspelmeier}}, \bibinfo {author} {\bibfnamefont {A.~J.}\ \bibnamefont {Bray}},\ and\ \bibinfo {author} {\bibfnamefont {M.~A.}\ \bibnamefont {Moore}},\ }\bibfield  {title} {\bibinfo {title} {Complexity of {{Ising Spin Glasses}}},\ }\href {https://doi.org/10.1103/PhysRevLett.92.087203} {\bibfield  {journal} {\bibinfo  {journal} {Physical Review Letters}\ }\textbf {\bibinfo {volume} {92}},\ \bibinfo {pages} {087203} (\bibinfo {year} {2004})}\BibitemShut {NoStop}%
\bibitem [{\citenamefont {Crisanti}\ \emph {et~al.}(2004{\natexlab{a}})\citenamefont {Crisanti}, \citenamefont {Leuzzi}, \citenamefont {Parisi},\ and\ \citenamefont {Rizzo}}]{crisantiSpinGlassComplexity2004}%
  \BibitemOpen
  \bibfield  {author} {\bibinfo {author} {\bibfnamefont {A.}~\bibnamefont {Crisanti}}, \bibinfo {author} {\bibfnamefont {L.}~\bibnamefont {Leuzzi}}, \bibinfo {author} {\bibfnamefont {G.}~\bibnamefont {Parisi}},\ and\ \bibinfo {author} {\bibfnamefont {T.}~\bibnamefont {Rizzo}},\ }\bibfield  {title} {\bibinfo {title} {Spin-{{Glass Complexity}}},\ }\href {https://doi.org/10.1103/PhysRevLett.92.127203} {\bibfield  {journal} {\bibinfo  {journal} {Physical Review Letters}\ }\textbf {\bibinfo {volume} {92}},\ \bibinfo {pages} {127203} (\bibinfo {year} {2004}{\natexlab{a}})}\BibitemShut {NoStop}%
\bibitem [{\citenamefont {Parisi}\ and\ \citenamefont {Rizzo}(2004)}]{parisiSupersymmetryBreakingComputation2004}%
  \BibitemOpen
  \bibfield  {author} {\bibinfo {author} {\bibfnamefont {G.}~\bibnamefont {Parisi}}\ and\ \bibinfo {author} {\bibfnamefont {T.}~\bibnamefont {Rizzo}},\ }\bibfield  {title} {\bibinfo {title} {On {{Supersymmetry Breaking}} in the {{Computation}} of the {{Complexity}}},\ }\href {https://doi.org/10.1088/0305-4470/37/33/001} {\bibfield  {journal} {\bibinfo  {journal} {Journal of Physics A: Mathematical and General}\ }\textbf {\bibinfo {volume} {37}},\ \bibinfo {pages} {7979} (\bibinfo {year} {2004})},\ \Eprint {https://arxiv.org/abs/cond-mat/0401509} {arXiv:cond-mat/0401509} \BibitemShut {NoStop}%
\bibitem [{\citenamefont {Crisanti}\ \emph {et~al.}(2005)\citenamefont {Crisanti}, \citenamefont {Leuzzi},\ and\ \citenamefont {Rizzo}}]{crisantiComplexityMeanfieldSpinglass2005}%
  \BibitemOpen
  \bibfield  {author} {\bibinfo {author} {\bibfnamefont {A.}~\bibnamefont {Crisanti}}, \bibinfo {author} {\bibfnamefont {L.}~\bibnamefont {Leuzzi}},\ and\ \bibinfo {author} {\bibfnamefont {T.}~\bibnamefont {Rizzo}},\ }\bibfield  {title} {\bibinfo {title} {Complexity in mean-field spin-glass models: {{Ising}} \$p\$-spin},\ }\href {https://doi.org/10.1103/PhysRevB.71.094202} {\bibfield  {journal} {\bibinfo  {journal} {Physical Review B}\ }\textbf {\bibinfo {volume} {71}},\ \bibinfo {pages} {094202} (\bibinfo {year} {2005})}\BibitemShut {NoStop}%
\bibitem [{\citenamefont {Rizzo}(2005)}]{rizzoTAPComplexityCavity2005}%
  \BibitemOpen
  \bibfield  {author} {\bibinfo {author} {\bibfnamefont {T.}~\bibnamefont {Rizzo}},\ }\bibfield  {title} {\bibinfo {title} {{{TAP}} complexity, the cavity method and supersymmetry},\ }\href {https://doi.org/10.1088/0305-4470/38/15/005} {\bibfield  {journal} {\bibinfo  {journal} {Journal of Physics A: Mathematical and General}\ }\textbf {\bibinfo {volume} {38}},\ \bibinfo {pages} {3287} (\bibinfo {year} {2005})}\BibitemShut {NoStop}%
\bibitem [{\citenamefont {Aspelmeier}\ \emph {et~al.}(2006)\citenamefont {Aspelmeier}, \citenamefont {Blythe}, \citenamefont {Bray},\ and\ \citenamefont {Moore}}]{aspelmeierFreeenergyLandscapesDynamics2006}%
  \BibitemOpen
  \bibfield  {author} {\bibinfo {author} {\bibfnamefont {T.}~\bibnamefont {Aspelmeier}}, \bibinfo {author} {\bibfnamefont {R.~A.}\ \bibnamefont {Blythe}}, \bibinfo {author} {\bibfnamefont {A.~J.}\ \bibnamefont {Bray}},\ and\ \bibinfo {author} {\bibfnamefont {M.~A.}\ \bibnamefont {Moore}},\ }\bibfield  {title} {\bibinfo {title} {Free-energy landscapes, dynamics, and the edge of chaos in mean-field models of spin glasses},\ }\href {https://doi.org/10.1103/PhysRevB.74.184411} {\bibfield  {journal} {\bibinfo  {journal} {Physical Review B}\ }\textbf {\bibinfo {volume} {74}},\ \bibinfo {pages} {184411} (\bibinfo {year} {2006})}\BibitemShut {NoStop}%
\bibitem [{\citenamefont {Fan}\ \emph {et~al.}(2020)\citenamefont {Fan}, \citenamefont {Mei},\ and\ \citenamefont {Montanari}}]{fanTAPFreeEnergy2020}%
  \BibitemOpen
  \bibfield  {author} {\bibinfo {author} {\bibfnamefont {Z.}~\bibnamefont {Fan}}, \bibinfo {author} {\bibfnamefont {S.}~\bibnamefont {Mei}},\ and\ \bibinfo {author} {\bibfnamefont {A.}~\bibnamefont {Montanari}},\ }\bibfield  {title} {\bibinfo {title} {{{TAP}} free energy, spin glasses, and variational inference},\ }\href@noop {} {\bibfield  {journal} {\bibinfo  {journal} {arXiv:1808.07890 [math-ph, stat]}\ } (\bibinfo {year} {2020})},\ \Eprint {https://arxiv.org/abs/1808.07890} {arXiv:1808.07890 [math-ph, stat]} \BibitemShut {NoStop}%
\bibitem [{\citenamefont {M{\"u}ller}\ \emph {et~al.}(2006)\citenamefont {M{\"u}ller}, \citenamefont {Leuzzi},\ and\ \citenamefont {Crisanti}}]{mullerMarginalStatesMeanfield2006}%
  \BibitemOpen
  \bibfield  {author} {\bibinfo {author} {\bibfnamefont {M.}~\bibnamefont {M{\"u}ller}}, \bibinfo {author} {\bibfnamefont {L.}~\bibnamefont {Leuzzi}},\ and\ \bibinfo {author} {\bibfnamefont {A.}~\bibnamefont {Crisanti}},\ }\bibfield  {title} {\bibinfo {title} {Marginal states in mean-field glasses},\ }\href {https://doi.org/10.1103/PhysRevB.74.134431} {\bibfield  {journal} {\bibinfo  {journal} {Physical Review B}\ }\textbf {\bibinfo {volume} {74}},\ \bibinfo {pages} {134431} (\bibinfo {year} {2006})}\BibitemShut {NoStop}%
\bibitem [{\citenamefont {Fyodorov}(2004)}]{fyodorovComplexityRandomEnergy2004}%
  \BibitemOpen
  \bibfield  {author} {\bibinfo {author} {\bibfnamefont {Y.~V.}\ \bibnamefont {Fyodorov}},\ }\bibfield  {title} {\bibinfo {title} {Complexity of {{Random Energy Landscapes}}, {{Glass Transition}}, and {{Absolute Value}} of the {{Spectral Determinant}} of {{Random Matrices}}},\ }\href {https://doi.org/10.1103/PhysRevLett.92.240601} {\bibfield  {journal} {\bibinfo  {journal} {Physical Review Letters}\ }\textbf {\bibinfo {volume} {92}},\ \bibinfo {pages} {240601} (\bibinfo {year} {2004})}\BibitemShut {NoStop}%
\bibitem [{\citenamefont {Fyodorov}(2005)}]{fyodorovCountingStationaryPoints2005}%
  \BibitemOpen
  \bibfield  {author} {\bibinfo {author} {\bibfnamefont {Y.~V.}\ \bibnamefont {Fyodorov}},\ }\bibfield  {title} {\bibinfo {title} {Counting {{Stationary Points}} of {{Random Landscapes}} as a {{Random Matrix Problem}}},\ }\href@noop {} {\bibfield  {journal} {\bibinfo  {journal} {arXiv:cond-mat/0507059}\ } (\bibinfo {year} {2005})},\ \Eprint {https://arxiv.org/abs/cond-mat/0507059} {arXiv:cond-mat/0507059} \BibitemShut {NoStop}%
\bibitem [{\citenamefont {Bray}\ and\ \citenamefont {Dean}(2007)}]{brayStatisticsCriticalPoints2007}%
  \BibitemOpen
  \bibfield  {author} {\bibinfo {author} {\bibfnamefont {A.~J.}\ \bibnamefont {Bray}}\ and\ \bibinfo {author} {\bibfnamefont {D.~S.}\ \bibnamefont {Dean}},\ }\bibfield  {title} {\bibinfo {title} {Statistics of {{Critical Points}} of {{Gaussian Fields}} on {{Large}}-{{Dimensional Spaces}}},\ }\href {https://doi.org/10.1103/PhysRevLett.98.150201} {\bibfield  {journal} {\bibinfo  {journal} {Physical Review Letters}\ }\textbf {\bibinfo {volume} {98}},\ \bibinfo {pages} {150201} (\bibinfo {year} {2007})}\BibitemShut {NoStop}%
\bibitem [{\citenamefont {Fyodorov}\ and\ \citenamefont {Williams}(2007)}]{fyodorovReplicaSymmetryBreaking2007}%
  \BibitemOpen
  \bibfield  {author} {\bibinfo {author} {\bibfnamefont {Y.~V.}\ \bibnamefont {Fyodorov}}\ and\ \bibinfo {author} {\bibfnamefont {I.}~\bibnamefont {Williams}},\ }\bibfield  {title} {\bibinfo {title} {Replica {{Symmetry Breaking Condition Exposed}} by {{Random Matrix Calculation}} of {{Landscape Complexity}}},\ }\href {https://doi.org/10.1007/s10955-007-9386-x} {\bibfield  {journal} {\bibinfo  {journal} {Journal of Statistical Physics}\ }\textbf {\bibinfo {volume} {129}},\ \bibinfo {pages} {1081} (\bibinfo {year} {2007})}\BibitemShut {NoStop}%
\bibitem [{\citenamefont {Fyodorov}(2008)}]{fyodorovStatisticalMechanicsSingle2008}%
  \BibitemOpen
  \bibfield  {author} {\bibinfo {author} {\bibfnamefont {Y.~V.}\ \bibnamefont {Fyodorov}},\ }\bibfield  {title} {\bibinfo {title} {On statistical mechanics of a single particle in high-dimensional random landscapes},\ }\href@noop {} {\bibfield  {journal} {\bibinfo  {journal} {arXiv:0801.0732 [cond-mat]}\ } (\bibinfo {year} {2008})},\ \Eprint {https://arxiv.org/abs/0801.0732} {arXiv:0801.0732 [cond-mat]} \BibitemShut {NoStop}%
\bibitem [{\citenamefont {Cavagna}\ \emph {et~al.}(1999)\citenamefont {Cavagna}, \citenamefont {Garrahan},\ and\ \citenamefont {Giardina}}]{cavagnaQuenchedComplexityMeanfieldpspin1999}%
  \BibitemOpen
  \bibfield  {author} {\bibinfo {author} {\bibfnamefont {A.}~\bibnamefont {Cavagna}}, \bibinfo {author} {\bibfnamefont {J.~P.}\ \bibnamefont {Garrahan}},\ and\ \bibinfo {author} {\bibfnamefont {I.}~\bibnamefont {Giardina}},\ }\bibfield  {title} {\bibinfo {title} {Quenched complexity of the mean-field p-spin spherical model with external magnetic field},\ }\href {https://doi.org/10.1088/0305-4470/32/5/004} {\bibfield  {journal} {\bibinfo  {journal} {Journal of Physics A: Mathematical and General}\ }\textbf {\bibinfo {volume} {32}},\ \bibinfo {pages} {711} (\bibinfo {year} {1999})}\BibitemShut {NoStop}%
\bibitem [{\citenamefont {Cavagna}\ \emph {et~al.}(1998)\citenamefont {Cavagna}, \citenamefont {Giardina},\ and\ \citenamefont {Parisi}}]{cavagnaStationaryPointsThoulessAndersonPalmer1998}%
  \BibitemOpen
  \bibfield  {author} {\bibinfo {author} {\bibfnamefont {A.}~\bibnamefont {Cavagna}}, \bibinfo {author} {\bibfnamefont {I.}~\bibnamefont {Giardina}},\ and\ \bibinfo {author} {\bibfnamefont {G.}~\bibnamefont {Parisi}},\ }\bibfield  {title} {\bibinfo {title} {Stationary points of the {{Thouless}}-{{Anderson}}-{{Palmer}} free energy},\ }\href {https://doi.org/10.1103/PhysRevB.57.11251} {\bibfield  {journal} {\bibinfo  {journal} {Physical Review B}\ }\textbf {\bibinfo {volume} {57}},\ \bibinfo {pages} {11251} (\bibinfo {year} {1998})}\BibitemShut {NoStop}%
\bibitem [{\citenamefont {Crisanti}\ \emph {et~al.}(2003{\natexlab{a}})\citenamefont {Crisanti}, \citenamefont {Leuzzi}, \citenamefont {Parisi},\ and\ \citenamefont {Rizzo}}]{crisantiComplexitySherringtonKirkpatrickModel2003}%
  \BibitemOpen
  \bibfield  {author} {\bibinfo {author} {\bibfnamefont {A.}~\bibnamefont {Crisanti}}, \bibinfo {author} {\bibfnamefont {L.}~\bibnamefont {Leuzzi}}, \bibinfo {author} {\bibfnamefont {G.}~\bibnamefont {Parisi}},\ and\ \bibinfo {author} {\bibfnamefont {T.}~\bibnamefont {Rizzo}},\ }\bibfield  {title} {\bibinfo {title} {Complexity in the {{Sherrington}}-{{Kirkpatrick}} model in the annealed approximation},\ }\href {https://doi.org/10.1103/PhysRevB.68.174401} {\bibfield  {journal} {\bibinfo  {journal} {Physical Review B}\ }\textbf {\bibinfo {volume} {68}},\ \bibinfo {pages} {174401} (\bibinfo {year} {2003}{\natexlab{a}})}\BibitemShut {NoStop}%
\bibitem [{\citenamefont {Crisanti}\ \emph {et~al.}(2003{\natexlab{b}})\citenamefont {Crisanti}, \citenamefont {Leuzzi},\ and\ \citenamefont {Rizzo}}]{crisantiComplexitySphericalMathsfp2003}%
  \BibitemOpen
  \bibfield  {author} {\bibinfo {author} {\bibfnamefont {A.}~\bibnamefont {Crisanti}}, \bibinfo {author} {\bibfnamefont {L.}~\bibnamefont {Leuzzi}},\ and\ \bibinfo {author} {\bibfnamefont {T.}~\bibnamefont {Rizzo}},\ }\bibfield  {title} {\bibinfo {title} {The complexity of the spherical \$\textbackslash mathsf\{p\}\$ -spin spin glass model, revisited},\ }\href {https://doi.org/10.1140/epjb/e2003-00325-x} {\bibfield  {journal} {\bibinfo  {journal} {The European Physical Journal B - Condensed Matter}\ }\textbf {\bibinfo {volume} {36}},\ \bibinfo {pages} {129} (\bibinfo {year} {2003}{\natexlab{b}})}\BibitemShut {NoStop}%
\bibitem [{\citenamefont {Fyodorov}(2013)}]{fyodorovHighDimensionalRandomFields2013}%
  \BibitemOpen
  \bibfield  {author} {\bibinfo {author} {\bibfnamefont {Y.~V.}\ \bibnamefont {Fyodorov}},\ }\bibfield  {title} {\bibinfo {title} {High-{{Dimensional Random Fields}} and {{Random Matrix Theory}}},\ }\href@noop {} {\bibfield  {journal} {\bibinfo  {journal} {arXiv:1307.2379 [cond-mat, physics:math-ph]}\ } (\bibinfo {year} {2013})},\ \Eprint {https://arxiv.org/abs/1307.2379} {arXiv:1307.2379 [cond-mat, physics:math-ph]} \BibitemShut {NoStop}%
\bibitem [{\citenamefont {Auffinger}\ and\ \citenamefont {Arous}(2013)}]{auffingerComplexityRandomSmooth2013}%
  \BibitemOpen
  \bibfield  {author} {\bibinfo {author} {\bibfnamefont {A.}~\bibnamefont {Auffinger}}\ and\ \bibinfo {author} {\bibfnamefont {G.~B.}\ \bibnamefont {Arous}},\ }\bibfield  {title} {\bibinfo {title} {Complexity of random smooth functions on the high-dimensional sphere},\ }\href {https://doi.org/10.1214/13-AOP862} {\bibfield  {journal} {\bibinfo  {journal} {Annals of Probability}\ }\textbf {\bibinfo {volume} {41}},\ \bibinfo {pages} {4214} (\bibinfo {year} {2013})}\BibitemShut {NoStop}%
\bibitem [{\citenamefont {Auffinger}\ \emph {et~al.}(2013)\citenamefont {Auffinger}, \citenamefont {Arous},\ and\ \citenamefont {{\v C}ern{\'y}}}]{auffingerRandomMatricesComplexity2013}%
  \BibitemOpen
  \bibfield  {author} {\bibinfo {author} {\bibfnamefont {A.}~\bibnamefont {Auffinger}}, \bibinfo {author} {\bibfnamefont {G.~B.}\ \bibnamefont {Arous}},\ and\ \bibinfo {author} {\bibfnamefont {J.}~\bibnamefont {{\v C}ern{\'y}}},\ }\bibfield  {title} {\bibinfo {title} {Random {{Matrices}} and {{Complexity}} of {{Spin Glasses}}},\ }\href {https://doi.org/10.1002/cpa.21422} {\bibfield  {journal} {\bibinfo  {journal} {Communications on Pure and Applied Mathematics}\ }\textbf {\bibinfo {volume} {66}},\ \bibinfo {pages} {165} (\bibinfo {year} {2013})}\BibitemShut {NoStop}%
\bibitem [{\citenamefont {Ros}\ \emph {et~al.}(2019{\natexlab{a}})\citenamefont {Ros}, \citenamefont {Ben~Arous}, \citenamefont {Biroli},\ and\ \citenamefont {Cammarota}}]{rosComplexEnergyLandscapes2019}%
  \BibitemOpen
  \bibfield  {author} {\bibinfo {author} {\bibfnamefont {V.}~\bibnamefont {Ros}}, \bibinfo {author} {\bibfnamefont {G.}~\bibnamefont {Ben~Arous}}, \bibinfo {author} {\bibfnamefont {G.}~\bibnamefont {Biroli}},\ and\ \bibinfo {author} {\bibfnamefont {C.}~\bibnamefont {Cammarota}},\ }\bibfield  {title} {\bibinfo {title} {Complex {{Energy Landscapes}} in {{Spiked}}-{{Tensor}} and {{Simple Glassy Models}}: {{Ruggedness}}, {{Arrangements}} of {{Local Minima}}, and {{Phase Transitions}}},\ }\href {https://doi.org/10.1103/PhysRevX.9.011003} {\bibfield  {journal} {\bibinfo  {journal} {Physical Review X}\ }\textbf {\bibinfo {volume} {9}},\ \bibinfo {pages} {011003} (\bibinfo {year} {2019}{\natexlab{a}})}\BibitemShut {NoStop}%
\bibitem [{\citenamefont {Ros}\ \emph {et~al.}(2019{\natexlab{b}})\citenamefont {Ros}, \citenamefont {Biroli},\ and\ \citenamefont {Cammarota}}]{rosComplexityEnergyBarriers2019}%
  \BibitemOpen
  \bibfield  {author} {\bibinfo {author} {\bibfnamefont {V.}~\bibnamefont {Ros}}, \bibinfo {author} {\bibfnamefont {G.}~\bibnamefont {Biroli}},\ and\ \bibinfo {author} {\bibfnamefont {C.}~\bibnamefont {Cammarota}},\ }\bibfield  {title} {\bibinfo {title} {Complexity of energy barriers in mean-field glassy systems},\ }\href {https://doi.org/10.1209/0295-5075/126/20003} {\bibfield  {journal} {\bibinfo  {journal} {EPL (Europhysics Letters)}\ }\textbf {\bibinfo {volume} {126}},\ \bibinfo {pages} {20003} (\bibinfo {year} {2019}{\natexlab{b}})}\BibitemShut {NoStop}%
\bibitem [{\citenamefont {Becker}\ \emph {et~al.}(2020)\citenamefont {Becker}, \citenamefont {Zhang},\ and\ \citenamefont {Lee}}]{beckerGeometryEnergyLandscapes2020}%
  \BibitemOpen
  \bibfield  {author} {\bibinfo {author} {\bibfnamefont {S.}~\bibnamefont {Becker}}, \bibinfo {author} {\bibfnamefont {Y.}~\bibnamefont {Zhang}},\ and\ \bibinfo {author} {\bibfnamefont {A.~A.}\ \bibnamefont {Lee}},\ }\bibfield  {title} {\bibinfo {title} {Geometry of {{Energy Landscapes}} and the {{Optimizability}} of {{Deep Neural Networks}}},\ }\href {https://doi.org/10.1103/PhysRevLett.124.108301} {\bibfield  {journal} {\bibinfo  {journal} {Physical Review Letters}\ }\textbf {\bibinfo {volume} {124}},\ \bibinfo {pages} {108301} (\bibinfo {year} {2020})}\BibitemShut {NoStop}%
\bibitem [{\citenamefont {Adler}\ \emph {et~al.}(2007)\citenamefont {Adler}, \citenamefont {Taylor} \emph {et~al.}}]{adler2007random}%
  \BibitemOpen
  \bibfield  {author} {\bibinfo {author} {\bibfnamefont {R.~J.}\ \bibnamefont {Adler}}, \bibinfo {author} {\bibfnamefont {J.~E.}\ \bibnamefont {Taylor}}, \emph {et~al.},\ }\href@noop {} {\emph {\bibinfo {title} {Random fields and geometry}}},\ Vol.~\bibinfo {volume} {80}\ (\bibinfo  {publisher} {Springer},\ \bibinfo {year} {2007})\BibitemShut {NoStop}%
\bibitem [{\citenamefont {Potters}\ and\ \citenamefont {Bouchaud}(2020)}]{Potters2020-ta}%
  \BibitemOpen
  \bibfield  {author} {\bibinfo {author} {\bibfnamefont {M.}~\bibnamefont {Potters}}\ and\ \bibinfo {author} {\bibfnamefont {J.-P.}\ \bibnamefont {Bouchaud}},\ }\href@noop {} {\emph {\bibinfo {title} {A First Course in Random Matrix Theory: for Physicists, Engineers and Data Scientists}}}\ (\bibinfo  {publisher} {Cambridge University Press},\ \bibinfo {year} {2020})\BibitemShut {NoStop}%
\bibitem [{\citenamefont {Cavagna}\ \emph {et~al.}(2003)\citenamefont {Cavagna}, \citenamefont {Giardina}, \citenamefont {Parisi},\ and\ \citenamefont {Mezard}}]{cavagnaFormalEquivalenceTAP2003}%
  \BibitemOpen
  \bibfield  {author} {\bibinfo {author} {\bibfnamefont {A.}~\bibnamefont {Cavagna}}, \bibinfo {author} {\bibfnamefont {I.}~\bibnamefont {Giardina}}, \bibinfo {author} {\bibfnamefont {G.}~\bibnamefont {Parisi}},\ and\ \bibinfo {author} {\bibfnamefont {M.}~\bibnamefont {Mezard}},\ }\bibfield  {title} {\bibinfo {title} {On the formal equivalence of the {{TAP}} and thermodynamic methods in the {{SK}} model},\ }\href {https://doi.org/10.1088/0305-4470/36/5/301} {\bibfield  {journal} {\bibinfo  {journal} {Journal of Physics A: Mathematical and General}\ }\textbf {\bibinfo {volume} {36}},\ \bibinfo {pages} {1175} (\bibinfo {year} {2003})}\BibitemShut {NoStop}%
\bibitem [{\citenamefont {Annibale}\ \emph {et~al.}(2003{\natexlab{a}})\citenamefont {Annibale}, \citenamefont {Cavagna}, \citenamefont {Giardina}, \citenamefont {Parisi},\ and\ \citenamefont {Trevigne}}]{annibaleRoleBecchiRouet2003}%
  \BibitemOpen
  \bibfield  {author} {\bibinfo {author} {\bibfnamefont {A.}~\bibnamefont {Annibale}}, \bibinfo {author} {\bibfnamefont {A.}~\bibnamefont {Cavagna}}, \bibinfo {author} {\bibfnamefont {I.}~\bibnamefont {Giardina}}, \bibinfo {author} {\bibfnamefont {G.}~\bibnamefont {Parisi}},\ and\ \bibinfo {author} {\bibfnamefont {E.}~\bibnamefont {Trevigne}},\ }\bibfield  {title} {\bibinfo {title} {The role of the {{Becchi}}\textendash{{Rouet}}\textendash{{Stora}}\textendash{{Tyutin}} supersymmetry in the calculation of the complexity for the {{Sherrington}}\textendash{{Kirkpatrick}} model},\ }\href {https://doi.org/10.1088/0305-4470/36/43/018} {\bibfield  {journal} {\bibinfo  {journal} {Journal of Physics A: Mathematical and General}\ }\textbf {\bibinfo {volume} {36}},\ \bibinfo {pages} {10937} (\bibinfo {year} {2003}{\natexlab{a}})}\BibitemShut {NoStop}%
\bibitem [{\citenamefont {Annibale}\ \emph {et~al.}(2003{\natexlab{b}})\citenamefont {Annibale}, \citenamefont {Cavagna}, \citenamefont {Giardina},\ and\ \citenamefont {Parisi}}]{annibaleSupersymmetricComplexitySherringtonKirkpatrick2003}%
  \BibitemOpen
  \bibfield  {author} {\bibinfo {author} {\bibfnamefont {A.}~\bibnamefont {Annibale}}, \bibinfo {author} {\bibfnamefont {A.}~\bibnamefont {Cavagna}}, \bibinfo {author} {\bibfnamefont {I.}~\bibnamefont {Giardina}},\ and\ \bibinfo {author} {\bibfnamefont {G.}~\bibnamefont {Parisi}},\ }\bibfield  {title} {\bibinfo {title} {Supersymmetric complexity in the {{Sherrington}}-{{Kirkpatrick}} model},\ }\href {https://doi.org/10.1103/PhysRevE.68.061103} {\bibfield  {journal} {\bibinfo  {journal} {Physical Review E}\ }\textbf {\bibinfo {volume} {68}},\ \bibinfo {pages} {061103} (\bibinfo {year} {2003}{\natexlab{b}})}\BibitemShut {NoStop}%
\bibitem [{\citenamefont {Annibale}\ \emph {et~al.}(2004)\citenamefont {Annibale}, \citenamefont {Gualdi},\ and\ \citenamefont {Cavagna}}]{annibaleCoexistenceSupersymmetricSupersymmetrybreaking2004}%
  \BibitemOpen
  \bibfield  {author} {\bibinfo {author} {\bibfnamefont {A.}~\bibnamefont {Annibale}}, \bibinfo {author} {\bibfnamefont {G.}~\bibnamefont {Gualdi}},\ and\ \bibinfo {author} {\bibfnamefont {A.}~\bibnamefont {Cavagna}},\ }\bibfield  {title} {\bibinfo {title} {Coexistence of supersymmetric and supersymmetry-breaking states in spherical spin-glasses},\ }\href {https://doi.org/10.1088/0305-4470/37/47/001} {\bibfield  {journal} {\bibinfo  {journal} {Journal of Physics A: Mathematical and General}\ }\textbf {\bibinfo {volume} {37}},\ \bibinfo {pages} {11311} (\bibinfo {year} {2004})}\BibitemShut {NoStop}%
\bibitem [{\citenamefont {Cavagna}\ \emph {et~al.}(2005)\citenamefont {Cavagna}, \citenamefont {Giardina},\ and\ \citenamefont {Parisi}}]{cavagnaCavityMethodSupersymmetrybreaking2005}%
  \BibitemOpen
  \bibfield  {author} {\bibinfo {author} {\bibfnamefont {A.}~\bibnamefont {Cavagna}}, \bibinfo {author} {\bibfnamefont {I.}~\bibnamefont {Giardina}},\ and\ \bibinfo {author} {\bibfnamefont {G.}~\bibnamefont {Parisi}},\ }\bibfield  {title} {\bibinfo {title} {Cavity method for supersymmetry-breaking spin glasses},\ }\href {https://doi.org/10.1103/PhysRevB.71.024422} {\bibfield  {journal} {\bibinfo  {journal} {Physical Review B}\ }\textbf {\bibinfo {volume} {71}},\ \bibinfo {pages} {024422} (\bibinfo {year} {2005})}\BibitemShut {NoStop}%
\bibitem [{\citenamefont {Dyson}(1962)}]{Dyson1962-vn}%
  \BibitemOpen
  \bibfield  {author} {\bibinfo {author} {\bibfnamefont {F.~J.}\ \bibnamefont {Dyson}},\ }\bibfield  {title} {\bibinfo {title} {A {Brownian‐Motion} model for the eigenvalues of a random matrix},\ }\href@noop {} {\bibfield  {journal} {\bibinfo  {journal} {J. Math. Phys.}\ }\textbf {\bibinfo {volume} {3}},\ \bibinfo {pages} {1191} (\bibinfo {year} {1962})}\BibitemShut {NoStop}%
\bibitem [{\citenamefont {Madan~Lal}(2004)}]{madanlalRandomMatrices2004}%
  \BibitemOpen
  \bibfield  {author} {\bibinfo {author} {\bibfnamefont {M.}~\bibnamefont {Madan~Lal}},\ }\bibfield  {title} {\bibinfo {title} {Random {{Matrices}}},\ }in\ \href {https://doi.org/10.1016/S0079-8169(04)80091-6} {\emph {\bibinfo {booktitle} {Pure and {{Applied Mathematics}}}}},\ Vol.\ \bibinfo {volume} {142}\ (\bibinfo  {publisher} {{Elsevier}},\ \bibinfo {year} {2004})\ pp.\ \bibinfo {pages} {1--32}\BibitemShut {NoStop}%
\bibitem [{\citenamefont {Pastur}(1972)}]{pasturSpectrumRandomMatrices1972}%
  \BibitemOpen
  \bibfield  {author} {\bibinfo {author} {\bibfnamefont {L.~A.}\ \bibnamefont {Pastur}},\ }\bibfield  {title} {\bibinfo {title} {On the spectrum of random matrices},\ }\href {https://doi.org/10.1007/BF01035768} {\bibfield  {journal} {\bibinfo  {journal} {Theoretical and Mathematical Physics}\ }\textbf {\bibinfo {volume} {10}},\ \bibinfo {pages} {67} (\bibinfo {year} {1972})}\BibitemShut {NoStop}%
\bibitem [{\citenamefont {Aspelmeier}\ \emph {et~al.}(2008)\citenamefont {Aspelmeier}, \citenamefont {Billoire}, \citenamefont {Marinari},\ and\ \citenamefont {Moore}}]{aspelmeierFinitesizeCorrectionsSherrington2008}%
  \BibitemOpen
  \bibfield  {author} {\bibinfo {author} {\bibfnamefont {T.}~\bibnamefont {Aspelmeier}}, \bibinfo {author} {\bibfnamefont {A.}~\bibnamefont {Billoire}}, \bibinfo {author} {\bibfnamefont {E.}~\bibnamefont {Marinari}},\ and\ \bibinfo {author} {\bibfnamefont {M.~A.}\ \bibnamefont {Moore}},\ }\bibfield  {title} {\bibinfo {title} {Finite-size corrections in the {{Sherrington}}\textendash{{Kirkpatrick}} model},\ }\href {https://doi.org/10.1088/1751-8113/41/32/324008} {\bibfield  {journal} {\bibinfo  {journal} {Journal of Physics A: Mathematical and Theoretical}\ }\textbf {\bibinfo {volume} {41}},\ \bibinfo {pages} {324008} (\bibinfo {year} {2008})}\BibitemShut {NoStop}%
\bibitem [{\citenamefont {Crisanti}\ and\ \citenamefont {Rizzo}(2002)}]{crisantiAnalysisEnsuremathInfty2002}%
  \BibitemOpen
  \bibfield  {author} {\bibinfo {author} {\bibfnamefont {A.}~\bibnamefont {Crisanti}}\ and\ \bibinfo {author} {\bibfnamefont {T.}~\bibnamefont {Rizzo}},\ }\bibfield  {title} {\bibinfo {title} {Analysis of the \$\textbackslash ensuremath\{\textbackslash infty\}\$-replica symmetry breaking solution of the {{Sherrington}}-{{Kirkpatrick}} model},\ }\href {https://doi.org/10.1103/PhysRevE.65.046137} {\bibfield  {journal} {\bibinfo  {journal} {Physical Review E}\ }\textbf {\bibinfo {volume} {65}},\ \bibinfo {pages} {046137} (\bibinfo {year} {2002})}\BibitemShut {NoStop}%
\bibitem [{\citenamefont {Schmidt}(2008)}]{schmidtReplicaSymmetryBreaking2008}%
  \BibitemOpen
  \bibfield  {author} {\bibinfo {author} {\bibfnamefont {M.~J.}\ \bibnamefont {Schmidt}},\ }\href@noop {} {\bibinfo {title} {Replica {{Symmetry Breaking}} at {{Low Temperature}}}} (\bibinfo {year} {2008})\BibitemShut {NoStop}%
\bibitem [{\citenamefont {Montanari}(2019)}]{montanariOptimizationSherringtonKirkpatrickHamiltonian2019}%
  \BibitemOpen
  \bibfield  {author} {\bibinfo {author} {\bibfnamefont {A.}~\bibnamefont {Montanari}},\ }\bibfield  {title} {\bibinfo {title} {Optimization of the {{Sherrington}}-{{Kirkpatrick Hamiltonian}}},\ }\href@noop {} {\bibfield  {journal} {\bibinfo  {journal} {arXiv:1812.10897 [cond-mat]}\ } (\bibinfo {year} {2019})},\ \Eprint {https://arxiv.org/abs/1812.10897} {arXiv:1812.10897 [cond-mat]} \BibitemShut {NoStop}%
\bibitem [{\citenamefont {Morone}\ \emph {et~al.}(2014)\citenamefont {Morone}, \citenamefont {Caltagirone}, \citenamefont {Harrison},\ and\ \citenamefont {Parisi}}]{moroneReplicaTheorySpin2014}%
  \BibitemOpen
  \bibfield  {author} {\bibinfo {author} {\bibfnamefont {F.}~\bibnamefont {Morone}}, \bibinfo {author} {\bibfnamefont {F.}~\bibnamefont {Caltagirone}}, \bibinfo {author} {\bibfnamefont {E.}~\bibnamefont {Harrison}},\ and\ \bibinfo {author} {\bibfnamefont {G.}~\bibnamefont {Parisi}},\ }\bibfield  {title} {\bibinfo {title} {Replica {{Theory}} and {{Spin Glasses}}},\ }\href@noop {} {\bibfield  {journal} {\bibinfo  {journal} {arXiv:1409.2722 [cond-mat]}\ } (\bibinfo {year} {2014})},\ \Eprint {https://arxiv.org/abs/1409.2722} {arXiv:1409.2722 [cond-mat]} \BibitemShut {NoStop}%
\bibitem [{\citenamefont {Sommers}\ and\ \citenamefont {Dupont}(1984)}]{sommersDistributionFrozenFields1984}%
  \BibitemOpen
  \bibfield  {author} {\bibinfo {author} {\bibfnamefont {H.-J.}\ \bibnamefont {Sommers}}\ and\ \bibinfo {author} {\bibfnamefont {W.}~\bibnamefont {Dupont}},\ }\bibfield  {title} {\bibinfo {title} {Distribution of frozen fields in the mean-field theory of spin glasses},\ }\href {https://doi.org/10.1088/0022-3719/17/32/012} {\bibfield  {journal} {\bibinfo  {journal} {Journal of Physics C: Solid State Physics}\ }\textbf {\bibinfo {volume} {17}},\ \bibinfo {pages} {5785} (\bibinfo {year} {1984})}\BibitemShut {NoStop}%
\bibitem [{Note1()}]{Note1}%
  \BibitemOpen
  \bibinfo {note} {The Ising energy of random spin states is almost zero for the following reason. Flipping some of the Ising spins, we can make all spins $+1$. By flipping the sign of $J_{ij}$ properly at the same time, we can keep the Ising energy unchanged. This new $J$-matrix follows GOE as well. The Ising energy is given by the sum of all the elements of the new $J$-matrix. By the law of large numbers, this quantity converges to its mean $0$.}\BibitemShut {Stop}%
\bibitem [{Note2()}]{Note2}%
  \BibitemOpen
  \bibinfo {note} {We set the norm of the initialized principal eigenvector to match that of a Gaussian random initialization.}\BibitemShut {Stop}%
\bibitem [{\citenamefont {Stern}\ \emph {et~al.}(2014)\citenamefont {Stern}, \citenamefont {Sompolinsky},\ and\ \citenamefont {Abbott}}]{sternDynamicsRandomNeural2014}%
  \BibitemOpen
  \bibfield  {author} {\bibinfo {author} {\bibfnamefont {M.}~\bibnamefont {Stern}}, \bibinfo {author} {\bibfnamefont {H.}~\bibnamefont {Sompolinsky}},\ and\ \bibinfo {author} {\bibfnamefont {L.~F.}\ \bibnamefont {Abbott}},\ }\bibfield  {title} {\bibinfo {title} {Dynamics of random neural networks with bistable units},\ }\href {https://doi.org/10.1103/PhysRevE.90.062710} {\bibfield  {journal} {\bibinfo  {journal} {Physical Review E}\ }\textbf {\bibinfo {volume} {90}},\ \bibinfo {pages} {062710} (\bibinfo {year} {2014})}\BibitemShut {NoStop}%
\bibitem [{\citenamefont {Wainrib}\ and\ \citenamefont {Touboul}(2013)}]{Wainrib2013-of}%
  \BibitemOpen
  \bibfield  {author} {\bibinfo {author} {\bibfnamefont {G.}~\bibnamefont {Wainrib}}\ and\ \bibinfo {author} {\bibfnamefont {J.}~\bibnamefont {Touboul}},\ }\bibfield  {title} {\bibinfo {title} {Topological and dynamical complexity of random neural networks},\ }\href@noop {} {\bibfield  {journal} {\bibinfo  {journal} {Phys. Rev. Lett.}\ }\textbf {\bibinfo {volume} {110}},\ \bibinfo {pages} {118101} (\bibinfo {year} {2013})}\BibitemShut {NoStop}%
\bibitem [{\citenamefont {Biroli}\ \emph {et~al.}(2018)\citenamefont {Biroli}, \citenamefont {Bunin},\ and\ \citenamefont {Cammarota}}]{biroliMarginallyStableEquilibria2018}%
  \BibitemOpen
  \bibfield  {author} {\bibinfo {author} {\bibfnamefont {G.}~\bibnamefont {Biroli}}, \bibinfo {author} {\bibfnamefont {G.}~\bibnamefont {Bunin}},\ and\ \bibinfo {author} {\bibfnamefont {C.}~\bibnamefont {Cammarota}},\ }\bibfield  {title} {\bibinfo {title} {Marginally stable equilibria in critical ecosystems},\ }\href {https://doi.org/10.1088/1367-2630/aada58} {\bibfield  {journal} {\bibinfo  {journal} {New Journal of Physics}\ }\textbf {\bibinfo {volume} {20}},\ \bibinfo {pages} {083051} (\bibinfo {year} {2018})}\BibitemShut {NoStop}%
\bibitem [{\citenamefont {Bunin}(2017)}]{buninEcologicalCommunitiesLotkaVolterra2017}%
  \BibitemOpen
  \bibfield  {author} {\bibinfo {author} {\bibfnamefont {G.}~\bibnamefont {Bunin}},\ }\bibfield  {title} {\bibinfo {title} {Ecological communities with {{Lotka}}-{{Volterra}} dynamics},\ }\href {https://doi.org/10.1103/PhysRevE.95.042414} {\bibfield  {journal} {\bibinfo  {journal} {Physical Review E}\ }\textbf {\bibinfo {volume} {95}},\ \bibinfo {pages} {042414} (\bibinfo {year} {2017})}\BibitemShut {NoStop}%
\bibitem [{\citenamefont {Ipsen}\ and\ \citenamefont {Forrester}(2018)}]{ipsenKacRiceFixed2018}%
  \BibitemOpen
  \bibfield  {author} {\bibinfo {author} {\bibfnamefont {J.~R.}\ \bibnamefont {Ipsen}}\ and\ \bibinfo {author} {\bibfnamefont {P.~J.}\ \bibnamefont {Forrester}},\ }\bibfield  {title} {\bibinfo {title} {Kac\textendash{{Rice}} fixed point analysis for single- and multi-layered complex systems},\ }\href {https://doi.org/10.1088/1751-8121/aae76d} {\bibfield  {journal} {\bibinfo  {journal} {Journal of Physics A: Mathematical and Theoretical}\ }\textbf {\bibinfo {volume} {51}},\ \bibinfo {pages} {474003} (\bibinfo {year} {2018})}\BibitemShut {NoStop}%
\bibitem [{\citenamefont {Acebr{\'o}n}\ \emph {et~al.}(2005)\citenamefont {Acebr{\'o}n}, \citenamefont {Bonilla}, \citenamefont {P{\'e}rez~Vicente}, \citenamefont {Ritort},\ and\ \citenamefont {Spigler}}]{Acebron2005-vi}%
  \BibitemOpen
  \bibfield  {author} {\bibinfo {author} {\bibfnamefont {J.~A.}\ \bibnamefont {Acebr{\'o}n}}, \bibinfo {author} {\bibfnamefont {L.~L.}\ \bibnamefont {Bonilla}}, \bibinfo {author} {\bibfnamefont {C.~J.}\ \bibnamefont {P{\'e}rez~Vicente}}, \bibinfo {author} {\bibfnamefont {F.}~\bibnamefont {Ritort}},\ and\ \bibinfo {author} {\bibfnamefont {R.}~\bibnamefont {Spigler}},\ }\bibfield  {title} {\bibinfo {title} {The kuramoto model: A simple paradigm for synchronization phenomena},\ }\href@noop {} {\bibfield  {journal} {\bibinfo  {journal} {Rev. Mod. Phys.}\ }\textbf {\bibinfo {volume} {77}},\ \bibinfo {pages} {137} (\bibinfo {year} {2005})}\BibitemShut {NoStop}%
\bibitem [{\citenamefont {Wu}\ \emph {et~al.}(2011)\citenamefont {Wu}, \citenamefont {Jiao}, \citenamefont {Li},\ and\ \citenamefont {Chen}}]{Wu2011-ev}%
  \BibitemOpen
  \bibfield  {author} {\bibinfo {author} {\bibfnamefont {J.}~\bibnamefont {Wu}}, \bibinfo {author} {\bibfnamefont {L.}~\bibnamefont {Jiao}}, \bibinfo {author} {\bibfnamefont {R.}~\bibnamefont {Li}},\ and\ \bibinfo {author} {\bibfnamefont {W.}~\bibnamefont {Chen}},\ }\bibfield  {title} {\bibinfo {title} {Clustering dynamics of nonlinear oscillator network: Application to graph coloring problem},\ }\href@noop {} {\bibfield  {journal} {\bibinfo  {journal} {Physica D}\ }\textbf {\bibinfo {volume} {240}},\ \bibinfo {pages} {1972} (\bibinfo {year} {2011})}\BibitemShut {NoStop}%
\bibitem [{\citenamefont {Albertsson}\ \emph {et~al.}(2021)\citenamefont {Albertsson}, \citenamefont {Zahedinejad}, \citenamefont {Houshang}, \citenamefont {Khymyn}, \citenamefont {{\AA}kerman},\ and\ \citenamefont {Rusu}}]{Albertsson2021-yw}%
  \BibitemOpen
  \bibfield  {author} {\bibinfo {author} {\bibfnamefont {D.~I.}\ \bibnamefont {Albertsson}}, \bibinfo {author} {\bibfnamefont {M.}~\bibnamefont {Zahedinejad}}, \bibinfo {author} {\bibfnamefont {A.}~\bibnamefont {Houshang}}, \bibinfo {author} {\bibfnamefont {R.}~\bibnamefont {Khymyn}}, \bibinfo {author} {\bibfnamefont {J.}~\bibnamefont {{\AA}kerman}},\ and\ \bibinfo {author} {\bibfnamefont {A.}~\bibnamefont {Rusu}},\ }\bibfield  {title} {\bibinfo {title} {Ultrafast ising machines using spin torque nano-oscillators},\ }\href@noop {} {\bibfield  {journal} {\bibinfo  {journal} {Appl. Phys. Lett.}\ }\textbf {\bibinfo {volume} {118}} (\bibinfo {year} {2021})}\BibitemShut {NoStop}%
\bibitem [{\citenamefont {Wang}\ and\ \citenamefont {Roychowdhury}(2019)}]{wang2019oim}%
  \BibitemOpen
  \bibfield  {author} {\bibinfo {author} {\bibfnamefont {T.}~\bibnamefont {Wang}}\ and\ \bibinfo {author} {\bibfnamefont {J.}~\bibnamefont {Roychowdhury}},\ }\bibfield  {title} {\bibinfo {title} {Oim: Oscillator-based ising machines for solving combinatorial optimisation problems},\ }in\ \href@noop {} {\emph {\bibinfo {booktitle} {Unconventional Computation and Natural Computation: 18th International Conference, UCNC 2019, Tokyo, Japan, June 3--7, 2019, Proceedings 18}}}\ (\bibinfo {organization} {Springer},\ \bibinfo {year} {2019})\ pp.\ \bibinfo {pages} {232--256}\BibitemShut {NoStop}%
\bibitem [{\citenamefont {Inui}\ and\ \citenamefont {Yamamoto}(2020)}]{InuiYamamoto2020}%
  \BibitemOpen
  \bibfield  {author} {\bibinfo {author} {\bibfnamefont {Y.}~\bibnamefont {Inui}}\ and\ \bibinfo {author} {\bibfnamefont {Y.}~\bibnamefont {Yamamoto}},\ }\bibfield  {title} {\bibinfo {title} {Entanglement and quantum discord in optically coupled coherent ising machines},\ }\href {https://doi.org/10.1103/PhysRevA.102.062419} {\bibfield  {journal} {\bibinfo  {journal} {Phys. Rev. A}\ }\textbf {\bibinfo {volume} {102}},\ \bibinfo {pages} {062419} (\bibinfo {year} {2020})}\BibitemShut {NoStop}%
\bibitem [{\citenamefont {Blais}\ \emph {et~al.}(2021)\citenamefont {Blais}, \citenamefont {Grimsmo}, \citenamefont {Girvin},\ and\ \citenamefont {Wallraff}}]{BlaisGrimsmoGirvinWallraff2021}%
  \BibitemOpen
  \bibfield  {author} {\bibinfo {author} {\bibfnamefont {A.}~\bibnamefont {Blais}}, \bibinfo {author} {\bibfnamefont {A.~L.}\ \bibnamefont {Grimsmo}}, \bibinfo {author} {\bibfnamefont {S.~M.}\ \bibnamefont {Girvin}},\ and\ \bibinfo {author} {\bibfnamefont {A.}~\bibnamefont {Wallraff}},\ }\bibfield  {title} {\bibinfo {title} {Circuit quantum electrodynamics},\ }\href {https://doi.org/10.1103/RevModPhys.93.025005} {\bibfield  {journal} {\bibinfo  {journal} {Rev. Mod. Phys.}\ }\textbf {\bibinfo {volume} {93}},\ \bibinfo {pages} {025005} (\bibinfo {year} {2021})}\BibitemShut {NoStop}%
\bibitem [{\citenamefont {Hopfield}(1982)}]{Hopfield1982-fx}%
  \BibitemOpen
  \bibfield  {author} {\bibinfo {author} {\bibfnamefont {J.~J.}\ \bibnamefont {Hopfield}},\ }\bibfield  {title} {\bibinfo {title} {{Neural networks and physical systems with emergent collective computational abilities}},\ }\href@noop {} {\bibfield  {journal} {\bibinfo  {journal} {Proc. Natl. Acad. Sci. U. S. A.}\ }\textbf {\bibinfo {volume} {79}},\ \bibinfo {pages} {2554} (\bibinfo {year} {1982})}\BibitemShut {NoStop}%
\bibitem [{\citenamefont {Marsh}\ \emph {et~al.}(2021)\citenamefont {Marsh}, \citenamefont {Guo}, \citenamefont {Kroeze}, \citenamefont {Gopalakrishnan}, \citenamefont {Ganguli}, \citenamefont {Keeling},\ and\ \citenamefont {Lev}}]{Marsh2021-mz}%
  \BibitemOpen
  \bibfield  {author} {\bibinfo {author} {\bibfnamefont {B.~P.}\ \bibnamefont {Marsh}}, \bibinfo {author} {\bibfnamefont {Y.}~\bibnamefont {Guo}}, \bibinfo {author} {\bibfnamefont {R.~M.}\ \bibnamefont {Kroeze}}, \bibinfo {author} {\bibfnamefont {S.}~\bibnamefont {Gopalakrishnan}}, \bibinfo {author} {\bibfnamefont {S.}~\bibnamefont {Ganguli}}, \bibinfo {author} {\bibfnamefont {J.}~\bibnamefont {Keeling}},\ and\ \bibinfo {author} {\bibfnamefont {B.~L.}\ \bibnamefont {Lev}},\ }\href@noop {} {\bibinfo {title} {Enhancing associative memory recall and storage capacity using confocal cavity {QED}}} (\bibinfo {year} {2021})\BibitemShut {NoStop}%
\bibitem [{\citenamefont {Marsh}\ \emph {et~al.}(2023)\citenamefont {Marsh}, \citenamefont {Kroeze}, \citenamefont {Ganguli}, \citenamefont {Gopalakrishnan}, \citenamefont {Keeling},\ and\ \citenamefont {Lev}}]{Marsh2023-wy}%
  \BibitemOpen
  \bibfield  {author} {\bibinfo {author} {\bibfnamefont {B.~P.}\ \bibnamefont {Marsh}}, \bibinfo {author} {\bibfnamefont {R.~M.}\ \bibnamefont {Kroeze}}, \bibinfo {author} {\bibfnamefont {S.}~\bibnamefont {Ganguli}}, \bibinfo {author} {\bibfnamefont {S.}~\bibnamefont {Gopalakrishnan}}, \bibinfo {author} {\bibfnamefont {J.}~\bibnamefont {Keeling}},\ and\ \bibinfo {author} {\bibfnamefont {B.~L.}\ \bibnamefont {Lev}},\ }\bibfield  {title} {\bibinfo {title} {Entanglement and replica symmetry breaking in a driven-dissipative quantum spin glass},\ }\href@noop {} {\  (\bibinfo {year} {2023})},\ \Eprint {https://arxiv.org/abs/2307.10176} {arXiv:2307.10176 [quant-ph]} \BibitemShut {NoStop}%
\bibitem [{\citenamefont {Grimaldi}\ \emph {et~al.}(2023)\citenamefont {Grimaldi}, \citenamefont {Mazza}, \citenamefont {Raimondo}, \citenamefont {Tullo}, \citenamefont {Rodrigues}, \citenamefont {Camsari}, \citenamefont {Crupi}, \citenamefont {Carpentieri}, \citenamefont {Puliafito},\ and\ \citenamefont {Finocchio}}]{grimaldi2023evaluating}%
  \BibitemOpen
  \bibfield  {author} {\bibinfo {author} {\bibfnamefont {A.}~\bibnamefont {Grimaldi}}, \bibinfo {author} {\bibfnamefont {L.}~\bibnamefont {Mazza}}, \bibinfo {author} {\bibfnamefont {E.}~\bibnamefont {Raimondo}}, \bibinfo {author} {\bibfnamefont {P.}~\bibnamefont {Tullo}}, \bibinfo {author} {\bibfnamefont {D.}~\bibnamefont {Rodrigues}}, \bibinfo {author} {\bibfnamefont {K.~Y.}\ \bibnamefont {Camsari}}, \bibinfo {author} {\bibfnamefont {V.}~\bibnamefont {Crupi}}, \bibinfo {author} {\bibfnamefont {M.}~\bibnamefont {Carpentieri}}, \bibinfo {author} {\bibfnamefont {V.}~\bibnamefont {Puliafito}},\ and\ \bibinfo {author} {\bibfnamefont {G.}~\bibnamefont {Finocchio}},\ }\bibfield  {title} {\bibinfo {title} {Evaluating spintronics-compatible implementations of ising machines},\ }\href {https://arxiv.org/pdf/2304.04177.pdf} {\bibfield  {journal} {\bibinfo  {journal} {arXiv preprint arXiv:2304.04177}\ } (\bibinfo {year} {2023})}\BibitemShut {NoStop}%
\bibitem [{\citenamefont {English}\ \emph {et~al.}(2022)\citenamefont {English}, \citenamefont {Zampetaki}, \citenamefont {Kalinin}, \citenamefont {Berloff},\ and\ \citenamefont {Kevrekidis}}]{english2022ising}%
  \BibitemOpen
  \bibfield  {author} {\bibinfo {author} {\bibfnamefont {L.}~\bibnamefont {English}}, \bibinfo {author} {\bibfnamefont {A.}~\bibnamefont {Zampetaki}}, \bibinfo {author} {\bibfnamefont {K.}~\bibnamefont {Kalinin}}, \bibinfo {author} {\bibfnamefont {N.}~\bibnamefont {Berloff}},\ and\ \bibinfo {author} {\bibfnamefont {P.~G.}\ \bibnamefont {Kevrekidis}},\ }\bibfield  {title} {\bibinfo {title} {An ising machine based on networks of subharmonic electrical resonators},\ }\href {https://www.nature.com/articles/s42005-022-01111-x} {\bibfield  {journal} {\bibinfo  {journal} {Communications Physics}\ }\textbf {\bibinfo {volume} {5}},\ \bibinfo {pages} {333} (\bibinfo {year} {2022})}\BibitemShut {NoStop}%
\bibitem [{\citenamefont {Pierangeli}\ \emph {et~al.}(2020)\citenamefont {Pierangeli}, \citenamefont {Marcucci},\ and\ \citenamefont {Conti}}]{pierangeli2020adiabatic}%
  \BibitemOpen
  \bibfield  {author} {\bibinfo {author} {\bibfnamefont {D.}~\bibnamefont {Pierangeli}}, \bibinfo {author} {\bibfnamefont {G.}~\bibnamefont {Marcucci}},\ and\ \bibinfo {author} {\bibfnamefont {C.}~\bibnamefont {Conti}},\ }\bibfield  {title} {\bibinfo {title} {Adiabatic evolution on a spatial-photonic ising machine},\ }\href {https://opg.optica.org/optica/fulltext.cfm?uri=optica-7-11-1535&id=442147} {\bibfield  {journal} {\bibinfo  {journal} {Optica}\ }\textbf {\bibinfo {volume} {7}},\ \bibinfo {pages} {1535} (\bibinfo {year} {2020})}\BibitemShut {NoStop}%
\bibitem [{\citenamefont {Wang}\ \emph {et~al.}(2019)\citenamefont {Wang}, \citenamefont {Wu},\ and\ \citenamefont {Roychowdhury}}]{wang2019new}%
  \BibitemOpen
  \bibfield  {author} {\bibinfo {author} {\bibfnamefont {T.}~\bibnamefont {Wang}}, \bibinfo {author} {\bibfnamefont {L.}~\bibnamefont {Wu}},\ and\ \bibinfo {author} {\bibfnamefont {J.}~\bibnamefont {Roychowdhury}},\ }\bibfield  {title} {\bibinfo {title} {New computational results and hardware prototypes for oscillator-based ising machines},\ }in\ \href {https://dl.acm.org/doi/abs/10.1145/3316781.3322473} {\emph {\bibinfo {booktitle} {Proceedings of the 56th Annual Design Automation Conference 2019}}}\ (\bibinfo {year} {2019})\ pp.\ \bibinfo {pages} {1--2}\BibitemShut {NoStop}%
\bibitem [{\citenamefont {Dauphin}\ \emph {et~al.}(2014)\citenamefont {Dauphin}, \citenamefont {Pascanu}, \citenamefont {Gulcehre}, \citenamefont {Cho}, \citenamefont {Ganguli},\ and\ \citenamefont {Bengio}}]{dauphin2014identifying}%
  \BibitemOpen
  \bibfield  {author} {\bibinfo {author} {\bibfnamefont {Y.~N.}\ \bibnamefont {Dauphin}}, \bibinfo {author} {\bibfnamefont {R.}~\bibnamefont {Pascanu}}, \bibinfo {author} {\bibfnamefont {C.}~\bibnamefont {Gulcehre}}, \bibinfo {author} {\bibfnamefont {K.}~\bibnamefont {Cho}}, \bibinfo {author} {\bibfnamefont {S.}~\bibnamefont {Ganguli}},\ and\ \bibinfo {author} {\bibfnamefont {Y.}~\bibnamefont {Bengio}},\ }\bibfield  {title} {\bibinfo {title} {Identifying and attacking the saddle point problem in high-dimensional non-convex optimization},\ }\href@noop {} {\bibfield  {journal} {\bibinfo  {journal} {Advances in neural information processing systems}\ }\textbf {\bibinfo {volume} {27}} (\bibinfo {year} {2014})}\BibitemShut {NoStop}%
\bibitem [{\citenamefont {Cavagna}\ \emph {et~al.}(2000)\citenamefont {Cavagna}, \citenamefont {Garrahan},\ and\ \citenamefont {Giardina}}]{cavagnaIndexDistributionRandom2000}%
  \BibitemOpen
  \bibfield  {author} {\bibinfo {author} {\bibfnamefont {A.}~\bibnamefont {Cavagna}}, \bibinfo {author} {\bibfnamefont {J.~P.}\ \bibnamefont {Garrahan}},\ and\ \bibinfo {author} {\bibfnamefont {I.}~\bibnamefont {Giardina}},\ }\bibfield  {title} {\bibinfo {title} {Index distribution of random matrices with an application to disordered systems},\ }\href {https://doi.org/10.1103/PhysRevB.61.3960} {\bibfield  {journal} {\bibinfo  {journal} {Physical Review B}\ }\textbf {\bibinfo {volume} {61}},\ \bibinfo {pages} {3960} (\bibinfo {year} {2000})}\BibitemShut {NoStop}%
\bibitem [{\citenamefont {Crisanti}\ \emph {et~al.}(2004{\natexlab{b}})\citenamefont {Crisanti}, \citenamefont {Leuzzi}, \citenamefont {Parisi},\ and\ \citenamefont {Rizzo}}]{crisantiQuenchedComputationDependence2004}%
  \BibitemOpen
  \bibfield  {author} {\bibinfo {author} {\bibfnamefont {A.}~\bibnamefont {Crisanti}}, \bibinfo {author} {\bibfnamefont {L.}~\bibnamefont {Leuzzi}}, \bibinfo {author} {\bibfnamefont {G.}~\bibnamefont {Parisi}},\ and\ \bibinfo {author} {\bibfnamefont {T.}~\bibnamefont {Rizzo}},\ }\bibfield  {title} {\bibinfo {title} {Quenched computation of the dependence of complexity on the free energy in the {{Sherrington}}-{{Kirkpatrick}} model},\ }\href {https://doi.org/10.1103/PhysRevB.70.064423} {\bibfield  {journal} {\bibinfo  {journal} {Physical Review B}\ }\textbf {\bibinfo {volume} {70}},\ \bibinfo {pages} {064423} (\bibinfo {year} {2004}{\natexlab{b}})}\BibitemShut {NoStop}%
\end{thebibliography}%

\end{document}